\documentclass[11pt,a4paper]{article}
\usepackage{latexsym}
\usepackage{graphicx}
\usepackage{psfrag}
\usepackage{amsmath}
\usepackage{amssymb}
\usepackage{color}
\usepackage{rotating}
\usepackage{multirow}
\usepackage{colortbl}
\usepackage{lscape,epsfig}
\usepackage{a4wide}
\usepackage[]{xcolor}

\usepackage[dvips,colorlinks=true,a4paper=true,backref=false,linktocpage=true,citecolor=blue,urlcolor=blue,linkcolor=blue,pdfpagemode=UseOutlines]{hyperref}
\usepackage[sort&compress,square,comma,numbers]{natbib}
\usepackage{hypernat}

\definecolor{orange}{rgb}{1.0,.6,0}
\newcommand{\tbr}{\hspace{0.25mm}{\color{red}$\blacksquare$}} 
\newcommand{\tbg}{{\color{green}$\bigstar$}}

\newcommand{\rC}{C}
\newcommand{\gA}{A}
\newcommand{\oP}{P}

\newcommand{\bda}{\begin{\displaymath}\begin{array}{rl}}
\newcommand{\eda}{\end{array}\end{displaymath}}
\newcommand{\be}{\begin{equation}}
\newcommand{\ee}{\end{equation}}
\newcommand{\bdm}{\begin{displaymath}}
\newcommand{\edm}{\end{displaymath}}
\newcommand{\bea}{\begin{eqnarray}}
\newcommand{\eea}{\end{eqnarray}}

\newcommand{\fs}{\,.}
\newcommand{\co}{\,,}

\newcommand{\ind}{\scriptscriptstyle}

\newcommand{\qbar}{\overline{\rule[0.42em]{0.4em}{0em}}\hspace{-0.45em}q}
\newcommand{\ubar}{\overline{\rule[0.42em]{0.4em}{0em}}\hspace{-0.5em}u}
\newcommand{\dbar}{\,\overline{\rule[0.65em]{0.4em}{0em}}\hspace{-0.6em}d}
\newcommand{\sbar}{\,\overline{\rule[0.42em]{0.4em}{0em}}\hspace{-0.5em}s}
\newcommand{\lbar}{\bar{\ell}}

\newcommand{\gsim}{\,\raisebox{-0.3em}{$\stackrel{\raisebox{-0.1em}{$>$}}{\sim}$
}\,}
\newcommand{\lvac}{\langle 0|\,}
\newcommand{\rvac}{\,|0\rangle}

\newcommand{\al}{&\!\!\!}

\newcommand{\Ch}{$\chi$} 

\newcommand{\Mpibar}{\rule{0.05cm}{0cm}\overline{\hspace{-0.08cm}M}_{\hspace{-0.04cm}\pi}}
\newcommand{\MKbar}{\rule{0.05cm}{0cm}\overline{\hspace{-0.08cm}M}_{\hspace{-0.04cm}K}}

\newcommand{\bi}{\begin{itemize}}
\newcommand{\ei}{\end{itemize}}

\newcommand{\beq}{\begin{equation}}
\newcommand{\eeq}{\end{equation}}

\newcommand{\Mpi}{M_\pi}
\newcommand{\Fpi}{F_\pi}
\newcommand{\Mka}{M_K}
\newcommand{\Fka}{F_K}

\newcommand{\<}{\langle}
\renewcommand{\>}{\rangle}

\newcommand{\lonebar}{\ln\frac{\Lambda_1^2}{M_\pi^2}}
\newcommand{\ltwobar}{\ln\frac{\Lambda_2^2}{M_\pi^2}}
\newcommand{\lthreebar}{\ln\frac{\Lambda_3^2}{M_\pi^2}}
\newcommand{\lfourbar}{\ln\frac{\Lambda_4^2}{M_\pi^2}}
\newcommand{\lsixbar}{\ln\frac{\Lambda_6^2}{M_\pi^2}}
\newcommand{\lMbar}{\ln\frac{\Omega_M^2}{M_\pi^2}}
\newcommand{\lFbar}{\ln\frac{\Omega_F^2}{M_\pi^2}}

\newcommand{\MeV}{\,\mathrm{MeV}}
\newcommand{\GeV}{\,\mathrm{GeV}}
\newcommand{\fm}{\,\mathrm{fm}}

\newcommand{\et}{\eta}

\newcommand{\rmO}{{\rm O}}
\newcommand{\msbar}{{\overline{{\rm MS}}}}

\def\mev{{\rm MeV}}
\def\gev{{\rm GeV}}
\def\tev{{\rm TeV}}
\def\fm{{\rm fm}}
 
\def\qbar{\bar{q}}
\def\psibar{\bar{\psi}}

\def\ubar{\bar{u}} 

\def\csw{c_{\rm sw}}

\def\gbar{\bar{g}}

\newcommand{\bd}{\begin{displaymath}}
\newcommand{\ed}{\end{displaymath}}

\newcommand{\eq}[1]{eq.\,(\ref{#1})}

\newcommand{\figurebox}[2]{\fbox{\vbox to#2in{\hbox to #1in{\hfil}\vfil}}}
\newcommand{\gtaeq}{\raisebox{-.6ex}{$\stackrel{\textstyle{>}}{\sim}$}}

\newcommand{\Nf}{N_{\hspace{-0.08 em} f}}
\newcommand{\Tr}{{\rm Tr}\,}

\newcommand{\mK}{m_{\rm K}}

\newcommand{\fK}{f_{\rm K}}

\newcommand{\half}{\textstyle{1\over2}}

\newcommand{\abar}{\overline{a}}

\newcommand{\rb}[1]{\raisebox{1.5ex}[-1.5ex]{#1}}

\newcommand{\Lo}{\stackrel{\rule[-0.1cm]{0cm}{0cm}\mbox{\tiny LO}}{=}}
\newcommand{\NLo}{\stackrel{\rule[-0.1cm]{0cm}{0cm}\mbox{\tiny NLO}}{=}} 
\newcommand{\epsilonD}{\epsilon}

\def\good{\raisebox{0.35mm}{{\color{green}$\bigstar$}}}
\def\bad{\raisebox{0.35mm}{\hspace{0.65mm}{\color{red}\tiny$\blacksquare$}}} 
\def\soso{\hspace{0.25mm}\raisebox{-0.2mm}{{\color{orange}\Large\textbullet}}}

\def\half{{1\over2}}

\def\Tr{\,\mathrm{Tr}}

\def\fm{\mathrm{fm}}
\def\ev{\mathrm{e\kern-0.1em V}}
\def\kev{\mathrm{ke\kern-0.1em V}}
\def\mev{\mathrm{Me\kern-0.1em V}}
\def\gev{\mathrm{Ge\kern-0.1em V}}
\def\tev{\mathrm{Te\kern-0.1em V}}

\let\Re=\re \let\Im=\im

\def\n#1e#2n{{#1}\times 10^{#2}}

\def\bea{\begin{eqnarray}}
\def\eea{\end{eqnarray}}
\def\nn{\nonumber}

\def\ods2{\mathcal{O}_{\Delta S=2}}
\def\zds2{Z_{\Delta S=2}}

\def\msbar{{\overline{\mathrm{MS}}}}

\makeatletter
\def\slash#1{{\mathpalette\c@ncel{#1}}} 
\def\big#1{{\hbox{$\left#1\vbox to1.012\ht\strutbox{}\right.\n@space$}}}
\def\Big#1{{\hbox{$\left#1\vbox to1.369\ht\strutbox{}\right.\n@space$}}}
\def\bigg#1{{\hbox{$\left#1\vbox to1.726\ht\strutbox{}\right.\n@space$}}}
\def\Bigg#1{{\hbox{$\left#1\vbox
to2.083\ht\strutbox{}\right.\n@space$}}}
\makeatother

\begin{document}

\title{Review of lattice results\\ concerning low energy particle physics}
\author{FLAG working group$^\star$ of FLAVIANET}
\date{\today}

\maketitle

\begin{center}
 $\rule{0cm}{0cm}^\star$ G.~Colangelo$^a\!$, S.~D\"urr$^{b,c}\!$, A.~J\"uttner$^d\!$,
L.~Lellouch$^e\!$, H.~Leutwyler$^a\!$, V.~Lubicz$^f\!$, \\S.~Necco$^d\!$,
C.~T.~Sachrajda$^g\!$, S.~Simula$^h\!$, 
A.~Vladikas$^i\!$, U.~Wenger$^a\!$, H.~Wittig$^j\!$\\[6mm]
{\footnotesize $^a\,$Albert Einstein Center for Fundamental Physics, Institut
  f\"ur Theoretische Physik,}\\
{\footnotesize Universit\"at Bern,
 Sidlerstr. 5, CH-3012 Bern, Switzerland }\\
{\footnotesize $^b$ Bergische Universit\"at Wuppertal, Gaussstr.\,20,
D-42119 Wuppertal, Germany}\\
{\footnotesize $^c$ J\"ulich Supercomputing Centre, Forschungszentrum J\"ulich,
D-52425 J\"ulich, Germany}\\
{\footnotesize $^d$ CERN, Physics Department, TH Unit, CH-1211 Geneva 23,
  Switzerland}\\
{\footnotesize $^e$ Centre de Physique Th\'eorique\footnote{CPT is research unit UMR 6207 of the CNRS and
of the universities Aix-Marseille I, Aix-Marseille II and Sud
Toulon-Var, and is affiliated with the FRUMAM.}, Case 907,
Campus de Luminy, F-13288 Marseille, France}\\
{\footnotesize $^f$ Dipartimento di Fisica, Universit\`a  Roma Tre, and INFN, Via della
  Vasca Navale 84, I-00146 Roma, Italy}\\
{\footnotesize $^g$ School of Physics and Astronomy, University of Southampton,
Southampton SO17 1BJ, UK}\\
{\footnotesize $^h$ INFN, Sezione di Roma Tre, Via della Vasca Navale 84, I-00146
  Roma, Italy}\\
{\footnotesize $^i$ INFN, Sezione di Tor Vergata, c/o Dipartimento di Fisica,
  Universit\`a di Roma Tor Vergata,\\ Via della Ricerca Scientifica 1,
  I-00133 Rome, Italy}\\
{\footnotesize $^j$ Institut f\"ur Kernphysik and Helmholtz Institute
  Mainz, University of Mainz,\\ Becher Weg 45, 55099 Mainz, Germany}
\end{center}

\begin{abstract}
\noindent
We review lattice results relevant for pion and kaon physics with the aim
of making them easily accessible to the particle physics community.
Specifically, we review the determination of the light-quark masses, the
form factor $f_+(0)$, relevant for the semileptonic $K\to\pi$ transition at
zero momentum transfer as well as the ratio $f_K/f_\pi$ of decay constants
and discuss the consequences for the elements $V_{us}$ and $V_{ud}$ of the
CKM matrix. Furthermore, we describe the results obtained on the lattice
for some of the low-energy constants of SU(2)$_L\times$SU(2)$_R$ and
SU(3)$_L\times$SU(3)$_R$ Chiral Perturbation Theory and review the
determination of the $B_K$ parameter of neutral kaon mixing. We introduce
quality criteria and use these when forming averages. Although subjective
and imperfect, these criteria may help the reader to judge different
aspects of current lattice computations. Our main results are summarized in
section \ref{sec:conclusion}, but we stress the importance of the detailed
discussion that underlies these results and constitutes the bulk of the
present review.
\end{abstract}

\vspace{-22 cm}
\begin{flushright}
CERN-PH-TH/2010-218\\ CPT-P056-2010 \\ SHEP-1039 
\end{flushright}

\clearpage
\tableofcontents

\clearpage
\section{Introduction}

Flavour physics provides an important opportunity for exploring the limits
of the Standard Model of particle physics and in constraining possible
extensions of theories ``Beyond the Standard Model''. As the LHC explores a
new energy frontier, the importance of flavour physics will grow still
further, in searches for signatures of new physics through precision
measurements and/or in helping to unravel the theoretical framework behind
direct discoveries of new particles. The major theoretical limitation
arises from the precision with which strong interaction effects can be
quantified and large-scale numerical simulations of lattice QCD provide the
opportunity of computing these effects from first principles. In this paper
we review the current status of lattice results for a variety of physical
quantities in low-energy physics; our aim is to provide the answer to the
frequently posed question ``What is currently the best lattice value for a
particular quantity?'' in a way which is readily accessible to
non-lattice-experts.  This is generally not an easy question to answer;
different collaborations use different lattice actions (discretizations of
QCD) with a variety of lattice spacings and volumes, and with a range of
masses for the $u$ and $d$ quarks. Not only are the systematic errors
therefore different, but also the methodology used to estimate these
uncertainties vary between collaborations. Below we summarize the main
features of each of the calculations and provide a framework for judging
and combining the different results. Sometimes it is a single result which
provides the ``best'' value; more often it is a combination of results from
different collaborations. Indeed, the consistency of values obtained using
different formulations adds significantly to our confidence in the results.

\subsection{Scope of the present review\label{sec:scope}}
The Flavianet Lattice Averaging Group (FLAG) was constituted
in November 2007, at a general meeting of the European Network on Flavour
Physics (Flavianet), with the remit of providing the current lattice results
to the network's working groups and to the wider community. In this paper
we focus on physically important quantities in kaon and pion
physics, specifically the masses of the light quarks, the CKM matrix
element $V_{us}$ determined from leptonic and semileptonic kaon decays, the
$B_K$ parameter of $K^0$\,--\,$\bar{K}^0$ mixing and the low-energy
constants of the SU(2)$_{\textrm L}\times$ SU(2)$_{\textrm R}$ and
SU(3)$_{\textrm L}\times$ SU(3)$_{\textrm R}$ chiral Lagrangians.  In the
future, we plan to extend the range of topics covered and to update the
material regularly. Below, we review papers which appeared before the
closing date of the present edition, 28 February 2011. The averages are based
on the results quoted in the tables, which exclusively list articles that
appeared before this date. Most of
the data concern simulations for which the masses of the two lightest quarks 
are set equal. This is indicated by the notation: 
$\Nf=2+1+1$, for instance, denotes a lattice calculation with four dynamical
quark flavours and $m_u=m_d\neq m_s\neq m_c$. Our
review is also available on the FLAG webpage \cite{FLAG:webpage}, which
will be updated as new lattice results appear. A compilation of some of the
results discussed below can also be found on the web page of Laiho, Lunghi
and Van de Water \cite{Laiho:webpage,Laiho:2009eu}.  These authors restrict
themselves to simulations with $\Nf=2+1$ dynamical flavours, but in
addition cover lattice results for bound states containing charmed or
beauty quarks, which are beyond the scope of the present review.
Our analysis does include simulations with $\Nf=2$. The significance of these 
data for the physics conclusions to be drawn from the work done on the lattice
is discussed in section \ref{sec:averages}.   

It is by now generally accepted that, taken together, QCD and QED provide a
very accurate approximation for the laws governing the low energy
properties of nature -- other degrees of freedom, such as those of the $W$
and $Z$ bosons, appear to be frozen at low energies. Moreover, since QED 
is infrared stable, the electromagnetic interaction can be dealt with 
perturbatively at low energy, but for QCD, this is not the case. The
present review deals with the information obtained about the low energy 
structure of that theory by simulating it on a lattice.

Lattice QCD is a mature field today as a result of a 30-year history of
theoretical and computational developments, many of which were tested on
quenched simulations. Indeed the remarkable recent progress in the
precision of lattice calculations is due to improved algorithms, better
computing resources and, last but not least, conceptual developments, such
as improved actions that reduce lattice artifacts, regularizations which
preserve (remnants of) chiral symmetry, understanding the finite size
effects, non-perturbative renormalization, etc.  For recent discussions of
these developments, we refer to \cite{Jansen:2008vs,Jung:2010jt}. A concise
characterization of the various discretizations that underlie the results
reported in the present review is given in appendix \ref{sec_quark_actions}
(cf.~\cite{Bazavov:2009bb} for a more extended overview).  The recent
developments in the domain of computer algorithms were triggered by
\cite{Hasenbusch:2001ne,Luscher:2005rx,Urbach:2005ji,Clark:2006fx}. For an
outline of the progress in computer architecture, we refer to
\cite{Ishikawa:2008pf}, which also contains an overview of recent
developments in simulation-algorithms.

Each discretization has its merits, but also its shortcomings. One of the
main points emerging from this review is that the large variety of
discretizations employed by the various collaborations lead to consistent
results, confirming universality within the accuracy reached. In our
opinion, only the eventual convergence of the various lattice results,
obtained with different discretizations and methods, can lead to a reliable
determination, for which all systematic errors can be claimed to be fully
under control.

The lattice spacings reached in recent simulations go down to 0.05 fm or
even smaller. In that region, growing autocorrelation times may slow down
the sampling of the configurations
\cite{Antonio:2008zz,Bazavov:2010xr,Schaefer:2010hu,Luscher:2010we,Schaefer:2010qh}.  
Many groups check for autocorrelations in a number of observables, including the
topological charge, for which a rapid growth of the autocorrelation time is
observed if the lattice spacing becomes small. In the following, we assume
that the continuum limit can be reached by extrapolating the existing
simulations.

Lattice simulations of QCD currently involve at most four dynamical quark flavours. To connect these with the QCD sector of the Standard Model, which is characterized by $\Nf = 6$, one considers a sequence of effective theories, obtained from full QCD by sending one of the quark masses after the other to infinity: QCD$_5$, QCD$_4$, \ldots The series terminates with gluodynamics, QCD$_0$. The physical value of the coupling constant is conventionally specified in terms of the value in QCD$_5$, in the $\msbar$ scheme at scale $\mu =M_Z$. We characterize the masses of the light quarks by their running $\msbar$ values in QCD$_4$ at scale $\mu=2$ GeV.\footnote{Note that this convention is not universally adopted -- the scale at which QCD$_3$ is matched to QCD$_4$ is not always taken below 2 GeV. The difference between the numerical values in QCD$_3$ and QCD$_4$ is small, but with the increase of precision on the lattice, it does become necessary to specify which one of these effective theories the quoted running masses refer to (see section \ref{sec:qmass}).} 

In this language, lattice calculations with $\Nf=3$ concern QCD$_3$. This theory deviates from full QCD by Zweig-suppressed corrections of $O(1/m_c^2)$. Note that calculations with $\Nf=2$ dynamical flavours often include strange valence quarks interacting with gluons, so that bound states with the quantum numbers of the kaons can be studied -- what these simulations share with QCD$_2$ is that strange sea quark fluctuations are neglected. Likewise, calculations done in the quenched approximation, in which sea quark effects are treated as a mean field, can be made to involve light valence quarks and mesons as well as baryons, while in QCD$_0$, the spectrum exclusively contains glueballs. 

For $\Nf\leq 3$, the connection between the observables and the running coupling constant and quark masses of the effective theory is now understood at the non-perturbative level. In the case of QCD$_3$, for instance, which is obtained from the QCD sector of the Standard Model by sending $m_t,m_b$ and $m_c$ to infinity, the running quark masses $m_u(\mu)$, $m_d(\mu)$, $m_s(\mu)$ and the running coupling constant $g(\mu)$ can unambiguously be related to the masses $M_{\pi^+}$, $M_{K^+}$, $M_{K^0}$ and the mass of one of the  baryons that occur in the spectrum, so that the simulation yields parameter free results for all of the observables of the effective theory in terms of these masses. The lattice results obtained so far, which concern the isospin limit of QCD$_3$, agree with experiment, within the accuracy of the calculation. The results reported for the masses of the long-lived baryons in BMW 08 \cite{Durr:2008zz}, for instance, have a precision of better than 4\%. In particular, if the scale is fixed with $M_\Xi$, the mass of the $\Omega$ can be calculated to an accuracy of 1.5\%. The fact that the result agrees with experiment indicates that  the residual effects generated by the neglected heavy degrees of freedom are small.

We concentrate on pion and kaon observables and assume that, at the precision indicated, simulations which do not account for the heavy degrees of freedom represent a sufficiently accurate approximation to full QCD. In particular, for $f_K/f_\pi$ and $f_+(0)$, where flavour symmetry suppresses the dynamics, we expect the effects generated by the presence of heavy sea quarks to be too small to be seen, despite the fact that the level of accuracy reached is impressive indeed.  We emphasize that the estimates quoted below and labeled $\Nf=2$ or $\Nf=2+1$ concern the values of the relevant quantities in full QCD -- the label merely indicates the method by means of which these estimates are obtained.

At low energies, the fact that QCD has an approximate, spontaneously
broken {\it chiral symmetry} plays a crucial role. In the limit where
the masses of the three lightest quarks are set to zero, this symmetry
becomes exact, so that the spectrum contains a massless octet of
Nambu-Goldsone bosons.  There are lattice formulations of the theory
that preserve this symmetry and we refer to appendix
\ref{sec_quark_actions} for an overview.  We rely on the understanding
of the dependence on the lattice spacing reached with Symanzik's
effective theory \cite{Symanzik:1983dc,Symanzik:1983gh}. This framework 
can be combined with Chiral Perturbation Theory ({\Ch}PT),
to cope with the Nambu-Goldstone nature of the lowest excitations that
occur in the presence of light quarks
\cite{Bernard:1993sv,Sharpe:1997by,Golterman:1997st,Sharpe:1998xm,
Lee:1999zxa,Sharpe:2000bc,Rupak:2002sm,Bar:2002nr, Aubin:2003mg,
Bar:2003mh,Aubin:2003uc,Aoki:2003yv,Aoki:2004ta,Sharpe:2004is,
Sharpe:2004ny,Bar:2005tu,Golterman:2005xa,Chen:2006wf,Chen:2007ug,Chen:2009su,Bar:2010jk}. The form of the effective theory depends on the discretization used -- see appendix \ref{sec_ChiPT} for a brief characterization of the different variants in use. 

As the calculations become more precise, it is important to emphasize the
need for continued close collaboration between the lattice and {\Ch}PT
communities. Lattice computations are generally performed at $u$ and $d$
quark masses which are heavier than the physical ones and the results are
then extrapolated to the physical point. Although these masses are
decreasing very significantly with time (see, in particular, the most
recent calculations by BMW \cite{BMW10} and PACS-CS \cite{Aoki:2009ix}), it
remains true that this extrapolation is one of the most significant sources
of systematic error. Indeed several lattice collaborations have found that
the expansion of $M_\pi^2$, $M_K^2$, $f_\pi$, $f_K$ to first non-leading
order in the quark masses $m_u,m_d,m_s$ [SU(3) {\Ch}PT to NLO] does not
yield an adequate parametrization of the data, and so cannot reliably be
used for guiding the chiral extrapolation. In part, the problem may be due
to the fact that in the simulations, the mean mass of $u$ and $d$ is
usually significantly heavier than in nature, while $m_s$ is taken in the
vicinity of the physical mass of the strange quark, so that the kaon mass
is then too heavy for {\Ch}PT to be of use, but it may also indicate that
contributions of NNLO need to be accounted for to have sufficient accuracy
(an analytic approximation for the latter that goes beyond the double-log
approximation \cite{Bijnens:1998yu} is proposed in \cite{Ecker:2010nc,Ecker:2010xz}).
One method to circumvent the impasse is to (a) work at a smaller value of
$m_s$, so that the data taken concern the region where the kaon mass does
not exceed its physical value \cite{Bietenholz:2010jr,Bietenholz:2010si} and (b) use NNLO
{\Ch}PT for the extrapolation to the quark masses of physical interest.
Alternatively, one may try to apply brute force, lower the mean mass of $u$
and $d$ to the physical value and avoid using an extrapolation altogether
\cite{BMW10,Aoki:2009ix}. Which one of these approaches yields better
results yet remains to be seen, but all of them can teach us something
about the low-energy regime of QCD and can contribute to a better
determination of the low-energy constants of {\Ch}PT.

\subsection{Summary of the main results\label{sec:conclusion}}

We end this introduction with a summary of our main results, but we stress
that they should be considered together with the many comments and
observations which accompany their determination. The lattice results 
for the running masses of the three lightest quarks are discussed in
detail in section \ref{sec:qmass}. Using the $\msbar$ scheme and fixing
the running scale at 2 $\gev$, the lattice data with $\Nf=2+1$ dynamical
flavours lead to 
\begin{eqnarray}\label{eq:ms+mud_Nf=3}
m_{ud}= 3.43\pm 0.11\,\mev\co \hspace{0.5cm}
m_s = 94\pm 3\,\mev\co\hspace{0.5cm}
\frac{m_s}{m_{ud}}=27.4\pm 0.4\co
\end{eqnarray}
with $m_{ud}\equiv \frac{1}{2}(m_u+m_d)$. We emphasize that these results represent a very significant improvement of our knowledge: as compared to the estimates for the light quark masses given by the Particle Data Group \cite{Manohar_and_Sachrajda}, the uncertainties in the above estimates for $m_s$, $m_{ud}$ and $m_s/m_{ud}$ are reduced by about an order of magnitude, in all three cases ! Note also that
the estimates (\ref{eq:ms+mud_Nf=3}) agree with the only $\Nf=2$
determination that satisfies our quality criteria~\cite{Blossier:2007vv}:
\begin{eqnarray}
m_{ud} = 3.6  \pm 0.2  \,\mbox{MeV}  \co \hspace{0.5cm}
m_s= 95  \pm 6 \,\mbox{MeV}\co  \hspace{0.5cm}
\frac{m_s}{m_{ud}} = 27.4 \pm 0.4 \fs  
\end{eqnarray} 
The lattice simulations
are performed with $m_u=m_d$, so some additional input is required to
obtain $m_u$ and $m_d$ separately (see the discussion in section
\ref{subsec:mumd}). We quote
\begin{eqnarray}
m_u = 2.19\pm 0.15\,\mev\co\hspace{0.5cm}
m_d= 4.67\pm 0.20\,\mev\co\hspace{0.5cm}
\frac{m_u}{m_d}= 0.47\pm 0.04\fs 
\end{eqnarray}

In the determination of the CKM matrix elements $V_{us}$ and $V_{ud}$, the
term $f_+(0)$ (form factor relevant for the semileptonic transition
$K^0\to\pi^-$, evaluated at zero momentum transfer) and the ratio
$f_K/f_\pi$ of the kaon and pion decay constants play a central role. 
Both of these quantities can now be determined rather precisely on the lattice. 
Adding statistical and systematic errors in quadrature, the results for $f_+(0)$ read
 \bea 
f_+(0)\al=\al 0.959\pm 0.005\,, \hspace{1cm}(\mbox{direct},\,\Nf=2+1),  \\
f_+(0)\al=\al 0.956\pm 0.008\,,
\hspace{1cm}(\mbox{direct},\,\Nf=2).\nonumber \eea
On the basis of the detailed discussion in section \ref{sec:Direct}, we conclude that the value 
\be
\label{eq:f+0}
f_+(0) = 0.956\pm 0.008  \hspace{1cm}\mbox{(direct)}
\ee
represents a conservative estimate for the range permitted by the presently available
direct determinations of $f_+(0)$ in lattice QCD. 

For $f_K/f_\pi$, the results are
\bea
\label{eq:direct} 
f_K/f_\pi \al=\al 1.193 \pm
0.005 \,,  \hspace{1cm}(\mbox{direct}\,,\;\Nf=2+1)\,,\\
  f_K/f_\pi\al =\al 1.210 \pm 0.018\,,
  \hspace{1cm}(\mbox{direct}\,,\;\Nf=2)\,, \nonumber
\eea 
where the first number is the average of three calculations whereas the
second stems from a single one. We refer to these results, which do
not rely on the Standard Model, as {\it direct} determinations. 

Two precise experimental results constrain the determination of $V_{us}$
and $V_{ud}$: the observed rate for the semileptonic $K^0\to\pi^-$ decay
determines the product $|V_{us}|f_+(0)$, while the ratios of the kaon's and
pion's leptonic widths determine the ratio $|V_{us}|f_K/|V_{ud}|f_\pi$. In
section \ref{sec:testing}, these results are combined with the lattice
determinations of $f_K/f_\pi$ and $f_+(0)$ to test the unitarity of the CKM
matrix. We find that the first row obeys this property within errors:
$|V_{ud}|^2+|V_{us}|^2+|V_{ub}|^2=1.002\pm 0.015$ and $1.036\pm 0.037$ for $\Nf =
2+1$ and $\Nf=2$, respectively.\footnote{The experimental information on
  $|V_{ub}|$ shows that the contribution from this term is negligibly
  small.} In obtaining these results, we have not made use of the recent
determinations of $|V_{ud}|$ from super-allowed nuclear $\beta$-decays 
\cite{Hardy:2008gy}, but these are perfectly consistent with the outcome of the
lattice calculations. In fact, the unitarity test sharpens considerably if
the results obtained on the lattice are combined with those found in
$\beta$-decay: the interval allowed for the sum
$|V_{ud}|^2+|V_{us}|^2+|V_{ub}|^2$ then shrinks by more than an order of
magnitude. Since this interval includes unity, CKM unitarity passes the test.

Within the Standard Model of particle physics, the CKM matrix is
unitary. As explained in section \ref{sec:SM}, this condition and the
experimental results for $|V_{us}|f_+(0)$, $|V_{us}|f_K/|V_{ud}|f_\pi$  
imply three equations for the four quantities $|V_{ud}|$,
$|V_{us}|$, $f_K/f_\pi$, $f_+(0)$. Thus it is sufficient to add one
additional constraint, such as the lattice result for $f_K/f_\pi$ or
$f_+(0)$, to determine all four of these quantities. Averaging over the
numbers obtained for $f_K/f_\pi$ and $f_+(0)$ with $\Nf=2+1$ 
dynamical quark flavours, we find 
\begin{eqnarray}\label{eq:final3}
|V_{ud}|\al=\al 0.97427 \pm 0.00021 \,, \qquad |V_{us}|=0.2254\pm 0.0009 \,,
 \\
f_+(0) \al=\al 0.9597\pm 0.0038 \,, \,\,\,\, \qquad \frac{f_K}{f_\pi}= 1.1925 \pm 0.0050 \,.
\nonumber\end{eqnarray} 
The results obtained from the data with $\Nf =2$ are the same within errors:
\begin{eqnarray}
\label{eq:final2}
|V_{ud}|\al=\al 0.97433 \pm 0.0042 \,, \qquad |V_{us}|=0.2251 \pm 0.0018 \,,\\
f_+(0) \al=\al 0.9604 \pm 0.0075 \,, \, \qquad \frac{f_K}{f_\pi}= 1.194 \pm 0.010 \,.
\nonumber\end{eqnarray}
Since the lattice results are consistent with CKM unitarity, it was to be
expected that the above values for $f_+(0)$ and $f_K/f_\pi$ are consistent
with those derived from the direct determinations, listed in equation
(\ref{eq:direct}).  The main effect of the unitarity constraint is a
reduction of the uncertainties.

As far as the low-energy constants are concerned, there are many results
coming from lattice calculations which are of interest. A coherent picture 
does not emerge yet for all cases, but for two quantities related to the SU(2)
chiral expansion, the lattice provides results which are competitive with
what has been obtained from phenomenology. These are
\begin{equation}
\lbar_3 = 3.2 \pm 0.8 \qquad \mbox{and} \qquad \frac{F_\pi}{F}= 1.073 \pm 0.015
\end{equation}
which are estimates based on $\Nf =2$, $\Nf =2+1$ and even $\Nf
=2+1+1$ calculations. We refer the interested reader
to Section~\ref{sec:LECs} for further results and all details about the
calculations.

For the $B_K$ parameter of $K$-$\bar{K}$ mixing we quote ($B_K$ refers to the 
$\msbar$-scheme at scale 2 GeV, while $\hat{B}_K$ denotes the corresponding
renormalization group invariant parameter):
\bea
B_K \al=\al 0.536\pm 0.017\co
\quad\hat{B}_K= 0.738\pm 0.020 \hspace{1cm} (\Nf=2+1)\co\\
 B_K \al=\al  0.516\pm 0.022\co
\quad\hat{B}_K= 0.729\pm 0.030 \hspace{1cm} (\Nf=2)\fs\nonumber
\eea
 
\subsection{Plan of the paper\label{sec:plan}}
The plan for the remainder of this paper is as follows. In the next section
we suggest criteria by which the quality of lattice calculations can be
judged and compared. We recognize of course that these are necessarily
subjective and generally give an incomplete picture of the quality of a
given simulation. Nevertheless we feel that some framework of quality
standards is necessary. There then follow five sections with physics
results: quark masses (section \ref{sec:qmass}), $V_{us}$ as determined
from leptonic and semileptonic Kaon decays (section \ref{sec:vusvud}), the
low energy constants of SU(2)$_{\textrm{L}}\times$SU(2)$_{\textrm{R}}$ and
SU(3)$_{\textrm{L}}\times$SU(3)$_{\textrm{R}}$ chiral perturbation theory
(sections \ref{sec:su2} and \ref{sec:su3} respectively) and the $B_K$
parameter which contains the non-perturbative strong interaction effects in
$K^0$--$\bar{K}^0$ mixing (section \ref{sec:BK}).

The final point we wish to raise in this introduction is a particularly
delicate one. As stated above, our aim is to make lattice QCD results
easily accessible to non-lattice-experts and we are well aware that it is
likely that some readers will only consult the present paper and not the
original lattice literature. We consider it very important that this paper
is not the only one which gets cited when the lattice results which are
discussed and analyzed here are quoted. Readers who find the review and
compilations offered in this paper to be useful are therefore kindly
requested also to cite the original sources -- the bibliography at the end
of this paper should make this task easier. Indeed we hope that the
bibliography will be one of the most widely used elements of the whole
paper.


\section{Quality criteria}
\label{sec:color-code}

In this article we present a collection of results and references
concerning lattice results of physical quantities in low energy particle
physics. In addition however, we aim to help the reader in assessing the
reliability of particular lattice results without necessarily studying the
original article in depth.
We understand that this is a delicate issue and are well aware of
the risk of making things ``simpler than they are''.  On the other hand, it
is rather common nowadays that theorists and experimentalists use
the results of lattice calculations in order to draw the physics
conclusions coming from a new measurement or a new analysis of a set of
measurements, without a critical assessment of the
quality of the various calculations. This may lead to a substantial
underestimate of systematic errors or in the worst case to a distorted
physics conclusion. We believe that despite the risks mentioned, it is
important to provide some compact information about the quality of a
calculation.
\subsection{Colour code}
In the past, at lattice conferences, some speakers have used a ``colour
code'' to provide such information, for example when showing plots
\cite{Pena:2006tw}. We find that this way of summarizing the quality of a
lattice calculation serves well its purpose and adopt it in what follows.
We identify a number of sources of systematic errors for which a systematic
improvement is possible and assign to each calculation a colour with
respect to each of
these:\\
\hspace{-0.1em}\good \hspace{0.2cm} when the systematic error has been
estimated in a 
satisfactory manner and convincingly\\
\rule{0.7cm}{0cm}shown to be under control;\\
\rule{0.1em}{0em}\soso \hspace{0.2cm} when a reasonable attempt at
estimating the systematic 
error has been made, although\\ \rule{0.7cm}{0cm}this could be improved;\\
\rule{0.1em}{0em}\bad \hspace{0.2cm} when no or a clearly unsatisfactory
attempt at 
estimating the systematic error has been\\\rule{0.7cm}{0cm}made.\\
The precise criteria used in determining the colour coding is unavoidably
time-dependent because as lattice calculations become more accurate the
standards against which they are measured become tighter. 
Today, our definition of the colour code related to the
systematic error coming from {\em i)} the chiral extrapolation, {\em ii)}
the continuum extrapolation, {\em iii)} the finite-volume effects, and
(where applicable) {\em iv)} the renormalization and {\em v)} the
running of operators and their matrix elements are as follows:
\begin{itemize}
\item Chiral extrapolation:\\
\good \hspace{0.2cm}  $M_{\pi,\mathrm{min}}< 250$ MeV  \\
\rule{0.05em}{0em}\soso \hspace{0.2cm}  250 MeV $\le M_{\pi,{\mathrm{min}}} \le$ 400 MeV \\
\rule{0.05em}{0em}\bad \hspace{0.2cm}  $M_{\pi,\mathrm{min}}> 400$ MeV \\
It is assumed that the chiral extrapolation is done with at least a
three-point analysis -- otherwise this will be explicitly mentioned in a
footnote. In case of nondegeneracies among the different pion states
$M_{\pi,\mathrm{min}}$ stands for a root-mean-squared (RMS) pion mass.

\item Continuum extrapolation:\\
\good \hspace{0.2cm}  3 or more lattice spacings, at least 2 points below
0.1 fm\\ 
\rule{0.05em}{0em}\soso \hspace{0.2cm}  2 or more lattice spacings, at least 1 point below 0.1
fm \\ 
\rule{0.05em}{0em}\bad \hspace{0.2cm}  otherwise\\
It is assumed that the action is $O(a)$-improved, i.e. the discretization
errors vanish quadratically with the lattice spacing -- otherwise this will be
explicitly mentioned in a footnote. Moreover the colour coding criteria for
non-improved actions change as follows: one lattice spacing more needed. 

\item Finite-volume effects:\\
\good \hspace{0.2cm}  $M_{\pi,\mathrm{min}} L > 4$ or at least 3 volumes \\
\rule{0.05em}{0em}\soso \hspace{0.2cm}  $M_{\pi,\mathrm{min}} L > 3$ and at least 2 volumes \\
\rule{0.05em}{0em}\bad \hspace{0.2cm}  otherwise\\
These criteria apply to calculations in the $p$-regime, and it is
assumed that $L_\mathrm{min}\ge $ 2 fm, otherwise this will be
explicitly mentioned in a footnote and a red square will be assigned. In
case of nondegeneracies among the different pion states
$M_{\pi,\mathrm{min}}$ stands for a root-mean-squared (RMS) pion mass.

\item Renormalization (where applicable):\\
\good \hspace{0.2cm}  non-perturbative\\
\rule{0.05em}{0em}\soso \hspace{0.2cm}  2-loop perturbation theory  \\
\rule{0.05em}{0em}\bad \hspace{0.2cm}  otherwise
\newpage
\item Running (where applicable): 
 
For scale-dependent quantities, such as quark masses or $B_K$, it is
essential that contact with continuum perturbation theory can be established. 
Various different methods are used for
this purpose (cf.~Appendix \ref{sec_match}): Regularization-independent
Momentum Subtraction (RI/MOM), Schr\"odinger functional, direct comparison
with (resummed) perturbation theory. In the case of the quark masses, a
further approach has been proposed recently: determination of $m_s$ via the
ratio $m_c/m_s$. Quite irrespective of the particular method used, the uncertainty 
associated with the choice of intermediate renormalization scales in the construction of 
physical observables must be brought under control. This is best achieved by performing 
comparisons between non-perturbative and perturbative running over a reasonably 
large range of scales. These comparisons were initially only made in the Schr\"odinger 
functional (SF) approach, but are now also being performed in RI/MOM schemes.

In the framework of the Schr\"odinger functional, the comparison of the
lattice results for the relevant renormalization factors with perturbation
theory has thoroughly been explored. Among the calculations relying on the
RI/MOM framework, the most recent ones are aiming for a level of control
over running and matching which is of comparable quality. However, since
these approaches are new, we postpone the formulation of quantitative
criteria until the systematics associated with their use is better
understood. We mark those data for which information about non-perturbative
running checks is available and give some details, but do not attempt to
translate this into a colour-code.

\end{itemize}

The mass of the pion plays an important role in our colour coding, but there
are fermion action discretizations in which pion states are non-degenerate.
This is the case in particular for the twisted-mass formulation of
lattice QCD. In that discretization, isospin symmetry and parity are traded
for a chiral symmetry which simplifies the renormalization of the
theory~\cite{Frezzotti:2000nk} and, at ``maximal twist'', eliminates
leading $O(a)$ discretization errors from all
observables~\cite{Frezzotti:2003ni}. As a consequence of the loss of
isospin symmetry, the charged and neutral pions have different masses, even
when the $u$ and $d$ quarks are mass degenerate.  This mass difference is
an $O(a^2)$ discretization effect, whose actual size depends on the
specific choice of the lattice regularization (in both the gauge and
fermionic part of the action). In the $\Nf = 2$ simulation
of~\cite{Boucaud:2007uk} it is found $1-M_{\pi^0}/M_{\pi^+}\sim 0.2$ for
$a\sim 0.09\,\fm$. Since the charged pion mass turns out also to be larger
than the mass of the neutral pion, when applying the criterion adopted for
assessing the quality of the chiral extrapolation to twisted mass fermion
simulations we will conservatively choose as $M_{\pi,\mathrm{min}}$ the
mass of the charged pion.

In staggered fermion calculations,\footnote{We refer the interested reader
  to a number of good reviews written about the
  subject~\cite{Durr:2005ax,Sharpe:2006re,Kronfeld:2007ek,Golterman:2008gt,
    Bazavov:2009bb}.}
the Nambu-Goldstone pion has ``taste'' partners whose masses are those of
the Nambu-Goldstone boson plus a discretization error of order $a^2$, which
can be significant at larger lattice spacings. For instance, in MILC's
simulations, at $a=0.15\,\fm$ and for a Nambu-Goldstone mass of 241~MeV,
the largest taste partner (i.e. the taste singlet) mass is 673~MeV and the
RMS taste-partner mass is 542~MeV~\cite{Bazavov:2009fk}. The situation
improves significantly as the lattice spacing is reduced: the singlet and
RMS masses become 341~MeV and 334~MeV at $a=0.045\,\fm$ for a
Nambu-Goldstone of mass 324~MeV.  While it is possible to pick out the
Nambu-Goldstone pion in the valence sector, the contributions of the sea
correspond to some complicated admixture of the 16 taste partners. MILC
reduces to one the contribution of the four quark tastes associated with a
single quark flavor by taking the fourth root of the quark determinant, and
hence the name rooted staggered fermions. In any event, the effective pion
mass in the sea can be significantly larger than that of the valence
Nambu-Goldstone and, in comparing staggered fermion calculations with those
performed using other discretizations, it makes sense to quote as an
$M_{\pi,\mathrm{min}}$ a value which is characteristic of the sea pion mass
contributions. For most quantities, the RMS taste-partner pion mass is a
reasonable choice.

Of course any colour coding has to be treated with caution. First of all
we repeat that the criteria are subjective and evolving. Sometimes a single
source of systematic error dominates the systematic uncertainty and it is
more important to reduce this uncertainty than to aim for green stars for
other sources of error. In spite of these caveats we hope that our attempt
to introduce quality measures for lattice results will prove to be a useful
guide. In addition we would like to underline that the agreement of lattice
results obtained using different actions and procedures evident in many of
the tables presented below provides further reassurance.

Finally it is important to stress that the colour coding does not imply any
credit for the development of theoretical or technical ideas used in the
simulations. Frequently new ideas and techniques are first tested on
smaller lattices, perhaps on quenched configurations or at heavier quark
masses, to demonstrate their feasibility. They may then be applied on more
modern lattices by other collaborations to obtain physical results, but
since our primary aim is to review the latest results concerning quantities 
of physical interest, we do not try to attribute credit to the underlying 
ideas or techniques.

For a coherent assessment of the present situation, the quality of the data plays a key role, but the colour code cannot be made visible in the figures. Simply showing all data on equal footing would give the misleading impression that the overall consistency of the information available on the lattice is questionable. As a way out, the figures do indicate the quality in a rudimentary way: data for which the colour code is free of red tags are shown as green symbols, otherwise they are depicted in red. Note that the pictures do not distinguish updates from earlier data obtained by the same collaboration -- the reader needs to compare the legends to identify closely correlated data.
 
\subsection{Averages and estimates}\label{sec:averages}
For some observables there may be enough independent lattice calculations
of good quality that it makes sense to average them and propose such an
{\em average} as the best current lattice number. In order to decide if
this situation is realized for a certain observable we rely on the colour
coding. Unless special reasons are given for making an exception, we
restrict the averages to data for which the colour code does not contain
any red tags. In some cases, the averaging procedure nevertheless leads to a 
result which in our opinion does not cover all uncertainties -- this happens, 
in particular, if some of the data have comparatively
small systematic errors. In such cases, we may provide our {\em estimate}
as the best current lattice number. This estimate is not obtained with a prescribed 
mathematical procedure, but is based on our own assessment of the 
information collected on the lattice.

There are two other important criteria which also play a role in this
respect, but which cannot be represented with the colour coding, because a
systematic improvement is not possible. These are: {\em i)} the publication
status, and {\em ii)} the number of flavours. As far as the former is
concerned we adopt the following policy: we consider in our averages only
calculations which have been published, {\em i.e.} which have been endorsed
by a referee. The only exception to this rule is an obvious update of numbers 
appearing in a previous publication. Other cases will be listed and
identified as such also by a (coloured) symbol, but not used in averages:
\begin{itemize}
\item Publication status:\\
\gA  \hspace{0.2cm}published or plain update of published results\\
\oP  \hspace{0.2cm}preprint\\ 
\rC  \hspace{0.2cm}conference contribution
\end{itemize}

Several active lattice collaborations do their calculations with only two
dynamical flavours ($\Nf=2$). We are not aware of visible differences
between results obtained in this framework and QCD with $\Nf=3$. At the
present level of accuracy, it is thus perfectly meaningful to compare
results obtained for $\Nf=2$ with experiment. On the other hand, the two
theories are intrinsically different; the manner in which quantities like
the coupling constant or the quark masses depend on the running scale
depends on $\Nf$.

Since we are not aware of an {\em a priori} way to quantitatively
estimate the difference between $\Nf=3$ and $\Nf=2$ calculations we
will present separate averages (where applicable) for the two sets of
calculations.  Averages of $\Nf=3$ and $\Nf=2$ calculations will not
be provided.  A first lattice calculation with $\Nf=4$ dynamical
flavours has recently appeared \cite{Baron:2010bv}. While in principle
it would be cleaner to keep $\Nf=4$ separate from the rest, we believe
that, for the quantities under discussion in this review, the effects
of the $c$-quark are below the precision of current lattice
calculations: if appropriate, we will average $\Nf=4$ and $\Nf=3$ results.

We stress that $\Nf=2$ calculations have in several cases a better control
over systematic effects than $\Nf=3$ calculations, as will be clear in
colour coding tables below, and therefore play a very important role today
in the comparison with experiment. Several calculations discussed
below provide examples of the relevance of $\Nf=2$ calculations today, and
in particular they enter in some of our estimates. In the future, as the
accuracy and the control over systematic effects in lattice calculations
will increase, it will hopefully be possible to see a difference between
$\Nf=2$ and $\Nf=3$ calculations and so determine the size of the
Zweig-rule violations related to strange quark loops. This is a very
interesting issue {\em per se}, and one which can be quantitatively
addressed only with lattice calculations.

One of the problems that arises when forming averages is that not all of 
the data sets are independent -- in fact, some rely on the same ensembles.
For this reason, we do not average updates with earlier results. Note that,
in this respect, the figures give a somewhat distorted picture: they include
all of the data listed in the tables and, moreover, indicate their quality only in a very
rudimentary way. Specific problems encountered in connection with correlations 
between different data sets are mentioned in the text. In our opinion, the 
variety of algorithms and configurations used and the fact that we attach 
conservative errors to our estimates ensure that these correlations do not 
distort our results in a significant manner.

We take the mean values from the standard procedure, where $\chi^2$ is
evaluated by adding the statistical and systematic errors in quadrature 
(data sets for which the error estimate only accounts for the statistical 
uncertainties are not taken into consideration).
The standard procedure also offers an estimate for the net uncertainty $\delta$ to
be attached to the mean: if the fit is of good quality
($\chi^2_{min}/\mbox{dof}\leq 1$), calculate $\delta$ from
$\chi^2=\chi^2_{min}+1$; otherwise, stretch the result obtained in this way
by the factor $S=
\sqrt{\rule[-0.07cm]{0cm}{0.37cm}\rule{1.5cm}{0cm}}\hspace{-1.5cm}
\chi^2_{min}/\mbox{dof}$. The problem with this recipe is that the
systematic errors are not stochastic. In particular, applying it to the
lattice data on $m_{ud}$, $m_s$, $m_s/m_{ud}$ and $f_K/f_\pi$, one arrives
at a total error that is smaller than the smallest systematic error of the
individual data sets. In our opinion, the latter represents a lower limit
for the total error to be attached to the average. It is not a trivial
matter, however, to split the total error of the average into a statistical
and a systematic part. We are not aware of a generally accepted
prescription for doing this and refrain from proposing one in this review.
In application to the lattice data to be discussed below, where the
statistical errors are small compared to the systematic ones and thus
barely affect the result, replacing the quantity $\delta$ by the smallest
systematic error of the individual data sets does not lead to a significant
underestimate of the total error. In the following, we adopt this
prescription, which has the advantage of being simple and transparent.  For
a more sophisticated discussion of the problem, we refer to
\cite{Lellouch:2009fg}. At the precision used in the present review, the
numerical results obtained for the total errors with the method proposed
there are the same.


\section{Quark masses}
\label{sec:qmass}
Quark masses are fundamental parameters of the Standard Model. An accurate
determination of these parameters is important for both phenomenological
and theoretical applications. The charm and bottom masses, for instance,
enter the theoretical expressions of several cross sections and decay
rates, in heavy quark expansions. The up, down and strange quark masses
govern the amount of explicit chiral symmetry breaking in QCD. From a
theoretical point of view, the values of quark masses provide information
about the flavour structure of physics beyond the Standard Model. The Review
of Particle Physics of the Particle Data Group contains a review of quark
masses \cite{Manohar_and_Sachrajda}, which covers light as well as heavy
flavours.  The present summary only deals with the light quark masses (those
of the up, down and strange quarks), but discusses the lattice results for
these in more detail.

Quark masses cannot be measured directly with experiment, because
quarks cannot be isolated, as they are confined inside hadrons. On the
other hand, quark masses are free parameters of the theory and, as
such, cannot be obtained on the basis of purely theoretical
considerations. Their values can only be determined by comparing the
theoretical prediction for an observable, which depends on the quark
mass of interest, with the corresponding experimental value. What
makes light quark masses particularly difficult to determine is the
fact that they are very small (for the up and down) or small (for the
strange) compared to typical hadronic scales. Thus, their impact on
typical hadronic observables is minute and it is difficult to isolate
their contribution accurately.

Fortunately, the spontaneous breaking of SU(3)$_L\otimes$SU(3)$_R$
chiral symmetry provides observables which are particularly sensitive
to the light quark masses: the masses of the resulting Nambu-Goldstone
bosons (NGB), i.e.~pions, kaons and etas. Indeed, the
Gell-Mann-Oakes-Renner relation~\cite{GellMann:1968rz} predicts that
the squared mass of a NGB is directly proportional to the sum of the
masses of the quark and antiquark which compose it, up to higher order
mass corrections. Moreover, because these NGBs are light and are
composed of only two valence particles, their masses have a
particularly clean statistical signal in lattice QCD calculations. In
addition to which, the experimental uncertainties on these meson
masses are negligible.

\medskip

Three flavour QCD has four free parameters: the strong coupling,
$\alpha_s$ (alternatively $\Lambda_\mathrm{QCD}$) and the up, down and
strange quark masses, $m_u$, $m_d$ and $m_s$. However, present day
lattice calculations are performed in the isospin limit, and the up
and down quark masses get replaced by a single parameter: the isospin
averaged up and down quark mass, $m_{ud}=\frac12(m_u+m_d)$. A lattice
determination of these parameters requires two steps:
\begin{enumerate}
\item Calculations of three experimentally measurable quantities are
 used to fix the three bare parameters. As already discussed, NGB
 masses are particularly appropriate for fixing the light quark
 masses. Another observable, such as the mass of a member of the
 baryon octet, can be used to fix the overall scale. It is important
 to note that most of present day calculations are performed at values of 
 $m_{ud}$ which are still substantially larger than its physical
 value, typically four times as large. Thus, reaching the physical up
 and down quark mass point still requires a significant
 extrapolation. The only exceptions are the
 PACS-CS 08, 09, 10~\cite{Aoki:2008sm,Aoki:2009ix,Aoki:2010wm} and BMW
 08, 10A~\cite{Durr:2008zz,Durr:2010vn} 
 calculations discussed below, where masses very close to the physical
 value are reached. The situation is radically different for the
 strange quark.  Indeed, modern simulations can include strange
 quarks whose masses bracket its physical value, and only
 interpolations are needed.

\item Renormalizations of these bare parameters must be performed to relate
 them to the corresponding cutoff-independent, renormalized
 parameters.\footnote{Throughout this review, the quark masses $m_u$,
   $m_d$ and $m_s$ refer to the $\msbar$ scheme at running scale
   $\mu=2\,\gev$ and the numerical values are given in MeV units.} These
 are short distance calculations, which may be performed perturbatively.
 Experience shows that one-loop calculations are particularly unreliable
 for the renormalization of quark masses, in all discretizations. Thus, at
 least two loops are required to have trustworthy results. Nevertheless, it
 is best to perform the renormalizations non-perturbatively, to avoid
 potentially large perturbative uncertainties due to neglected higher
 order terms.
\end{enumerate}
Of course, in quark mass ratios the renormalization factor cancels, so that
this second step is no longer relevant.

\subsection{Contributions from the electromagnetic interaction}
\label{subsec:electromagnetic interaction}
As mentioned in section \ref{sec:color-code}, the present review
relies on the hypothesis that, at low energies, the Lagrangian ${\cal
 L}_{\mbox{\tiny QCD}}+{\cal L}_{\mbox{\tiny QED}}$ describes
nature to a high degree of precision. Moreover, we assume that, at the
accuracy reached by now and for the quantities discussed here, the
difference between the results obtained from simulations with three
dynamical flavours and full QCD is small in comparison with the quoted
systematic uncertainties. The electromagnetic interaction, on the
other hand, cannot be ignored.  Quite generally, when comparing QCD
calculations with experiment, radiative corrections need to be
applied. In lattice simulations, where the QCD parameters are fixed in
terms of the masses of some of the hadrons, the electromagnetic
contributions to these masses must be accounted for.\footnote{Since
 the decomposition of the sum ${\cal L}_{\mbox{\tiny QCD}}+{\cal
   L}_{\mbox{\tiny QED}}$ into two parts is not unique, specifying
 the QCD part requires a convention. In order to give results for the
 quark masses in the Standard Model at scale $\mu=2\,\mbox{GeV}$, on
 the basis of a calculation done within QCD, it is convenient to
 match the two theories at that scale. We use this convention
 throughout the present review.  Note that a different convention is
 used in the analysis of the precision measurements carried out in
 low energy pion physics. The result $a_0\, M_{\pi^+} =
 0.220(5)(2)(6)$ of the NA48/2 collaboration
 \cite{BlochDevaux:2008zz} for the $I=0$ S-wave $\pi\pi$ scattering
 length, for instance, also concerns QCD in the isospin limit, but
 refers to the convention where $\hat{M}_\pi$ is identified with
 $M_{\pi^+}$. When comparing  lattice results with experiment, it is
 important to fix the QCD parameters in accordance with the convention used
 in the analysis of the experimental data (for a more detailed discussion,
 see
 \cite{Gasser:2003hk,Rusetsky:2009ic,Gasser:2007de,Leutwyler:2009jg}). } 

The electromagnetic interaction plays a crucial role in determinations
of the ratio $m_u/m_d$, because the isospin breaking effects generated
by this interaction are comparable to those from $m_u\neq m_d$ (see
subsection \ref{subsec:mumd}). In determinations of the ratio
$m_s/m_{ud}$, the electromagnetic interaction is less important, but
at the accuracy reached, it cannot be neglected. The reason is that,
in the determination of this ratio, the pion mass enters as an input
parameter. Because $M_\pi$ represents a small symmetry breaking
effect, it is rather sensitive to the perturbations generated by QED.

We distinguish the physical mass $M_P$, $P\in\{\pi^+,$ $\pi^0$, $K^+$,
$K^0\}$, from the mass $\hat{M}_P$ within QCD alone. The e.m.\ self-energy
is the difference between the two, $M_P^\gamma\equiv
M_P-\hat{M}_P$. Since the self-energy of the Nambu-Goldstone bosons
diverges in the chiral limit, it is convenient to replace it by the
contribution of the e.m.~interaction to the {\it square} of the mass,
\be \label{eq:DeltaP}
\Delta_{P}^\gamma\equiv M_P^2-\hat{M}_P^2= 2\,M_P M_P^\gamma+O(e^4)\,.\ee 
The main
effect of the e.m.\ interaction is an increase in the mass of the
charged particles, generated by the photon cloud that surrounds
them. The self-energies of the neutral ones are comparatively small,
particularly for the Nambu-Goldstone bosons, which do not have a
magnetic moment. Dashen's theorem \cite{Dashen:1969eg} confirms this
picture, as it states that, to leading order (LO) of the chiral expansion,
the self-energies of the neutral NGBs vanish, while the charged ones
obey $\Delta_{K^+}^\gamma = \Delta_{\pi^+}^\gamma $. It is convenient
to express the self-energies of the neutral particles as well as the
mass difference between the charged and neutral pions within QCD in
units of the observed mass difference, $\Delta_\pi\equiv
M_{\pi^+}^2-M_{\pi^0}^2$:
\be\label{eq:epsilon1}
\Delta_{\pi^0}^\gamma \equiv
\epsilon_{\pi^0}\,\Delta_\pi\co\hspace{0.2cm}\Delta_{K^0}^\gamma \equiv
\epsilon_{K^0}\,\Delta_\pi\co\hspace{0.2cm}\hat{M}_{\pi^+}^2-
\hat{M}_{\pi^0}^2\equiv
\epsilon_m\,\Delta_\pi\fs\ee
In this notation, the self-energies of the charged particles are given
by 
\be\label{eq:epsilon2}
\Delta_{\pi^+}^\gamma=(1+\epsilon_{\pi^0}-\epsilon_m)\,\Delta_\pi\co\hspace{0.5cm}
\Delta_{K^+}^\gamma=(1+\epsilonD+\epsilon_{K^0}-\epsilon_m)\,\Delta_\pi\co\ee
where the dimensionless coefficient $\epsilonD$ parametrizes the
violation of Dashen's theorem, 
\be\label{eq:epsilon3}
\Delta_{K^+}^\gamma-\Delta_{K^0}^\gamma-
\Delta_{\pi^+}^\gamma+\Delta_{\pi^0}^\gamma\equiv\epsilonD\,\Delta_\pi\fs\ee
Any determination of the light quark masses based on a calculation of
the masses of $\pi^+,K^+$ and $K^0$ within QCD requires an estimate
for the coefficients $\epsilonD$, $\epsilon_{\pi^0}$, $\epsilon_{K^0}$
and $\epsilon_m$.

The first determination of the self-energies on the lattice was
carried out by Duncan, Eichten and Thacker \cite{Duncan:1996xy}. Using
the quenched approximation, they arrived at
$M_{K^+}^\gamma-M_{K^0}^\gamma= 1.9\,\mbox{MeV}$. Actually, the
parametrization of the masses given in that paper yields an estimate
for all but one of the coefficients introduced above (since the mass
splitting between the charged and neutral pions in QCD is neglected,
the parametrization amounts to setting $\epsilon_m=0$ ab
initio). Evaluating 
the differences between the masses obtained at the
physical value of the electromagnetic coupling constant and at $e=0$,
we obtain $\epsilonD = 0.50(8)$, $\epsilon_{\pi^0} = 0.034(5)$ and
$\epsilon_{K^0} = 0.23(3)$. The errors quoted are statistical only: an
estimate of lattice systematic errors is not possible from the
limited results of \cite{Duncan:1996xy}. The result for $\epsilonD$
indicates that the violations of Dashen's theorem are sizable:
according to this calculation, the non-leading contributions to the
self-energy difference of the kaons amount to 50\% of the leading
term. The result for the self-energy of the neutral pion cannot be
taken at face value, because it is small, comparable with the
neglected mass difference $\hat{M}_{\pi^+}-\hat{M}_{\pi^0}$. To
illustrate this, we note that the numbers quoted above are obtained by
matching the parametrization with the physical masses for $\pi^0$,
$K^+$ and $K^0$. This gives a mass for the charged pion that is too
high by 0.32 MeV. Tuning the parameters instead such that $M_{\pi^+}$
comes out correctly, the result for the self-energy of the neutral
pion becomes larger: $\epsilon_{\pi^0}=0.10(7)$. Also, the
uncertainties due to the systematic errors of the lattice calculation
need to be taken into account.

In an update of this calculation by the RBC collaboration
\cite{Blum:2007cy} (RBC 07), the electromagnetic interaction is still
treated in the quenched approximation, but the strong interaction is
simulated with $\Nf=2$ dynamical quark flavours. The quark masses are fixed
with the physical masses of $\pi^0$, $K^+$ and $K^0$. The outcome for the
difference in the electromagnetic self-energy of the kaons reads
$M_{K^+}^\gamma-M_{K^0}^\gamma= 1.443(55)\,\mbox{MeV}$. This corresponds to
a remarkably small violation of Dashen's theorem. Indeed, a recent
extension of this work to $\Nf=2+1$ dynamical flavours \cite{Blum:2010ym}
leads to a significantly larger self-energy difference:
$M_{K^+}^\gamma-M_{K^0}^\gamma= 1.87(10)\,\mbox{MeV}$, in good agreement
with the estimate of Eichten et al. Expressed in terms of the coefficient
$\epsilonD$ that measures the size of the violation of Dashen's theorem, it
corresponds to $\epsilonD=0.5(1)$.

More recently, the BMW collaboration has reported preliminary results for
the electromagnetic self-energies, based on simulations in which a U(1)
degree of freedom is superimposed on their $\Nf = 2+1$ QCD configurations
\cite{Portelli:2010yn}. The result for the kaon self-energy difference
reads $M_{K^+}^\gamma-M_{K^0}^\gamma= 2.2(2)\,\mbox{MeV}$, that is
$\epsilonD=0.66(14)$, where the error indicates the statistical
uncertainties.

The input for the electromagnetic corrections used by the MILC
collaboration is specified in \cite{Aubin:2004he}. In their analysis of the
lattice data, $\epsilon_{\pi^0}$, $\epsilon_{K^0}$ and $\epsilon_m$ are set
equal to zero. For the remaining coefficient, which plays a crucial role in
determinations of the ratio $m_u/m_d$, the very conservative range
$\epsilonD=1\pm1$ was used in MILC 04 \cite{Aubin:2004fs}, while in more
recent work, in particular in MILC 09 \cite{Bazavov:2009bb} and MILC
09A~\cite{Bazavov:2009fk}, this input is replaced by $\epsilonD=1.2\pm0.5$,
as suggested by phenomenological estimates for the corrections to Dashen's
theorem \cite{Bijnens:1996kk,Donoghue:1996zn}. First results of an
evaluation of the electromagnetic self-energies based on $\Nf=2+1$
dynamical quarks in the QCD sector and on the quenched approximation in the
QED sector are also reported by the MILC collaboration
\cite{Basak:2008na}. Further calculations using staggered quarks and
quenched photons are underway \cite{Bernard:2010qd}.

The effective Lagrangian that governs the self-energies to next-to-leading
order (NLO) of the chiral expansion was set up in \cite{Urech:1994hd}. The
estimates in \cite{Bijnens:1996kk,Donoghue:1996zn} are obtained by
replacing QCD with a model, matching this model with the effective theory
and assuming that the effective coupling constants obtained in this way
represent a decent approximation to those of QCD. For alternative model
estimates and a detailed discussion of the problems encountered in models
based on saturation by resonances, see
\cite{Baur:1995ig,Baur:1996ya,Moussallam:1997xx}.  In the present review of
the information obtained on the lattice, we avoid the use of models
altogether.

There is an indirect phenomenological determination of $\epsilonD$, which
is based on the decay $\eta\rightarrow 3\pi$ and does not rely on models.
The result for the quark mass ratio $Q$ obtained from a dispersive analysis
of this decay implies $\epsilonD = 0.70(28)$ (see section
\ref{subsec:mumd}).  While the values found on the lattice
\cite{Duncan:1996xy,Blum:2007cy,Blum:2010ym,Portelli:2010yn} are lower, the
phenomenological estimate used by MILC 09A \cite{Bazavov:2009fk} is on the
high side. In the following, we take the central value from $\eta$-decay,
but stretch the error, so that the numbers given in
\cite{Bazavov:2009fk,Duncan:1996xy,Blum:2007cy,Blum:2010ym,Portelli:2010yn}
are all within one standard deviation: $\epsilonD=0.7(5)$.

We add a few comments concerning the physics of the self-energies and then
specify the estimates used as an input in our analysis of the data. The
Cottingham formula \cite{Cottingham} represents the self-energy of a
particle as an integral over electron scattering cross sections; elastic as
well as inelastic reactions contribute. For the charged pion, the term due
to elastic scattering, which involves the square of the e.m.~form factor,
makes a substantial contribution. In the case of the $\pi^0$, this term is
absent, because the form factor vanishes on account of charge conjugation
invariance. Indeed, the contribution from the form factor to the
self-energy of the $\pi^+$ roughly reproduces the observed mass difference
between the two particles. Furthermore, the numbers given in
\cite{Socolow:1965zz,Gross:1979ur,Gasser:1982ap} indicate that the
inelastic contributions are significantly smaller than the elastic
contributions to the self-energy of the $\pi^+$. The low energy theorem of
Das, Guralnik, Mathur, Low and Young \cite{Das:1967it} ensures that, in the
limit $m_u,m_d\rightarrow 0$, the e.m.~self-energy of the $\pi^0$ vanishes,
while the one of the $\pi^+$ is given by an integral over the difference
between the vector and axial spectral functions. The estimates for
$\epsilon_{\pi^0}$ obtained in \cite{Duncan:1996xy} are consistent with the
suppression of the self-energy of the $\pi^0$ implied by chiral
SU(2)$\times$SU(2). In our opinion, $\epsilon_{\pi^0}=0.07(7)$ is a
conservative estimate for this coefficient. The self-energy of the $K^0$ is
suppressed less strongly, because it remains different from zero if $m_u$
and $m_d$ are taken massless and only disappears if $m_s$ is turned off as
well. Note also that, since the e.m.~form factor of the $K^0$ is different
from zero, the self-energy of the $K^0$ does pick up an elastic
contribution. The lattice result for $\epsilon_{K^0}$ indicates that the
violation of Dashen's theorem is smaller than in the case of $\epsilonD$.
In the following, we use $\epsilon_{K^0}=0.3(3)$. 

Finally, we consider the mass splitting between the charged and neutral
pions in QCD. This effect is known to be very small, because it is of
second order in $m_u-m_d$. There is a parameter-free prediction, which
expresses the difference $\hat{M}_{\pi^+}^2-\hat{M}_{\pi^0}^2$ in terms of
the physical masses of the pseudoscalar octet and is valid to NLO of the
chiral perturbation series. Numerically, the relation yields
$\epsilon_m=0.04$ \cite{Gasser:1984gg}, indicating that this contribution
does not play a significant role at the present level of accuracy. We
attach a conservative error also to this coefficient: $\epsilon_m=0.04(2)$.
The lattice result for the self-energy difference of the pions, reported in
\cite{Blum:2010ym}, $M_{\pi^+}^\gamma-M_{\pi^0}^\gamma=
4.50(23)\,\mbox{MeV}$, agrees with this estimate: expressed in terms of the
coefficient $\epsilon_m$ that measures the pion mass splitting in QCD, the
result corresponds to $\epsilon_m=0.04(5)$. The corrections of
next-to-next-to-leading order (NNLO) have been worked out
\cite{Amoros:2001cp}, but the numerical evaluation of the 
formulae again meets with the problem that the relevant effective coupling
constants are not reliably known. 

In summary, we use the following estimates for the e.m.~corrections:
\be\label{eq:epsilon num}\epsilonD=0.7(5)
\co\hspace{0.5cm}\epsilon_{\pi^0}=0.07(7)\co\hspace{0.5cm}
\epsilon_{K^0}=0.3(3)\co\hspace{0.5cm}\epsilon_m=0.04(2)\fs\ee
While the range used for the coefficient $\epsilonD$ affects our
analysis in a significant way, the numerical values of the other
coefficients only serve to set the scale of these contributions. The
range given for $\epsilon_{\pi^0}$ and $\epsilon_{K^0}$ may be overly
generous, but because of the exploratory nature of the lattice
determinations, we consider it advisable to use a conservative
estimate.

Treating the uncertainties in the four coefficients as statistically
independent and adding errors in quadrature, the numbers in equation
(\ref{eq:epsilon num}) yield the following estimates for the
e.m.~self-energies,
\bea\label{eq:Mem}&&\hspace{-1cm} M_{\pi^+}^\gamma= 4.7(3)\,
\mbox{MeV}\co\hspace{0.45cm} M_{\pi^0}^\gamma = 0.3(3)\,\mbox{MeV}
\co\hspace{0.5cm} M_{\pi^+}^\gamma-M_{\pi^0}^\gamma=4.4(1)\, \mbox{MeV}\co\\
&&\hspace{-1cm} M_{K^+}^\gamma= 2.5(7)\,\mbox{MeV}\co\hspace{0.35cm}
M_{K^0}^\gamma
=0.4(4)\,\mbox{MeV}\co\hspace{0.3cm}M_{K^+}^\gamma-M_{K^0}^\gamma= 2.1(6)\,
\mbox{MeV}\,,\nonumber \eea
and for the pion and kaon masses occurring in the QCD sector of the
Standard Model, 
\bea\label{eq:MQCD}&&\hspace{-1cm} \hat{M}_{\pi^+}= 134.8(3)\,
\mbox{MeV}\co\hspace{0.2cm} \hat{M}_{\pi^0} = 134.6(3)\,\mbox{MeV}
\co\hspace{0.5cm} \hat{M}_{\pi^+}-\hat{M}_{\pi^0}=\hspace{0.25cm}0.2(1)\,
\mbox{MeV}\co\\ 
&&\hspace{-1.1cm} \hat{M}_{K^+}= 491.2(7)\,\mbox{MeV}\co\hspace{0.12cm}
\hat{M}_{K^0}
=497.2(4)\,\mbox{MeV}\co\hspace{0.3cm}\hat{M}_{K^+}-\hat{M}_{K^0}=-6.1(6)\,
\mbox{MeV}\fs\nonumber \eea
The self-energy difference between the charged and neutral pion involves
the same coefficient $\epsilon_m$ that describes the mass difference in QCD
-- this is why the estimate for $ M_{\pi^+}^\gamma-M_{\pi^0}^\gamma$ is so
sharp.

\subsection{Pion and kaon masses in the isospin limit}
\label{subsec:mu not equal md}

As mentioned above, most of the lattice calculations concerning the
properties of the light mesons are performed in the isospin limit of QCD 
($m_u-m_d\rightarrow0$ at fixed $m_u+m_d$). We
denote the pion and kaon masses in that limit by $\Mpibar$ and
$\MKbar$, respectively. Their numerical values can be estimated as
follows. Since the operation $u\leftrightarrow d$ interchanges $\pi^+$
with $\pi^-$ and $K^+$ with $K^0$, the expansion of the quantities
$\hat{M}_{\pi^+}^2$ and $\frac{1}{2}(\hat{M}_{K^+}^2+\hat{M}_{K^0}^2)$
in powers of $m_u-m_d$ only contains even powers. As shown in
\cite{Gasser:1983yg}, the effects generated by $m_u-m_d$ in the mass
of the charged pion are strongly suppressed: the difference
$\hat{M}_{\pi^+}^2-\Mpibar^{\,2}$ represents a quantity of
$O[(m_u-m_d)^2(m_u+m_d)]$ and is therefore small compared to the
difference $\hat{M}_{\pi^+}^2-\hat{M}_{\pi^0}^2$, for which an
estimate was given above. In the case of 
$\frac{1}{2}(\hat{M}_{K^+}^2+\hat{M}_{K^0}^2)-\MKbar^{\,2}$, the
expansion does contain a contribution at NLO, determined by the
combination $2L_8-L_5$ of low energy constants, but the lattice
results for that combination show that this contribution is very
small, too. Numerically, the effects generated by $m_u-m_d$ in
$\hat{M}_{\pi^+}^2$ and in
$\frac{1}{2}(\hat{M}_{K^+}^2+\hat{M}_{K^0}^2)$ are negligible compared
to the uncertainties in the electromagnetic self-energies. The
estimates for these given in equation (\ref{eq:epsilon num}) thus imply
\be
\Mpibar =134.8(3)\,\mev\ ,\hspace{1cm} \MKbar= 494.2(5)\,\mev\ . 
\label{eq:MpiMKiso}
\ee
This shows that, for the convention used above to specify the QCD sector of
the Standard Model, and within the accuracy to which this convention can
currently be implemented, the mass of the pion in the isospin limit agrees
with the physical mass of the neutral pion: $\Mpibar-M_{\pi^0}=-0.2(3)$
MeV.

\subsection{Lattice determination of $m_s$ and $m_{ud}$}

We now turn to a review of the lattice calculations of the light quark
masses and begin with $m_s$, the isospin averaged up and down quark
mass, $m_{ud}$, and their ratio. Most groups quote only $m_{ud}$, not
the individual up and down quark masses. We then discuss the ratio
$m_u/m_d$ and the individual determination of $m_u$ and $m_d$.

Quark masses have been calculated on the lattice since the
mid-nineties. However early calculations were performed in the quenched
approximation, leading to unquantifiable systematics. Thus in the following,
we only review modern, unquenched calculations, which include the effects of
light sea quarks.

\begin{table}[!ht]
{\footnotesize{
\begin{tabular*}{\textwidth}[l]{l@{\extracolsep{\fill}}rllllllll}
Collaboration & Ref. & \hspace{0.15cm}\begin{rotate}{60}{publication status}\end{rotate}\hspace{-0.15cm} &
 \hspace{0.15cm}\begin{rotate}{60}{chiral extrapolation}\end{rotate}\hspace{-0.15cm} &
 \hspace{0.15cm}\begin{rotate}{60}{continuum  extrapolation}\end{rotate}\hspace{-0.15cm}  &
 \hspace{0.15cm}\begin{rotate}{60}{finite volume}\end{rotate}\hspace{-0.15cm}  &  
 \hspace{0.15cm}\begin{rotate}{60}{renormalization}\end{rotate}\hspace{-0.15cm} &  
 \hspace{0.15cm}\begin{rotate}{60}{running}\end{rotate}\hspace{-0.15cm}  & 
\rule{0.6cm}{0cm}$m_{ud} $ & \rule{0.6cm}{0cm}$m_s $ \\
&&&&&&&&& \\[-0.1cm]
\hline
\hline
&&&&&&&&& \\[-0.1cm]
{ETM 10B}& \cite{Blossier:2010cr} & \gA & \soso & \good & \soso & \good & $\,a$
 & 3.6(1)(2) & 95(2)(6) \\ 

{JLQCD/TWQCD 08A}& \cite{Noaki:2008iy} & \gA& \soso&\bad&\bad&\good& $-$& 4.452(81)(38)$\binom{+0}{-227}$ &\rule{0.6cm}{0cm}--\\                  

{RBC 07$^\dagger$} & \cite{Blum:2007cy} & \gA & \bad & \bad & \good  & \good &
$-$       & $4.25(23)(26)$        & 119.5(5.6)(7.4)              \\

{ETM 07} & \cite{Blossier:2007vv} & \gA &  \soso & \bad & \soso & \good &$-$
& $3.85(12)(40)$        & $105(3)(9)$                  \\

\hspace{-0.2cm}{\begin{tabular}{l}QCDSF/\\
UKQCD 06\end{tabular}} & \cite{Gockeler:2006jt} & \gA &  \bad  & \good & \bad &
\good &$-$      & $4.08(23)(19)(23)$ &  $111(6)(4)(6)$ \\

{SPQcdR 05} & \cite{Becirevic:2005ta} & \gA & \bad & \soso & \soso & \good &
$-$& $4.3(4)(^{+1.1}_{-0.0})$ & $101(8)(^{+25}_{-0})$        \\

{ALPHA 05} & \cite{DellaMorte:2005kg} & \gA &  \bad & \soso & \good  & \good &
$\,b$  &                      & 97(4)(18)$^\S$           \\

\hspace{-0.2cm}{\begin{tabular}{l}QCDSF/\\
UKQCD 04\end{tabular}} & \cite{Gockeler:2004rp} & \gA &  \bad  & \good & \bad &
\good & $-$       & $4.7(2)(3)$ & $119(5)(8)$    \\

{JLQCD 02} & \cite{Aoki:2002uc} & \gA &  \bad  & \bad & \soso & \bad & $-$   
& $3.223(^{+46}_{-69})$ & $84.5(^{+12.0}_{-1.7})$        \\

{CP-PACS 01} & \cite{AliKhan:2001tx} & \gA & \bad & \bad & \good & \bad &$-$ &
$3.45(10)(^{+11}_{-18})$ & $89(2)(^{+2}_{-6})^\star$    \\
&&&&&&&&& \\[-0.1cm] 
\hline
\end{tabular*}
\begin{tabular*}{\textwidth}[l]{l@{\extracolsep{\fill}}lllllllll}
\multicolumn{10}{l}{\vbox{\begin{flushleft} 
$^\dagger$ The calculation includes quenched e.m. effects.\\
$^\S$ The data used to obtain the bare value of $m_s$ are from RBC/UKQCD 04
 \cite{Gockeler:2004rp}\\
 $^\star$ This value of $m_s$ was obtained
 using the kaon mass as input. If the $\phi$ meson mass is used instead, the\\\hspace{0.3cm}authors find 
 $m_s =90^{+5}_{-11}.$\\\rule{0cm}{0.3cm}\hspace{-0.1cm}
$a$ The masses are renormalized non-perturbatively at scales $1/a\sim 2\div3\,\gev$ in the $N_f=2$ RI/MOM \\\hspace{0.3cm}scheme.  In this
scheme, non-perturbative and N$^3$LO running for the quark masses
are shown to agree\\\hspace{0.3cm}from 4~GeV down 2 GeV to better than 3\%
\cite{Constantinou:2010gr}.\\\rule{0cm}{0.3cm}\hspace{-0.1cm}
$b$ The masses are renormalized and run non-perturbatively up to
a scale of $100\,\gev$ in the $N_f=2$ SF\\\hspace{0.3cm}scheme. In this
scheme, non-perturbative and NLO running for the quark masses are
shown to agree\\\hspace{0.3cm}well from 100 GeV all the way down to 2
GeV \cite{DellaMorte:2005kg}.
\end{flushleft}}}
\end{tabular*}
}}

\vspace{-0.3cm}
\caption{\label{tab:masses2} $\Nf=2$ lattice results for the masses $m_{ud}$ and $m_s$,
  together with the colour coding of the calculation used to obtain
  these. If information about non-perturbative running is available, this is indicated in
  the column "running", with details given at the bottom of the table.}
\end{table}

\begin{table}[!ht]
{\footnotesize{
\begin{tabular*}{\textwidth}[l]{l@{\extracolsep{\fill}}rllllllll}
Collaboration & Ref. & \hspace{0.15cm}\begin{rotate}{60}{publication status}\end{rotate}\hspace{-0.15cm} &
 \hspace{0.15cm}\begin{rotate}{60}{chiral extrapolation}\end{rotate}\hspace{-0.15cm} &
 \hspace{0.15cm}\begin{rotate}{60}{continuum  extrapolation}\end{rotate}\hspace{-0.15cm}  &
 \hspace{0.15cm}\begin{rotate}{60}{finite volume}\end{rotate}\hspace{-0.15cm}  &  
 \hspace{0.15cm}\begin{rotate}{60}{renormalization}\end{rotate}\hspace{-0.15cm} &  
 \hspace{0.15cm}\begin{rotate}{60}{running}\end{rotate}\hspace{-0.15cm}  & 
\rule{0.6cm}{0cm}$m_{ud} $ & \rule{0.6cm}{0cm}$m_s $ \\
&&&&&&&&& \\[-0.1cm]
\hline
\hline
&&&&&&&&& \\[-0.1cm]
{PACS-CS 10}& \protect{\cite{Aoki:2010wm}} & \oP & \good & \bad & \bad & \good & $\,a$
&  2.78(27) &  86.7(2.3) \\

{MILC 10A}& \cite{Bazavov:2010yq} & \rC & \soso  & \good & \good &
\soso  &$-$& 3.19(4)(5)(16)&\rule{0.6cm}{0cm}-- \\

{HPQCD~10}&  \cite{McNeile:2010ji} &\gA & \soso & \good & \good & \good 
&$-$& 3.39(6)$^\ast $ & 92.2(1.3) \\

{BMW 10A, 10B$^+$} & \cite{Durr:2010vn,Durr:2010aw} & \oP & \good & \good & \good & \good &
$\,b$ & 3.469(47)(48)& 95.5(1.1)(1.5)\\

{RBC/UKQCD 10A}& \cite{Aoki:2010dy} & \oP & \soso & \soso & \good &
\good & $\,c$  &  3.59(13)(14)(8) & 96.2(1.6)(0.2)(2.1)\\

{Blum~10$^\dagger$}&\cite{Blum:2010ym}& \oP& \soso & \bad & \soso & \good &
$-$ &3.44(12)(22)&97.6(2.9)(5.5)\\

{PACS-CS 09}& \cite{Aoki:2009ix}& \gA &\good   &\bad   & \bad & \good  &  $\,a$
 & 2.97(28)(3) &92.75(58)(95)\\

{HPQCD 09}&  \cite{Davies:2009ih}&\gA & \soso & \good & \good & \good 
& $-$& 3.40(7) & 92.4(1.5) \\

{MILC 09A} & \cite{Bazavov:2009fk} & \rC &  \soso & \good & \good & \soso &
$-$ & 3.25 (1)(7)(16)(0) & 89.0(0.2)(1.6)(4.5)(0.1)\\

{MILC 09} & \cite{Bazavov:2009bb} & \gA & \soso & \good & \good & \soso & $-$
& 3.2(0)(1)(2)(0) & 88(0)(3)(4)(0)\\

{PACS-CS 08} & \cite{Aoki:2008sm} &  \gA & \good & \bad & \bad  & \bad & $-$ &
2.527(47) & 72.72(78)\\

{RBC/UKQCD 08} & \cite{Allton:2008pn} & \gA & \soso & \bad & \good & \good &
$-$ &$3.72(16)(33)(18)$ & $107.3(4.4)(9.7)(4.9)$\\

\hspace{-0.2cm}{\begin{tabular}{l}CP-PACS/\\JLQCD 07\end{tabular}} 
& \cite{Ishikawa:2007nn}& \gA & \bad & \good & \good  & \bad & $-$ &
$3.55(19)(^{+56}_{-20})$ & $90.1(4.3)(^{+16.7}_{-4.3})$ \\

{HPQCD 05}
& 
\cite{Mason:2005bj}& \gA & \soso & \soso & \soso & \soso &$-$&
$3.2(0)(2)(2)(0)^\ddagger$ & $87(0)(4)(4)(0)^\ddagger$\\

\hspace{-0.2cm}{\begin{tabular}{l}MILC 04, HPQCD/\\MILC/UKQCD 04\end{tabular}} 
& \cite{Aubin:2004fs,Aubin:2004ck} & \gA & \soso & \soso & \soso & \bad & $-$ &
$2.8(0)(1)(3)(0)$ & $76(0)(3)(7)(0)$\\

\hline
\hline\\[-0.7cm]
\end{tabular*}
\begin{tabular*}{\textwidth}[l]{l@{\extracolsep{\fill}}lllllllll}
\multicolumn{10}{l}{\vbox{\begin{flushleft}
$^\ast$ Value obtained by combining the HPQCD 10 result for $m_s$ with the MILC 09 result for $m_s/m_{ud}$.
\\
\hspace{-0.04cm}$^+$\hspace{-0.05cm} The fermion action used is tree-level improved.\\
$^\dagger$ The calculation includes quenched e.m. effects.\\
$^\ddagger$ The bare numbers are those of MILC 04. The masses are simply rescaled, using the
ratio of the 2-loop to\\\hspace{0.3cm}1-loop renormalization factors.\\\rule{0cm}{0.3cm}\hspace{-0.1cm}
$a$ The masses are renormalized and run non-perturbatively up to
a scale of $40\,\gev$ in the $N_f=3$ SF\\ \hspace{0.3cm}scheme. In this
scheme, non-perturbative and NLO running for the quark masses are
shown to agree well\\ \hspace{0.3cm}from 40 GeV all the way down to 3 GeV\cite{Aoki:2010wm}.\\\rule{0cm}{0.3cm}\hspace{-0.1cm}
$b$ The masses are renormalized and run non-perturbatively up to
a scale of 4 GeV in the $N_f=3$ RI/MOM\\\hspace{0.3cm}scheme.  In this
scheme, non-perturbative and N$^3$LO running for the quark masses
are shown to agree\\\hspace{0.3cm}from 6~GeV down to 3~GeV to better than 1\%
\cite{Durr:2010aw}.\\\rule{0cm}{0.3cm}\hspace{-0.1cm}
$c$ The masses are renormalized non-perturbatively at a scale of 2~GeV 
in a couple of $N_f=3$ RI/SMOM\\ \hspace{0.3cm}schemes. A careful study of
perturbative matching uncertainties has been performed by comparing
results\\\hspace{0.3cm}in the two schemes in the region of 2~GeV to 3~GeV
\cite{Aoki:2010dy}.
\end{flushleft}}}
\end{tabular*}
}}
\vspace{-0.7cm}
\caption{$\Nf=2+1$ lattice results for the masses $m_{ud}$ and $m_s$,
  together with the colour coding of the calculation used to obtain
  these. If information about non-perturbative running is available, this is indicated in
  the column "running", with details given at the bottom of the table.}
\label{tab:masses3}
\end{table}
\begin{table}[!ht]
\vspace{1cm}
{\footnotesize{
\begin{tabular*}{\textwidth}[l]{l@{\extracolsep{\fill}}rlllll}
Collaboration & Ref. &  \hspace{0.15cm}\begin{rotate}{60}{publication status}\end{rotate}\hspace{-0.15cm}  &
 \hspace{0.15cm}\begin{rotate}{60}{chiral extrapolation}\end{rotate}\hspace{-0.15cm} &
 \hspace{0.15cm}\begin{rotate}{60}{continuum  extrapolation}\end{rotate}\hspace{-0.15cm}  &
 \hspace{0.15cm}\begin{rotate}{60}{finite volume}\end{rotate}\hspace{-0.15cm}  & \rule{0.1cm}{0cm} 
$m_s/m_{ud}$ \\
&&&&&& \\[-0.1cm]
\hline
\hline
&&&&&& \\[-0.1cm]
{BMW 10A, 10B$^+$}& \cite{Durr:2010vn,Durr:2010aw} & \oP & \good & \good & \good & 27.53(20)(8) \\

{RBC/UKQCD 10A}& \cite{Aoki:2010dy} & \oP & \soso & \soso & \good & 26.8(0.8)(1.1) \\

{Blum~10$^\dagger$}&\cite{Blum:2010ym}& \oP& \soso & \bad & \soso & 
28.31(0.29)(1.77)\\

{PACS-CS 09}  & \cite{Aoki:2009ix}   &  \gA &\good   &\bad   & \bad    &
31.2(2.7)  \\

{MILC 09A}      & \cite{Bazavov:2009fk}       & \rC & \soso & \good & \good &  
27.41(5)(22)(0)(4)  \\

{MILC 09}      & \cite{Bazavov:2009bb}       & \gA & \soso & \good & \good  &
27.2(1)(3)(0)(0)  \\

{PACS-CS 08}   & \cite{Aoki:2008sm}          & \gA & \good & \bad  & \bad    &
28.8(4)\\

{RBC/UKQCD 08} & \cite{Allton:2008pn}        & \gA & \soso & \bad  & \good   &
$28.8(0.4)(1.6)$ \\

\hspace{-0.2cm}{\begin{tabular}{l}MILC 04, HPQCD/\\MILC/UKQCD 04\end{tabular}} 
& \cite{Aubin:2004fs,Aubin:2004ck} & \gA & \soso & \soso & \soso    &
27.4(1)(4)(0)(1)  \\

&&&&&& \\[-0.1cm]
\hline
&&&&&& \\[-0.1cm]
{ETM 10B}& \cite{Blossier:2010cr} & \gA & \soso & \good  & \soso &   27.3(5)(7) \\

{RBC 07}$^\dagger$       & \cite{Blum:2007cy}          & \gA & \bad  & \bad  & \good   &
28.10(38) \\

{ETM 07}       & \cite{Blossier:2007vv}      & \gA & \soso & \bad  & \soso  & 
$27.3(0.3)(1.2)$ \\

{QCDSF/UKQCD 06}
                 & \cite{Gockeler:2006jt}   & \gA &  \bad & \good & \bad   &
27.2(3.2)\\
&&&&&& \\[-0.1cm]
\hline
\hline
\end{tabular*}
\begin{tabular*}{\textwidth}[l]{l@{\extracolsep{\fill}}lllllll}
\multicolumn{7}{l}{\vbox{\begin{flushleft}
$^+$ The fermion action used is tree-level improved.\\
$^\dagger$
The calculation includes quenched e.m. effects.\\
\end{flushleft}}}
\end{tabular*}
}}
\vspace{-0.7cm}
\caption{Lattice results for $m_s/m_{ud}$, together with the
 colour coding of the calculation used to obtain them. The upper half of the
table corresponds
 to $\Nf=2+1$ calculations, the lower half to $\Nf=2$ results.}
\label{tab:ratio_msmud}
\end{table}
Tables~\ref{tab:masses2} and \ref{tab:masses3} list the results of $\Nf=2$ and $\Nf=2+1$
lattice calculations of $m_s$ and $m_{ud}$. These results are given in the
$\msbar$ scheme at $2\,\gev$, which is standard nowadays. The tables also show
the colour coding of the calculations leading to these results. The corresponding
results for $m_s/m_{ud}$ are given in Table~\ref{tab:ratio_msmud}.
As indicated in the Introduction, we treat $\Nf=2$ and $\Nf=2+1$
calculations separately. The latter include the effects of a strange sea
quark, but the former do not. 

\subsubsection{$\Nf=2$ lattice calculations}

We begin with $\Nf=2$ calculations. A quick inspection of
Table~\ref{tab:masses2} indicates that only the
most recent calculation, ETM 10B~\cite{Blossier:2010cr}, 
controls all systematic effects. Only ETM
10B~\cite{Blossier:2010cr} and ETM 07~\cite{Blossier:2007vv} really enter the 
chiral regime, with pion masses down to about
300~MeV. All the others have significantly more massive pions, the lightest
being about 430~MeV, in the calculation by CP-PACS 01~\cite{AliKhan:2001tx}.
Moreover, the latter calculation is performed on very coarse lattices, with
lattice spacings $a\ge 0.11\,\fm$ and only one loop perturbation theory is
used to renormalize the results.

One calculation of $m_s$ which performs quite well in terms of colour
coding is that of ALPHA 05~\cite{DellaMorte:2005kg}. This is due to the
fact that both non-perturbative running and non-perturbative
renormalization are performed in a controlled fashion, using
Schr\"odinger functional methods. However, the lightest sea pion in
the calculation is 600~MeV, significantly above our thresholds for a
controlled chiral extrapolation. In that sense, this calculation
should be viewed more as a means of giving a first estimate of
possible sea quark corrections to the equally precise quenched
calculation ALPHA 99~\cite{Garden:1999fg}, than a full fledged dynamical
fermion calculation.

The conclusion of our analysis of $\Nf=2$ calculations is that the results
of ETM 10B~\cite{Blossier:2010cr} (which updates and extends ETM
07~\cite{Blossier:2007vv}), is the only one to date which satisfies our
selection criteria. Thus we do not offer an average of
$\Nf=2$ results for the strange and isospin averaged quark masses and their
ratios, but simply quote the results of  ETM 10B~\cite{Blossier:2010cr}:
\be
\label{eq:quark masses Nf=2} \Nf=2 :\hspace{0.4cm}
m_s= 95(2)(6) \,\mbox{MeV}\co \hspace{0.4cm}
m_{ud}= 3.6(1)(2)  \,\mbox{MeV}  \co\hspace{0.4cm}
\frac{m_s}{m_{ud}} = 27.3(5)(7)\fs\ee
The combined statistical and systematic errors on these results are 7\%,
6\% and 3\% respectively. The errors are smaller in the ratio because many
systematics cancel, most notably those associated with renormalization and the
setting of the scale. Moreover, because of statistical correlations between
$m_s$ and $m_{ud}$, those errors are also reduced in the ratio.

\subsubsection{$\Nf=2+1$ lattice calculations}\label{sec:Nf=2+1}

We turn now to $\Nf=2+1$ calculations. These and the corresponding results
are summarized in Table~\ref{tab:masses3} and
\ref{tab:ratio_msmud}. Somewhat paradoxically, these calculations are more
mature than those with $\Nf=2$. This is thanks, in large part, to the head
start and sustained effort of MILC, which has been performing $\Nf=2+1$
rooted staggered fermion calculations for the past ten or so years. They
have covered an impressive range of parameter space, with lattice spacings
which, today, go down to 0.045~fm and valence pion masses down to
approximately 180~MeV~\cite{Bazavov:2009fk}.  The most recent updates, MILC
10A~\cite{Bazavov:2010yq} and MILC 09A~\cite{Bazavov:2009fk}, include
significantly more data and use two loop renormalization.
Since these data sets subsume those of their previous calculations, these
latest results are the only ones that must be kept in any world average.

We recall that valence and sea quarks are treated differently in
the rooted staggered approach, as discussed in section
\ref{sec:color-code}, and that the lightest effective sea-pion mass
of all of their ensembles is about 260~MeV rather than
180~MeV.  Regarding the specific subject of light quark masses, a weak point of
present day staggered fermion calculations is the use of perturbative
renormalization.  As already noted, it is preferable, a priori, to
renormalize non-perturbatively, to avoid introducing uncontrolled
systematic errors from neglected, higher order contributions. That
this is a problem, in practice, for the staggered calculations can be
seen from the following observation. In
MILC 04 \cite{Aubin:2004fs} and HPQCD/MILC/UKQCD 
~04~\cite{Aubin:2004ck}, only one loop perturbation theory
was used. $O(\alpha_s^2)$ corrections were included in the
HPQCD~05~\cite{Mason:2005bj} update. The only difference between this
update and the earlier calculation is the inclusion of these
corrections: the bare lattice results are the same in both. Yet, the
$O(\alpha_s^2)$ correction increased the renormalized value of $m_s$
by 11~MeV, i.e.~14\%. In light of this, one may question whether the 
perturbative expansion is under control. Moreover, preliminary results
including non-perturbative renormalization suggest that the omitted terms
may contribute an additional increase of $\sim 20\%$ in $m_s$ 
\cite{Lytle:2009xm}. 

Of all the entries in Tables~\ref{tab:masses3} 
and \ref{tab:ratio_msmud}, Blum~10~\cite{Blum:2010ym} is the only one
which provides an improved determination of the electromagnetic
self-energies on the lattice and points the way for future promising
developments: still, at present, these data do not pass the
selection criteria formulated in section \ref{sec:averages} as the authors 
did not investigate all sources of systematic error. There are four other
calculations with $\Nf = 2 + 1$, however, that do and that we discuss now.

The most recent one is BMW 10A, 10B~\cite{Durr:2010vn,Durr:2010aw}
which, as shown by the color coding in Tables~\ref{tab:masses3}
and \ref{tab:ratio_msmud} has addressed all sources of systematic
effects. One of the most impressive aspects of this calculation is
that they have reached the physical up and down quark mass by {\it
  interpolation} and did not need any chiral extrapolation. Moreover,
their calculation was performed at five lattice spacings ranging from
0.054 to 0.116~fm, with full non-perturbative renormalization and
running and in volumes of up to (6~fm)$^3$ guaranteeing that all
sources of systematic error are controlled. However, according to the
rules set forth in section~\ref{sec:averages}, these results will have
to be published before they can be included in our averages.

The second calculation that we discuss is PACS-CS
10~\cite{Aoki:2010wm}, which updates 09~\cite{Aoki:2009ix}: these
calculations have been performed with $O(a)$-improved Wilson fermions
on lattices with $a\simeq 0.09\,\fm$. The important aspect of this
calculation is that it also probes pion masses down to $M_\pi\simeq
135\,\mev$, i.e.\ down to the physical mass point. This is achieved by
reweighting the simulations performed in PACS-CS 08~\cite{Aoki:2008sm}
at $M_\pi\simeq 160\,\mev$. If adequately controlled, this procedure
eliminates the need to extrapolate to the physical mass point and,
hence, the corresponding systematic error.  Moreover, renormalization
of quark masses is implemented non-perturbatively, through the
Schr\"odinger functional method~\cite{Luscher:1992an}. As it stands,
the main drawback of the calculation, which makes the inclusion of its
results in a world average of lattice results inappropriate at this
stage, is that for the lightest quark mass the volume is very small,
corresponding to $LM_\pi\simeq 2.0$, a value for which finite volume
effects will be difficult to control. Another problem is that the
calculation was performed at a single lattice spacing, forbidding a
continuum extrapolation. Furthermore, it is unclear at this point what
might be the systematic errors associated with the reweighting
procedure. The PACS-CS collaboration plans to perform physical point
simulations with larger and finer lattices.

The third calculation of particular interest is RBC/UKQCD
10A~\cite{Aoki:2010dy} (which updates RBC/UKQCD
08~\cite{Allton:2008pn}).  This calculation uses the theoretically
appealing domain wall fermions. It reaches down to 330~MeV pion masses
in the sea and to 242~MeV valence pions. Moreover, it makes use of
non-perturbative renormalization \`a la
Rome-Southampton~\cite{Martinelli:1994ty}. The latest update removes
the main drawback of the previous calculation by adding a second,
finer lattice spacing of about 0.09 fm to the first rather coarse
lattice spacing of 0.11~fm, and allows a first estimate of the
systematic error from the continuum extrapolation.

Finally we come to the fourth calculation which satisfies our selection
criteria, HPQCD~10~\cite{McNeile:2010ji} (which updates HPQCD
09~\cite{Davies:2009ih}). The strange quark mass is computed using a
precise determination of the charm quark mass, $m_c(m_c)=1.273(6)$
GeV~\cite{Allison:2008xk,McNeile:2010ji}, whose accuracy is better than
0.5\%, and a calculation of the quark mass ratio
$m_c/m_s=11.85(16)$~\cite{Davies:2009ih}, which achieves a precision
slightly above 1\%.  The determination of $m_s$ via the ratio $m_c/m_s$
circumvents the problem of running by calculating a scale-invariant ratio
non-perturbatively. The MILC 09 determination of the quark mass ratio
$m_s/m_{ud}$~\cite{Bazavov:2009bb} is also used in
HPQCD~10~\cite{McNeile:2010ji} to calculate $m_{ud}$.

\begin{figure}[t]
\psfrag{V 78}{\sffamily\tiny  \hspace{-0.03cm}\cite{Vainshtein:1978nn}}
\psfrag{N 06}{\sffamily\tiny  \hspace{-0.03cm}\cite{Narison:2005ny}}
\psfrag{J 06}{\sffamily\tiny  \hspace{-0.03cm}\cite{Jamin:2006tj}}
\psfrag{C 06}{\sffamily\tiny\hspace{-0.03cm}\cite{Chetyrkin:2005kn}}
\psfrag{D 09}{\sffamily\tiny \hspace{-0.03cm}\cite{Dominguez:2008jz}}
\psfrag{P 10}{\sffamily\tiny \hspace{-0.03cm}\cite{Nakamura:2010zzi}}
\psfrag{y}{$\dagger$}
\begin{center}
\psfig{file=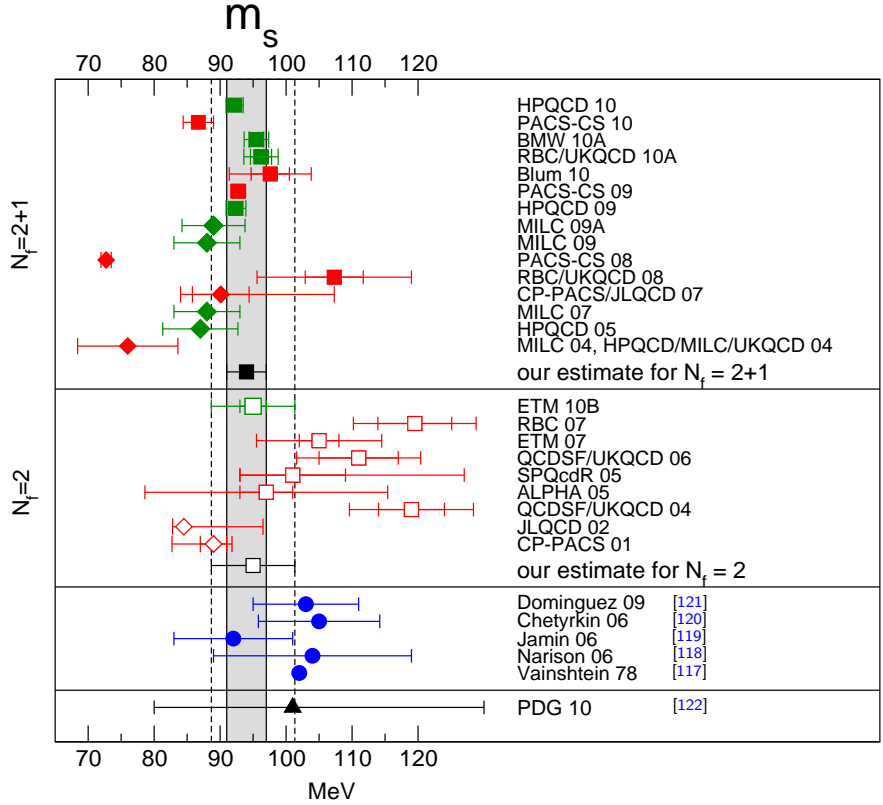,width=11.5cm}

\caption{\label{fig:ms}Mass of the strange quark ($\msbar$ scheme, running
 scale 2 GeV). The upper part shows the lattice results listed in Tables
 \ref{tab:masses2} and \ref{tab:masses3}. Full and empty symbols correspond to
 simulations with $\Nf=2+1$ and $\Nf=2$, respectively. Diamonds represent
 results based on perturbative renormalization, while squares indicate
 that, in the relation between the lattice-regularized and renormalized
 $\msbar$ masses, non-perturbative effects are accounted for. The
 vertical bands indicate our  estimates (\ref{eq:quark masses Nf=2}) and (\ref{eq:nf3msmud}). 
 The lower part shows recent determinations obtained from the evaluation of 
 sum rules, together with the earliest result based on this method, as
well as the most recent estimate of the Particle Data Group.}
\end{center}
\end{figure}

\begin{figure}[t]
\psfrag{M 01}{\sffamily\tiny  \hspace{-0.03cm}\cite{Maltman:2001nx}}
\psfrag{D 09}{\sffamily\tiny \hspace{-0.03cm}\cite{Dominguez:2008jz}}
\psfrag{N 06}{\sffamily\tiny \hspace{-0.03cm}\cite{Narison:2005ny}}
\psfrag{P 10}{\sffamily\tiny\hspace{-0.03cm}\cite{Nakamura:2010zzi}}

\begin{center}
\psfig{file=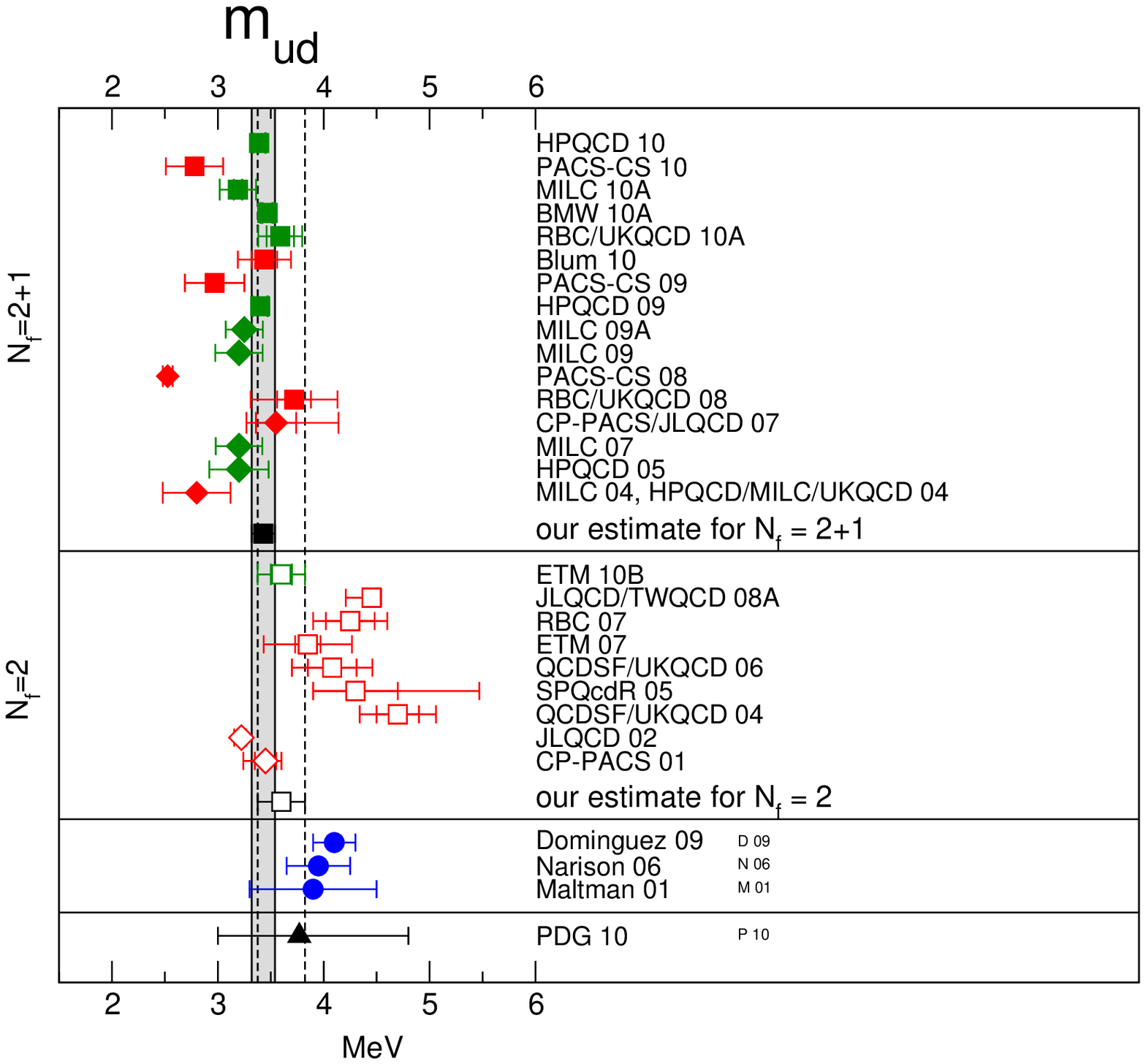,width=11.5cm}

\caption{\label{fig:mud} Mean mass of the two lightest quarks,
 $m_{ud}=\frac{1}{2}(m_u+m_d)$. The meaning of the various symbols is
 explained in the caption of Fig.\ \ref{fig:ms}.}
\end{center}

\end{figure}

The high precision quoted by HPQCD~10 on the strange quark mass relies
in large part on the precision reached in the determination of the
charm quark mass~\cite{Allison:2008xk,McNeile:2010ji}. This
calculation uses a novel approach based on the lattice determination
of moments of charm-quark pseudoscalar, vector and axial-vector
correlators. These moments are then combined with four-loop results
from continuum perturbation theory to obtain a determination of the
charm quark mass in the $\msbar$ scheme. In the preferred case, in
which pseudoscalar correlators are used for the analysis, there are no
renormalization factors required, since the corresponding axial-vector
current is partially conserved in the staggered lattice
formalism. Overall, the 0.5\% accuracy claimed in the calculation of
$m_c$ is certainly impressive, since it implies that all sources of
uncertainty, coming from the scale determination, discretization
effects (the bare charm mass in lattice units is as large as $am_c
\simeq 0.85$), perturbation theory as well as non-perturbative effects
in the estimate of moments, are all controlled at the per mille
level.

This discussion leaves us with three results to average for $m_s$,
those of MILC 09A~\cite{Bazavov:2009fk}, RBC/UKQCD
10A~\cite{Aoki:2010dy} and HPQCD~10~\cite{McNeile:2010ji}. Treating them as
independent measurements and simply adding the statistical and
systematic errors in quadrature, we obtain a good fit, with
$\chi^2=2.5$ for 3 data points and 1 free parameter. With the
prescription specified in section \ref{sec:averages}, the average
becomes $m_s=92.8(1.3)\,\mev$. When repeating the exercise for $m_{ud}$, we replace MILC 09A by the more recent analysis reported in  MILC 10A~\cite{Bazavov:2010yq} and again obtain a good fit: $\chi^2=2.3$ for 2 degrees of freedom and $m_{ud}=3.38(6)\,\mev$ for the average. The outcome of the averaging procedure
thus amounts to a determination of $m_s$ and $m_{ud}$ of 1.4\%. and 1.8\%, respectively. 

The results in BMW 10A, 10B~\cite{Durr:2010vn,Durr:2010aw}, which are not included in the averaging procedure on account of our criteria concerning the publication status, offer an independent, remarkably sharp determination of $m_{ud}$ and $m_s$. These data are perfectly consistent with the averages quoted above.

The heavy sea quarks affect the determination of the light quark masses only through contributions of order $1/m_c^2$, which moreover are suppressed by the Zweig-rule. We expect these contributions to be small, but do not know of a reliable quantitative evaluation. The problem originates in the fact that the relation between the parameters of QCD$_3$ and those of full QCD can currently be analyzed only in the framework of perturbation theory. The $\beta$- and $\gamma$-functions, which control the renormalization of the coupling constants and quark masses, respectively, are known to four loops \cite{vanRitbergen:1997va,Chetyrkin:1997sg,Chetyrkin:1999pq,Bethke:2009jm}. The precision achieved in this framework for the decoupling of the $t$- and $b$-quarks is excellent, but the $c$-quark is not heavy enough: at the percent level, the corrections of order $1/m_c^2$ cannot be neglected and the decoupling formulae of perturbation theory do not provide a reliable evaluation, because the scale $m_c(m_c)\simeq 1.27\,\gev$ is too low for these formulae to be taken at face value. Consequently, the accuracy to which it is possible to identify the running masses of the light quarks of full QCD in terms of those occurring in QCD$_3$ is limited. For this reason, it is preferable to characterize the masses $m_u$, $m_d$, $m_s$ in terms of QCD$_4$,  where the connection with full QCD is under good control. The role of the $c$-quarks in the determination of the light quark masses will soon be studied in detail -- some simulations with 2+1+1 dynamical quarks have already been carried out~\cite{Baron:2010bv}.  
 
A crude upper bound on the size of the effects due to the neglected heavy quarks can be established within the $\Nf=2+1$ simulations themselves, without invoking perturbation theory. In  \cite{Durr:2008zz} it is found that when the scale is set by $M_\Xi$, the result for $M_\Lambda$ agrees well with experiment within the 2.3\% accuracy of the calculation. Because of the very strong correlations between the statistical and systematic errors of these two masses, we expect the uncertainty in the difference $M_\Xi-M_\Lambda$ to also be of order 2\%. To leading order in the chiral expansion, this mass difference is proportional to $m_s-m_{ud}$. We conclude that the agreement of $\Nf=2+1$ calculations with experiment yields an upper bound on the sensitivity of $m_s$ to heavy sea quarks of order 2\%.

In order to stay on the conservative side in this rapidly developing field, our final estimates for $m_{ud}$ and $m_s$ come with sizable uncertainties:
\bea\label{eq:nf3msmud}
\Nf=2+1 :\hspace{1cm} m_{ud}\al=\al 3.43(11)\;\mev 
\co\hspace{1cm}m_s=94(3)\;\mev\fs \eea
As discussed in section \ref{sec:averages}, these numbers are not obtained from a mathematical prescription, but are chosen so as to cover all of the high quality data discussed in the present section (note also that the central values are slightly higher than those of the averages discussed above).

The estimates (\ref{eq:nf3msmud}) represent the conclusions we draw from the information gathered on the lattice until now. They are shown as vertical bands in Figures \ref{fig:ms} and \ref{fig:mud}, together with the $\Nf=2$ results (\ref{eq:quark masses Nf=2}) and a few estimates from other sources. The values quoted in the 2010 edition of the Review of
Particle Properties~\cite{Nakamura:2010zzi} are $m_{ud}=3.9(9)$, $m_s = 101(^{+29}_{-21})$ MeV.  The errors attached to our estimates are smaller by an order of magnitude. In our opinion, the remarkable recent progress achieved with the simulation of light dynamical quarks and the fact that all of the data that pass our quality criteria are consistent with one another fully
justify this reduction of the uncertainties. 

The lower half of Figure~\ref{fig:ms} shows that the sum rule results
for $m_s$ agree with the $\Nf=2+1$ data, but Figure~\ref{fig:mud}
indicates that in the case of $m_{ud}$, the sum rule estimates are on
the high side. For the most recent sum rule
evaluation~\cite{Dominguez:2008jz}, the origin of this apparent
discrepancy is readily identified. In that work, the sum rules are
used to estimate the value of $m_s+m_{ud}$. In order to disentangle
the two terms, the authors take the ratio $m_s/m_{ud}$ from the early
literature on {\Ch}PT. If the numerical value of this ratio is updated
to the range given in (\ref{eq:msovmud}), the value of $m_s$
practically stays put, but the outcome for $m_{ud}$ is lowered
substantially, so that the difference shrinks to about 1 $\sigma$. In
other words, there is no discrepancy to speak of between the lattice
and sum rule results for the masses of the light quarks -- the size of
$m_s$ and $m_{ud}$ is well understood.
 
In the ratio $m_s/m_{ud}$, one of the sources of systematic error -- the
uncertainties in the renormalization factors -- drops out. Also, we can
compare the lattice results with the leading order formula of {\Ch}PT, 
\be\label{eq:LO1}\frac{m_s}{m_{ud}}\Lo\frac{\hat{M}_{K^+}^2+
\hat{M}_{K^0}^2-\hat{M}_{\pi^+}^2}{\hat{M}_{\pi^+}^2}\co\ee
which relates the quantity $m_s/m_{ud}$ to a ratio of meson masses in QCD.
Expressing these in terms of the physical masses and the four coefficients
introduced in (\ref{eq:epsilon1})-(\ref{eq:epsilon3}), linearizing the
result with respect to the corrections and inserting the observed mass
values, we obtain 
\be\label{eq:LO1 num} \frac{m_s}{m_{ud}} \Lo 25.9 - 0.1\,
\epsilonD + 1.9\, \epsilon_{\pi^0} - 0.1\, \epsilon_{K^0} -1.8
\,\epsilon_m\fs\ee 
If the coefficients $\epsilonD$, $\epsilon_{\pi^0}$,
$\epsilon_{K^0}$ and $\epsilon_m$ are set equal to zero, the right hand
side reduces to the value $m_s/m_{ud}=25.9$ that follows from Weinberg's
leading order formulae for $m_u/m_d$ and $m_s/m_d$ \cite{Weinberg:1977hb},
in accordance with the fact that these do account for the e.m.\ interaction
at leading order, but neglect the mass difference between the charged and
neutral pions in QCD.  Inserting the estimates (\ref{eq:epsilon num}), the
LO prediction becomes 
\be\label{eq:LO ms/mud}\frac{m_s}{m_{ud}}\Lo 25.9(1)\fs\ee 
The quoted uncertainty does not include an estimate for the
higher order contributions, but only accounts for the error bars in the
coefficients, which is dominated by the one in the estimate given for
$\epsilon_{\pi^0}$. The result shows that, at the accuracy reached in
lattice determinations of the ratio $m_s/m_{ud}$, the uncertainties due to
the electromagnetic corrections is smaller than the systematic errors
from other sources.

\begin{figure}[t]
\psfrag{W 77}{\sffamily\tiny  \hspace{-0.03cm}\cite{Weinberg:1977hb}} 
\psfrag{N 06}{\sffamily\tiny  \hspace{-0.03cm}\cite{Narison:2005ny}}
\psfrag{L 96}{\sffamily\tiny  \hspace{-0.03cm}\cite{Leutwyler:1996qg}}
\psfrag{K 98}{\sffamily\tiny  \hspace{-0.03cm}\cite{Kaiser}}
\psfrag{O 07}{\sffamily\tiny  \hspace{-0.03cm}\cite{Oller:2006xb}}
\psfrag{P 10}{\sffamily\tiny \hspace{-0.03cm}\cite{Nakamura:2010zzi}}

\begin{center}
\psfig{file=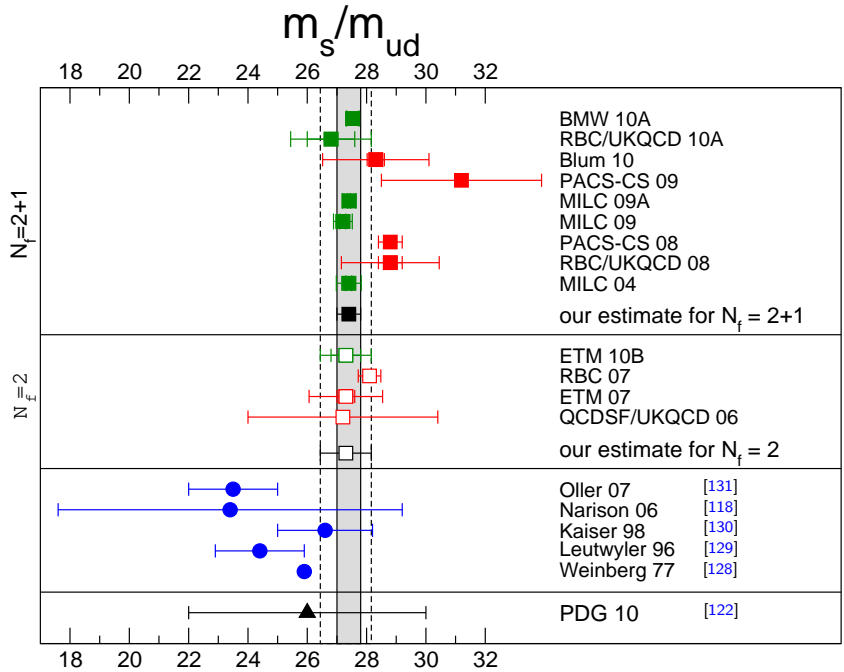,width=11cm}
\caption{\label{fig:msovmud}Results for the ratio $m_s/m_{ud}$, in
  which the renormalization factors cancel. The upper part of the
  figure shows the lattice results listed in Table
  \ref{tab:ratio_msmud} (full and empty symbols correspond to
  $\Nf=2+1$ and $\Nf=2$). The vertical bands indicate our estimates
  (\ref{eq:quark masses Nf=2}) and (\ref{eq:msovmud}). The lower part
  shows results obtained on the basis of {\Ch}PT or from QCD sum
  rules, as well as the most recent estimate of the Particle Data
  Group.}
\end{center}
\end{figure}

The lattice results in Table \ref{tab:ratio_msmud} indicate that the
corrections generated by the non-leading terms of the chiral perturbation
series are remarkably small, in the range 5--11\%.  Despite the fact that
the SU(3) flavour symmetry breaking effects in the Nambu-Goldstone boson
masses are very large ($M_K^2\simeq 13\, M_\pi^2$), the mass spectrum of
the pseudoscalar octet obeys the SU(3)$\times$SU(3) formula (\ref{eq:LO1})
very well. 

Our average for $m_s/m_{ud}$ is based on the results of MILC 09A and RBC/UK\-QCD 10A  -- the value quoted by HPQCD~10 does not represent independent information as it relies on the result for $m_s/m_{ud}$ obtained by the MILC collaboration. Adding the statistical and systematic errors in quadrature, we obtain a good fit with $m_s/m_{ud}=27.39(23)$ and $\chi^2=0.2$ for 2 data points and one free parameter. The fit is dominated by MILC 09A. Since the errors associated with renormalization drop out in the ratio, the uncertainties are even smaller than in the case of the quark masses themselves: the above number for $m_s/m_{ud}$ amounts to an accuracy of 0.8\%. Again, the results of the BMW collaboration are not included in the fit because they are
not yet published. The result reported in BMW 10A, 10B, $m_s/m_{ud}=27.53(22)$,  is perfectly consistent with the above average and reaches comparable accuracy.

At this level of precision, the uncertainties in the electromagnetic and strong isospin breaking corrections are not completely negligible. The error estimate in the LO result  (\ref{eq:LO ms/mud}) indicates the expected order of magnitude. The uncertainties in $m_s$ and $m_{ud}$ associated with the heavy sea quarks cancel at least partly. In view of this, we are convinced that our  final estimate,  
\be\label{eq:msovmud} \mbox{$N_f=2+1$ :}
\hspace{1cm}\frac{m_s}{m_{ud}}=27.4(4)\co \ee
is on the conservative side -- it corresponds to an accuracy of 1.5\%.

The lattice results show that the LO prediction of {\Ch}PT in (\ref{eq:LO ms/mud}) receives only small corrections from higher orders of the chiral expansion: according to (\ref{eq:msovmud}), these generate a shift of $5.8\pm 1.5 \%$. Our estimate does therefore not represent a very sharp determination of the higher order contributions. Note however, that in comparison with the value $m_s/m_{ud}=26(4)$ quoted by the Particle Data Group \cite{Nakamura:2010zzi}, the uncertainty is reduced by a factor of 10. 

The ratio $m_s/m_{ud}$ can also be extracted from the masses of the neutral
Nambu-Goldstone bosons: neglecting effects of order $(m_u-m_d)^2$ also
here, the leading order formula reads 
$m_s/m_{ud}\Lo\frac{3}{2}\hat{M}_\eta^2/\hat{M}_\pi^2-\frac{1}{2}$.
Numerically, this gives $m_s/m_{ud}\Lo 24.2$. The relation has the advantage
that the e.m.\ corrections are expected to be much smaller here, but it is
more difficult to calculate the $\eta$-mass on the lattice. The comparison with
(\ref{eq:msovmud}) shows that, in this case, the contributions of NLO are
somewhat larger: $13\pm 2$\%.

\subsection{Lattice determination of $m_u$ and $m_d$}\label{subsec:mumd}

The determination of $m_u$ and $m_d$ separately requires additional
input.  MILC 09A \cite{Bazavov:2009fk} uses the mass difference between 
$K^0$ and $K^+$, from which they subtract electromagnetic effects, using 
Dashen's theorem with corrections, as discussed in section 
\ref{subsec:electromagnetic interaction}.  The up and down sea quarks remain degenerate 
in their calculation, fixed to the value of $m_{ud}$ obtained from $M_{\pi^0}$.
Instead of subtracting electromagnetic effects using phenomenology,
RBC~07~\cite{Blum:2007cy} and Blum~10~\cite{Blum:2010ym} actually 
include a quenched electromagnetic field in their calculation. This means 
that their results include corrections to Dashen's theorem, albeit only in the 
presence of quenched electromagnetism. Since the up and down quarks in the  
sea are treated as degenerate, very small isospin corrections are neglected, 
as in MILC's calculation. The results are summarized in Table~\ref{tab:mu_md_grading}.

In order to discuss these results, we consider the LO formula
\be\label{eq:LO2}\frac{m_u}{m_d}\Lo\frac{\hat{M}_{K^+}^2-\hat{M}_{K^0}^2+\hat{M}_{\pi^+}^2}
{\hat{M}_{K^0}^2-\hat{M}_{K^+}^2+\hat{M}_{\pi^+}^2} \fs\ee
Using equations (\ref{eq:epsilon1})--(\ref{eq:epsilon3}) to express the
meson masses in QCD in terms of the physical ones and linearizing in the
corrections, this relation takes the form \be\label{eq:LO2
  num}\frac{m_u}{m_d}\Lo 0.558 - 0.084\, \epsilonD - 0.02\,
\epsilon_{\pi^0} + 0.11\, \epsilon_m \fs\ee Inserting the estimates
(\ref{eq:epsilon num}) and adding errors in quadrature, the LO prediction
becomes \be\label{eq:mu/md LO}\frac{m_u}{m_d}\Lo 0.50(4)\fs\ee Again, the
quoted error exclusively accounts for the errors attached to the estimates
(12) Ð- contributions of non-leading order are ignored. The uncertainty in
the leading order prediction is dominated by the one in the coefficient
$\epsilonD$, which specifies the difference between the mass splittings
generated by the e.m. interaction in the kaon and pion multiplets.
\begin{table}[t]
\vspace{1cm}
{\footnotesize{
\begin{tabular*}{\textwidth}[l]{l@{\extracolsep{\fill}}rlllllllll}
Collaboration \al  Ref. \al \hspace{0.15cm}\begin{rotate}{60}{publication status}\end{rotate}\hspace{-0.15cm}  \al 
\hspace{0.15cm}\begin{rotate}{60}{chiral extrapolation}\end{rotate}\hspace{-0.15cm} \al 
\hspace{0.15cm}\begin{rotate}{60}{continuum  extrapolation}\end{rotate}\hspace{-0.15cm}  \al 
\hspace{0.15cm}\begin{rotate}{60}{finite volume}\end{rotate}\hspace{-0.15cm}  \al   
\hspace{0.15cm}\begin{rotate}{60}{renormalization}\end{rotate}\hspace{-0.15cm} \al   
\hspace{0.15cm}\begin{rotate}{60}{running}\end{rotate}\hspace{-0.15cm}  \al  
\rule{0.6cm}{0cm}$m_u$\al 
\rule{0.6cm}{0cm}$m_d$ \al \rule{0.3cm}{0cm} $m_u/m_d$\\
\al \al \al \al \al \al \al \al \al \al  \\[-0.1cm]
\hline
\hline
\al \al \al \al \al \al \al \al \al \al  \\[-0.1cm]
{HPQCD~10$^\ddagger$}\al \cite{McNeile:2010ji} \al \oP \al \soso \al \good \al \good \al \good \al
$-$ \al 2.01(14) \al 4.77(15) \al  \\

{BMW 10A, 10B$^+$}\al \cite{Durr:2010vn,Durr:2010aw} \al \oP \al \good \al \good \al \good \al \good \al
$\,a$ \al 2.15(03)(10) \al 4.79(07)(12) \al 0.448(06)(29) \\

{Blum~10$^\dagger$}\al\cite{Blum:2010ym}\al\oP\al \soso \al \bad \al \soso \al \good \al $-$ \al 2.24(10)(34)\al 4.65(15)(32)\al 0.4818(96)(860)\\

{MILC 09A} \al  \cite{Bazavov:2009fk} \al  \rC \al  \soso \al \good \al \good \al \soso \al $-$
\al 1.96(0)(6)(10)(12)
\al  4.53(1)(8)(23)(12)  \al   0.432(1)(9)(0)(39) \\

{MILC 09} \al  \cite{Bazavov:2009bb} \al  \gA \al  \soso \al  \good \al  \good \al  \soso \al 
$-$\al  1.9(0)(1)(1)(1)
\al  4.6(0)(2)(2)(1) \al  0.42(0)(1)(0)(4) \\

\hspace{-0.2cm}{\begin{tabular}{l}MILC 04, HPQCD/\rule{0.1cm}{0cm}\\MILC/UKQCD 04\end{tabular}} \al  \cite{Aubin:2004fs,Aubin:2004ck} \al  \gA \al  \soso \al  \soso \al  \soso \al 
\bad \al$-$\al  1.7(0)(1)(2)(2)
\al  3.9(0)(1)(4)(2)  \al  0.43(0)(1)(0)(8) \\

\al \al \al \al \al \al \al \al \al \al  \\[-0.1cm]
\hline
\al \al \al \al \al \al \al \al \al \al  \\[-0.1cm]
{RBC 07$^\dagger$} \al  \cite{Blum:2007cy} \al  \gA \al  \bad \al  \bad \al  \good  \al  \good \al  $-$
\al  3.02(27)(19) \al  5.49(20)(34)  \al  0.550(31)\\

\al \al \al \al \al \al \al \al \al \al  \\[-0.1cm]
\hline
\hline\\[-0.7cm]
\end{tabular*}
\begin{tabular*}{\textwidth}[l]{l@{\extracolsep{\fill}}lllllll}
\multicolumn{7}{l}{\vbox{\begin{flushleft}
$^\ddagger$ Values obtained by combining the HPQCD 10 result for $m_s$ with the MILC 09 results for $m_s/m_{ud}$ and\\\hspace{0.3cm}$m_u/m_d$.\\
\hspace{-0.04cm}$^+$\hspace{-0.05cm} The fermion action used is tree-level improved.\\
$^\dagger$
The calculation includes quenched e.m. effects.\\\rule{0cm}{0.3cm}\hspace{-0.1cm}
$a$ The masses are renormalized and run non-perturbatively up to
a scale of 4 GeV in the $N_f=3$ RI/MOM\\\hspace{0.3cm}scheme.  In this
scheme, non-perturbative and N$^3$LO running for the quark masses
are shown to agree\\\hspace{0.3cm}from 6~GeV down to 3~GeV to better than 1\%
\cite{Durr:2010aw}. 
\end{flushleft}}}
\end{tabular*}
}}
\vspace{-0.7cm}
\caption{Lattice results for $m_u$, $m_d$ and $m_u/m_d$, together with the
  colour coding of the calculation used to obtain them.  The upper part
  of the table lists results obtained with $\Nf=2+1$, while the lowest row 
  represents a calculation with $\Nf=2$.}
\label{tab:mu_md_grading}
\end{table}
Since the input $\epsilonD=1.2(5)$ used in MILC 09A \cite{Bazavov:2009fk}
differs from the results $\epsilonD=0.13(4)$ and $\epsilonD=0.5(1)$
obtained by RBC 07 \cite{Blum:2007cy} and Blum~10~\cite{Blum:2010ym},
respectively, it does not come as a surprise that the values for $m_u/m_d$
are different. Evaluating the relation (\ref{eq:LO2 num}) for these values
of $\epsilonD$ and neglecting the self-energies of the neutral particles,
we obtain $m_u/m_d=0.46(4)$, 0.547(3) and 0.52(1) for MILC 09A, RBC 07 and
Blum 10, respectively. Within errors, these numbers agree with the values
for $m_u/m_d$ quoted in the last column of Table \ref{tab:mu_md_grading}.
This indicates that, as in the case of $m_s/m_{ud}$, the corrections to the
leading order prediction are small.

The uncertainties in the lattice results discussed so far for the
ratio $m_u/m_d$ are dominated by those in the e.m.\,contributions to
the meson masses. Given the exploratory nature of the RBC 07
calculation, we do not draw a conclusion concerning the values of
$m_u$ and $m_d$ obtained with $\Nf=2$.  As discussed in section
\ref{sec:Nf=2+1}, the results of Blum 10 do not pass our selection
criteria either. We therefore resort to the phenomenological
estimates of the electromagnetic self-energies discussed in section
\ref{subsec:electromagnetic interaction} and use the MILC data for
$\Nf=2+1$ to assess the size of the corrections from higher orders of
the chiral expansion. With this input, we can calculate the mass ratio
of the two lightest quarks, with the result $m_u/m_d=0.47(4)$. The two
individual masses can then be worked out from the estimate
(\ref{eq:nf3msmud}) for their mean. The result reads
\be
\label{eq:mumd} \hspace{0cm}\Nf = 2+1:\hspace{0.2cm}m_u
=2.19(15)\,\mev\co\hspace{0.2cm} m_d = 4.67(20) \,\mev\co\hspace{0.2cm}
\frac{m_u}{m_d} = 0.47(4)\fs\ee 
These estimates have uncertainties of order 7\%, 4\% and 9\%,
respectively.  

In contrast to the situation with the ratio $m_s/m_{ud}$, the error in
the result for $m_u/m_d$ is totally dominated by the uncertainties in
the input used for the electromagnetic corrections. In fact, the
comparison of equations (\ref{eq:mu/md LO}) and (\ref{eq:mumd})
indicates that more than half of the difference between the prediction
$m_u/m_d=0.558$ obtained from Weinberg's mass formulae
\cite{Weinberg:1977hb} and the result for $m_u/m_d$ obtained on the
lattice stems from electromagnetism -- the higher orders in the chiral
perturbation series only generate a small correction.

To determine $m_u/m_d$, BMW 10A, 10B~\cite{Durr:2010vn,Durr:2010aw}
follow a slightly different strategy. They obtain this ratio from
their result for $m_s/m_{ud}$ combined with a phenomenological
determination of the isospin breaking quark mass ratio $Q=22.3(8)$
from $\eta\to3\pi$ decays~\cite{Leutwyler:2009jg}. In fact, as
discussed in section \ref{sec:RandQ}, the central value of the e.m.~parameter
$\epsilon$ in (\ref{eq:epsilon num}) is taken from the same source. Their results for $m_u/m_d$, $m_u$ and $m_d$ are thus correlated
with our estimates in (\ref{eq:mumd}) and are compatible with them, but are
more precise.  

In view of the fact that a {\it massless up quark} would solve the
strong CP-problem, many authors have considered this an attractive
possibility, but the results presented above exclude this possibility:
the value of $m_u$ in (\ref{eq:mumd}) differs from zero by 15 standard deviations. We conclude that nature solves the strong CP-problem differently. This conclusion relies on lattice calculations of kaon masses and on
the phenomenological estimates of the e.m.~self-energies discussed in
section \ref{subsec:electromagnetic interaction}. The uncertainties
therein currently represent the limiting factor in determinations of
$m_u$ and $m_d$. As demonstrated in
\cite{Duncan:1996xy,Blum:2007cy,Blum:2010ym}, lattice methods can be
used to calculate the e.m.~self-energies. Further progress on the
determination of the light quark masses hinges on an improved
understanding of the e.m.~effects.

\subsection{Estimates for $R$ and $Q$}\label{sec:RandQ}

The quark mass ratios
\be\label{eq:Qm}
R\equiv \frac{m_s-m_{ud}}{m_d-m_u}\hspace{0.5cm} \mbox{and}\hspace{0.5cm}Q^2\equiv\frac{m_s^2-m_{ud}^2}{m_d^2-m_u^2}
\ee
compare SU(3) breaking  with isospin breaking. The quantity $Q$ is of
particular interest because of a low energy theorem \cite{Gasser:1984pr},
which relates it to a ratio of meson masses,  
\begin{equation}\label{eq:QM}
 Q^2_M\equiv \frac{\hat{M}_K^2}{\hat{M}_\pi^2}\cdot\frac{\hat{M}_K^2-\hat{M}_\pi^2}{\hat{M}_{K^0}^2-
   \hat{M}_{K^+}^2}\co\hspace{1cm}\hat{M}^2_\pi\equiv\mbox{$\frac{1}{2}$}( \hat{M}^2_{\pi^+}+ \hat{M}^2_{\pi^0})
 \co\hspace{0.5cm}\hat{M}^2_K\equiv\mbox{$\frac{1}{2}$}( \hat{M}^2_{K^+}+ \hat{M}^2_{K^0})\fs\end{equation}
Chiral symmetry implies that the expansion of $Q_M^2$ in powers of the
quark masses (i) starts with $Q^2$ and (ii) does not receive any
contributions at NLO:
\be\label{eq:LET Q}Q_M\NLo Q \fs\ee

Inserting the estimates for the mass ratios $m_s/m_{ud}$, and $m_u/m_d$
given in equations (\ref{eq:msovmud}) and (\ref{eq:mumd}), respectively, we
obtain
\be\label{eq:RQres} R=36.6(3.8)\co\hspace{2cm}Q=22.8(1.2)\fs\ee 
As is the case with the ratio $m_u/m_d$, the errors are dominated by the
uncertainties in the electromagnetic corrections. To investigate the
sensitivity to the latter, we use {\Ch}PT to express the quark mass
ratios in terms of the pion and kaon masses in QCD and then again use
equations (\ref{eq:epsilon1})--(\ref{eq:epsilon3}) to relate the QCD
masses to the physical ones. Linearizing in the corrections, this
leads to \bea\label{eq:R epsilon}R\al \Lo\al 43.9 - 10.8\, \epsilonD +
0.2\, \epsilon_{\pi^0} - 0.2\, \epsilon_{K^0}- 10.6\, \epsilon_m\co\\
\label{eq:Q epsilon} Q\al\NLo\al 24.3 - 3.0\, \epsilonD + 0.9\,
\epsilon_{\pi^0} - 0.1\, \epsilon_{K^0} + 2.1 \,\epsilon_m \fs\nonumber \eea While
the first relation only holds to LO of the chiral perturbation series, the
second remains valid at NLO, on account of the low energy theorem mentioned
above. The first terms on the right hand side represent the values of $R$
and $Q$ obtained with the Weinberg leading order formulae for the quark
mass ratios \cite{Weinberg:1977hb}. Inserting the estimates
(\ref{eq:epsilon num}), we find that the e.m.~corrections lower the
Weinberg values to $R\Lo 36.7(7.6)$ and $Q\NLo 22.3(2.1)$, respectively.
The comparison with the full results quoted above shows that there is
little room for the higher order terms in the chiral expansion.

As mentioned in section \ref{subsec:electromagnetic interaction}, there is
a phenomenological determination of $Q$ based on the decay $\eta\rightarrow
3\pi$ \cite{Kambor:1995yc,Anisovich:1996tx}. The key point is that the
transition $\eta\rightarrow 3\pi$ violates isospin conservation. The
dominating contribution to the transition amplitude stems from the mass
difference $m_u-m_d$. At NLO of {\Ch}PT, the QCD part of the amplitude can
be expressed in a parameter free manner in terms of $Q$.  It is well-known
that the electromagnetic contributions to the transition amplitude are
suppressed (a thorough recent analysis is given in \cite{Ditsche:2008cq}).
This implies that the result for $Q$ is less sensitive to the
electromagnetic uncertainties than the value obtained from the masses of
the Nambu-Goldstone bosons.  For a recent update of this determination and
for further references to the literature, we refer to
\cite{Colangelo:2009db}. Using dispersion theory to pin down the momentum
dependence of the amplitude, the observed decay rate implies $Q=22.3(8)$
(since the uncertainty quoted in \cite{Colangelo:2009db} does not include
an estimate for all sources of error, we have retained the error estimate
given in \cite{Leutwyler:1996qg}, which is twice as large). The formulae
for the corrections of NNLO are available also in this case
\cite{Bijnens:2007pr} -- the poor knowledge of the effective coupling
constants, particularly of those that are relevant for the dependence on
the quark masses, is currently the limiting factor encountered in the
application of these formulae.

As was to be expected, the central value of $Q$ obtained from
$\eta$-decay agrees exactly with the central value obtained from the low
energy theorem: we have used that theorem to estimate the coefficient
$\epsilonD$, which dominates the e.m.~corrections. Using the numbers for
$\epsilon_m$, $\epsilon_{\pi^0}$ and $\epsilon_{K^0}$ in (\ref{eq:epsilon
  num}) and adding the corresponding uncertainties in quadrature to those
in the phenomenological result for $Q$, we obtain \be\label{eq:epsilon eta}
\epsilonD\NLo 0.70(28)\fs\ee The estimate (\ref{eq:epsilon num}) for the 
size of the coefficient $\epsilonD$ is taken from here, except that
the error bar is stretched, to account for the fact that the
above relation is valid only up to NNLO contributions.

The MILC data offer a welcome check on the size of the neglected higher
orders: with the input $\epsilonD=1.2(5)$ and
$\epsilon_{\pi^0}=\epsilon_{K^0}=\epsilon_m=0$ used in the MILC analysis,
equation (\ref{eq:Q epsilon}) implies $Q_M=21.3(1.0)$. On the other hand,
inserting the MILC results for the quark mass ratios $m_s/m_{ud}$ and
$m_u/m_d$ (see Tables \ref{tab:ratio_msmud} and \ref{tab:mu_md_grading}) in
the definition of $Q$, equation (\ref{eq:Qm}), and ignoring a possible
correlation between the results for the two ratios, we get $Q=21.7(1.1)$.
For MILC 09A, the central values of $Q_M$ and $Q$ thus agree to an accuracy
of 2\%, indicating that the corrections of NNLO or higher are smaller than
the uncertainty of the result.

In contrast to this, the picture which follows from the numbers obtained
with the $\Nf=2$ simulation of RBC 07 \cite{Blum:2007cy} is difficult to
understand. Repeating the above calculation with the estimate
$\epsilonD=0.13(4)$ that corresponds to the kaon self-energy difference
obtained in this reference and again neglecting the other terms in equation
(\ref{eq:Q epsilon}), we get $Q_M=23.9(1)$, while the value of $Q$ that
follows from the quark mass ratios $m_s/m_{ud}$ and $m_u/m_d$ given in the
same reference, implies $Q=26.1(1.2)$. A difference of this size calls for
large contributions of NNLO, in conflict with the data analysis, which
assumes that it is meaningful to truncate the chiral series at NLO.

The recent extension of this calculation to $\Nf=2+1$ dynamical flavours
described in \cite{Blum:2010ym} appears to solve the puzzle. The values
obtained for the kaon and pion self-energies obtained in this reference
imply $\epsilonD=0.5(1)$, $\epsilon_m=0.04(5)$. Inserting this in equation
(\ref{eq:Q epsilon}) and neglecting the self-energies of the neutral
particles, we obtain $Q_M=22.8(3)$, to be compared with the value
$Q=23.9(3.4)$ that follows from the quark mass ratios given in the same
reference. This implies that the NNLO corrections to the low energy theorem
(\ref{eq:LET Q}) are too small to be seen at the accuracy of this
calculation: $Q_M/Q=0.96(12)$.

The determination of $R$ and $Q$ in BMW 10A, 10B \cite{Durr:2010vn,Durr:2010aw}
relies on the phenomenological value of $Q$ discussed above~\cite{Leutwyler:2009jg} -- their results do not bring new information on $Q$. The value of $R$ which follows from their result for the ratio $m_s/m_{ud}$ and the input used for $Q$ is $R=34.9(2.5)$, slightly different from our estimate in (\ref{eq:RQres}), but perfectly compatible with it.\footnote{Repeating the calculation with the same input for $Q$, but replacing the BMW result for $m_s/m_{ud}$ by our estimate in equation (\ref{eq:msovmud}), we obtain $R=35.0(2.6)$.}  

Our final results for the masses $m_u$, $m_d$, $m_s$ and the mass ratios
$m_u/m_d$, $m_s/m_{ud}$, $R$, $Q$ are collected in the table below.

\begin{table}[!thb]
\begin{tabular*}{\textwidth}[l]{@{\extracolsep{\fill}}ccccccccc}
$\Nf$ & $m_u  $ & $m_d $ & $ m_s $&  $m_{ud}$ & $m_u/m_d$ & $m_s/m_{ud}$ & $R$ & $Q$\\ 
&&&&&&&& \\[-2ex]
\hline\hline
&&&&&&&& \\[-2ex]
2+1 & 2.19(15) & 4.67(20) & 94(3) & 3.43(11) & 0.47(4) & 27.4(4) & 36.6(3.8) & 22.8(1.2) \\ 
&&&&&&&& \\[-2ex]
\hline\rule[-0.1cm]{0cm}{0.5cm}
&&&&&&&& \\[-2ex]
2 & -- & -- & 95(6) & 3.6(2) & -- & 27.3(9) & -- & --\\ 
&&&&&&&& \\[-2ex]
\hline
\hline
\end{tabular*}
\caption{\label{tab:collection qm} Our estimates for the masses of the
  three lightest quarks, and related ratios (the masses refer to the
  $\msbar$ scheme at running scale $\mu=2\,\gev$ for $N_f=3$, the
  numerical values are given in MeV units). }
\end{table}
\newpage

\section{$|V_{ud}|$ and $|V_{us}|$}\label{sec:vusvud}
\subsection{Experimental information concerning $|V_{ud}|$, $|V_{us}|$,
$f_+(0)$ and $f_K/f_\pi$}\label{sec:Exp} 
The following review relies on the fact that precision data on kaon decays
very accurately determine the product $|V_{us}|f_+(0)$ and the ratio
$|V_{us}f_K|/|V_{ud}f_\pi|$ \cite{Antonelli:2010yf}: \be\label{eq:products}
|V_{us}| f_+(0) = 0.2163(5)\co \hspace{1cm} \rule[-0.4cm]{0.02cm}{1cm}\;
\frac{V_{us} f_K}{V_{ud} f_\pi} \;
\rule[-0.4cm]{0.02cm}{1cm}=0.2758(5)\fs\ee $V_{ud}$ and $V_{us}$ are
elements of the Cabibbo-Kobayashi-Maskawa matrix and $f_+(t)$ represents
one of the form factors relevant for the semileptonic decay
$K^0\rightarrow\pi^-\ell\,\nu$, which depends on the momentum transfer $t$
between the two mesons.  What matters here is the value at $t=0$:
$f_+(0)\equiv
f_+^{K^0\pi^-}\hspace{-0.1cm}(t)\,\rule[-0.15cm]{0.02cm}{0.5cm}_{\;t\rightarrow
  0}$. The pion and kaon decay constants are defined by\footnote{The pion
  decay constant represents a QCD-matrix-element -- in the full Standard
  Model, the one-pion state is not a meaningful notion: the correlation
  function of the charged axial current does not have a pole at
  $p^2=M_{\pi^+}^2$, but a branch cut extending from $M_{\pi^+}^2$ to
  $\infty$. The analytic properties of the correlation function and the
  problems encountered in the determination of $f_\pi$ are thoroughly
  discussed in \cite{Gasser:2010wz}. The "experimental" value of $f_\pi$
  depends on the convention used when splitting the sum ${\cal
    L}_{\mbox{\tiny QCD}}+{\cal L}_{\mbox{\tiny QED}}$ into two parts
  (compare section \ref{subsec:electromagnetic interaction}).  The lattice
  determinations of $f_\pi$ do not yet reach the accuracy where this is of
  significance, but at the precision claimed by the Particle Data Group
  \cite{Rosner}, the numerical value does depend on the convention used
  \cite{Gasser:2003hk,Rusetsky:2009ic,Gasser:2007de,Gasser:2010wz}. A recent analysis of the electromagnetic and strong isospin breaking effects in the leptonic decays of the pseudoscalar mesons is given in \cite{Cirigliano:2011tm}.}  \bdm
\lvac \dbar\gamma_\mu\gamma_5 \hspace{0.05cm}u|\pi^+(p)\rangle=i
\hspace{0.05cm}p_\mu f_\pi\co\hspace{1cm} \lvac \sbar\gamma_\mu\gamma_5
\hspace{0.05cm} u|K^+(p)\rangle=i \hspace{0.05cm}p_\mu f_K\fs\edm In this
normalization, $f_\pi \simeq 130$ MeV, $f_K\simeq 155$ MeV.
 
The measurement of $|V_{ud}|$ based on superallowed nuclear $\beta$
transitions has now become remarkably precise. The result of the recent
update of Hardy and Towner \cite{Hardy:2008gy}, which is based on 20
different superallowed transitions, reads\footnote{It is not a trivial
  matter to perform the data analysis at this precision. In particular,
  isospin-breaking effects need to be properly accounted for
  \cite{Towner:2007np,Miller:2008my,Auerbach:2008ut,Liang:2009pf,Miller:2009cg}.
  For a review of recent work on this issue, we refer to
  \cite{Towner:2010bx}.}
\be\label{eq:Vud beta}
|V_{ud}| = 0.97425(22)\fs\ee 

The matrix element $|V_{us}|$ can be measured in $\tau$ decays
\cite{Gamiz:2002nu,Gamiz:2004ar,Maltman:2008na,Pich_Kass}. Separating the
inclusive decay $\tau\rightarrow \mbox{hadrons}+\nu$ into non-strange and
strange final states, Gamiz et al.~\cite{Gamiz:2007qs} obtain
\be\label{eq:Vus tau}|V_{us}|=0.2165(26)_{exp}(5)_{th}\fs \ee Maltman et
al.~\cite{Maltman:2008ib} arrive at very similar values.

As recently pointed out by Maltman \cite{Maltman:2008na}, the theoretical
uncertainties of the analysis which underlies the result (\ref{eq:Vus tau})
can be reduced by invoking the experimental information about the spectral
function of the electromagnetic current, but the experimental uncertainties
then play a more important role. Applying this method, the outcome for
$|V_{us}|$ reads \cite{Maltman:2009bh}
\be\label{eq:Maltman}|V_{us}|=0.2208(39)\fs\ee

In principle, $\tau$ decay offers a clean measurement of $|V_{us}|$, but a
number of open issues yet remain to be clarified. In particular, the
measured exclusive decay rates for $\tau\rightarrow \pi\nu$ and
$\tau\rightarrow K\nu$ are below the Standard Model predictions, which
determine these rates in terms of the same matrix elements $|V_{ud}f_\pi|$
and $|V_{us}f_K|$ that govern the leptonic decays $\pi\rightarrow\mu\nu$
and $K\rightarrow \mu\nu$ (violation of $\tau/\mu$ universality). On the
other hand, the value obtained for $|V_{us} f_K|/|V_{ud}f_\pi|$ from the
ratio of the tau decay rates is perfectly consistent with the one from the
leptonic $\pi$ and $K$-decays quoted in (\ref{eq:products}), but this does
not shed any light on the value of $|V_{us}|$. It is important to pursue
the measurement of modes that have previously been studied only with low
statistics (especially the large mode $K^0 \pi^0 \pi^-$), as well as those
of higher multiplicity (final states with more than two pions, for
instance). The recent developments on the theoretical side
\cite{Beneke:2008ad,Caprini:2009vf,Menke:2009vg} also need to be pursued.
The most interesting possibility is that $\tau$ decay involves new physics,
but more work is required before $\tau$ decay becomes a competitive source
of information about the CKM matrix elements.
\subsection{Lattice results for $f_+(0)$ and $f_K/f_\pi$}
\begin{table}[t]
\centering 
\vspace{2.8cm}
{\footnotesize\noindent
\begin{tabular*}{\textwidth}[l]{@{\extracolsep{\fill}}llllllll}
Collaboration & Ref. & $\Nf$ & 
\hspace{0.15cm}\begin{rotate}{60}{publication status}\end{rotate}\hspace{-0.15cm}&
\hspace{0.15cm}\begin{rotate}{60}{chiral extrapolation}\end{rotate}\hspace{-0.15cm}&
\hspace{0.15cm}\begin{rotate}{60}{continuum extrapolation}\end{rotate}\hspace{-0.15cm}&
\hspace{0.15cm}\begin{rotate}{60}{finite volume errors}\end{rotate}\hspace{-0.15cm}&\rule{0.3cm}{0cm}
$f_+(0)$ \\
&&&&&&& \\[-0.1cm]
\hline
\hline&&&&&&& \\[-0.1cm]
RBC/UKQCD 10              & \cite{RBCUKQCD10}    &2+1  &\gA&\soso&\tbr&\tbg& 0.9599(34)($^{+31}_{-47}$)(14)\rule{0cm}{0.4cm}\\ 
RBC/UKQCD 07              & \cite{RBCUKQCD07}    &2+1  &\gA&\soso&\tbr&\tbg& 0.9644(33)(34)(14)\\
&&&&&&& \\[-0.1cm]
\hline
&&&&&&& \\[-0.1cm]
ETM 10D                   & \cite{Lubicz:2010bv}  &2 &\rC&\soso&\tbg&\soso& 0.9544(68)$_{stat}$\\
ETM 09A 	          & \cite{Lubicz:2009ht}  &2 &\gA&\soso&\soso&\soso& 0.9560(57)(62)\\	
QCDSF 07	          & \cite{Brommel:2007wn} &2 &\rC&\tbr&\tbr&\tbg& 0.9647(15)$_{stat}$ \\
RBC 06  	          & \cite{Dawson:2006qc}  &2 &\gA&\tbr&\tbr&\tbg& 0.968(9)(6)\\	
JLQCD 05 	          & \cite{Tsutsui:2005cj} &2 &\rC&\tbr&\tbr&\tbg& 0.967(6), 0.952(6)\\ 
 &&&&&&& \\[-0.1cm]
\hline
\hline
\end{tabular*}}
\caption{Colour code for the data on $f_+(0)$.\hfill}\label{tab:f+(0)}
\end{table}
The traditional way of determining $|V_{us}|$ relies on using theory for
the value of $f_+(0)$, invoking the Ademollo-Gatto theorem
\cite{Ademollo_Gatto}.  Since this theorem only holds to leading order of
the expansion in powers of $m_u$, $m_d$ and $m_s$, theoretical models are
used to estimate the corrections. Lattice methods have now reached the
stage where quantities like $f_+(0)$ or $f_K/f_\pi$ can be determined to
good accuracy. As a consequence, the uncertainties inherent in the
theoretical estimates for the higher order effects in the value of $f_+(0)$
do not represent a limiting factor any more and we shall therefore not
invoke those estimates. Also, we will use the experimental results based on
nuclear $\beta$ decay and $\tau$ decay exclusively for comparison -- the
main aim of the present review is to assess the information gathered with
lattice methods and to use it for testing the consistency of the SM and its
potential to provide constraints for its extensions.

The database underlying the present review of $f_+(0)$ and $f_K/f_\pi$ is
listed in Tables \ref{tab:f+(0)} and \ref{tab:FKFpi}. The properties of the
lattice data play a crucial role for the conclusions to be drawn from these
results: range of $M_\pi$, size of $L M_\pi$, continuum extrapolation,
extrapolation in the quark masses, finite size effects, etc. The key
features of the various data sets are characterized by means of the colour
code specified in section \ref{sec:color-code}.

The quantity $f_+(0)$ represents a matrix element of a strangeness changing
null plane charge, $f_+(0)=(K|Q^{us}|\pi)$. The vector charges obey the
commutation relations of the Lie algebra of SU(3), in particular
$[Q^{us},Q^{su}]=Q^{uu-ss}$. This relation implies the sum rule $\sum_n
|(K|Q^{us}|n)|^2-\sum_n |(K|Q^{su}|n)|^2=1$. Since the contribution from
the one-pion intermediate state to the first sum is given by $f_+(0)^2$,
the relation amounts to an exact representation for this quantity
\cite{Furlan}: \be \label{eq:Ademollo-Gatto} f_+(0)^2=1-\sum_{n\neq \pi}
|(K|Q^{us}|n)|^2+\sum_n |(K|Q^{su}|n)|^2\fs\ee While the first sum on the
right extends over non-strange intermediate states, the second runs over
exotic states with strangeness $\pm 2$ and is expected to be small compared
to the first.

The expansion of $f_+(0)$ in powers of $m_u$, $m_d$ and $m_s$ starts with
$f_+(0)=1+f_2+f_4+\ldots\,$ \cite{Gasser:1984gg}.  Since all of the low
energy constants occurring in $f_2$ can be expressed in terms of $M_\pi$,
$M_K$, $M_\eta$ and $f_\pi$ \cite{Gasser:1984ux}, the correction of NLO is
known. In the language of the sum rule (\ref{eq:Ademollo-Gatto}), $f_2$
stems from non-strange intermediate states with three mesons. Like all
other non-exotic intermediate states, it lowers the value of $f_+(0)$:
$f_2=-0.023$.  The corresponding expressions have also been derived in
quenched or partially quenched chiral perturbation theory
\cite{Becirevic:2005py}.  At the same order in the SU(2) expansion
\cite{Flynn:2008tg}, $f_+(0)$ is parametrized in terms of $M_\pi$ and two
\textit{a priori} unknown parameters. The latter can be determined from the
dependence of the lattice results on the masses of the quarks.  Note that
any calculation that relies on the {\Ch}PT formula for $f_2$ is subject to
the uncertainties inherent in NLO results: instead of using the physical
value of the pion decay constant $f_\pi$, one may, for instance, work with
the constant $f_0$ that occurs in the effective Lagrangian and represents
the value of $f_\pi$ in the chiral limit. Although trading $f_\pi$ for
$f_0$ in the expression for the NLO term affects the result only at NNLO,
it may make a significant numerical difference in calculations where the
latter are not explicitly accounted for (the lattice results concerning the
value of the ratio $f_\pi/f_0$ are reviewed in section \ref{sec:su3}).
\begin{table}[t]
\centering
\vspace{1.5cm}{\footnotesize\noindent
\begin{tabular*}{\textwidth}[l]{@{\extracolsep{\fill}}lrllllll}
Collaboration & Ref. & $\Nf$ &
\hspace{0.15cm}\begin{rotate}{60}{publication status}\end{rotate}\hspace{-0.15cm}&
\hspace{0.15cm}\begin{rotate}{60}{chiral extrapolation}\end{rotate}\hspace{-0.15cm}&
\hspace{0.15cm}\begin{rotate}{60}{continuum extrapolation}\end{rotate}\hspace{-0.15cm}&
\hspace{0.15cm}\begin{rotate}{60}{finite volume errors}\end{rotate}\hspace{-0.15cm}&\rule{0.2cm}{0cm}
$f_K/f_\pi$ \\  
&&&&&&& \\[-0.1cm]
\hline
\hline
&&&&&&& \\[-0.1cm]
ETM 10E       &\cite{Farchioni:2010tb}&2+1+1&\rC&\soso&\soso&\soso&1.224(13)$_{stat}$\\
&&&&&&& \\[-0.1cm]
\hline
&&&&&&& \\[-0.1cm]
MILC 10        &\cite{Bazavov:2010hj}&2+1&\rC&\soso&\good&\good&1.197(2)($^{+3}_{-7}$)\\
RBC/UKQCD 10A  &\cite{Aoki:2010dy}   &2+1&\oP&\soso&\soso&\good&1.204(7)(25)\\
BMW 10         &\cite{BMW10}         &2+1&\gA&\tbg &\tbg&\tbg& 1.192(7)(6)\\
JLQCD/TWQCD 09A&\cite{JLQCD:2009sk}  &2+1&\rC&\soso&\tbr&\tbr& $1.210(12)_{\rm stat}$\\
MILC 09A       &\cite{Bazavov:2009fk}&2+1&\rC&\soso&\tbg&\tbg& 1.198(2)($^{\hspace{0.01cm}+6}_{-8}$)\\
MILC 09        &\cite{Bazavov:2009bb}&2+1&\gA&\soso&\tbg&\tbg& 1.197(3)($^{\;+6}_{-13}$)\\
Aubin 08       &\cite{Aubin:2008ie}  &2+1&\rC&\soso&\soso&\soso& 1.191(16)(17)\\
PACS-CS 08, 08A&\cite{Aoki:2008sm, Kuramashi:2008tb} &2+1&\gA&\tbg&\tbr&\tbr& 1.189(20)\\
RBC/UKQCD 08   &\cite{Allton:2008pn} &2+1&\gA&\soso&\tbr&\tbg& 1.205(18)(62)\\
HPQCD/UKQCD 07 &\cite{Follana:2007uv}&2+1&\gA&\soso&\tbg&\soso& 1.189(2)(7)\\
NPLQCD 06      &\cite{Beane:2006kx}  &2+1&\gA&\soso&\tbr&\tbr& 1.218(2)($^{+11}_{-24}$)\\
MILC 04 &\cite{Aubin:2004fs}&2+1&\gA&\good&\soso&\soso&1.210(4)(13)\\
&&&&&&& \\[-0.1cm]
\hline
&&&&&&& \\[-0.1cm]
ETM 10D        &\cite{Lubicz:2010bv} &2  &\rC&\soso&\tbg&\soso& 1.190(8)$_{stat}$ \\
ETM 09         &\cite{ETM09}         &2  &\gA&\soso&\tbg&\soso& 1.210(6)(15)(9)\\
QCDSF/UKQCD 07 &\cite{QCDSFUKQCD}    &2  &\rC&\soso&\soso&\tbg& 1.21(3)\\
&&&&&&& \\[-0.1cm]
\hline
\hline
&&&&&&& \\[-0.1cm]
\end{tabular*}}

\vspace{-0.5cm}
\caption{Colour code for the data on $f_K/f_\pi$.\hfill}\label{tab:FKFpi}

\end{table}
\begin{figure}[ht]
\psfrag{L 84}{\sffamily\tiny \hspace{-0.24cm}\cite{Leutwyler:1984je}}
\psfrag{B 03}{\sffamily\tiny \hspace{-0.24cm}\nocite{Post:2001si}\cite{Bijnens:2003uy}}
\psfrag{J 04}{\sffamily\tiny \hspace{-0.24cm}\cite{Jamin:2004re}}
\psfrag{C 05}{\sffamily\tiny \hspace{-0.24cm}\cite{Cirigliano:2005xn}}
\psfrag{K 08}{\sffamily\tiny \hspace{-0.24cm}\cite{Kastner:2008ch}} 
\psfrag{y}{\tiny $\star$}
\includegraphics[height=6.3cm]{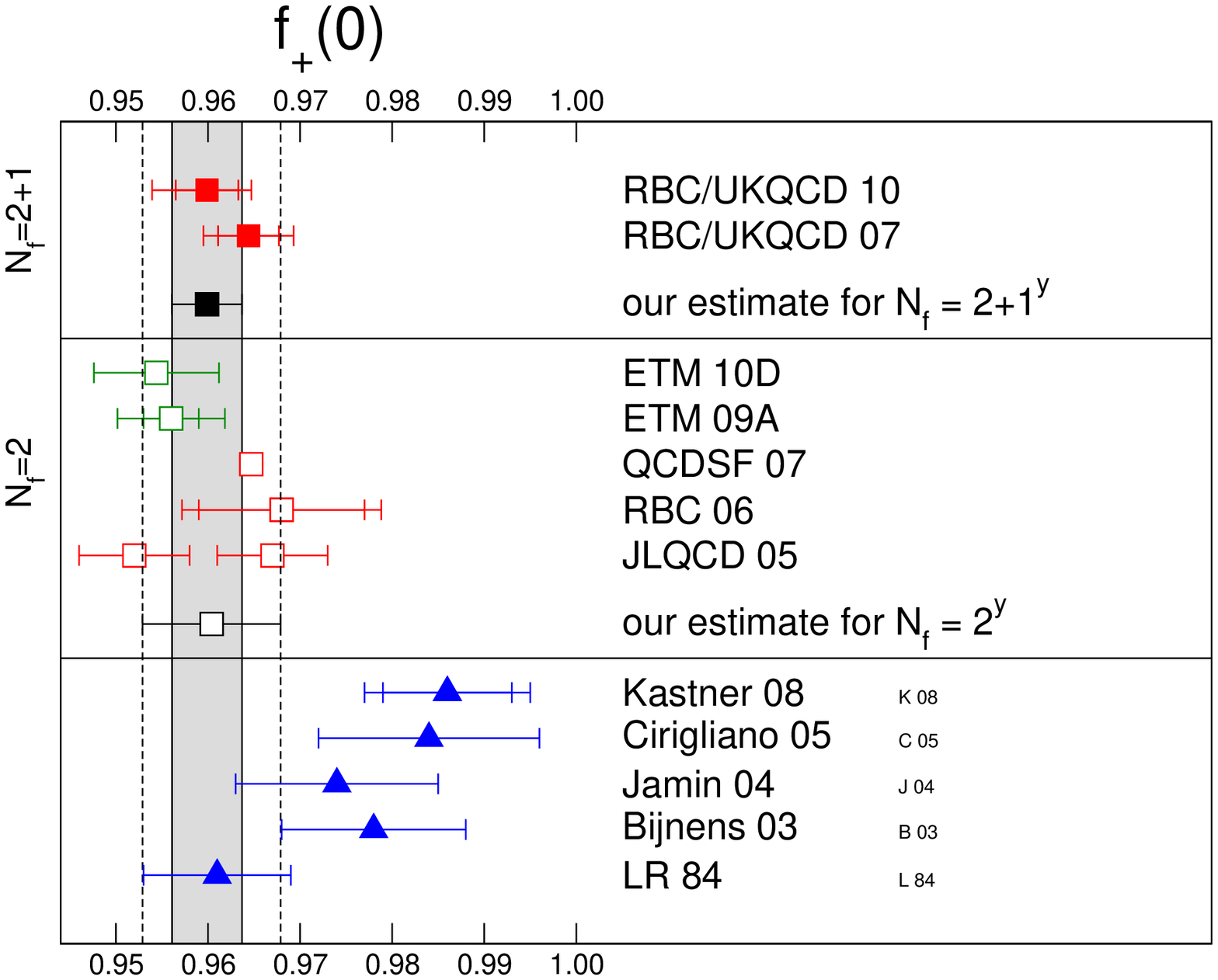}
\hfill 
\includegraphics[height=6.3cm]{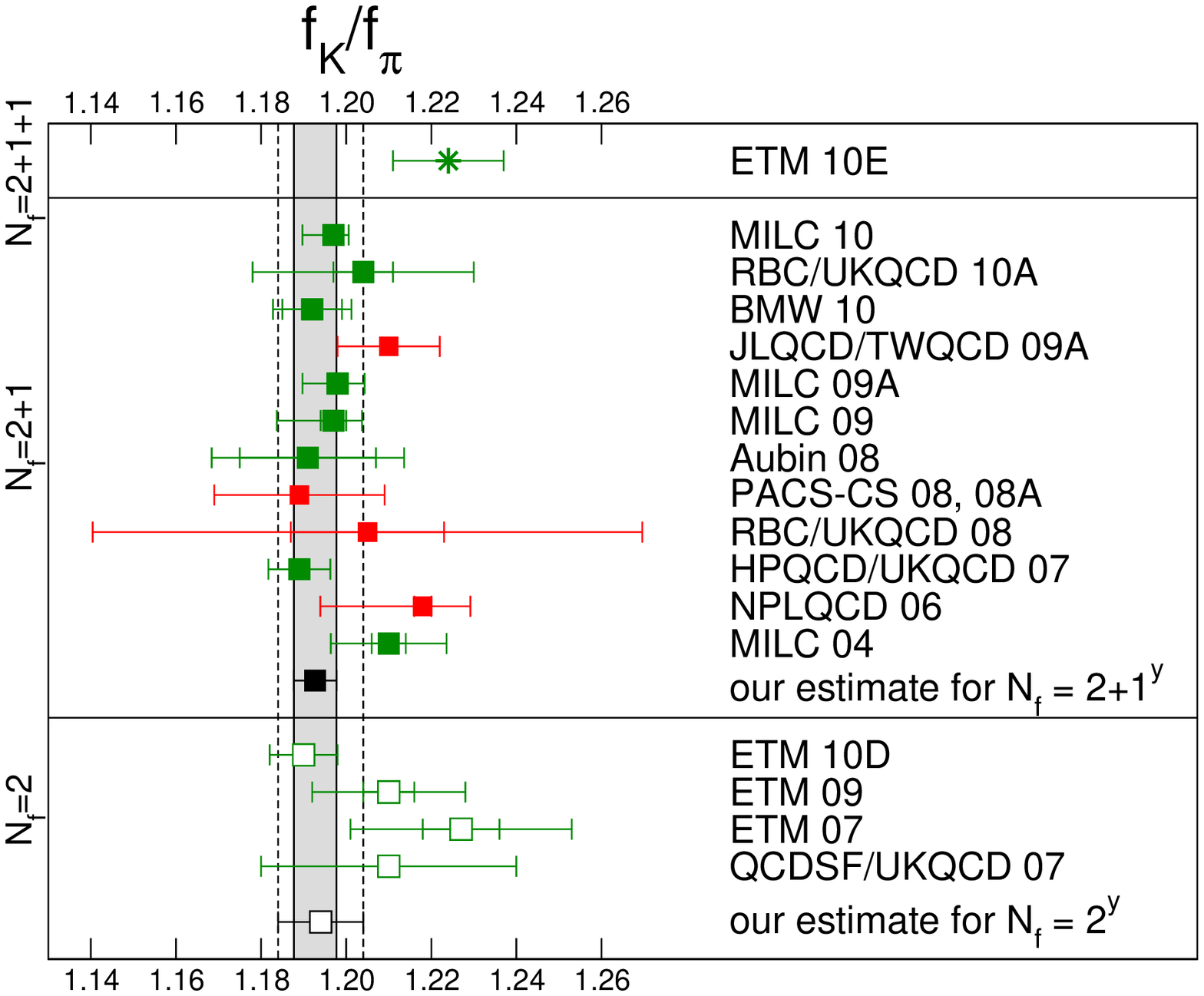}
  
  \vspace{0.3cm}
 {\footnotesize $^\star$ Estimates obtained from an analysis of the
  lattice data within the Standard Model, see section \ref{sec:SM}.}

\caption{\label{fig:lattice data}Comparison of lattice results (red squares) 
  for $f_+(0)$ and $f_K/ f_\pi$ with various model estimates based on {\Ch}PT (blue 
  triangles).  Full and empty squares represent simulations with $\Nf=2+1$
  and $\Nf=2$, respectively. The vertical bands indicate our estimates.$^\star$ }
\end{figure}

The lattice results shown in the left panel of Figure \ref{fig:lattice
  data} indicate that the higher order contributions $\Delta f\equiv
f_+(0)-1-f_2$ are negative and thus amplify the effect generated by $f_2$.
This confirms the expectation that the exotic contributions are small. The
entries in the lower part of the left panel represent various model
estimates for $f_4$. In \cite{Leutwyler:1984je} the symmetry breaking
effects are estimated in the framework of the quark model. The more recent
calculations are more sophisticated, as they make use of the known explicit
expression for the $K_{\ell3}$ form factors to NNLO of {\Ch}PT
\cite{Post:2001si,Bijnens:2003uy}. The corresponding formula for $f_4$
accounts for the chiral logarithms occurring at NNLO and is not subject to
the ambiguity mentioned above. The numerical result, however, depends on
the model used to estimate the low energy constants occurring in $f_4$
\cite{Bijnens:2003uy,Jamin:2004re,Cirigliano:2005xn,Kastner:2008ch}. The
figure indicates that the most recent numbers obtained in this way
correspond to a positive rather than a negative value for $\Delta f$.

A lattice determination of the low energy constants that occur in $f_4$
would be very useful to identify the origin of the problem. Lattice results
concerning the dependence of the scalar form factor on the momentum
transfer would also very useful, in particular they might help sorting out
the discrepancies in the experimental results for the slope of this form
factor \cite{Bernard:2009zm,Bernard:2009ds,
  PassemarCERN2009,Passemar:2010cc}.  Note that when analyzing lattice data
on the basis of the {\Ch}PT formulae, the loop integrals occurring therein
need to be evaluated for the masses at which the Nambu-Goldstone bosons
propagate in a given lattice simulation.\footnote{Fortran programs for the
  numerical evaluation of the form factor representation in
  \cite{Bijnens:2003uy} are available on request from Johan Bijnens.}
 
\subsection{Direct determination of $f_+(0)$ and $f_K/f_\pi$}\label{sec:Direct} 
Figure \ref{fig:lattice data} shows that the lattice results for $f_+(0)$
and $f_K/f_\pi$ obtained by the various collaborations which do a
comprehensive error analysis are consistent with each 
other.
We now proceed to form the corresponding averages, separately
for the data with $\Nf=2+1$ and $\Nf=2$ dynamical flavours.  As will be
discussed in detail in section \ref{sec:SM}, CKM unitarity and experiment
correlate $f_+(0)$ with $f_K/f_\pi$. Indeed, the lattice data are
consistent also with this correlation. First, however, we analyze the
lattice information available separately for $f_+(0)$ and for $f_K/f_\pi$.
We refer to results obtained in this way as "direct" determinations.

The colour code in Table 5 shows that for $f_+(0)$, presently only the results of
the ETM collaboration with $\Nf = 2$ dynamical flavours of fermions are without a 
red tag. In the following, we rely on the published work ETM 09A 
\cite{Lubicz:2009ht}. The table contains two data sets with $\Nf=2+1$, but since one
represents an update of the other, we only discuss RBC/UKQCD 10. These data
carry a red tag, because a systematic study of cut-off effects is missing.
These two computations of $f_+(0)$ are the most advanced ones and we quote
their results: \bea\label{eq:fplus_direct}
f_+(0)\al=\al 0.9599(34)(^{+31}_{-47})(14)\,, \hspace{1cm}(\mbox{direct},\,\Nf=2+1),  \\
f_+(0)\al=\al 0.9560(57)(62)\,,
\hspace{1.86cm}(\mbox{direct},\,\Nf=2).\nonumber \eea

The first error in the result quoted for $\Nf=2+1$ is statistical, the
second is due to the uncertainties in the chiral extrapolation of the
lattice data and the third is an estimate of potential discretization
effects. Flavour symmetry implies that if $m_{ud}$ is set equal to $m_s$,
the lattice data yield $f_+(0)=1$, irrespective of the lattice spacing or
the size of the box and for any value of $m_s$. Cut-off effects can
therefore only affect the difference $1-f_+(0)$, which turns out to be
about 0.04. Indeed, the estimate provided by RBC/UKQCD 10 for the
uncertainties due to discretization effects shows that these are
sub-dominant: inflating the corresponding error by a factor of up to 2
barely affects the net systematic uncertainty.

In the result quoted for $\Nf=2$, the brackets indicate the statistical and
systematic errors, respectively. The ETM collaboration provides a more
comprehensive study of the systematics by presenting results for three
lattice spacings \cite{DiVita:2009by} and simulating at lighter pion masses
(down to $M_\pi=260$ MeV). This allows to better constrain the chiral
extrapolation, using both SU(3) \cite{Gasser:1984ux} and SU(2)
\cite{Flynn:2008tg} chiral perturbation theory. Moreover, a rough estimate
for the size of the effects due to quenching the strange quark is given,
based on the comparison of the result for $\Nf=2$ dynamical quark flavours
\cite{ETM09} with the one in the quenched approximation, obtained earlier
by the SPQcdR collaboration
\cite{Becirevic:2004ya}.
  
The quality criteria laid out in section \ref{sec:color-code} require a
systematic study of lattice artifacts. As indicated by the colour code in
Table \ref{tab:f+(0)}, the errors due to the continuum extrapolation yet
need to be investigated in more detail for the data with $\Nf=2+1$, while
for $\Nf=2$, where the quoted uncertainties are larger, these errors are
under somewhat better control. The value 
\be 
f_+(0)=0.956(8)  \hspace{1.5cm} (\mbox{our estimate, direct})
\ee 
covers both
results in equation (\ref{eq:fplus_direct}). In our opinion, it represents
a conservative estimate for the range permitted by the presently available
direct determinations of $f_+(0)$ in lattice QCD, not only for $\Nf=2$, but
also for $\Nf=2+1$.

For $f_K/f_\pi$, Table \ref{tab:FKFpi} contains several simulations with
$\Nf=2+1$ dynamical quark flavours. The latest update of the MILC program
is reported in MILC 10 \cite{Bazavov:2010hj}. We use the results quoted
there when forming averages. Three further data sets meet the criteria
formulated in the introduction: BMW 10 \cite{BMW10} and HPQCD/UKQCD 07
\cite{Follana:2007uv} with $\Nf=2+1$ and ETM 09 \cite{ETM09} with $\Nf=2$
dynamical flavours.  We ignore possible correlations due to the fact that
MILC 10 and HPQCD/UKQCD 07 have partly used the same set of gauge
configurations and apply the procedure outlined in section
\ref{sec:averages} to the three sets with $\Nf=2+1$. The resulting fit is
of good quality, with $f_K/f_\pi=1.193(4)$ and $\chi^2=0.4$ for 3 data
points and 1 free parameter. The systematic errors of the individual data
sets are larger: 0.005, 0.007 and 0.006 for MILC 10, HPQCD/UKQCD 07 and
BMW 10, respectively. Following the prescription of section
\ref{sec:averages}, we replace the error by the smallest one of these
numbers. Together with the ETM 09 result for $\Nf=2$, our estimates thus
read
\bea 
\label{eq:fKfpi_direct} 
f_K / f_\pi\al=\al  1.193(5)\,, \hspace{1.18cm}(\mbox{direct},\, \Nf=2+1)\,,\\
f_K/f_\pi\al=\al 1.210 (6)(17)\,, \hspace{0.5cm} (\mbox{direct},\,
\Nf=2)\,. \nonumber\eea
 
\begin{figure}[t]\vspace{0.2cm}\centering
\includegraphics[width=9cm]{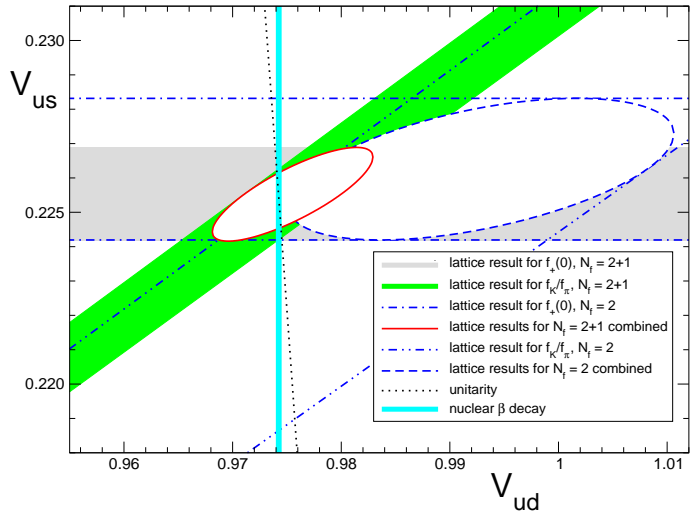}  
\caption{\label{fig:VusVersusVud}The plot compares the information for
  $|V_{ud}|$, $|V_{us}|$ obtained on the lattice with the experimental
  result extracted from nuclear $\beta$ transitions. The dotted arc
  indicates the correlation between $|V_{ud}|$ and $|V_{us}|$ that follows
  if the three-flavour CKM-matrix is unitary. }
\end{figure}

It is instructive to convert the above results for $f_+(0)$ and $f_K/f_\pi$
into a corresponding range for the CKM matrix elements $V_{ud}$ and
$V_{us}$, using the relations (\ref{eq:products}). Consider first the
results for $\Nf=2+1$.  The range for $f_+(0)$ in (\ref{eq:fplus_direct})
is mapped into the interval $V_{us}=0.2255(14)$, depicted as a horizontal
gray band in Figure \ref{fig:VusVersusVud}, while the one for $f_K/f_\pi$
in (\ref{eq:fKfpi_direct}) is converted into $V_{us}/V_{ud}= 0.2312(11)$,
shown as a green band. The red curve is the intersection of these two
bands.
More precisely, it represents the 68\% likelihood contour, obtained by
treating the above two results as independent measurements. A Gaussian in
$f_+(0)$ corresponds to a Gaussian in the variable $1/V_{us}$. Since the
width is small, the distribution in the variable $V_{us}$ is also
approximately Gaussian. The corresponding likelihood function is given by
$\chi_a^2=(V_{us}-0.2255)^2/0.0014^2$. Likewise, a Gaussian in $f_K/f_\pi$
is mapped into an approximately Gaussian distribution of the variable
$V_{us}/V_{ud}$, with $\chi_b^2= (V_{us}/V_{ud}-0.2312 )^2/0.0011^2$.
Expressed in terms of the CKM matrix elements, the $\Nf=2+1$ results for
$f_+(0)$ and $f_K/f_\pi$ are thus characterized by the likelihood function
$\chi^2=\chi_a^2+\chi_b^2$. The minimum occurs at the intersection of the
centers of the two bands, where $\chi^2$ vanishes. The contour shown is the
line where $\chi^2$ differs from the minimum by unity. Values of $V_{us}$,
$V_{ud}$ in the region enclosed by this contour are consistent with the
lattice data for $\Nf=2+1$, within one standard deviation. In particular,
the plot shows that the nuclear $\beta$ decay result for $V_{ud}$ is
perfectly consistent with these data.
 
Repeating the exercise with the $\Nf=2$ results leads to the dashed
ellipse. The figure thus indicates that the data are consistent within
errors. We conclude that, at the accuracy reached up to now for the
quantities $f_+(0)$, $f_K/f_\pi$, $V_{us}$ and $V_{ud}$, the distortions
generated by the quenching of the strange quark are too small to be
visible.

\subsection{Testing the Standard Model}\label{sec:testing}

\begin{table}[t]\centering
\noindent
\begin{tabular*}{\textwidth}[l]{@{\extracolsep{\fill}}lrlcl}
Collaboration & Ref. &$\Nf$&from&\rule{0.8cm}{0cm}$|V_{us}|$\\
&&&& \\[-2ex]
\hline \hline &&&&\\[-2ex]
MILC 10 &\cite{Bazavov:2010hj}&$2+1$&$f_K/f_\pi$ \rule{0cm}{0.45cm} &0.2245(5)($^{+12}_{-5}$)\\
RBC/UKQCD 10A  &\cite{Aoki:2010dy}   &$2+1$&$f_K/f_\pi$ \rule{0cm}{0.45cm} &0.2233(13)(44)\\ 
RBC/UKQCD 10 & \cite{RBCUKQCD10} & $2+1$ & $f_+(0)$ \rule{0cm}{0.45cm} &$0.2253(10)(^{+12}_{-8})$\\
BMW 10 &\cite{BMW10}  & $2+1$ \rule{0cm}{0.45cm}& $f_K/f_\pi$ & $0.2254(13)(11)$\\
HPQCD/UKQCD 07 &\cite{Follana:2007uv}\rule{0cm}{0.4cm}& $2+1$ & $f_K/f_\pi$&  $  0.2260(5)(13)$\\
&&&& \\[-2ex]
 \hline
&&&& \\[-2ex]
 ETM 09  &\cite{ETM09}\rule{0cm}{0.4cm}&2&$f_K/f_\pi$& $ 0.2222 (11) (31)$\\
 ETM 09A & \cite{Lubicz:2009ht}\rule{0cm}{0.4cm}&2&$f_+(0)$&   $ 0.2263 (14) (15)$\\
&&&& \\[-2ex]
 \hline \hline 
\end{tabular*}
\caption{\label{tab:Vus}Values of $|V_{us}|$ obtained from lattice
  determinations of $f_+(0)$ or $f_K/f_\pi$ with CKM unitarity. The first
  (second) number in brackets represents the statistical (systematic)
  error.} 
  \end{table} 
  
In the Standard Model, the CKM matrix is unitary.  In particular, the
elements of the first row obey \be\label{eq:CKM unitarity}|V_u|^2\equiv
|V_{ud}|^2 + |V_{us}|^2 + |V_{ub}|^2 = 1\fs\ee The tiny contribution from
$|V_{ub}|$ is known much better than needed in the present context: $
|V_{ub}| = 3.89(44)\cdot 10^{-3} $ \cite{Kowalewski}. In the following, we
first discuss the evidence for the validity of the relation (\ref{eq:CKM
  unitarity}) and only then use it to analyze the lattice data within the 
Standard Model.

In Figure \ref{fig:VusVersusVud}, the correlation between $V_{ud}$ and
$V_{us}$ imposed by the unitarity of the CKM matrix is indicated by a
dotted arc (more precisely, in view of the uncertainty in $V_{ub}$, the
correlation corresponds to a band of finite width, but the effect is too
small to be seen here). The plot shows that the data for $\Nf=2+1$ are
perfectly consistent with this constraint. Numerically, the outcome for the
sum of the squares of the first row of the CKM matrix reads $|V_u|^2
=1.002(15)$. The Standard Model thus passes a nontrivial test that
exclusively involves lattice data and well-established kaon decay branching
ratios. Combining the lattice results for $f_+(0)$ and $f_K/f_\pi$ in
(\ref{eq:fplus_direct}) and (\ref{eq:fKfpi_direct}) with the $\beta$-decay
value of $|V_{ud}|$ quoted in (\ref{eq:Vud beta}), the test sharpens
considerably: the lattice result for $f_+(0)$ leads to $|V_u|^2=
1.0000(7)$, while the one for $f_K/f_\pi$ implies $|V_u|^2=0.9999(6)$, thus
confirming CKM unitarity at the permille level.

Repeating the analysis for $\Nf=2$, we find $|V_u|^2=1.037(36)$ with the
lattice data alone. The number is somewhat larger than 1, in accordance
with the fact that the dotted curve passes just outside the blue contour,
but the difference barely exceeds one standard deviation. Moreover, it only
concerns the comparison of the $\Nf=2$ results for $f_+(0)$ with those for
$f_K/f_\pi$. Taken by themselves, these results are perfectly consistent
with the value of $|V_{ud}|$ found in nuclear $\beta$ decay: combining this
value with the data on $f_+(0)$ yields $|V_u|^2=1.0004(10)$, combining it
with the data on $f_K/f_\pi$ gives $|V_u|^2=0.9985(16)$.

\begin{figure}[t]
\psfrag{H 09}{\sffamily\scriptsize \hspace{0.13cm}\cite{Hardy:2008gy}}
\psfrag{G 08}{\sffamily\scriptsize \cite{Gamiz:2007qs}}
\psfrag{M 09}{\sffamily\scriptsize \cite{Maltman:2009bh}}
\psfrag{y}{\tiny $\star$}
\begin{center}
\vspace{0.5cm} 
\includegraphics[width=13cm]{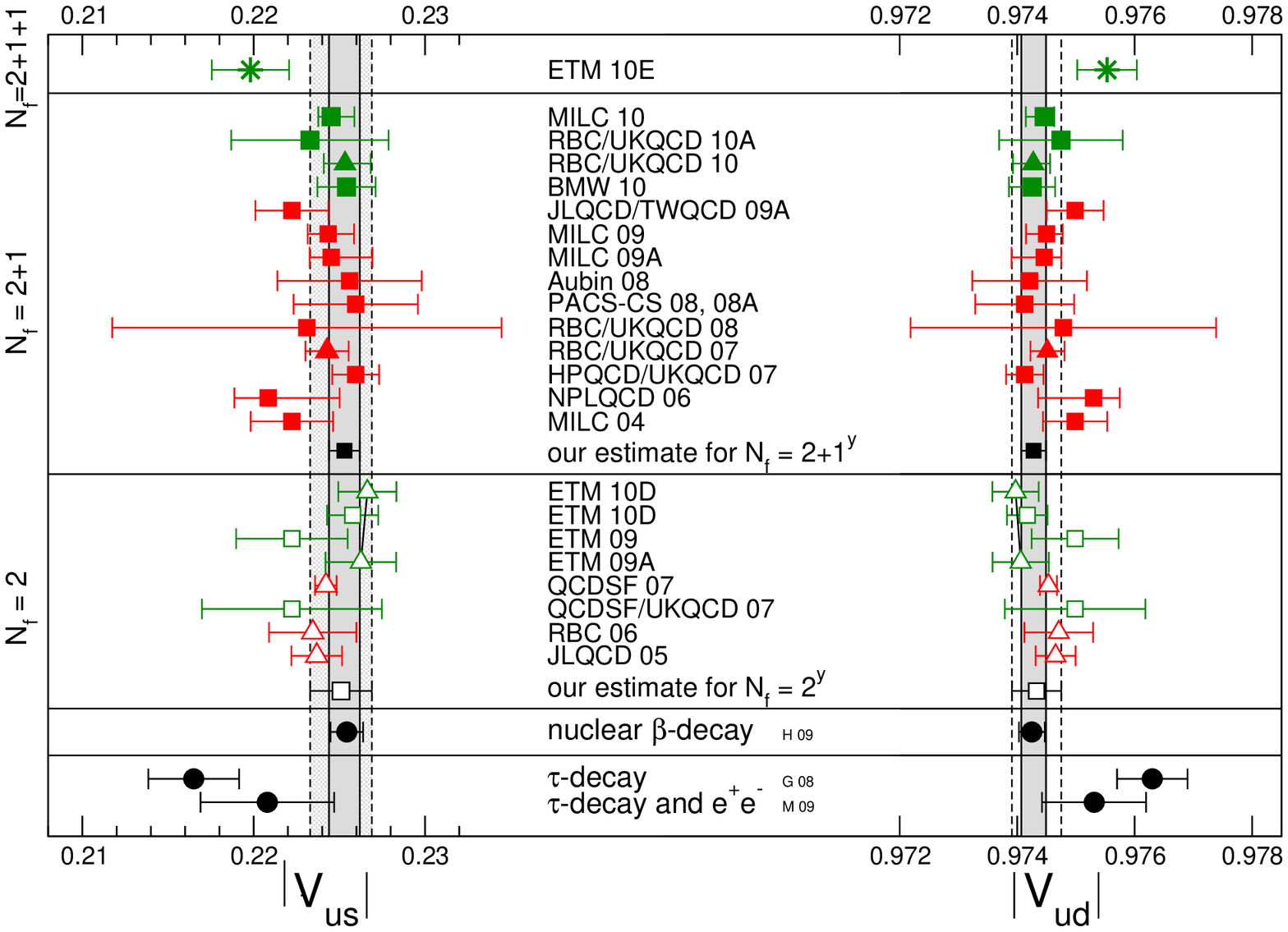}
\end{center}
\mbox{}\\[-4ex]
{\footnotesize $^\star$ Estimates obtained from an analysis of the
  lattice data within the Standard Model, see section \ref{sec:SM}.}

\caption{\label{fig:Vus Vud} Results for $|V_{us}|$ and $|V_{ud}|$ that
  follow from the lattice data for $f_+(0)$ (red triangles) and $f_K/f_\pi$
  (red squares), on the basis of the assumption that the CKM matrix is
  unitary. The black squares and the bands represent our estimates,
  obtained by combining these two different ways of measuring $|V_{us}|$
  and $|V_{ud}|$ on a lattice. For comparison, the figure also indicates
  the results obtained if the data on nuclear $\beta$ decay and $\tau$
  decay are analyzed within the Standard Model. }
\end{figure}  

Note that the above tests also offer a check of the basic hypothesis that
underlies our analysis: we are assuming that the weak interaction between
the quarks and the leptons is governed by the same Fermi constant as the
one that determines the strength of the weak interaction among the leptons
and determines the lifetime of the muon. In certain modifications of the
Standard Model, this is not the case. In those models it need not be true
that the rates of the decays $\pi\rightarrow \ell\nu$,
$K\rightarrow\ell\nu$ and $K\rightarrow \pi\ell \nu$ can be used to
determine the matrix elements $|V_{ud}f_\pi|$, $|V_{us}f_K|$ and
$|V_{us}f_+(0)|$, respectively and that $|V_{ud}|$ can be measured in
nuclear $\beta$ decay. The fact that the lattice data are consistent with
unitarity and with the value of $|V_{ud}|$ found in nuclear $\beta$ decay
indirectly also checks the equality of the Fermi constants.

\subsection{Analysis within the Standard Model} \label{sec:SM} 
 
The Standard Model implies that the CKM matrix is unitary. The precise
experimental constraints quoted in (\ref{eq:products}) and the unitarity
condition (\ref{eq:CKM unitarity}) then reduce the four quantities
$|V_{ud}|,|V_{us}|,f_+(0),f_K/f_\pi$ to a single unknown: any one of these
determines the other three within narrow uncertainties.
 
Figure \ref{fig:Vus Vud} shows that the results obtained for $|V_{us}|$ and
$|V_{ud}|$ from the data on $f_K/f_\pi$ (red squares) are quite consistent
with the determinations via $f_+(0)$ (red triangles). In order to calculate
the corresponding average values, we restrict ourselves to those
determinations that we have considered best in section \ref{sec:Direct}.
The corresponding results for $|V_{us}|$ are listed in Table \ref{tab:Vus}
(the noise in the experimental numbers used to convert the values of
$f_+(0)$ and $f_K/f_\pi$ into values for $|V_{us}|$ is included in the
statistical error).
  
We consider the fact that the results from the four $\Nf=2+1$ data sets
RBC/UKQCD 10 \cite{RBCUKQCD10}, BMW 10 \cite{BMW10}, MILC 10
\cite{Bazavov:2010hj} and HPQCD/UKQCD 07 \cite{Follana:2007uv} are
consistent with each other to be an important reliability test of the
lattice work. Treating the four sets as independent measurements, and
applying the standard averaging procedure, we obtain a fit of good quality,
with $|V_{us}| = 0.2253(6)$ and $\chi^2 = 0.6$ for 4 data points and 1 free
parameter. The results listed in the upper half of Table \ref{tab:Vus}
show, however, that the standard procedure underestimates the systematic
uncertainties also in this case. Applying the prescription of section
\ref{sec:averages}, we arrive at a somewhat larger error: $|V_{us}| =
0.2253(9)$. This result is indicated on the left hand side of
Fig.~\ref{fig:Vus Vud} by the narrow vertical band. The broader band shows
the corresponding value for $\Nf=2$ (standard error analysis, $|V_{us}|=
0.2253(17)$, with $\chi^2=1.2$ for 2 data points and 1 free parameter,
$S=1.09$). The figure shows that the result obtained for the data with
$\Nf=2$ is perfectly consistent with the one found for $\Nf=2+1$.
 
Alternatively, we can solve the relations for $|V_{ud}|$ instead of
$|V_{us}|$. Again, the result $|V_{ud}|=0.97428(21)$ which follows from the
lattice data with $\Nf=2+1$ is perfectly consistent with the value
$|V_{ud}|=0.97433(42)$ obtained from those with $\Nf=2$.  The reduction of
the uncertainties in the result for $|V_{ud}|$ due to CKM unitarity is to
be expected from Figure \ref{fig:VusVersusVud}: the unitarity condition
reduces the region allowed by the lattice results to a nearly vertical
interval.

Next, we determine the value of $f_+(0)$ that follows from the lattice data
within the Standard Model. Using CKM unitarity to convert the lattice
determinations of $f_K/f_\pi$ into corresponding values for $f_+(0)$ and
then combining these with the direct determinations of $f_+(0)$, we find
$f_+(0)= 0.9599(38)$ from the data with $\Nf=2+1$ and $f_+(0)= 0.9604(75)$
for $\Nf=2$. The results are shown in the left panel of
Fig.~\ref{fig:lattice data}.

Finally, we work out the analogous Standard Model fits for $f_K/f_\pi$,
converting the direct determinations of $f_+(0)$ into corresponding values
for $f_K/f_\pi$ and combining the outcome with the direct determinations of
that quantity. The results read $f_K/f_\pi=1.1927(50)$ for $\Nf=2+1$ and
$f_K/f_\pi= 1.194(10) $ for $\Nf=2$, respectively; they are shown in the
right panel of Fig.~\ref{fig:lattice data}.

\begin{table}[thb]
\centering
\begin{tabular*}{\textwidth}[l]{@{\extracolsep{\fill}}llllll}
\rule[-0.2cm]{0cm}{0.5cm}& Ref. & \rule{0.3cm}{0cm} $|V_{us}|$&\rule{0.3cm}{0cm} $|V_{ud}|$&\rule{0.25cm}{0cm} $f_+(0)$& \rule{0.2cm}{0cm} $f_K/f_\pi$\\
&&&& \\[-2ex]
\hline \hline
&&&& \\[-2ex]
$\Nf= 2+1$& &\rule{0cm}{0.4cm}0.2253(9)& 0.97428(21)  & 0.9599(38)   & 1.1927(50)\\
&&&& \\[-2ex]
\hline
&&&& \\[-2ex]
$\Nf=2$ & &\rule{0cm}{0.4cm}0.2251(18) &0.97433(42)  &0.9604(75) &1.194(10)\\
&&&& \\[-2ex]
\hline\hline
&&&& \\[-2ex]
$\beta$-decay &\cite{Hardy:2008gy}&0.22544(95)& 0.97425(22) & 0.9595(46)&
1.1919(57) \\ 
&&&& \\[-2ex]
$\tau$-decay &\cite{Gamiz:2007qs}&0.2165(26)&0.9763(6)& 0.999(12)&
1.244(16)\\ 
&&&& \\[-2ex]
$\tau$-decay  &\cite{Maltman:2009bh}&0.2208(39)&0.9753(9)& 0.980(18)&
1.218(23)\\ 
&&&& \\[-2ex]
\hline\hline
\end{tabular*}
\caption{\label{tab:Final results}The upper half of the table shows our
  final results for $|V_{us}|$, $|V_{ud}|$,  $f_+(0)$ and $f_K/f_\pi$,
  which are obtained by analyzing the lattice 
  data within the Standard Model. For comparison, the lower half lists the
  values that follow if the lattice results are replaced by the
  experimental results on nuclear $\beta$ decay and $\tau$ decay,
  respectively.}
\end{table} 

The results obtained by analyzing the lattice data in the framework of the
Standard Model are collected in the upper half of Table \ref{tab:Final
  results}.  In the lower half of this table, we list the analogous
results, found by working out the consequences of CKM-unitarity for the
experimental values of $|V_{ud}|$ and $|V_{us}|$ obtained from nuclear
$\beta$-decay and $\tau$-decay, respectively. The comparison shows that the
lattice result for $|V_{ud}|$ not only agrees very well with the totally
independent determination based on nuclear $\beta$ transitions, but is also
remarkably precise. On the other hand, the values of $|V_{ud}|$, $f_+(0)$
and $f_K/f_\pi$  which follow from the $\tau$-decay data if the Standard
Model is assumed to be valid, are not in good agreement with the lattice
results for these quantities. The disagreement is reduced considerably if
the analysis of the $\tau$ data is supplemented with experimental results
on electroproduction \cite{Maltman:2009bh}: the discrepancy then amounts to
little more than one standard deviation.


\section{Low-energy constants}\label{sec:LECs}

In studying the quark mass dependence of QCD observables calculated on
the lattice it is good practice to invoke {\Ch}PT.
For a given quantity this framework predicts the nonanalytic quark mass
dependence  and it provides symmetry relations among different observables.
These symmetry relations are best expressed with the help of a set of
universal and independent low-energy constants (LECs), which appear as
coefficients of the polynomial terms in different observables. 
If one expands around the SU(2) chiral limit, in the Chiral Effective
Lagrangian there are 2 LECs at order $p^2$ 
\be
F\equiv   F_\pi\,\rule[-0.3cm]{0.01cm}{0.7cm}_{\;m_u,m_d\rightarrow 0} \; , \qquad 
B\equiv -\frac{\lvac \ubar u \rvac}{F_\pi^2}\,\rule[-0.3cm]{0.01cm}{0.9cm}_{\;m_u,m_d\rightarrow 0} \; ,
\ee
and seven at order $p^4$, indicated by $\bar \ell_i$, $i=1, \ldots, 7$.
In the analysis of the SU(3) chiral limit, at order $p^2$ there are also
only two LECs
\be
F_0\equiv   F_\pi\,\rule[-0.3cm]{0.01cm}{0.7cm}_{\;m_u,m_d,m_s\rightarrow 0} \; , \qquad 
B_0\equiv -\frac{\lvac \ubar  u \rvac}{F_\pi^2}\,\rule[-0.3cm]{0.01cm}{0.9cm}_{\;m_u,m_d,m_s\rightarrow 0} \; ,\ee 
but ten at order $p^4$ indicated by the capital letter $L_i(\mu)$,
$i=1,\ldots,10$. These constants are independent of the quark masses,%
\footnote{More precisely, they are independent of the 2 or 3 light
quark masses which are explicitly considered in the respective framework.
However, all low-energy constants depend on the masses of the remaining quarks
$s,c,b,t$ or $c,b,t$ in the SU(2) and SU(3) framework, respectively.}
but they become scale dependent after renormalization (sometimes a superscript
$r$ is added). The SU(2) constants are scale independent, since they are
defined at $\mu=M_\pi$ (as indicated by the bar).
For the precise definition of these constants, their scale dependence
etc, we refer the reader to \cite{Gasser:1983yg,Gasser:1984gg}.

First of all, lattice calculations can be used to test if chiral symmetry is
indeed broken as SU$(\Nf)_L \times $SU$(\Nf)_R \to $SU$(\Nf)_{L+R}$ by measuring
non-zero chiral condensates and by veri\-fying, for instance, the validity of the
GMOR relation $M_\pi^2\propto m$ close to the chiral limit.
If the chiral extrapolation of quantities calculated on the lattice is made
with the help of {\Ch}PT, not only does one get the observable at the physical
value of the quark masses, but one also determines the relevant LECs. This
is a very important by-product for two reasons:
\begin{enumerate}
\item All the LECs up to order $p^4$ (with the exception of $B$ and $B_0$,
  since only the product of these times the quark masses can be estimated from
  phenomenology) have been either determined by comparison to experiments
  or estimated theoretically. A lattice determination of the better known
  ones provides therefore a stringent test of the {\Ch}PT approach.
\item The less well known LECs are those which describe the quark mass
  dependence of observables -- these cannot be determined from experiments,
  and therefore the lattice provides for these unique quantitative
  information. This information is essential for improving those {\Ch}PT
  predictions in which these LECs play a role.
\end{enumerate}
We stress that this program is based on the non-obvious assumption that
{\Ch}PT is valid in the region of masses used in the lattice simulation
under consideration.

The fact that, at large volume, the finite-size effects, which occur if a
system undergoes spontaneous symmetry breakdown, are controlled by the
Nambu-Goldstone modes, was first noted in solid state physics, in
connection with magnetic systems \cite{Fisher:1985zz,Brezin:1985xx}. As
pointed out in \cite{Gasser:1986vb}, in the context of QCD, the thermal
properties of such systems can be studied in a systematic and model
independent manner by means of the corresponding effective field theory,
provided only the temperature is low enough. While finite volumes are not
of physical interest in particle physics, lattice simulations are
necessarily carried out in a box. As shown in
\cite{Gasser:1987ah,Gasser:1987zq,Hasenfratz:1989pk}, the ensuing finite
size effects can also be studied on the basis of the effective theory --
{\Ch}PT in the case of QCD -- provided the simulation is close enough to
the continuum limit, the volume is sufficiently large and the explicit
breaking of chiral symmetry generated by the quark masses is sufficiently
small. Indeed, {\Ch}PT represents a useful tool also for the analysis of
the finite size effects in lattice simulations.

In the following two sections we will summarize the lattice results for the
SU(2) and SU(3) LECs at order $p^2$ and $p^4$. The $O(p^2)$ constants are
determined from the chiral extrapolation of masses and decay constants or,
alternatively, from a finite-size study of correlators in the $\epsilon$-regime
or from spectral observables.
At order $p^4$ some LECs influence two-point functions while others appear only
in three- or four-point functions; the latter need to be determined via a
chiral extrapolation of form factors or scattering amplitudes.
The {\Ch}PT analysis of the phenomenology is nowadays based on $O(p^6)$
formulae.%
\footnote{Some of the $O(p^6)$ formulae presented below have been derived in an
unpublished note by three of us (GC, SD and HL) and J\"urg Gasser. We thank him
for allowing us to publish them here.}
At this level the number of LECs explodes and we will not discuss any of these
here.
We will, however, discuss how comparing different orders and different
expansions ($x$ versus $\xi$, see below) can help to reliably assess the
theoretical uncertainties of the LECs determined on the lattice.

\subsection{SU(2) Low-Energy Constants}\label{sec:su2}

\subsubsection{Quark-mass dependence of pseudoscalar masses and
  decay constants}
\label{sec_MF}

The expansions of $M_\pi^2$ and $F_\pi$ in powers of the quark mass are
known to next-to-next-to-leading order in the SU(2) chiral effective theory.%
\footnote{Here and in the following, we stick to the notation used in the
papers where the {\Ch}PT formulae were established, i.e.\ we work with
$F_\pi\equiv f_\pi/\sqrt{2}=92.2(1)\MeV$ and $F_K\equiv f_K/\sqrt{2}$. The
occurrence of different normalization conventions is not convenient, but
avoiding it by reformulating the formulae in terms of $f_\pi$, $f_K$ is not a
good way out. Since we are using different symbols, confusion cannot arise.
\label{foot:fpi}}
In the isospin limit, $m_u=m_d=m$, the explicit expressions may be written
in the form \cite{Colangelo:2001df}
\begin{eqnarray}\label{eq:MF}
M_\pi^2 & = & M^2\left\{1-\frac{1}{2}x\ln\frac{\Lambda_3^2}{M^2}
  +\frac{17}{8}x^2 \left(\ln\frac{\Lambda_M^2}{M^2}  \right)^2 +x^2 k_M
  +O(x^3)             \right\},
\\
F_\pi & = & F\left\{1+x\ln\frac{\Lambda_4^2}{M^2} -\frac{5}{4}x^2
  \left(\ln\frac{\Lambda_F^2}{M^2}  \right)^2 +x^2k_F   +O(x^3)
\right\}.
 \nonumber
\end{eqnarray}
The expansion parameter is given by
\begin{equation}
x=\frac{M^2}{(4\pi F)^2},\;\;\;\;\;\;\;\;\;\;M^2=2Bm=\frac{2m \Sigma }{F^2}.
\label{eq:xM2}
\end{equation}
The logarithmic scales $\Lambda_3,\Lambda_4$ are related to the effective
coupling constants $\bar\ell_3,\bar\ell_4$ of the chiral Lagrangian at
running scale $M_\pi$:
\begin{equation}
\bar\ell_n=\ln\frac{\Lambda_n^2}{M_\pi^2},\;\;\;\;\;\;\;\;\;\;\;n=1,...,7.
\end{equation}
Note that in Eq.\,(\ref{eq:MF}), the logarithms are evaluated at $M^2$, and
not at $M_\pi^2$. The coupling constants $k_i$ in Eq.\,(\ref{eq:MF}) are
mass-independent.
The scales of the quadratic logarithms can be expressed in terms of the
coupling constants:
\begin{eqnarray}
  \ln\frac{\Lambda_M^2}{M^2} & = &
  \frac{1}{51}\left(28\ln\frac{\Lambda_1^2}{M^2}
    +32\ln\frac{\Lambda_2^2}{M^2}    -9 \ln\frac{\Lambda_3^2}{M^2}+49
  \right),    \\  
  \ln\frac{\Lambda_F^2}{M^2} & = &
  \frac{1}{30}\left(14\ln\frac{\Lambda_1^2}{M^2}
    +16\ln\frac{\Lambda_2^2}{M^2}    +6 \ln\frac{\Lambda_3^2}{M^2} 
    - 6 \ln\frac{\Lambda_4^2}{M^2}      +23  \right).    \nonumber
\end{eqnarray}

By analyzing the quark mass dependence of $M_\pi$ and $F_\pi$ with
Eq.\,(\ref{eq:MF}), possibly truncated at NLO, one can therefore determine%
\footnote{Notice that one could analyze the quark mass dependence entirely in
terms of the parameter $M$ defined in Eq.\,(\ref{eq:xM2}) and determine equally
well all other LECs. Using the determination of the quark masses described in
Sec.~\ref{sec:qmass} one can then extract $B$ or $\Sigma$.}
the $O(p^2)$ LECs $B$ and $F$, as well as the $O(p^4)$ LECs $\bar \ell_3$ and
$\bar \ell_4$.
The quark condensate in the chiral limit is determined by $\Sigma=F^2B$.
With precise enough data at several low pion masses, one could in principle
also determine $\Lambda_M$, $\Lambda_F$ and $k_M$, $k_F$.
This is not yet the case.
The results for the LO and NLO constants will be presented in
Sec.~\ref{sec:su2results}.

Alternatively, one can invert Eq.\,(\ref{eq:MF}) and express $M$ and
$F$ as an expansion in  
\be
\xi \equiv \frac{M_\pi^2}{16 \pi^2 F_\pi^2} \; \; .
\label{eq:xi}
\ee
The corresponding expressions read
\bea\label{eq:MpiFpi} M^2&=& M_\pi^2\,\left\{ 
1+\frac{1}{2}\,\xi\,\lthreebar-
\frac{5}{8}\,\xi^2 \left(\!\lMbar\!\right)^2+
\xi^2 c_{\ind M}+O(\xi^3)\right\} \co \\
F&=& F_\pi\,\left\{1-\xi\,\lfourbar-\frac{1}{4}\,\xi^2
\left(\!\lFbar\!\right)^2
+\xi^2 c_{\ind F}+O(\xi^3)\right\} \fs \nn
\eea
The scales of the quadratic logarithms are determined by $\Lambda_1,\ldots,\Lambda_4$: 
\bea\lMbar&=&\frac{1}{15}\left(28\,\lonebar+32\,\ltwobar-
33\,\lthreebar-12\,\lfourbar +52\right) \co \\
\lFbar&=&\frac{1}{3}\,\left(-7\,\lonebar-8\,\ltwobar+
18\,\lfourbar- \frac{29}{2}\right)\nonumber \fs
\eea

\subsubsection{Two-point correlation functions in the $\epsilon$-regime}\label{sec_eps} 
The finite-size effects encountered in lattice calculations can be used to
determine some of the Low-Energy Constants of QCD. In order to illustrate
the method, we focus on the two lightest quarks, take the isospin limit
$m_u=m_d=m$ and consider a box of size $L_s$ in the three space directions
and size $L_t$ in the time direction. If $m$ is sent to zero at fixed box
size, chiral symmetry is restored. The behaviour of the various observables
in the symmetry restoration region is controlled by the parameter
$\mu=m\,\Sigma\,V$, where $V=L_s^3L_t$ is the volume of the box. Up to a
sign and a factor of two, the parameter $\mu$ represents the minimum of the
classical action that belongs to the leading order effective Lagrangian of
QCD.

For $\mu\gg1$, the system behaves qualitatively as if the box was
infinitely large. In that region, the $p$-expansion, which counts $1/L_s$,
$1/L_t$ and $M$ as quantities of the same order, is adequate. In view of
$\mu=\frac{1}{2}F^2 M^2V $, this region includes configurations with $M
L\gsim\! 1$, where the finite-size effects due to pion loop diagrams are
suppressed by the factor $e^{-M L}$.

If $\mu$ is comparable to or smaller than 1, however, the chiral
perturbation series must be reordered. The $\epsilon$-expansion achieves
this by counting $1/L_s, 1/L_t$ as quantities of $O(\epsilon)$, while the
quark mass $m$ is booked as a term of $O(\epsilon^4)$. This ensures that
the symmetry restoration parameter $\mu$ represents a term of order
$O(\epsilon^0)$, so that the manner in which chiral symmetry is restored
can be worked out.

As an example, we consider the correlator of the axial charge carried by
the two lightest quarks, $q(x)=\{u(x),d(x)\}$.  The axial current and the
pseudoscalar density are given by \be\hspace{1cm} A_\mu^i(x)=
\qbar(x)\mbox{$\frac{1}{2}$} \tau^i\,\gamma_\mu\gamma_5\,q(x)\,,
\hspace{1cm}P^i(x) = \qbar(x)\mbox{$\frac{1}{2}$} \tau^i\, i \gamma_5\,q
(x)\,, \ee where $\tau^1, \tau^2,\tau^3$, are the Pauli matrices. In
euclidean space, the correlators of the axial charge and of the space
integral over the pseudoscalar density are given by
\begin{eqnarray}\label{eq:correlators}
\delta^{ik}C_{AA}(t)\al = \al L_s^3\int \hspace{-0.12cm}d^3\hspace{-0.04cm}\vec{x}\;\langle A_4^i(\vec{x},t)
A_4^k(0)\rangle\,,\\ 
\delta^{ik}C_{PP}(t)\al  =\al L_s^3\int \hspace{-0.12cm}d^3\hspace{-0.04cm}\vec{x}\;\langle P^i(\vec{x},t)
P^k(0)\rangle\,.\nonumber
\end{eqnarray}
{\Ch}PT yields explicit finite-size scaling formulae for these quantities \cite{Hasenfratz:1989pk,Hansen:1990un,Hansen:1990yg}. In the $\epsilon$-regime, the expansion starts with
\begin{eqnarray}  \label{aa-eps}
C_{AA}(t) \al = \al \frac{F^2L_s^3}{L_t}\left[a_A+
  \frac{L_t}{F^2L_s^3}\,b_A\,h_1\hspace{-0.1cm}\left(\frac{t}{L_t}  \right)
+O(\epsilon^4)\right],\\
C_{PP}(t) \al = \al
\Sigma^2L_s^6\left[a_P+\frac{L_t}{F^2L_s^3}\,b_P\,h_1\hspace{-0.1cm}\left(\frac{t}{L_t}  \right)
+O(\epsilon^4)\right],\nonumber 
\end{eqnarray}
where the coefficients $a_A$, $b_A$, $a_P$, $b_P$ stand for quantities of
$O(\epsilon^0)$. They can be expressed in terms of the variables $L_s$,
$L_t$ and $m$ and exclusively involve the two leading low-energy constants
$F$ and $\Sigma$. In fact, at leading order, only the combination
$\mu=m\,\Sigma\, L_s^3 L_t$ matters, the correlators are independent of $t$
and the dependence on $\mu$ is fully determined by the structure of the
groups involved in the spontaneous symmetry breakdown. In the case of
SU(2)$\times$SU(2) $\rightarrow$ SU(2), relevant for QCD in the symmetry
restoration region of the two lightest quarks, the coefficients can be
expressed in terms of Bessel functions. The $t$-dependence of the
correlators starts showing up at $O(\epsilon^2)$, in the form of a
parabola: $h_1(\tau)=\frac{1}{2}\left[\left(\tau-\frac{1}{2}
  \right)^2-\frac{1}{12} \right]$. Explicit expressions for $a_A$, $b_A$,
$a_P$, $b_P$ can be found in
\cite{Hasenfratz:1989pk,Hansen:1990un,Hansen:1990yg}, where some of the
correlation functions are worked out to NNLO. By matching the finite-size
scaling of correlators computed on the lattice with these predictions one
can extract $F$ and $\Sigma$.

The fact that the representation of the correlators to NLO is not
``contaminated'' by higher order unknown LECs, makes the $\epsilon$-regime
potentially convenient for a clean extraction of the LO couplings.  The
determination of the LECs is then affected by different systematic
uncertainties with respect to the standard case; simulations in this regime
yield complementary information which can serve as a valuable cross-check
to get a comprehensive picture of the low-energy properties of QCD.

The effective theory can also be used to study the distribution of the
topological charge in QCD \cite{Leutwyler:1992yt} and the various
quantities of interest may be defined for a fixed value of this charge. The
expectation values and correlation functions then not only depend on the
symmetry restoration parameter $\mu$, but also on the topological charge
$\nu$. The dependence on these two variables can explicitly be calculated.
It turns out that the two-point correlation functions considered above
retain the form (\ref{aa-eps}), but the coefficients $a_A$, $b_A$, $a_P$,
$b_P$ now depend on the topological charge as well as on the symmetry
restoration parameter (see
\cite{Damgaard:2001js,Damgaard:2002qe,Aoki:2009mx} for explicit
expressions).
 
A specific issue with $\epsilon$-regime calculations is the scale setting.
Ideally one would perform a $p$-regime study with the same bare parameters to 
measure a hadronic scale (e.g.\ the proton mass).
In the literature, sometimes a gluonic scale (e.g.\ $r_0$) is used instead; this introduces 
an extra uncertainty of the order of 5\%.

It is important to stress that in the $\epsilon$-expansion higher order
finite-volume corrections might be significant, and the physical box size (in
fm) should still be large in order to keep them under control.
The criteria for the chiral extrapolation and finite volume effects are
obviously different with respect to the $p$-regime.
For these reasons we have to adjust the colour coding defined in
Sect.\,\ref{sec:color-code} (see \ref{sec:su2results} for more details).

Recently, the chiral effective theory has been extended also to the so-called
mixed regime, where some quarks are in the $p$-regime and others in the
$\epsilon$-regime \cite{Bernardoni:2008ei}.
In \cite{Damgaard:2008zs} a technique is proposed to smoothly connect $p$- and
$\epsilon$-regimes.
These theoretical advances can be adopted for future extraction of the Low
Energy Couplings from lattice simulations.   
 
\subsubsection{Energy levels of the QCD Hamiltonian in a box, $\delta$-regime}\label{sec_su2_delta}

At low temperature, the properties of the partition function are governed
by the lowest eigenvalues of the Hamiltonian. In the case of QCD, the
lowest levels are due to the Nambu-Goldstone bosons and can be worked out
with {\Ch}PT \cite{Leutwyler:1987ak}. In the chiral limit, the level
pattern follows the one of a quantum-mechanical rotator: $E_\ell
=\ell(\ell+2)/(2\,\Theta)$, $\ell = 0, 1,2,\ldots$ For a cubic spatial box
and to leading order in the expansion in inverse powers of the box size
$L_s$, the moment of inertia is fixed by the value of the pion decay
constant in the chiral limit: $\Theta=F^2L_s^3$. In order to analyze the
dependence of the levels on the quark masses and on the parameters that
specify the size of the box, a reordering of the chiral series is required,
the so-called $\delta$-expansion; the region where the properties of the
system are controlled by this expansion is referred to as the
$\delta$-regime. Evaluating the chiral perturbation series in this regime,
one finds that the expansion of the partition function goes in even inverse
powers of $FL_s$, that the rotator formula for the energy levels holds up
to NNLO and the expression for the moment of inertia is now also known up
to and including order $(FL_s)^{-4}$
\cite{Hasenfratz:2009mp,Niedermayer:2010mx,Weingart:2010yv}. Since the
level spectrum is governed by the value of the pion decay constant in the
chiral limit, an evaluation of this spectrum on the lattice can be used to
measure $F$. More generally, the evaluation of various observables in the
$\delta$-regime offers an alternative method for a determination of some of
the low energy constants occurring in the effective Lagrangian. At present,
however, the numerical results obtained in this way
\cite{Hasenfratz:2006xi,Bietenholz:2010az} are not yet competitive with
those found in the $p$- or $\epsilon$-regimes.
 
\subsubsection{Other methods for the extraction of the Low-Energy Constants}\label{sec_su2_extra}

An observable that can be used to extract the LECs is the topological
susceptibility, which is defined as
\begin{equation}
\chi_t=\int \hspace{-0.12cm}d^4\hspace{-0.014cm} x\, \langle \omega(x) \omega(0)\rangle,
\end{equation}
where $\omega(x)$ is the topological charge density,
\begin{equation}
\omega(x)=\frac{1}{32\pi^2}\epsilon^{\mu\nu\rho\sigma}{\rm Tr}\left[F_{\mu\nu}(x)F_{\rho\sigma}(x)   \right].
\end{equation}
At infinite volume, the expansion of the topological susceptibility in
powers of the quark masses starts with \cite{DiVecchia:1980ve}
\begin{equation}\label{chi_t}
\chi_t=\overline{m}\,\Sigma \,\{1+O(m)\}\,,\hspace{2cm}\overline{m}\equiv\left(\frac{1}{m_u}+\frac{1}{m_d}+\frac{1}{m_s}+\ldots \right)^{-1}.
\end{equation}
The quark condensate $\Sigma$ can thus be extracted from the properties of
the topological susceptibility close to the chiral limit. The behaviour at
finite volume, in particular also in the region where the symmetry is
restored, are discussed in \cite{Hansen:1990yg}. The dependence on the
vacuum angle $\theta$ and the projection on sectors of given topological
charge have also been investigated \cite{Leutwyler:1992yt}. For a
discussion of the finite-size effects to NLO, including the dependence on
$\theta$, we refer to \cite{Mao:2009sy,Aoki:2009mx}.

Another method to compute directly the quark condensate has been proposed
in \cite{Giusti:2008vb}, where it is shown that starting from the
Banks-Casher relation \cite{Banks:1979yr}, it is possible to extract the
condensate from suitable (renormalizable) spectral observables, for
instance the number of Dirac operator modes
contained in a given interval. For those spectral observables higher order
corrections can be systematically computed in terms of the chiral effective
theory.\\
  
An alternative strategy is based on the fact that at LO in the
$\epsilon$-expansion, the partition function in a given topological sector
$\nu$ is equivalent to the one of a chiral Random Matrix Theory (RMT)
\cite{Shuryak:1992pi,Verbaarschot:1993pm,Verbaarschot:1994qf,Verbaarschot:2000dy}.
In RMT it is possible to extract the probability distributions of
individual eigenvalues
\cite{Nishigaki:1998is,Damgaard:2000ah,Basile:2007ki} in terms of two
dimensionless variables $\zeta=\lambda\Sigma V$ and $\mu=m\Sigma V$, where
$\lambda$ represents the eigenvalue of the massless Dirac operator and $m$
is the sea quark mass. Hence it is possible to match the QCD low-lying
spectrum of the Dirac operator with the RMT predictions in order to
extract\footnote{By introducing an imaginary isospin chemical potential,
  this framework can be extended such that the low spectrum of the Dirac
  operator is sensitive also to the pseudoscalar decay constant $F$ at LO
  \cite{Akemann:2006ru}.} the chiral condensate $\Sigma$.  The main issue
concerning this method is that for the distributions of individual
eigenvalues higher order corrections are still not known by means of the
chiral effective theory, and this may introduce systematic effects which
are not under control.\footnote{Higher order systematic effects in the
  matching with RMT have been recently investigated in
  \cite{Lehner:2010mv,Lehner:2011km}.}\hspace{0.5mm}Another open question is that,
while it is clear how the spectral density is renormalized
\cite{DelDebbio:2005qa}, this is not the case for the individual
eigenvalues, and one has to rely on assumptions. There have been many
lattice studies
\cite{Fukaya:2007yv,Lang:2006ab,DeGrand:2006nv,Hasenfratz:2007yj,DeGrand:2007tm}
where the matching of the low-lying Dirac spectrum with RMT has been
investigated.  In this review we don't include the results of the LECs
obtained in this way.%
\footnote{The results obtained for $\Sigma$ and $F$ lie in the same range
  as the determinations reported in Tables~\protect{\ref{tabsigma} and
    \ref{tabf}}.}
 
\subsubsection{Pion form factors}

The scalar and vector form factors of the pion are defined by the matrix elements
\bea
\langle \pi^i(p_2) |\, \qbar\, q  \, | \pi^j(p_1) \rangle \al = \al
\delta^{ij} F_S^\pi(t) \co  \\
\langle \pi^i(p_2) | \,\qbar\, \mbox{$\frac{1}{2}$}\tau^k \gamma^\mu q\,| \pi^j(p_1)
\rangle \al = \al 
i \,\epsilon^{ikj} (p_1^\mu + p_2^\mu) F_V^\pi(t) \co\nonumber
\eea
where the operators only contain the lightest two quark flavours, $\tau^1$,
$\tau^2$, $\tau^3$ are the Pauli matrices and $t\equiv (p_1-p_2)^2$ denotes
the momentum transfer. 
  
The vector form factor has been measured by several experiments for
timelike as well as for spacelike values of $t$. The scalar
form factor is not measurable but can be evaluated theoretically from data on
the $\pi \pi$ and $\pi K$ phase shifts \cite{Donoghue:1990xh} on the basis
of analyticity and unitarity, {\em i.e.} in a model-independent way.
Lattice calculations can be compared with data or model-independent
theoretical evaluations at different values of $t$. At present, however,
most lattice calculations concentrate on the region close to $t=0$ and on
the evaluation of the slope and curvature, defined as:
\begin{eqnarray}
  F^\pi_V(t) & = & 1+\mbox{$\frac{1}{6}$}\langle r^2 \rangle^\pi_V t + 
c_V\hspace{0.025cm} t^2+\ldots \;\co \\
  F^\pi_S(t) & = & F^\pi_S(0) \left[1+\mbox{$\frac{1}{6}$}\langle r^2
    \rangle^\pi_S t + c_S\, t^2+ \ldots \right] \; \; . \nn
\end{eqnarray}
The slopes are related to the mean quadratic vector and scalar radii
which are the quantities on which most experiments and lattice calculations
concentrate. 

In chiral perturbation theory, the form factors are known at NNLO
\cite{Bijnens:1998fm}. The corresponding formulae are available in fully
analytical form and are compact enough that they can be used for the chiral
extrapolation of the data (as done, for example in
\cite{Frezzotti:2008dr,Kaneko:2008kx}). The expressions for the
scalar and vector radii and for the $c_{S,V}$ coefficients at two-loop
level reads
\bea
\langle r^2 \rangle^\pi_S &=&  \xi \left\{6 \lfourbar-\frac{13}{2} 
-\frac{29}{3}\,\xi \left(\!\ln\frac{\Omega_{r_S}^2}{M_\pi^2} \!\right)^2+ 6
\xi \, k_{r_S}+O(\xi^2)\right\} \co\nn \\ 
\langle r^2 \rangle^\pi_V&=&\xi \left\{ \lsixbar-1 
+2\,\xi \left(\!\ln\frac{\Omega_{r_V}^2}{M_\pi^2} \!\right)^2+6 \xi \,k_{r_V}+O(\xi^2)\right\}
 \co  \\
c_S&=&\xi \left\{\frac{19}{120}  + \xi \left[ \frac{43}{36} \left(\!
      \ln\frac{\Omega_{c_S}^2}{M_\pi^2} \!\right)^2 + k_{c_S} \right]
\right\} \co \nn \\
c_V&=&\xi \left\{\frac{1}{60}+\xi \left[\frac{1}{72} \left(\!
      \ln\frac{\Omega_{c_V}^2}{M_\pi^2} \!\right)^2 + k_{c_V} \right]
\right\} \co \nn
\eea
where 
\bea
\ln\frac{\Omega_{r_S}^2}{M_\pi^2}&=&\frac{1}{29}\,\left(31\,\lonebar+34\,\ltwobar -36\,\lfourbar
  +\frac{145}{24}\right)  \co \nn\\
\ln\frac{\Omega_{r_V}^2}{M_\pi^2}&=&\frac{1}{2}\,\left(\lonebar-\ltwobar+\lfourbar+\lsixbar 
-\frac{31}{12}\right) \co  \\
\ln\frac{\Omega_{c_S}^2}{M_\pi^2}&=&\frac{43}{63}\,\left(11\,\lonebar+14\,\ltwobar+18\,\lfourbar
  -\frac{6041}{120}\right) 
\co \nn \\
\ln\frac{\Omega_{c_V}^2}{M_\pi^2}&=&\frac{1}{72}\,\left(2\lonebar-2\ltwobar-\lsixbar 
-\frac{26}{30}\right) \co \nn
\eea
and $k_{r_{S,V}}$ and $k_{c_{S,V}}$ are four constants independent of the quark masses.
Their expression in terms of the $\ell_i$ and of the $O(p^6)$ constants
$c_i$ is known but will not be reproduced here.

\subsubsection{Results}\label{sec:su2results}
 
\begin{figure}[!tb]
\vspace{-2cm}
\centering
\includegraphics[width=10.0cm]{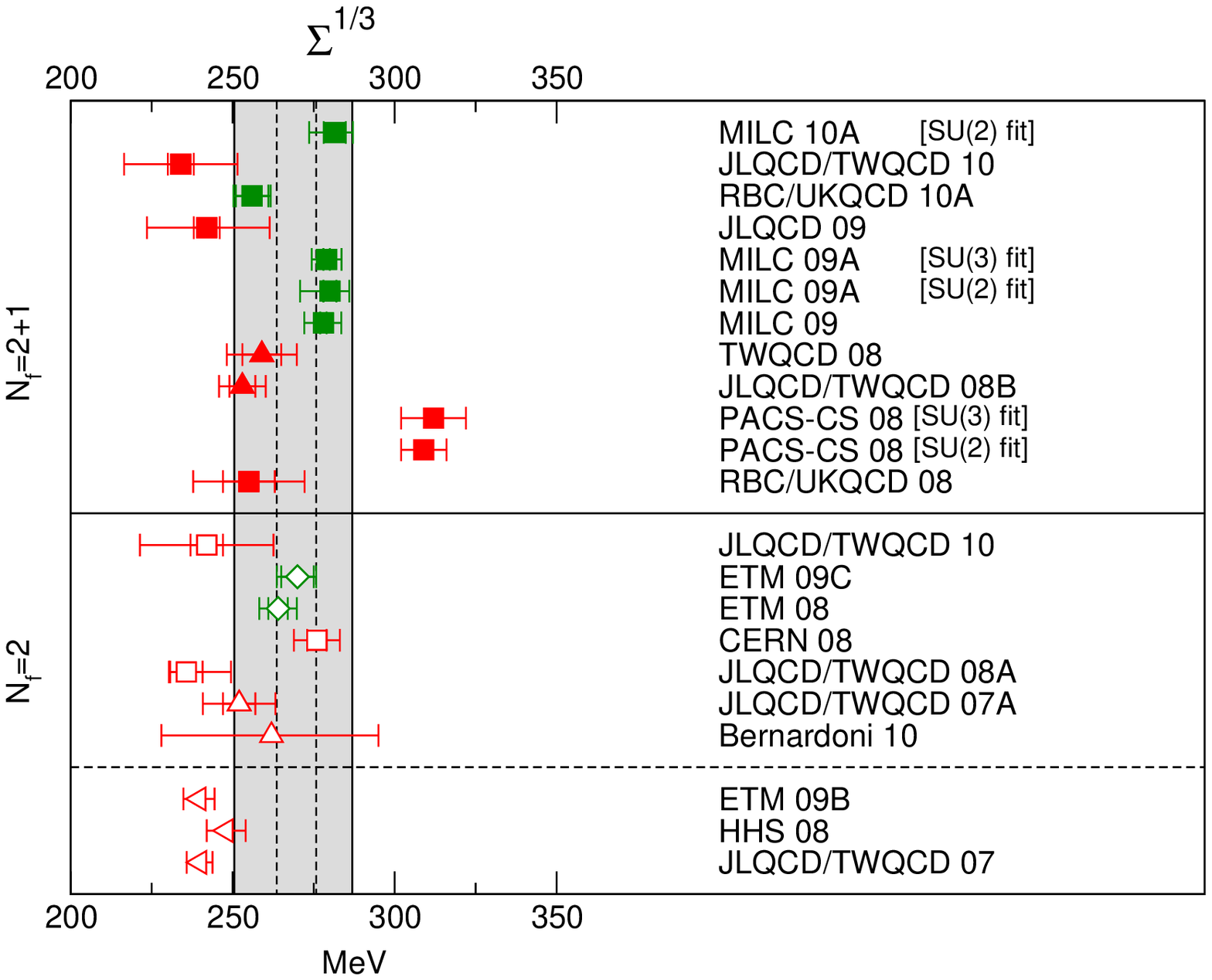}
\caption{Summary of lattice results for the quark condensate $\Sigma$ (in
  the $\overline{\rm MS}$ scheme at scale $\mu=2$ GeV). The filled symbols
  represent lattice simulations with $\Nf=2+1$ dynamical flavours, while
  the empty ones refer to $\Nf=2$ computations. Squares and left triangles
  indicate determinations from correlators in the $p$- and
  $\epsilon$-regimes, respectively. Up triangles refer to extractions from
  the topological susceptibility, diamonds to determinations from the pion
  form factor. The gray (dashed) band indicates our estimate for $\Nf = 2+1
  (\Nf = 2).$ }\label{fig_sigma}
\end{figure}

\begin{figure}[!t]
\vspace{-1cm}
\centering
\includegraphics[width=9.0cm]{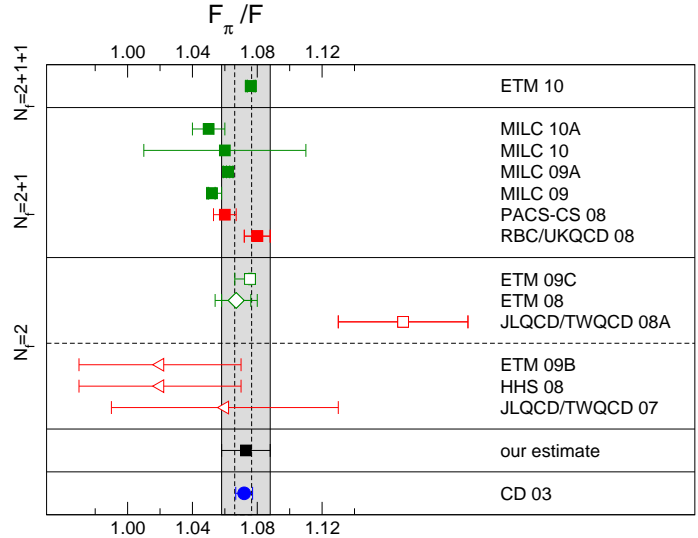}
\caption{Summary of lattice results for the ratio $F_\pi/F$, which compares
  the physical value of the pion decay constant with the value in the limit
  $m_u,m_d\rightarrow 0$. The meaning of the symbols is the same as in  
Fig.\,\ref{fig_sigma}. The gray band here represents our estimate, whereas
the dashed one the current $\Nf=2$ value.}
\label{fig_f} 
\end{figure}

\begin{figure}[!p]
\vspace{-3cm}
\centering
\psfrag{lbar3}{\sffamily $\lbar_3$}
\includegraphics[width=8.5cm]{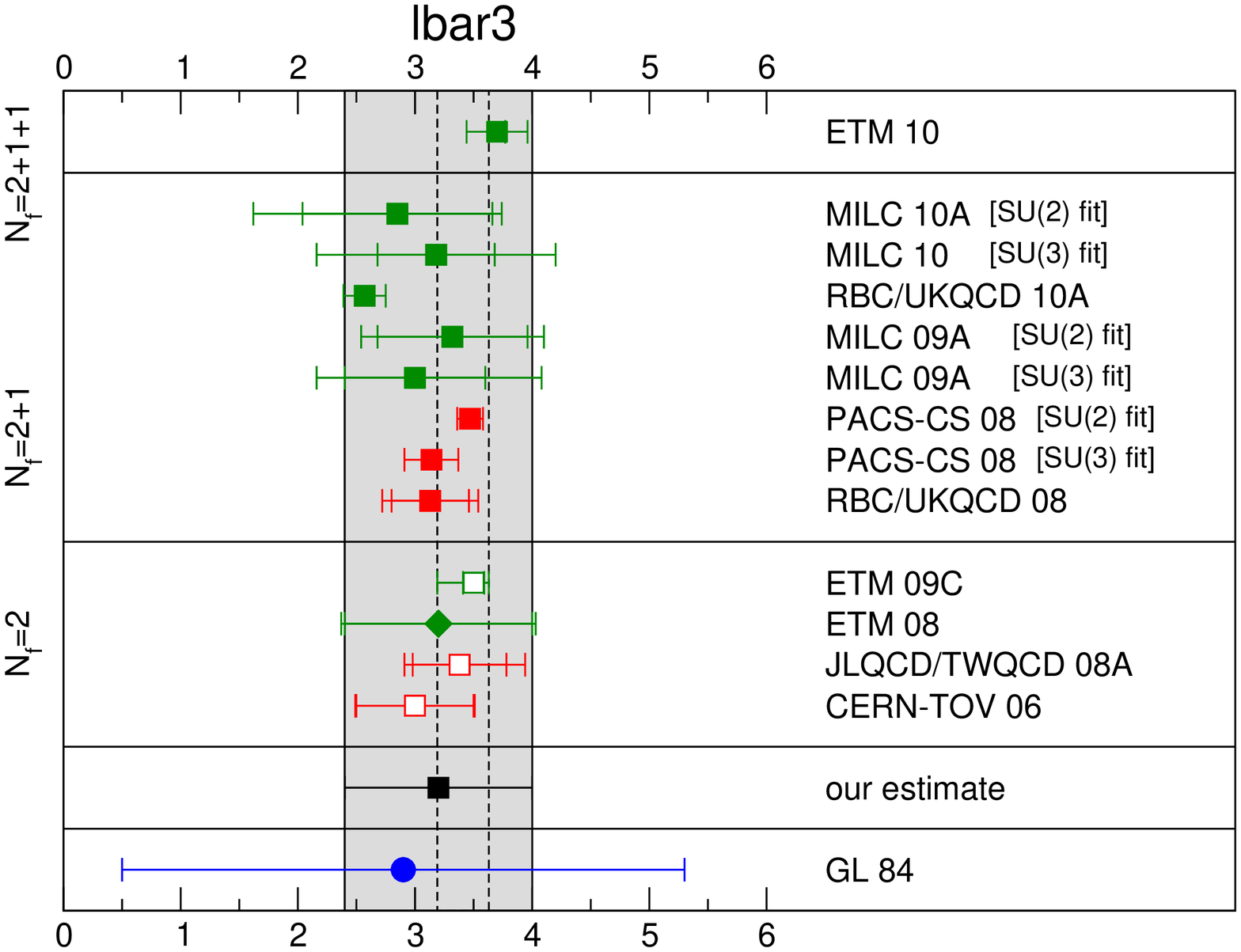}

\vspace{1cm}
\psfrag{lbar4}{\sffamily $\lbar_4$}
\includegraphics[width=8.5cm]{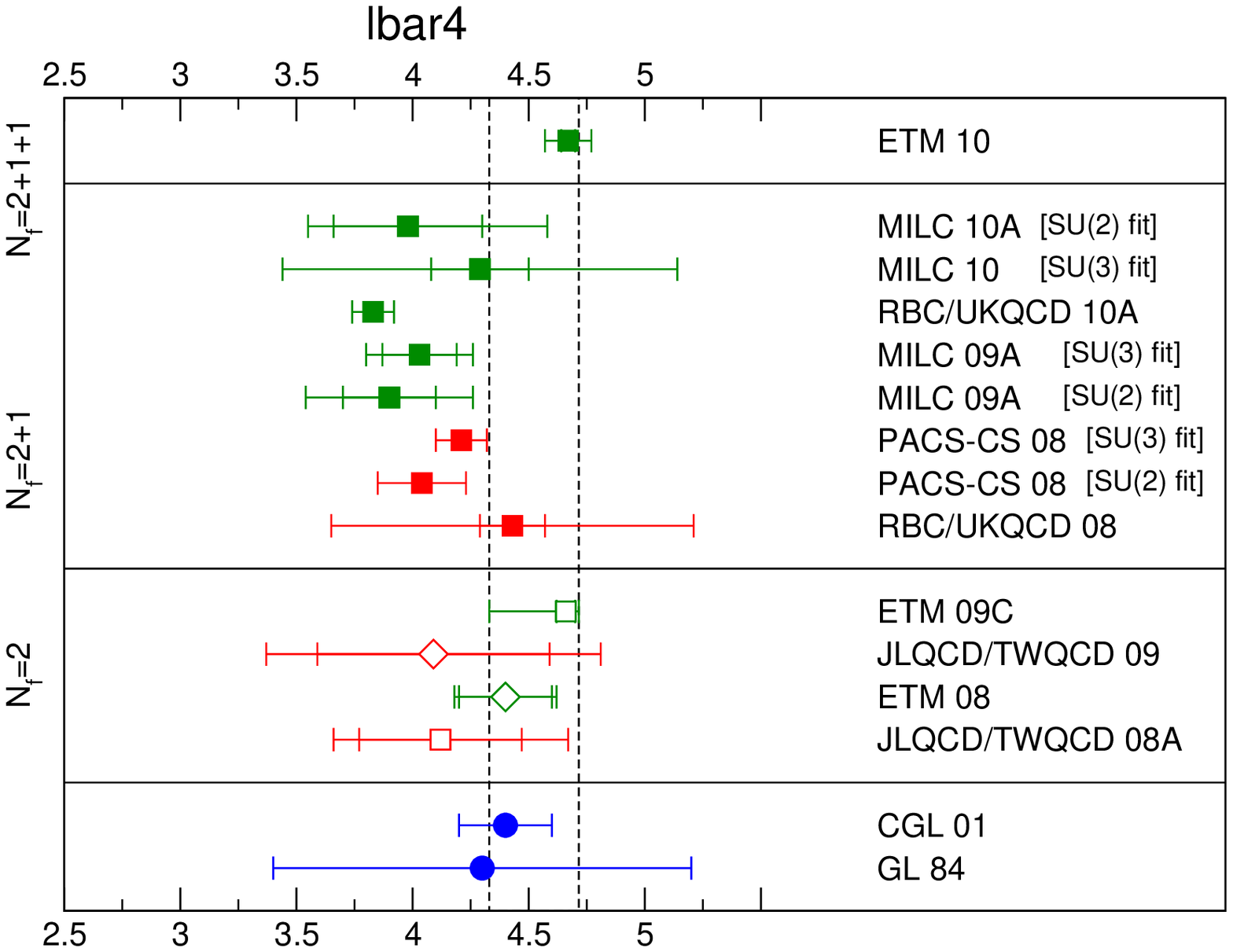}

\vspace{1cm}
\psfrag{lbar6}{\sffamily $\lbar_6$}
\includegraphics[width=8.5cm]{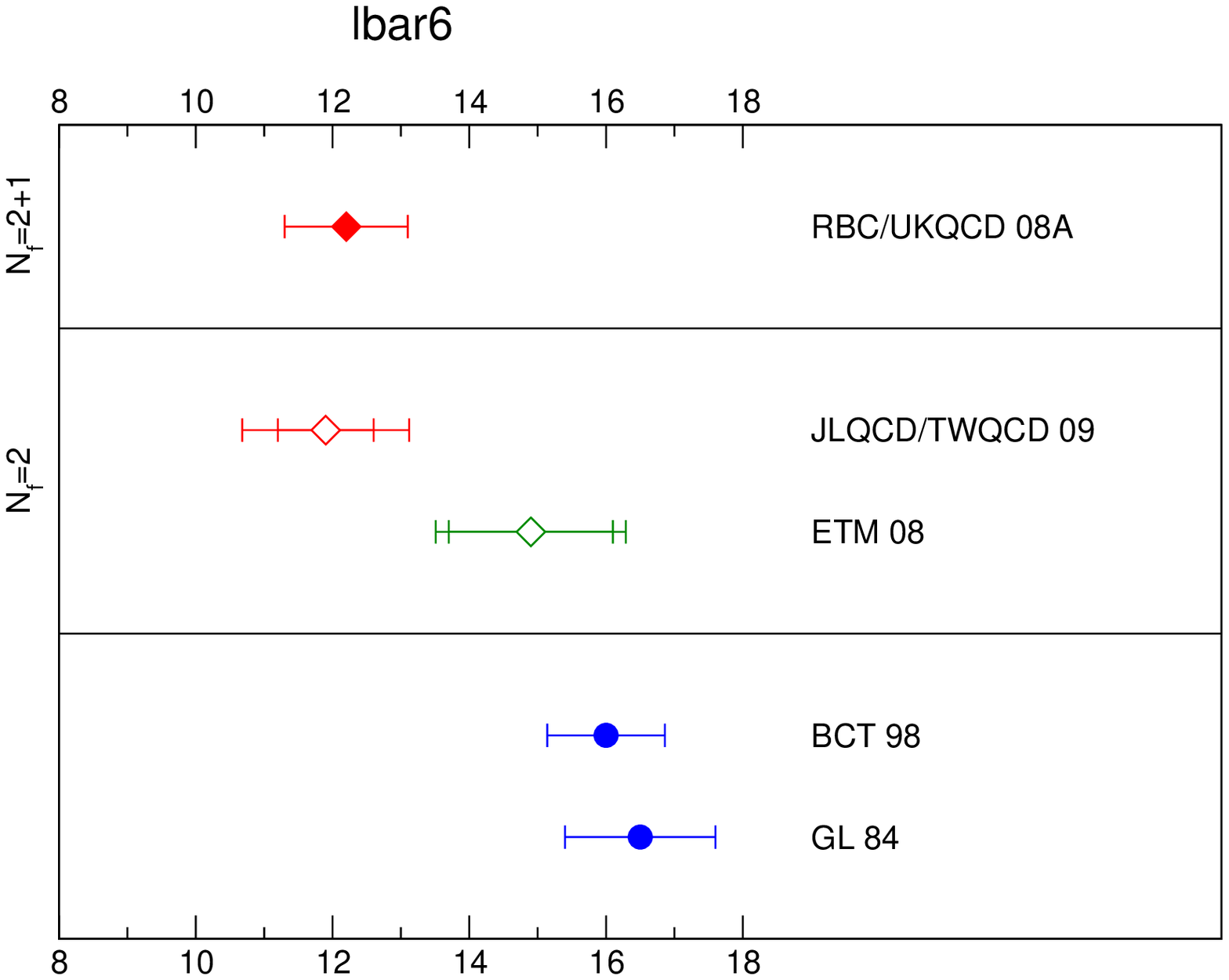}
\caption{Summary of lattice results for the effective coupling constants
  $\bar \ell_3,\bar\ell_4$ and $\bar\ell_6$. The filled symbols refer to
  lattice calculations with $\Nf=2+1$, while the empty ones represent
  simulations with $\Nf=2$. Squares indicate determinations from
  correlators in the $p$-regime, diamonds refer to determinations from the
  pion form factor. The gray band represents our estimate of $\lbar_3$, whereas the 
  dashed ones indicate the current $\Nf=2$ values of $\lbar_3$ and
  $\lbar_4$.}\label{fig_l3l4l6}  
\end{figure}

In this section we summarize the lattice results for the SU(2) couplings in
a set of tables (\ref{tabsigma}--\ref{tabl6}) and figures
(\ref{fig_sigma}--\ref{fig_l3l4l6}).
The tables present our usual colour coding which summarizes the main aspects
related to the treatment of the systematic errors of the various
calculations.

A delicate issue in the lattice determination of chiral LECs (in particular at
NLO) which cannot be reflected by our colour coding is a reliable assessment of
the theoretical error.
We add a few remarks on this point:
\begin{enumerate}
\item
Using \emph{both} the $x$ and the $\xi$ expansion is a good way to test how
the ambiguity of the chiral expansion (at a given order) affects the numerical
values of the LECs that are determined from a particular set of data.
For instance, to determine $\bar{\ell}_4$ (or $\Lambda_4$) from lattice data
for $\Fpi$ as a function of the quark mass, one may compare the fits based
on the parametrization 
$\Fpi=F\{1+x\ln(\Lambda_4^2/M^2)\}$ [see Eq.\,(\ref{eq:MF})] with those
obtained from  
$\Fpi=F/\{1-\xi\ln(\Lambda_4^2/\Mpi^2)\}$ [see Eq.\,(\ref{eq:MpiFpi})].
The difference between the two results provides an estimate
of the uncertainty due to the truncation of the chiral series.%
\footnote{Notice however that this difference could be accidentally small and
then lead to an underestimate of the true systematic error.}
Which central value one chooses is in principle arbitrary, but we find it
advisable to use the one obtained with the $\xi$ expansion,%
\footnote{There are theoretical arguments suggesting that the $\xi$ expansion
is preferable to the $x$ expansion, based on the observation that the
coefficients in front of the squared logs in (\ref{eq:MF}) are somewhat larger
than in (\ref{eq:MpiFpi}). This can be traced to the fact that a part of every
formula in the $x$ expansion is concerned with locating the position of the
pion pole (at the previous order) while in the $\xi$ expansion the knowledge of
this position is built in exactly. Numerical evidence supporting this view is
presented in \cite{Noaki:2008iy}.} in particular because it makes the
comparison with phenomenological determinations (where it is an established
practice to use the $\xi$ expansion) more meaningful. 
The fact that there is an apparent discrepancy in Tab.\,\ref{tabl6} between the
lattice and the phenomenological determination of $\bar\ell_1\!-\!\bar\ell_2$,
but not for $\<r^2\>_S^\pi$ (on which it is based), calls for a re-assessment
of the theore\-ti\-cal error which includes an $x$ versus $\xi$ expansion
analysis. 
\item As an alternative one could try to estimate the influence of higher
  chiral orders by reshuffling irrelevant higher-order terms.  For
  instance, in the example mentioned above one might use
  $\Fpi=F/\{1-x\ln(\Lambda_4^2/M^2)\}$ as a different functional form at
  NLO.  Another way to make this estimate is through introducing by hand
  ``analytic'' higher-order terms (e.g.\ ``analytic NNLO''), as done, in
  the past, by MILC \cite{Bazavov:2009bb}. In principle it would be
  preferable to include all NNLO terms or none, such that the structure of
  the chiral expansion is preserved at any order; this is what ETM
  \cite{Baron:2009wt} and JLQCD/TWQCD \cite{Noaki:2008iy} are doing now for
  SU(2) {\Ch}PT and MILC for SU(3) {\Ch}PT \cite{Bazavov:2009fk}.  In any case
  it is not advisable to include terms to which the data are not sensitive.
  The use of priors in fits has not yet found general consensus in the
  lattice community.
\item
Another issue concerns the $s$-quark mass dependence of the LECs
$\bar{\ell}_i$ or $\Lambda_i$ in the SU(2) framework. This has been studied
analytically in a series of papers
\cite{Gasser:1984gg,Gasser:2007sg,Gasser:2009hr}. An analysis of the
difference of the LECs determinations done in $\Nf=2$ or $\Nf=2+1$ lattice
simulations provides a unique means to analyze quantitatively this
dependence.
\item Finally, the determination of the LECs is in general affected by
  discretization effects, and it is important that these are corrected for
  with the help of the appropriate effective chiral Lagrangian (as done,
  {\em e.g.}  by MILC). For actions which respect chiral symmetry at finite
  lattice spacing (like the overlap formulation adopted by JLQCD/TWQCD or
  domain wall fermions used by RBC/UKQCD in the limit of large lattice size
  in the fifth dimension), this step is unnecessary.
\end{enumerate}

In the tables and figures we summarize the results of various lattice
collaborations for the SU(2) LECs at leading order ($F$ or $F/F_\pi$, $B$ or
$\Sigma$) and at NLO ($\bar\ell_1-\bar\ell_2$,
$\bar\ell_3$, $\bar\ell_4$, $\bar\ell_5$, $\bar\ell_6$).
The tables group the results into those which stem from $\Nf=2+1$
calculations and those which stem from $\Nf=2$ calculations.  Furthermore,
we make a distinction whether the results are obtained from simulations in
the $p$-regime or whether alternative methods ($\epsilon$-regime, spectral
quantities, topological susceptibility, etc.) have been used.
For comparison we add, in each case, a few phenomenological determinations
with high standing. 

There is only one set of $\Nf=2$ lattice results for the SU(2) LECs which
does not have any red tag, the one by the ETM collaboration
\cite{Frezzotti:2008dr,Baron:2009wt}. These results are at present the
lattice determinations of these quantities with the best control of
systematic effects (for $\Nf=2$ and in some cases even overall) and the
only ones which qualify for making an average according to our criteria. We
therefore offer no $\Nf=2$ averages, and the dashed bands in the plots
coincide with the lattice points of the ETM collaboration.

\begin{table}[!tb]
\vspace{3.0cm}
\footnotesize
\begin{tabular*}{\textwidth}[l]{l@{\extracolsep{\fill}}rlllllll}
Collaboration & Ref. & $\Nf$ &
\hspace{0.15cm}\begin{rotate}{60}{publication status}\end{rotate}\hspace{-0.15cm} &
\hspace{0.15cm}\begin{rotate}{60}{chiral extrapolation}\end{rotate}\hspace{-0.15cm}&
\hspace{0.15cm}\begin{rotate}{60}{continuum  extrapolation}\end{rotate}\hspace{-0.15cm} &
\hspace{0.15cm}\begin{rotate}{60}{finite volume}\end{rotate}\hspace{-0.15cm} &
\hspace{0.15cm}\begin{rotate}{60}{renormalization}\end{rotate}\hspace{-0.15cm} & \rule{0.2cm}{0cm}$\Sigma^{1/3}$ [MeV] \\
&&&&&&&& \\[-0.1cm]
\hline
\hline
&&&&&&&& \\[-0.1cm]
MILC 10A       & \cite{Bazavov:2010yq}    & 2+1 & \rC & \soso & \good & \good & \soso & 281.5(3.4)$\binom{+2.0}{-5.9}$(4.0) \\
JLQCD/TWQCD 10 & \cite{Fukaya:2010na}     & 2+1 & \gA & \good & \bad  & \soso & \good &  234(4)(17)   \\
RBC/UKQCD 10A  & \cite{Aoki:2010dy}       & 2+1 & \oP & \soso & \soso & \good & \good & 256(5)(2)(2)  \\
JLQCD 09       & \cite{Fukaya:2009fh}     & 2+1 & \gA & \good & \bad  & \soso & \good & 242(4)$\binom{+19}{-18}$\\
MILC 09A       & \cite{Bazavov:2009fk}    & 2+1 & \rC & \soso & \good & \good & \soso & 279(1)(2)(4)  \\
MILC 09A       & \cite{Bazavov:2009fk}    & 2+1 & \rC & \soso & \good & \good & \soso & 280(2)$\binom{+4}{-8}$(4)\\
MILC 09        & \cite{Bazavov:2009bb}    & 2+1 & \gA & \soso & \good & \good & \soso & 278(1)$\binom{+2}{-3}$(5)\\
TWQCD 08       & \cite{Chiu:2008jq}       & 2+1 & \gA & \soso & \bad  & \bad  & \good & 259(6)(9)     \\
JLQCD/TWQCD 08B& \cite{Chiu:2008kt}       & 2+1 & \rC & \soso & \bad  & \bad  & \good & 253(4)(6)     \\
PACS-CS 08     & \cite{Aoki:2008sm}       & 2+1 & \gA & \good & \bad  & \bad  & \bad  & 312(10)       \\
PACS-CS 08     & \cite{Aoki:2008sm}       & 2+1 & \gA & \good & \bad  & \bad  & \bad  & 309(7)        \\
RBC/UKQCD 08   & \cite{Allton:2008pn}     & 2+1 & \gA & \soso & \bad  & \good & \good & 255(8)(8)(13) \\
&&&&&&&& \\[-0.1cm]
\hline
&&&&&&&& \\[-0.1cm]
Bernardoni 10  & \cite{Bernardoni:2010nf} &  2  & \gA & \soso & \bad  & \bad  & \good & 262$\binom{+33}{-34}\binom{+4}{-5}$\\
JLQCD/TWQCD 10 & \cite{Fukaya:2010na}     &  2  & \gA & \good & \bad  & \bad  & \good & 242(5)(20)    \\
ETM 09C        & \cite{Baron:2009wt}      &  2  & \gA & \soso & \good & \soso & \good & 270(5)$\binom{+3}{-4}$ \\
ETM 08         & \cite{Frezzotti:2008dr}  &  2  & \gA & \soso & \soso & \soso & \good & 264(3)(5)     \\
CERN 08        & \cite{Giusti:2008vb}     &  2  & \gA & \soso & \bad  & \soso & \good & 276(3)(4)(5)  \\
JLQCD/TWQCD 08A& \cite{Noaki:2008iy}      &  2  & \gA & \soso & \bad  & \bad  & \good & 235.7(5.0)(2.0)$\binom{+12.7}{-0.0}$\\
JLQCD/TWQCD 07A& \cite{Aoki:2007pw}       &  2  & \gA & \soso & \bad  & \bad  & \good & 252(5)(10)    \\
&&&&&&&& \\[-0.1cm]
\hline
&&&&&&&& \\[-0.1cm]
ETM 09B        & \cite{Jansen:2009tt}     &  2  & \rC & \good & \soso & \bad  & \good & 239.6(4.8) \\
HHS 08         & \cite{Hasenfratz:2008ce} &  2  & \gA & \good & \bad  & \soso & \good & 248(6)     \\
JLQCD/TWQCD 07 & \cite{Fukaya:2007pn}     &  2  & \gA & \good & \bad  & \bad  & \good & 239.8(4.0) \\
&&&&&&&& \\[-0.1cm]
\hline
\hline
\end{tabular*}
\normalsize
\caption{Lattice results for the quark condensate $\Sigma$, in the
  $\overline{\rm MS}$ scheme at scale $\mu=2$ GeV. We separate $\Nf=2+1$
  from $\Nf=2$ results and, in the latter case, results obtained in the
  $p$-regime (middle) from those obtained in the $\epsilon$-regime
  (bottom). For MILC 09A and PACS-CS 08, converted values from SU(3) fits
  and values from direct SU(2) fits are included.
\label{tabsigma}}
\end{table}

The determination of the SU(2) LECs in $\Nf=2+1$ lattice calculations
requires more discussion: MILC has published values for $\Sigma$, $F$ and
$F_\pi/F$ in their summary paper \cite{Bazavov:2009bb}, but
not for the $\bar \ell_i$, which were instead extracted from a SU(2) fit
only in conference proceedings \cite{Bazavov:2009fk,Bazavov:2010yq}. As
discussed in the previous sections, these results have no red tag and those
presented in a regular article fully qualify for entering our averages. The
RBC/UKQCD collaboration has calculated all the SU(2) LECs discussed
here. After a first study with a single lattice spacing
\cite{Allton:2008pn}, a more recent publication with 2 lattice spacings
appeared \cite{Aoki:2010dy}.  
Those results have no red tag and are included in our averages.
PACS-CS \cite{Aoki:2008sm} has also performed a comprehensive analysis of
SU(2) LECs, and at remarkably small quark masses -- their calculation has
two red tags related to the continuum extrapolation and finite-size effects
and should also not be considered in averages. Finally JLQCD 
\cite{Fukaya:2009fh,Fukaya:2010na} has performed a calculation of $\Sigma$
(but no other SU(2) LECs) with $\Nf=2+1$ overlap dynamical quarks, but at a
single lattice spacings and has also a red tag. 

The ETM collaboration has recently started performing calculations with
$\Nf=2+1+1$ dynamical quarks and a first paper \cite{Baron:2010bv} on the
SU(2) LECs has appeared (a more recent paper \cite{Baron:2011sf} gives a
progress report but not yet an update of the LECs).
Assigning a color code to this calculation is not straightforward because
of the even larger splitting between the neutral and charged pion than
observed in $\Nf=2$ calculations, but as explained in the footnote in
Table~\ref{tabf} we did not assign it any red tag. We also refer to the
careful analysis by B\"ar \cite{Bar:2010jk} on the effect of this splitting
on the LEC determination. All this considered, we compare and take into
account their numbers when drawing our conclusions on $\Nf=2+1$
calculations of the SU(2) LECs.

A generic comment applies to the issue of the scale setting.  Since none of
the lattice studies involves simulations in the $p$-regime at the physical
value of $m_{ud}$, the setting of the scale $a^{-1}$ via an experimentally
measurable quantity involves a chiral extrapolation. As a result,
dimensionful quantities are particularly sensitive to this scale-setting
ambiguity, while in dimensionless ratios such as $F_\pi/F$, $F/F_0$,
$B/B_0$, $\Sigma/\Sigma_0$, the problem is much reduced, and often finite
lattice-to-continuum renormalization factors drop out.  In those cases
where the collaboration offers separate results for numerator and
denominator but not for the ratio, we have calculated the quotient
ourselves. Since the systematic errors are dominated by those due to the
renormalization of the scale and these cancel in $F_\pi/F$, we have dropped
the systematic errors altogether -- a lattice determination of the ratio
that properly accounts for the correlations would of course be preferable.
In the tables, results obtained in this way are shown in slanted fonts.
Since the difference between $F_\pi$ and $F$ manifests itself only at NNLO
of the $\epsilon$-expansion, the data taken in the $\epsilon$-regime do not
by themselves allow a determination of the ratio $F_\pi/F$.  The
corresponding entries in Tab.\,\ref{tabf} and Fig.\,\ref{fig_f} are
obtained by dividing the physical value of $F_\pi$ (cf.\ footnote
\ref{foot:fpi}) through the lattice result for $F$.

It is worth repeating here that the standard colour coding scheme of our
tables is necessarily schematic and cannot do justice to every calculation.
For instance, in the $\epsilon$-regime we routinely assign%
\footnote{Also for \cite{Fukaya:2009fh} and for \cite{Fukaya:2010na} (for
  $N_f=2+1,3$) the colour-coding criteria for the $\epsilon$-regime 
have been applied.}  a green star in the ``chiral
extrapolation'' column (since the pertinent expressions directly involve
$F$ and $B$, even if the calculation is done at finite quark mass), while
the ``infinite volume'' assessment is exclusively based on the number of
different volumes with $L>1.5\,\fm$ have been used (the $\Mpi L$ criterion
does not make sense here). Similarly, in the calculation of form factors
and charge radii the tables do not reflect whether an interpolation to the
desired $q^2$ has been performed or whether the evaluation has taken place
directly at the relevant $q^2$, or very close to it, by means of
``partially twisted boundary conditions'' \cite{Boyle:2008yd}.
Nevertheless, we feel that these tables give an adequate overview of the
qualities of the various calculations.

For most quantities our tables and figures show an overall reasonable
consistency among the various lattice determinations. One notable
exception is the case of $\Sigma$. The PACS-CS\,08 determination
\cite{Aoki:2008sm} of this quantity is in flat disagreement with all other
calculations.  A glimpse at Tab.\,\ref{tabsigma} reveals that this
calculation is the only one which received three red boxes.  In particular,
as was pointed out by the authors, for this quantity perturbative
renormalization may lead to theoretical errors which are hard to quantify. 
But even among the two $N_f=2+1$ calculations without red tags, MILC 10A
and RBC/UKQCD 10A there is some tension, at the level of 2.6 sigmas. As
pointed out above, the origin of the problem may lie in the scale setting
issue: the ratio $\Sigma/\Sigma_0$ would be a much more stringent test.
In any case, since it is not meaningful to average these two calculations,
the estimate shown in Figure~\ref{fig_sigma}, $\Sigma=269(18)$ is
constructed so as to cover both of them.
\begin{table}[!tb]
\footnotesize
\begin{tabular*}{\textwidth}[l]{l@{\extracolsep{\fill}}rllllllll}
Collaboration & Ref. & $\Nf$ & 
\hspace{0.15cm}\begin{rotate}{60}{publication status}\end{rotate}\hspace{-0.15cm}&
\hspace{0.15cm}\begin{rotate}{60}{chiral extrapolation}\end{rotate}\hspace{-0.15cm}&
\hspace{0.15cm}\begin{rotate}{60}{continuum  extrapolation}\end{rotate}\hspace{-0.15cm} &
\hspace{0.15cm}\begin{rotate}{60}{finite volume}\end{rotate}\hspace{-0.15cm} & 
\hspace{0.15cm}\begin{rotate}{60}{renormalization}\end{rotate}\hspace{-0.15cm} &\rule{0.1cm}{0cm}
 $F$ [MeV] &\rule{0.1cm}{0cm} $F_\pi/F$\\
&&&&&&&&& \\[-0.1cm]
\hline
\hline
&&&&&&&&& \\[-0.1cm]
ETM 10 & \cite{Baron:2010bv} &  2+1+1 & \gA & \soso & \soso &{\soso$^\dagger$\hspace{-0.2cm}}&\good & 85.66(6)(13)& 1.076(2)(2)  \\ 
&&&&&&&&& \\[-0.1cm]
\hline
&&&&&&&&& \\[-0.1cm]
MILC 10A       & \cite{Bazavov:2010yq}    & 2+1 & \rC & \soso & \good & \good & \soso & 87.5(1.0)($^{+0.7}_{-2.6}$) & {\sl 1.05(1)}\\
MILC 10        & \cite{Bazavov:2010hj}    & 2+1 & \rC & \soso & \good & \good & \soso & 87.0(4)(5)     & {\sl 1.06(5)}\\
MILC 09A       & \cite{Bazavov:2009fk}    & 2+1 & \rC & \soso & \good & \good & \soso & 86.8(2)(4)     & 1.062(1)(3)            \\
MILC 09        & \cite{Bazavov:2009bb}    & 2+1 & \gA & \soso & \good & \good & \soso &                & 1.052(2)($^{+6}_{-3}$) \\
PACS-CS 08     & \cite{Aoki:2008sm}       & 2+1 & \gA & \good & \bad  & \bad  & \bad  & 89.4(3.3)      & 1.060(7)               \\
RBC/UKQCD 08   & \cite{Allton:2008pn}     & 2+1 & \gA & \soso & \bad  & \good & \good & 81.2(2.9)(5.7) & 1.080(8)       \\ 
&&&&&&&&& \\[-0.1cm]
\hline
&&&&&&&&& \\[-0.1cm]
ETM 09C        & \cite{Baron:2009wt}      &  2  & \gA & \soso & \good & \soso & \good &                & 1.0755(6)($^{+8}_{-94}$) \\
ETM 08         & \cite{Frezzotti:2008dr}  &  2  & \gA & \soso & \soso & \soso & \good & 86.6(7)(7)     & 1.067(9)(9) \\
JLQCD/TWQCD 08A& \cite{Noaki:2008iy}      &  2  & \gA & \soso & \bad  & \bad  & \good & 79.0(2.5)(0.7)($^{+4.2}_{-0.0}$) & {\sl 1.17(4)} \\
&&&&&&&&& \\[-0.1cm]
\hline
&&&&&&&&& \\[-0.1cm]
ETM 09B        & \cite{Jansen:2009tt}     &  2  & \rC & \good & \soso & \bad  & \good & 90.2(4.8)$^\S$ & {\sl 1.02(5)} \\
HHS 08         & \cite{Hasenfratz:2008ce} &  2  & \gA & \good & \bad  & \soso & \good & 90(4)          & {\sl 1.02(5)} \\
JLQCD/TWQCD 07 & \cite{Fukaya:2007pn}     &  2  & \gA & \good & \bad  & \bad  & \good & 87.3(5.6)      & {\sl 1.06(7)} \\
&&&&&&&&& \\[-0.1cm]
\hline
&&&&&&&&& \\[-0.1cm]
CD 03          & \cite{Colangelo:2003hf}  &     &     &       &      &      &      & 86.2(5)        & 1.0719(52) \\
&&&&&&&&& \\[-0.1cm]
\hline
\hline
\end{tabular*}

\vspace{0.4cm}
$^\dagger$ The values of $M_{\pi^+} L$ correspond to a green tag, while
those of $M_{\pi^0} L$ imply a red one -- since both masses 
\vspace{-0.025cm}\hspace{0.3cm}play a role in finite-volume effects, we opt for orange.

\vspace{-0.05cm}
$^\S$ Result for $r_0 F$ converted into a value for $F$ with $r_0=0.49$ fm. 
\normalsize
\caption{\label{tabf}Results for the leading order SU(2) low energy
  constant $F$ and for the ratio $F_\pi/F$.   
  Numbers in slanted fonts have been calculated by us (see text for
  details). Horizontal lines establish the same grouping as in
  Tab.\,\ref{tabsigma}. The table shows that the precision reached on the
  lattice is now comparable to the accuracy of 
  the prediction, which is indicated in the last row. 
}
\end{table}

As far as $F_\pi/F$ is concerned,
there is some tension among the MILC and the RBC/UKQCD results, with the
latest ETM $\Nf=2+1+1$ agreeing with the latter. The tension is at the
level of about two sigmas, but it is worth stressing that the quoted
lattice uncertainties are all below one percent. We believe it is fair and
useful to quote an estimate which covers all three lattice determinations
and has an uncertainty of about 1.5\%,
\begin{equation}
\frac{F_\pi}{F}= 1.073(15)  \qquad \mbox{our estimate}.
\end{equation}
The ETM result \cite{Baron:2009wt} shows that the value obtained with
$\Nf=2$ dynamical flavours is the same within errors. \footnote{On the
  other hand, the result of JLQCD/TWQCD \cite{Noaki:2008iy} shows a 2-sigma
  deviation from the other results obtained with $N_f=2$. Since the quoted
  value for $F_\pi/F$ has been obtained by dividing the physical value of
  $F_\pi$ by the lattice result, we can't exclude that systematic errors
  are properly taken into account in this estimate.}  The crucial point
here is that the lattice results confirm the theoretical prediction
\cite{Colangelo:2003hf}: the breaking of chiral symmetry generated by the
small masses of the two lightest quarks is significantly larger than what
dimensional reasoning would indicate, because it is amplified by a chiral
logarithm with a large coefficient. The amplification is generated by the
propagation of Nambu-Goldstone boson pairs that necessarily accompany the
spontaneous breakdown of the symmetry.

At NLO, the ratio $F_\pi/F$ is determined by the effective coupling
$\lbar_4$. As we are dealing with an expansion in powers of the two
lightest quark masses here, the corrections of higher order are expected to
be very small (not necessarily at the quark masses used in the simulations,
but at their physical values). Unfortunately, however, some of the
collaborations appear to underestimate the uncertainties in the
determination of this coupling: in view of the small errors quoted, the
results are not coherent (see the middle panel in Fig.\,\ref{fig_l3l4l6}).
The dust needs to settle before a meaningful estimate can be given for
$\lbar_4$.

The leading term in the chiral expansion of the scalar radius $\< r^2
\>_S^\pi$ is also determined by the coupling $\lbar_4$. The bottom panel of
Table \ref{tabl6} shows that the result for the scalar radius obtained from
a lattice calculation of the scalar form factor by JLQCD/TWQCD 09
\cite{JLQCD:2009qn} is in good agreement with the result of CGL 01
\cite{Colangelo:2001df}, which is based on dispersion theory, but the red
tags indicate that some of the systematic uncertainties in that calculation
yet need to be studied.

The current situation with $\lbar_3$ is better than the one with $\lbar_4$:
the $\Nf=2+1$ determinations agree well with one another and the two ETM
calculations for $\Nf=2$ and $\Nf=2+1+1$ corroborate the result
further. There is some discrepancy between two among the most recent
determinations which have rather small error bars, namely ETM 10
\cite{Baron:2010bv} and RBC/UKQCD 10A  \cite{Aoki:2010dy}. 
We stress, however, that the value:
\begin{equation}
\lbar_3 = 3.2(8)  \qquad \mbox{our estimate} \; ,
\end{equation}
comfortably covers all available calculations and represents a substantial
progress with respect to earlier estimates.
The uncertainty is three times smaller than
the one attached to the old estimate in \cite{Gasser:1983yg}, which was based 
on analysis of the mass formulae for the pseudoscalar octet. 

Table \ref{tabl6} collects the results obtained for the electromagnetic
form factor of the pion: charge radius, curvature and the coupling constant
$\lbar_6$, that determines the charge radius to leading order of the chiral
expansion. The experimental information concerning the charge radius is
excellent and the curvature is also known very precisely, on the basis of
$e^+e^-$ data and dispersion theory. The form factor calculations thus
present an excellent testing ground for the lattice calculations. The table
shows that most of the available lattice results pass the test. Concerning
the value of $\lbar_6$, the situation is worse than for the charge radius
also because the extraction of this constant is done differently by
different groups (in particular some use the $x$ rather than the $\xi$
expansion and NLO rather than NNLO formulae): here the dust yet needs to
settle.

Perhaps the most important physics result of this
section is that the lattice simulations confirm the approximate validity of the
Gell-Mann-Oakes-Renner formula and show that the square of the pion mass
indeed grows in proportion to $m_{ud}$. The formula represents the leading
term of the chiral perturbation series and necessarily receives corrections
from higher orders. At first non-leading order, the correction is
determined by the effective coupling constant $\lbar_3$. The results
collected in Tab.\,\ref{tabl3and4} and in the top panel of
Fig.\,\ref{fig_l3l4l6} show that $\lbar_3$ is now known quite well. They
corroborate the conclusion drawn in the pioneering work of reference
\cite{Durr:2002zx}: the lattice confirms the early estimate of $\bar\ell_3$
derived in \cite{Gasser:1983yg}. In the graph of $M_\pi^2$ versus $m_{ud}$,
the values found on the lattice for $\lbar_3$ correspond to remarkably
little curvature: the Gell-Mann-Oakes-Renner formula represents a crude
first approximation out to values of $m_{ud}$ that exceed the physical
value by an order of magnitude.

\begin{table}[!tbp]
\vspace{0.5cm}
\footnotesize
\begin{tabular*}{\textwidth}[l]{l@{\extracolsep{\fill}}rlllllll}
Collaboration & Ref. & $\Nf$ & 
\hspace{0.15cm}\begin{rotate}{60}{publication status}\end{rotate}\hspace{-0.15cm} &
\hspace{0.15cm}\begin{rotate}{60}{chiral extrapolation}\end{rotate}\hspace{-0.15cm}&
\hspace{0.15cm}\begin{rotate}{60}{continuum  extrapolation}\end{rotate}\hspace{-0.15cm} &
\hspace{0.15cm}\begin{rotate}{60}{finite volume}\end{rotate}\hspace{-0.15cm} &\rule{0.3cm}{0cm} $\bar\ell_3$ & \rule{0.3cm}{0cm}$\bar\ell_4$\\
&&&&&&&& \\[-0.1cm]
\hline
\hline
&&&&&&&& \\[-0.1cm]
ETM 10 & \cite{Baron:2010bv} & 2+1+1& \gA & \soso & \soso & \soso & 3.70(7)(26) & 4.67(3)(10) \\
&&&&&&&& \\[-0.1cm]
\hline
&&&&&&&& \\[-0.1cm]
MILC 10A       & \cite{Bazavov:2010yq} & 2+1 & \rC & \soso & \good & \good & 2.85(81)$\binom{+37}{-92}$ & 3.98(32)$\binom{+51}{-28}$\\ 
MILC 10        & \cite{Bazavov:2010hj} & 2+1 & \rC & \soso & \good & \good & 3.18(50)(89) & 4.29(21)(82) \\
RBC/UKQCD 10A  &\cite{Aoki:2010dy}     & 2+1 & \oP & \soso & \soso& \good &  2.57(18) & 3.83(9) \\
MILC 09A       & \cite{Bazavov:2009fk}  & 2+1 & \rC & \soso & \good & \good & 3.32(64)(45)& 4.03(16)(17) \\
MILC 09A       & \cite{Bazavov:2009fk}  & 2+1 & \rC & \soso & \good & \good & 3.0(6)$\binom{+9}{-6}$& 3.9(2)(3)  \\
PACS-CS 08     & \cite{Aoki:2008sm}  & 2+1 & \gA & \good & \bad & \bad & 3.47(11)  & 4.21(11)  \\
PACS-CS 08     & \cite{Aoki:2008sm}  & 2+1 & \gA & \good & \bad & \bad & 3.14(23)  & 4.04(19)    \\
RBC/UKQCD 08   & \cite{Allton:2008pn}  & 2+1 & \gA & \soso & \bad & \good & 3.13(33)(24) & 4.43(14)(77)  \\
&&&&&&& \\[-0.1cm]
\hline
&&&&&&& \\[-0.1cm]
ETM 09C  & \cite{Baron:2009wt} &  2  & \gA & \soso & \good & \soso & 3.50(9)$\binom{+9}{-30}$&4.66(4)$\binom{+4}{-33}$ \\
JLQCD/TWQCD 09 & \cite{JLQCD:2009qn}      &  2  & \gA & \soso & \bad & \bad & & 4.09(50)(52)  \\
ETM 08         & \cite{Frezzotti:2008dr}  &  2  & \gA & \soso & \soso & \soso & 3.2(8)(2)  & 4.4(2)(1)      \\
JLQCD/TWQCD 08A& \cite{Noaki:2008iy}      &  2  & \gA & \soso & \bad & \bad & 3.38(40)(24)$\binom{+31}{-0}$& 4.12(35)(30)$\binom{+31}{-0}$ \\
CERN-TOV 06    & \cite{DelDebbio:2006cn}  &  2  & \gA & \soso & \soso & \bad & 3.0(5)(1)&      \\ &&&&&&& \\[-0.1cm]
\hline
&&&&&&& \\[-0.1cm]
CGL 01         & \cite{Colangelo:2001df}  &&&&&&& 4.4(2) \\
GL 84          & \cite{Gasser:1983yg}     &&&&&& 2.9(2.4)& 4.3(9) \\
&&&&&&& \\[-0.1cm]
\hline
\hline
\end{tabular*}
\normalsize
\caption{Results for the coupling constants $\bar\ell_3$ and $\lbar_4$ of
  the effective SU(2) Lagrangian. For MILC 09A and PACS-CS 08, converted 
values from SU(3) fits and values from direct SU(2) fits are included. The MILC 10 results are obtained by converting the SU(3) LECs, while the MILC 10A results are obtained with a direct SU(2) fit. For
comparison, the last two lines show the results obtained from a
phenomenological analysis. 
\label{tabl3and4}}
\end{table}

As emphasized by Stern and collaborators \cite{Fuchs:1991cq,Stern:1993rg,
  DescotesGenon:1999uh}, the analysis in the framework of $\chi$PT is
coherent only if (i) the leading term in the chiral expansion of $M_\pi^2$
dominates over the remainder and (ii) the ratio $m_s/m_{ud}$ is close to
the value $25.6$ that follows from Weinberg's leading order formulae. In
order to investigate the possibility that one or both of these conditions
might fail, the authors proposed a more general framework, referred to as
"Generalized {\Ch}PT", which includes {\Ch}PT as a special case. The
results found on the lattice demonstrate that QCD does satisfy both of the
above conditions -- in the context of QCD, the proposed generalization of
the effective theory does not appear to be needed.  There is a modified
version, however, referred to as "Resummed {\Ch}PT" \cite{Bernard:2010ex},
which is motivated by the possibility that the Zweig rule violating
couplings $L_4$ and $L_6$ might be larger than expected.  The available
lattice data do not exclude this possibility (see section \ref{sec:recent
  lattice determinations}).
  
\begin{table}[!tb]
\footnotesize
\begin{tabular*}{\textwidth}[l]{l@{\extracolsep{\fill}}rllllllll}
Collaboration & Ref. & $\Nf$ &
\hspace{0.15cm}\begin{rotate}{60}{publication status}\end{rotate}\hspace{-0.15cm} &
\hspace{0.15cm}\begin{rotate}{60}{chiral extrapolation}\end{rotate}\hspace{-0.15cm}&
\hspace{0.15cm}\begin{rotate}{60}{continuum  extrapolation}\end{rotate}\hspace{-0.15cm} &
\hspace{0.15cm}\begin{rotate}{60}{finite volume}\end{rotate}\hspace{-0.15cm} &
\rule{0.2cm}{0cm}$\< r^2 \>_V^\pi \, [\mathrm{fm}^2]$ & $c_V \, (\mathrm{GeV}^{-4})$ &\rule{0.3cm}{0cm}
 $\bar \ell_6$\\ 
&&&&&&&&& \\[-0.1cm]
\hline
\hline
&&&&&&&&& \\[-0.1cm]
RBC/UKQCD 08A  & \cite{Boyle:2008yd}     & 2+1 & \gA & \soso & \bad & \good & 0.418(31) & --- & 12.2(9)\\
LHP 04         & \cite{Bonnet:2004fr}    & 2+1 & \gA & \soso & \bad & \soso & 0.310(46) & --- & ---\\
&&&&&&&&& \\[-0.1cm]
\hline
&&&&&&&&& \\[-0.1cm]
JLQCD/TWQCD 09 & \cite{JLQCD:2009qn}     &  2  & \gA & \soso & \bad & \bad & 0.409(23)(37) & 3.22(17)(36) & 11.9(0.7)(1.0) \\
ETM 08         & \cite{Frezzotti:2008dr} &  2  & \gA & \soso & \soso & \soso & 0.456(30)(24) & 3.37(31)(27) & 14.9(1.2)(0.7) \\
QCDSF/UKQCD 06A  & \cite{Brommel:2006ww}   &  2  & \gA & \good & \soso & \soso & 0.441(19)(56)(29) & --- & --- \\[-0.1cm]
&&&&&&&&& \\[-0.1cm]
\hline
&&&&&&&&& \\[-0.1cm]
BCT 98         & \cite{Bijnens:1998fm} &&&&&& 0.437(16) & 3.85(60) & 16.0(0.5)(0.7) \\
NA7 86         & \cite{Amendolia:1986wj} &&&&&& 0.439(8)  &
&                \\ 
GL 84          & \cite{Gasser:1983yg}    &&&&&&           &
& 16.5(1.1)      \\ 
&&&&&&&&& \\[-0.1cm]
\hline
\hline
\end{tabular*}
\\[3cm]
\begin{tabular*}{\textwidth}[l]{l@{\extracolsep{\fill}}rlllllll}
Collaboration & Ref. & $\Nf$ &
\hspace{0.15cm}\begin{rotate}{60}{publication status}\end{rotate}\hspace{-0.15cm} &
\hspace{0.15cm}\begin{rotate}{60}{chiral extrapolation}\end{rotate}\hspace{-0.15cm}&
\hspace{0.15cm}\begin{rotate}{60}{continuum  extrapolation}\end{rotate}\hspace{-0.15cm} &
\hspace{0.15cm}\begin{rotate}{60}{finite volume}\end{rotate}\hspace{-0.15cm} &
\rule{0.1cm}{0cm}$\< r^2 \>_S^\pi \; [\mathrm{fm}^2]$ & \rule{0.3cm}{0cm}$\bar \ell_1-\bar \ell_2$ \\
&&&&&&&& \\[-0.1cm]
\hline
\hline
&&&&&&&& \\[-0.1cm]
JLQCD/TWQCD 09 & \cite{JLQCD:2009qn} &  2 &  \gA & \soso & \bad & \bad & 0.617(79)(66) & -2.9(0.9)(1.3) \\
&&&&&&&& \\[-0.1cm]
\hline
&&&&&&&& \\[-0.1cm]
CGL 01 & \cite{Colangelo:2001df} &&&&&& 0.61(4) & -4.7(6) \\
&&&&&&&& \\[-0.1cm]
\hline
\hline
&&&&&&&& \\[-0.1cm]
\end{tabular*}
\normalsize
\caption{{\it Top panel: vector form factor of the pion.} Lattice results
  for the charge radius $\< r^2 \>_V^\pi$, the curvature  $c_V$ and the
  effective coupling constant $\bar\ell_6$ are compared with the
  experimental value obtained by NA7 and some phenomenological estimates
  ($\lbar_6$ determines the leading contribution in the chiral expansion of
  the charge radius). {\it Bottom panel: scalar form factor of the pion.}
  Lattice results for the scalar radius $\< r^2 \>_S^\pi$ and the
  combination $\bar \ell_1-\bar \ell_2$ of effective coupling constants are
  compared with the outcome of a dispersive calculation of these quantities
  \cite{Colangelo:2001df} (the leading term in the chiral expansion of the
  scalar radius is determined by the coupling constant $\lbar_4$, for which
  the available information is collected in Tab.\,\ref{tabl3and4}; the lattice
  estimate for $\lbar_1-\lbar_2$ stems from an analysis of the momentum
  dependence of the vector and scalar form factors, based on the two-loop
  formulae of {\Ch}PT \cite{Bijnens:1998fm}). \label{tabl6}} 
\end{table}

\clearpage

\clearpage


\subsection{SU(3) Low-Energy Constants \label{sec:su3}}


\subsubsection{Quark-mass dependence of pseudoscalar masses and decay
constants}

In the isospin limit, the relevant SU(3) formulae take the form
\cite{Gasser:1984gg}
\bea
\Mpi^2\!\!&\!\!\NLo\!\!&\!\! 2B_0m_{ud}
\Big\{
1+\mu_\pi-\frac{1}{3}\mu_\et+\frac{B_0}{F_0^2}
\Big[16m_{ud}(2L_8\!-\!L_5)+16(m_s\!+\!2m_{ud})(2L_6\!-\!L_4)\Big]
\Big\}\;,\nn
\\
M_{\!K}^2\!\!&\!\!\NLo\!\!&\!\! B_0(m_s\!\!+\!m_{ud})
\Big\{
1\!+\!\frac{2}{3}\mu_\et\!+\!\frac{B_0}{F_0^2}
\Big[8(m_s\!\!+\!m_{ud})(2L_8\!-\!L_5)\!+\!16(m_s\!\!+\!2m_{ud})(2L_6\!-\!L_4)\Big]
\Big\}\;,\quad\nonumber
\\
\Fpi\!\!&\!\!\NLo\!\!&\!\!F_0
\Big\{
1-2\mu_\pi-\mu_K+\frac{B_0}{F_0^2}
\Big[8m_{ud}L_5+8(m_s\!+\!2m_{ud})L_4\Big]
\Big\}\;,
\\
\Fka\!\!&\!\!\NLo\!\!&\!\!F_0
\Big\{
1-\frac{3}{4}\mu_\pi-\frac{3}{2}\mu_K-\frac{3}{4}\mu_\et+\frac{B_0}{F_0^2}
\Big[4(m_s\!+\!m_{ud})L_5+8(m_s\!+\!2m_{ud})L_4\Big]
\Big\}\;,\nonumber
\eea
where $B_0=\Sigma_0/F_0^2$ and $F_0$ denote the condensate parameter and the
pseudoscalar decay constant in the SU(3) chiral limit, respectively, and
\beq
\mu_P=\frac{M_P^2}{32\pi^2F_0^2}
\ln\!\Big(\frac{M_P^2}{\mu^2}\Big)\;.
\label{def_muP}
\eeq
At the accuracy of these formulae, the quantities $\mu_\pi$, $\mu_K$,
$\mu_\eta$ can equally well be evaluated with the leading order expressions
for the masses,
\beq
M_\pi^2\Lo 2B_0\,m_{ud}\;,\quad
M_K^2\Lo B_0(m_s\!+\!m_{ud})\;,\quad
M_\et^2\Lo \mbox{$\frac{2}{3}$}B_0(2m_s\!+\!m_{ud})
\;.
\eeq 
Throughout, $L_i$ denotes the renormalized low-energy constant/coupling (LEC)
at scale $\mu$. The normalization used for the decay constants is specified in 
footnote \ref{foot:fpi}.


\subsubsection{Charge radius}

The SU(3) formula for the slope of the pion vector form factor reads
\cite{Gasser:1984ux}
\beq
\<r^2\>_{V,\pi\pi}\Lo-\frac{1}{32\pi^2F_0^2}
\Big\{
3+2\ln(\frac{\Mpi^2}{\mu^2})+\ln(\frac{\Mka^2}{\mu^2})
\Big\}
+\frac{12L_9}{F_0^2}
\;,
\eeq
while the expression $\<r^2\>_{S,\mathrm{oct}}$ for the octet part of the
scalar radius does not contain any NLO low-energy constant at the 1-loop order
\cite{Gasser:1984ux}.


\subsubsection{Partially quenched formulae}

The term ``partially quenched QCD'' is used in two ways.
For heavy quarks ($c,b$ and sometimes $s$) it usually means that these flavors
are included in the valence sector, but not to the functional determinant.
For the light quarks ($u,d$ and sometimes $s$) it means that they are
present in both the valence and the sea sector of the theory, but with
different masses (e.g.\ a series of valence quark masses is evaluated on an
ensemble with a fixed sea quark mass).

The program of extending the standard (unitary) SU(3) theory to the
(second version of) ``partially quenched QCD''
has been completed at the 2-loop (NNLO) level for masses and decay
constants~\cite{Bijnens:2006jv}.
These formulae tend to be complicated, with the consequence that a
state-of-the-art analysis with $O(2000)$ bootstrap samples on $O(20)$
ensembles with $O(5)$ masses each [and hence $O(200'000)$ different fits] will
require significant computational resources for the global fits.
For an up-to-date summary of recent developments in Chiral Perturbation Theory
relevant to lattice QCD we refer to~\cite{Bijnens:2009pq}.


\subsubsection{Recent lattice determinations}
\label{sec:recent lattice determinations}

\begin{table}[!t]
\centering\footnotesize
\vspace*{1cm}
\begin{tabular*}{\textwidth}[l]{l@{\extracolsep{\fill}}rclllllllllll}
 & Ref. & $\Nf$ &
\hspace{0.15cm}\begin{rotate}{60}{publication status}\end{rotate}\hspace{-0.15cm} &
\hspace{0.15cm}\begin{rotate}{60}{chiral extrapolation}\end{rotate}\hspace{-0.15cm} &
\hspace{0.15cm}\begin{rotate}{60}{continuum  extrapolation}\end{rotate}\hspace{-0.15cm} &
\hspace{0.15cm}\begin{rotate}{60}{finite volume effects}\end{rotate}\hspace{-0.15cm} &
\hspace{0.15cm}\begin{rotate}{60}{renormalization}\end{rotate}\hspace{-0.15cm} &
\rule{0.1cm}{0cm}$F_0\,[\mathrm{MeV}]$  &\rule{0.1cm}{0cm} $F/F_0$ & \rule{0.2cm}{0cm}$B/B_0$ & \hspace{2.5cm} \\ 
&&&&&&&&&&&& \\[-0.1cm]
\hline
\hline
&&&&&&&&&&&& \\[-0.1cm]
JLQCD/TWQCD 10 & \cite{Fukaya:2010na}                      & 3   & \gA & \bad  & \bad  & \bad  & \good & 71(3)(8) &  &  \\
&&&&&&&&&&&& \\[-0.2cm]
\hline \\[-0.1cm]
MILC 10        & \cite{Bazavov:2010hj}                     & 2+1 & \rC & \soso & \good & \good & \soso & 80.3(2.5)(5.4) & & \\
MILC 09A$^{\tiny\ref{foot:milc}}$  & \cite{Bazavov:2009fk} & 2+1 & \rC & \soso & \good & \good & \soso & 78.3(1.4)(2.9) & {\sl 1.104(3)(41)}    & {\sl 1.21(4)$\binom{+5}{-6}$}\  \\ 
MILC 09$^{\tiny\,\ref{foot:milc}}$ & \cite{Bazavov:2009bb} & 2+1 & \gA & \soso & \good & \good & \soso &                & 1.15(5)($^{+13}_{-3}$)& {\sl 1.15(16)$\binom{+39}{-13}$}\\ 
PACS-CS 08   & \cite{Aoki:2008sm}                          & 2+1 & \gA & \good & \bad  & \bad  & \bad  & 83.8(6.4)      & 1.078(44)             & 1.089(15)\\
RBC/UKQCD 08 & \cite{Allton:2008pn}                        & 2+1 & \gA & \soso & \bad  & \good & \good & 66.1(5.2)      & 1.229(59)             & 1.03(05) \\
&&&&&&&&&&& \\[-0.1cm]
\hline
\hline
\end{tabular*}
\newline
\vspace*{3cm}
\begin{tabular*}{\textwidth}[l]{l@{\extracolsep{\fill}}rclllllll}
 & Ref. & $\Nf$ &
\hspace{0.15cm}\begin{rotate}{60}{publication status}\end{rotate}\hspace{-0.15cm} &
\hspace{0.15cm}\begin{rotate}{60}{chiral extrapolation}\end{rotate}\hspace{-0.15cm} &
\hspace{0.15cm}\begin{rotate}{60}{continuum  extrapolation}\end{rotate}\hspace{-0.15cm} &
\hspace{0.15cm}\begin{rotate}{60}{finite volume effects}\end{rotate}\hspace{-0.15cm} &
\hspace{0.15cm}\begin{rotate}{60}{renormalization}\end{rotate}\hspace{-0.15cm} &
\rule{0.1cm}{0cm}$\Sigma_0^{1/3}\,[\mathrm{GeV}]$ &\rule{0.1cm}{0cm} 
$\Sigma/\Sigma_0$ \\ 
&&&&&&&&& \\[-0.1cm]
\hline
\hline
&&&&&&&&& \\[-0.1cm]
JLQCD/TWQCD 10                     & \cite{Fukaya:2010na}  & 3   & \gA & \bad  & \bad  & \bad  & \good & 0.214(6)(24) & {\sl 1.31(13)(52)} \\
&&&&&&&&& \\[-0.2cm]
\hline \\[-0.1cm] 
MILC 09A$^{\tiny\ref{foot:milc}}$  & \cite{Bazavov:2009fk} & 2+1 & \rC & \soso & \good & \good & \soso & 0.245(5)(4)(4)                & {\sl 1.48(9)(8)(10)}     \\
MILC 09$^{\tiny\,\ref{foot:milc}}$ & \cite{Bazavov:2009bb} & 2+1 & \gA & \soso & \good & \good & \soso & 0.242(9)($^{\;+5}_{-17}$)(4)  & 1.52(17)($^{+38}_{-15}$) \\
PACS-CS 08                         & \cite{Aoki:2008sm}    & 2+1 & \gA & \good & \bad  & \bad  & \bad  & 0.290(15)  & 1.245(10)     \\
RBC/UKQCD 08                       & \cite{Allton:2008pn}  & 2+1 & \gA & \soso & \bad  & \good & \good &            & 1.55(21)      \\
&&&&&&&&& \\[-0.1cm]
\hline
\hline
\end{tabular*}
\caption{\label{tab:SU3_overview}
  Summary of lattice results for the low-energy constants $F_0$, $B_0$ and
  $\Sigma_0\equiv F_0^2B_0$, which specify the effective SU(3) Lagrangian
  at leading order. The ratios $F/F_0$, $B/B_0$, $\Sigma/\Sigma_0$, which
  compare these with their SU(2) counterparts, indicate the strength of the
  Zweig-rule violations in these quantities: in the large-$N_c$ limit, they
  tend to unity. Numbers in slanted fonts are calculated by us, from the
  information given in the quoted references.}
\end{table}

To date, there are three comprehensive papers with results based on lattice QCD
with $\Nf\!=\!2\!+\!1$ dynamical flavors
\cite{Allton:2008pn,Aoki:2008sm,Bazavov:2009bb}.
It is an open issue whether the data collected at
$m_s\!\simeq\!m_s^\mathrm{phys}$ allow for an unambiguous determination of
SU(3) low-energy constants (cf.\ the discussion in \cite{Allton:2008pn}).
So far only MILC has some data at considerably smaller $m_s$
\cite{Bazavov:2009bb}.
Furthermore, we are aware of a few papers with a result on one SU(3)
low-energy constant each \cite{Beane:2006kx,Boyle:2008yd,Shintani:2008qe}.
Some particulars of the comprehensive papers are shown in
Tab.\,\ref{tab:SU3_overview}.

\begin{table}[!p]
\centering\footnotesize
\begin{tabular*}{\textwidth}[l]{l@{\extracolsep{\fill}}rcllllllll}
 & Ref. & $\Nf$ &
\hspace{0.15cm}\begin{rotate}{60}{publication status}\end{rotate}\hspace{-0.15cm} &
\hspace{0.15cm}\begin{rotate}{60}{chiral extrapolation}\end{rotate}\hspace{-0.15cm} &
\hspace{0.15cm}\begin{rotate}{60}{continuum  extrapolation}\end{rotate}\hspace{-0.15cm} &
\hspace{0.15cm}\begin{rotate}{60}{finite volume effects}\end{rotate}\hspace{-0.15cm} &
\rule{0.1cm}{0cm}$10^3L_4$ &$\rule{0.1cm}{0cm}10^3L_6$ 
& \hspace{-0.3cm} $10^3(2L_6\!-\!L_4)$ \\
&&&&&&&&&& \\[-0.1cm]
\hline
\hline
&&&&&&&&&& \\[-0.1cm]
JLQCD/TWQCD 10 & \cite{Fukaya:2010na}                      & 3   & \gA & \bad  & \bad  & \bad  &                           & 0.03(7)(17)  & \\
&&&&&&&&&& \\[-0.2cm]
\hline \\[-0.1cm] 
MILC 10        & \cite{Bazavov:2010hj}                     & 2+1 & \rC & \soso & \good & \good & -0.08(22)($^{+57}_{-33}$) & {\sl-0.02(16)($^{+33}_{-21}$)} & 0.03(24) ($^{+32}_{-27}$) \\
MILC 09A$^{\tiny\ref{foot:milc}}$  & \cite{Bazavov:2009fk} & 2+1 & \rC & \soso & \good & \good & 0.04(13)(4)               & 0.07(10)(3)          & 0.10(12)(2)          \\
MILC 09$^{\tiny\,\ref{foot:milc}}$ & \cite{Bazavov:2009bb} & 2+1 & \gA & \soso & \good & \good & 0.1(3)($^{+3}_{-1}$)      & 0.2(2)($^{+2}_{-1}$) & 0.3(1)($^{+2}_{-3}$) \\
PACS-CS 08   & \cite{Aoki:2008sm}                          & 2+1 & \gA & \good & \bad  & \bad  & -0.06(10)(-)              & {\sl0.02(5)(-)}      & 0.10(2)(-)           \\
RBC/UKQCD 08 & \cite{Allton:2008pn}                        & 2+1 & \gA & \soso & \bad  & \good & 0.14(8)(-)                & 0.07(6)(-)           & 0.00(4)(-)           \\
&&&&&&&&&& \\[-0.1cm]
\hline
&&&&&&&&&& \\[-0.1cm]
Bijnens\,09  & \cite{Bijnens:2009pq} &     &     &      &      &      & 0.0(-)(5)            & 0.0(-)(1)            & {\sl0.0(-)(5)}       \\
 GL\,85      & \cite{Gasser:1984gg}  &     &     &      &      &      & -0.3(5)              & -0.2(3)              & {\sl-0.1(8)}         \\
&&&&&&&&&& \\[-0.1cm]
\hline
\hline
\end{tabular*}
\newline
\vspace*{3cm}
\begin{tabular*}{\textwidth}[l]{l@{\extracolsep{\fill}}rlllllllll}
 & Ref. & $\Nf$ &
\hspace{0.15cm}\begin{rotate}{60}{publication status}\end{rotate}\hspace{-0.15cm} &
\hspace{0.15cm}\begin{rotate}{60}{chiral extrapolation}\end{rotate}\hspace{-0.15cm} &
\hspace{0.15cm}\begin{rotate}{60}{continuum  extrapolation}\end{rotate}\hspace{-0.15cm} &
\hspace{0.15cm}\begin{rotate}{60}{finite volume effects}\end{rotate}\hspace{-0.15cm} &
\rule{0.1cm}{0cm} $10^3L_5$ &\rule{0.05cm}{0cm}  $10^3L_8$ &\hspace{-0.3cm} $10^3(2L_8\!-\!L_5)$ \\
&&&&&&&&&& \\[-0.1cm]
\hline
\hline
&&&&&&&&&& \\[-0.1cm]
MILC 10        & \cite{Bazavov:2010hj}                     & 2+1 & \rC & \soso & \good & \good & 0.98(16)($^{+28}_{-41}$) & {\sl0.42(10)($^{+27}_{-23}$)} & -0.15(11)($^{+45}_{-19}$)\\
MILC 09A$^{\tiny\ref{foot:milc}}$  & \cite{Bazavov:2009fk} & 2+1 & \rC & \soso & \good & \good & 0.84(12)(36)         & 0.36(5)(7)      & -0.12(8)(21)     \\
MILC 09$^{\tiny\,\ref{foot:milc}}$ & \cite{Bazavov:2009bb} & 2+1 & \gA & \soso & \good & \good & 1.4(2)($^{+2}_{-1}$) & 0.8(1)(1)       & 0.3(1)(1)        \\
PACS-CS 08   & \cite{Aoki:2008sm}                          & 2+1 & \gA & \good & \bad  & \bad  & 1.45(7)(-)           & {\sl0.62(4)(-)} & -0.21(3)(-)      \\
RBC/UKQCD 08 & \cite{Allton:2008pn}                        & 2+1 & \gA & \soso & \bad  & \good & 0.87(10)(-)          & 0.56(4)(-)      & 0.24(4)(-)       \\
&&&&&&&&&& \\[-0.1cm]
\hline
&&&&&&&&&& \\[-0.1cm]
Bijnens\,09  & \cite{Bijnens:2009pq} &     &     &      &      &                       & 0.97(11)(-)          & 0.60(18)(-)     & {\sl0.23(38)(-)} \\
GL\,85       & \cite{Gasser:1984gg}  &     &     &      &      &                       & 1.4(5)               & 0.9(3)          & {\sl0.4(8)}      \\
&&&&&&&&&& \\[-0.1cm]
\hline
\hline
\end{tabular*}
\newline
\vspace*{3cm}
\begin{tabular*}{\textwidth}[l]{l@{\extracolsep{\fill}}lllllllll}
 & Ref. & $\Nf$ &
\hspace{0.15cm}\begin{rotate}{60}{publication status}\end{rotate}\hspace{-0.15cm} &
\hspace{0.15cm}\begin{rotate}{60}{chiral extrapolation}\end{rotate}\hspace{-0.15cm} &
\hspace{0.15cm}\begin{rotate}{60}{continuum  extrapolation}\end{rotate}\hspace{-0.15cm} &
\hspace{0.15cm}\begin{rotate}{60}{finite volume effects}\end{rotate}\hspace{-0.15cm} &
\rule{0.1cm}{0cm} $10^3L_5$ &\rule{0.1cm}{0cm} $10^3L_9$ &\rule{0.1cm}{0cm} $10^3L_{10}$ \\
&&&&&&&& \\[-0.1cm]
\hline
\hline
&&&&&&&& \\[-0.1cm]
JLQCD 08A    & \cite{Shintani:2008qe}  & 2   & \gA & \soso & \bad & \bad  &                         &              & -5.2(2)($^{+5}_{-3}$) \\
RBC/UKQCD 08A& \cite{Boyle:2008yd}     & 2+1 & \gA & \soso & \bad & \good &                         & 3.08(23)(51) &                       \\
NPLQCD 06    & \cite{Beane:2006kx}     & 2+1 & \gA & \soso & \bad & \bad  & 1.42(2)($^{+18}_{-54}$) &              &                       \\
&&&&&&&& \\[-0.1cm]
\hline
&&&&&&&& \\[-0.1cm]
Bijnens\,09  & \cite{Bijnens:2009pq}   &     &     &      &      &      & 0.97(11)(-)             & 5.93(43)(-)  &         \\
GL\,85       & \cite{Gasser:1984gg}    &     &     &      &      &      & 1.4(5)                  & 6.9(7)       & -5.5(7) \\
&&&&&&&& \\[-0.1cm]
\hline
\hline
\end{tabular*}
\caption{\label{tab:SU3_NLO}Summary of lattice results for some of the
  coupling constants that enter the effective SU(3) Lagrangian at NLO
  (running scale $\mu\!=\!770\MeV$ -- the values in
  \cite{Bazavov:2010hj,Bazavov:2009bb,Bazavov:2009fk,Gasser:1984gg} are evolved
  accordingly). The PACS-CS entry for $L_6$ is obtained from their results
  for $2L_6\!-\!L_4$ and $L_4$ (and similarly for other entries in slanted
  fonts). }

\end{table}
\begin{figure}[!tbh]
\centering
\includegraphics[width=7.5cm]{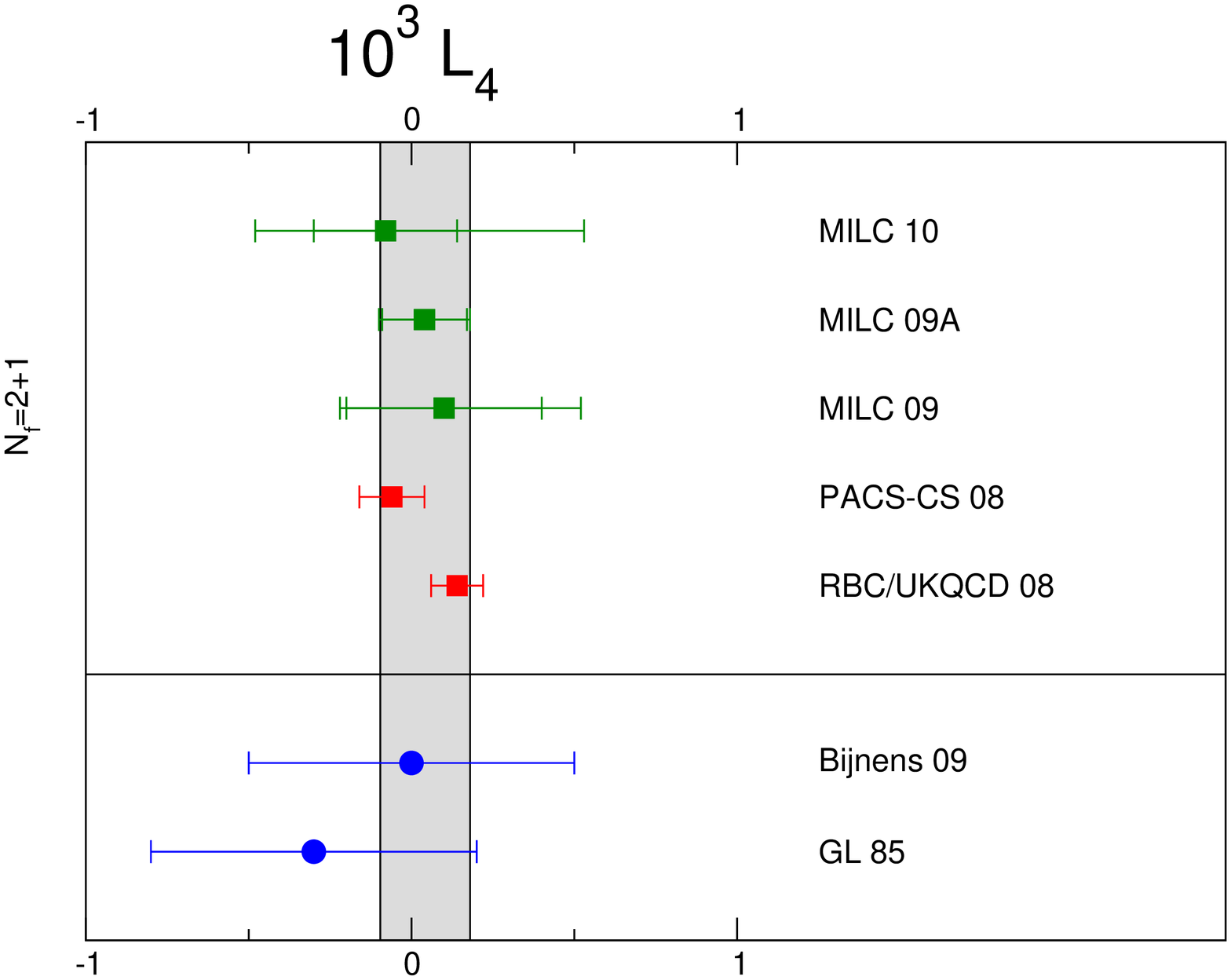} \hfill
\includegraphics[width=7.5cm]{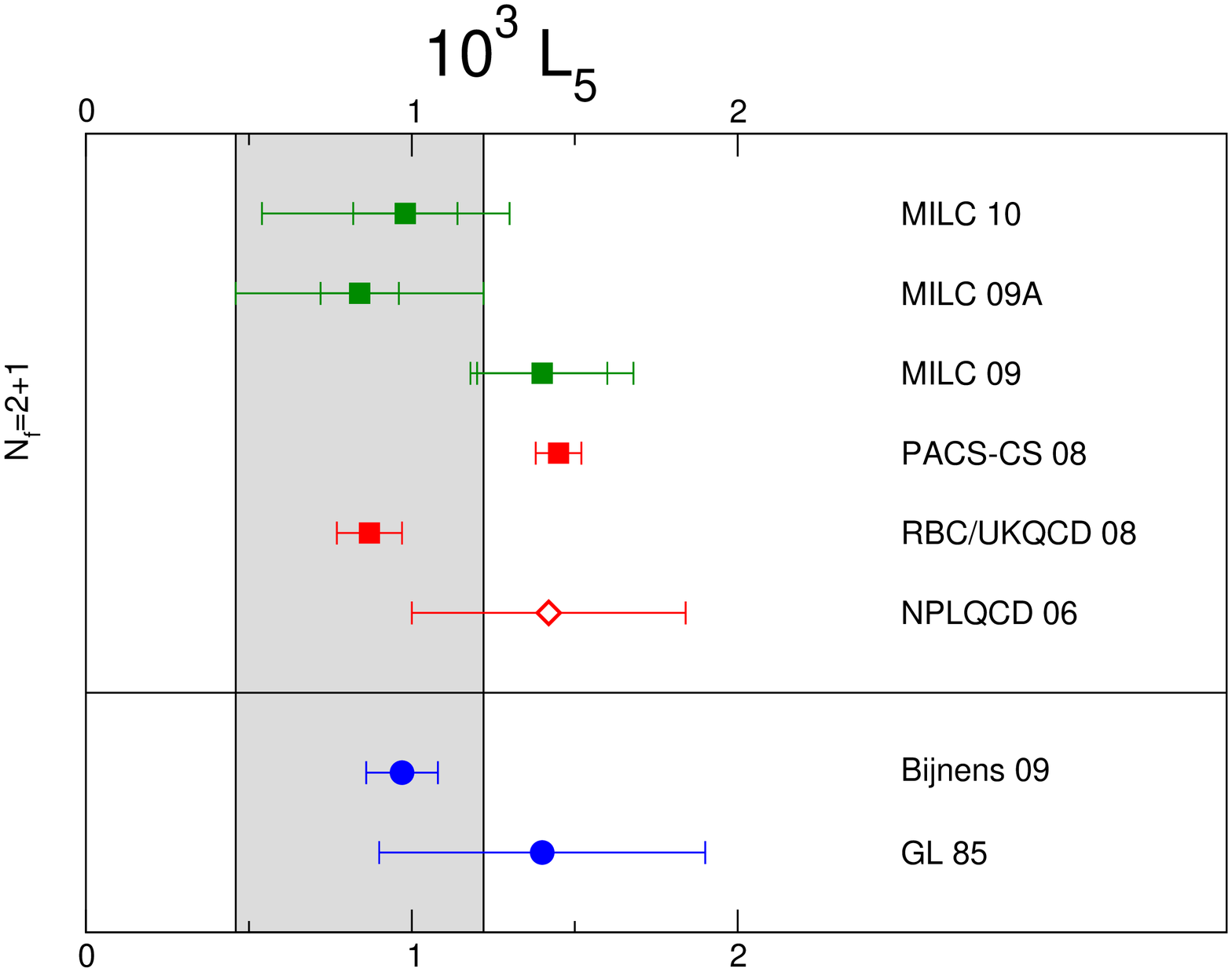} \\
\vspace{0.5cm}
\includegraphics[width=7.5cm]{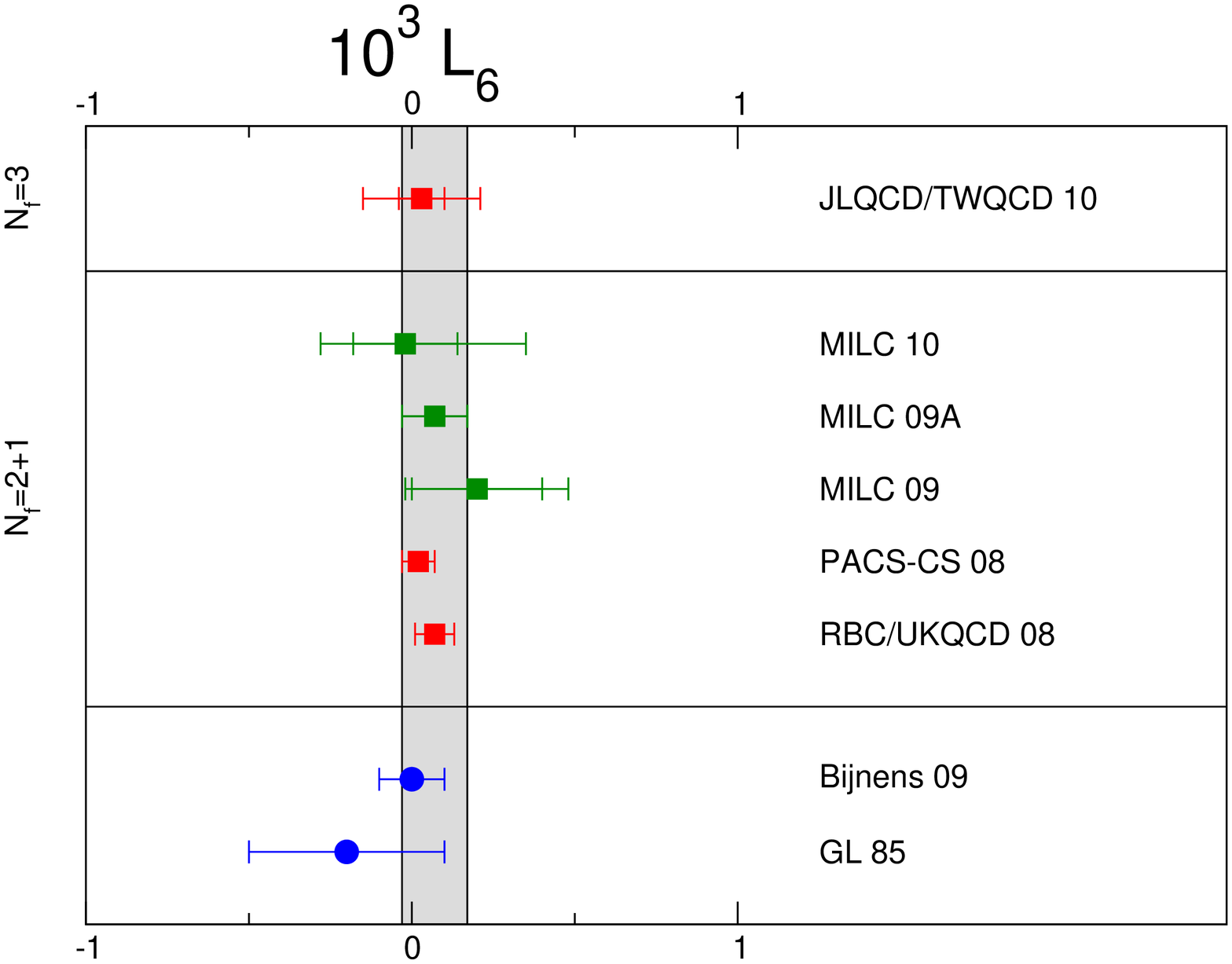} \hfill
\includegraphics[width=7.5cm]{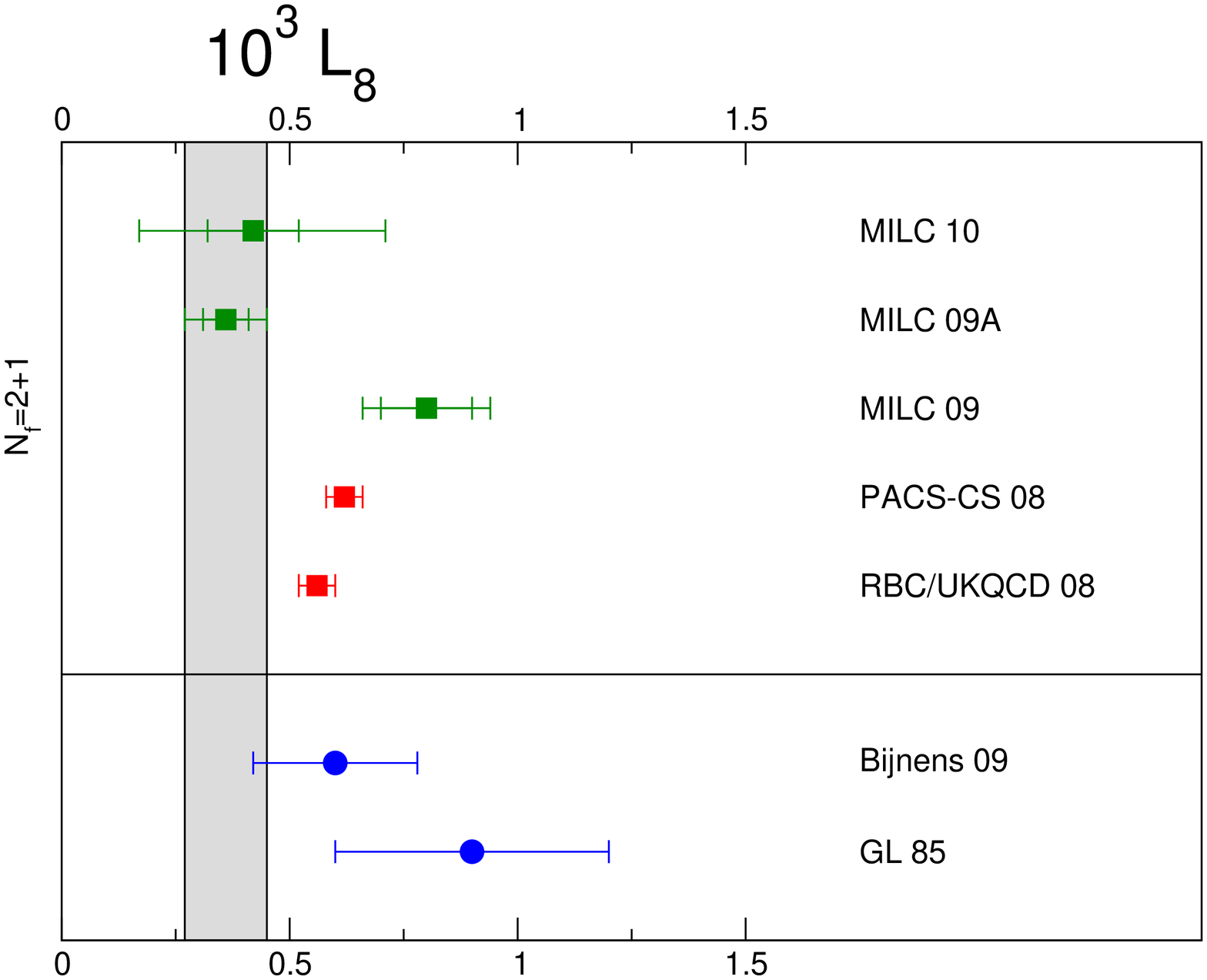} \\
\vspace{0.5cm}
\includegraphics[width=7.5cm]{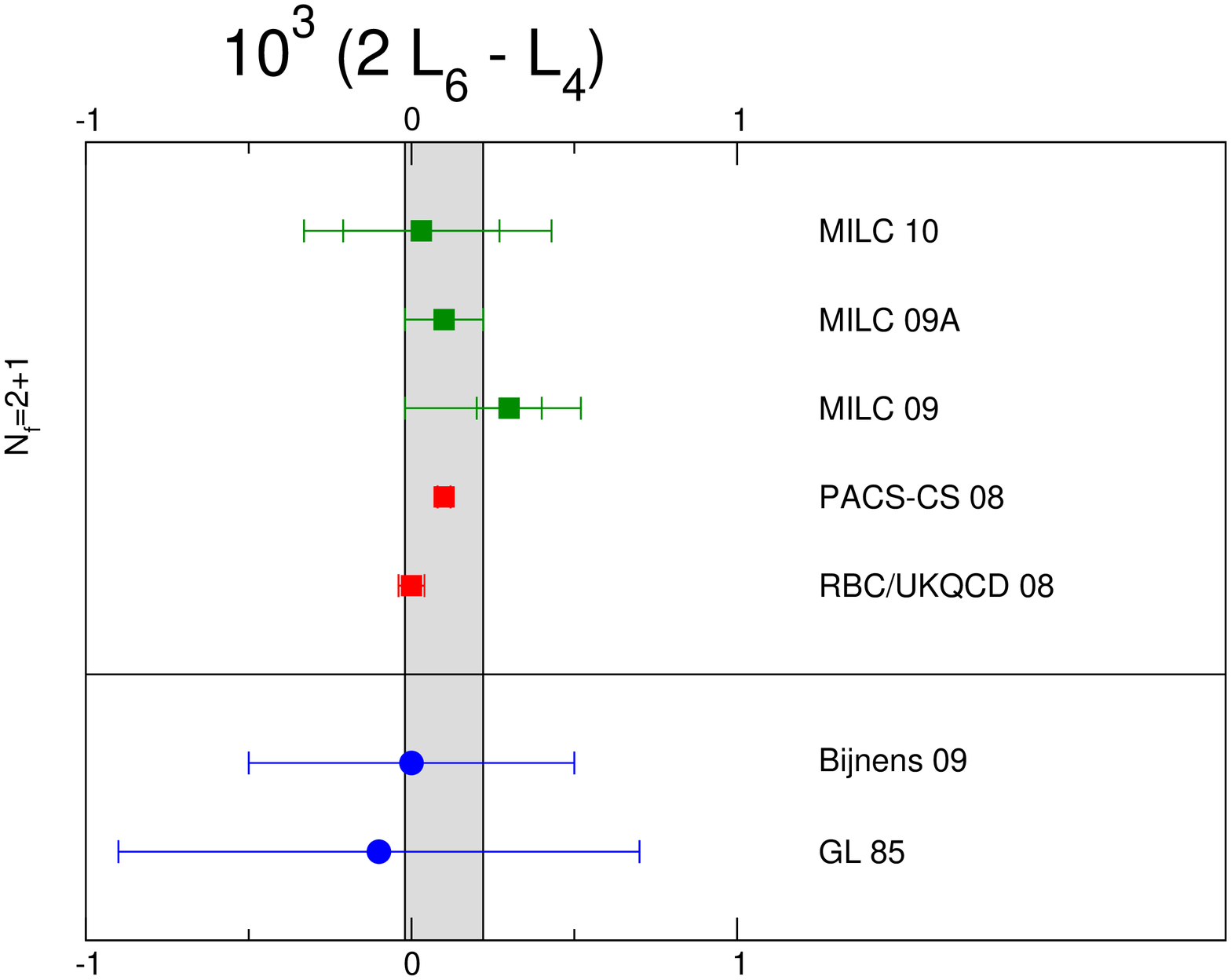} \hfill
\includegraphics[width=7.5cm]{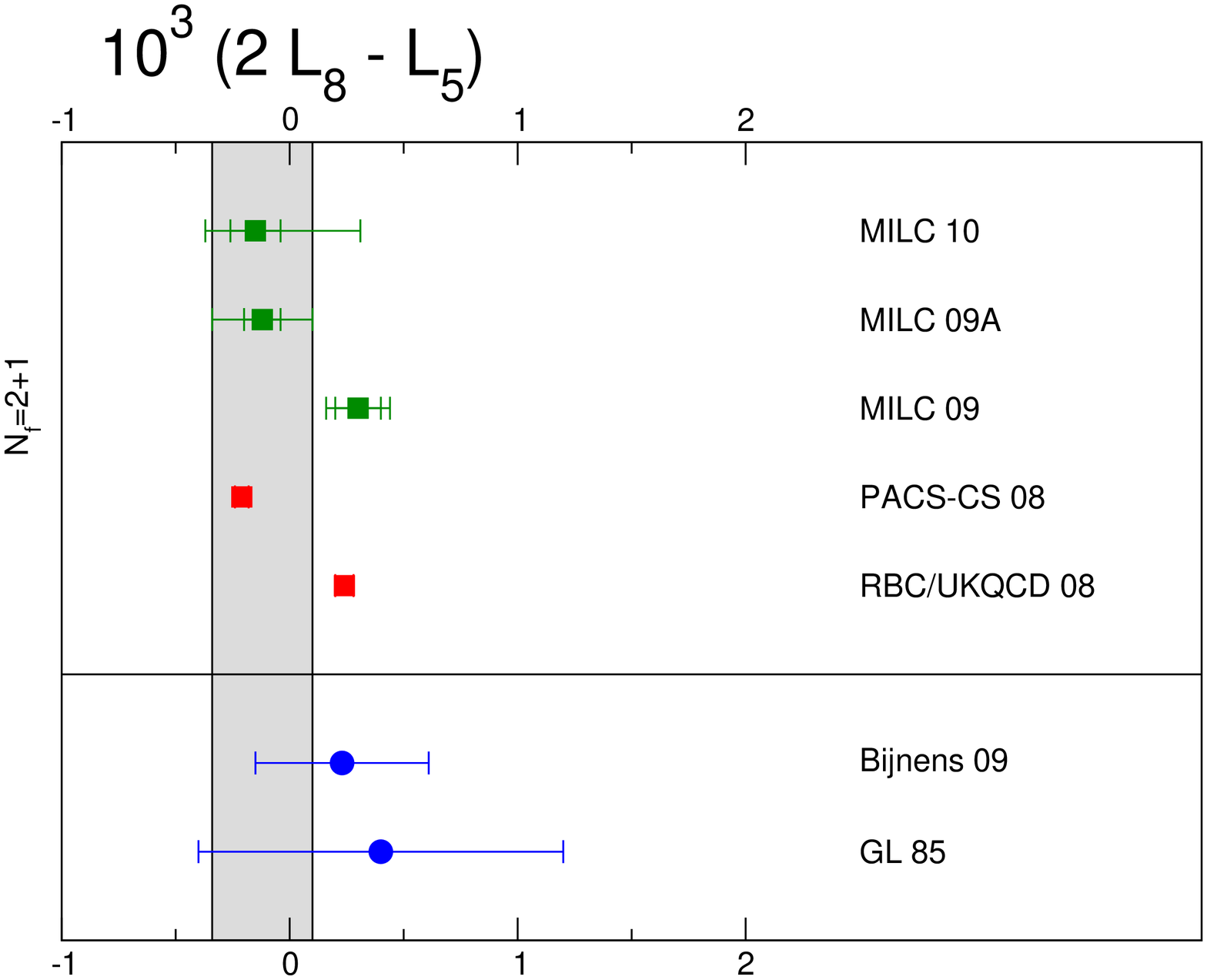}  
\caption{Summary of various lattice determinations of SU(3) low-energy
constants. \label{fig:plot_LECs_SU3}}
\end{figure}

Results for the SU(3) low-energy constants of leading order are found in
Tab.\,\ref{tab:SU3_overview} and analogous results for some of the
effective coupling constants that enter the chiral SU(3) Lagrangian at NLO
are collected in Tab.\,\ref{tab:SU3_NLO}.  From PACS-CS \cite{Aoki:2008sm}
only those results are quoted which have been \emph{corrected} for
finite-size effects (misleadingly labeled ``w/FSE'' in their tables).  For
staggered data our ``chiral'' colour coding criterion is slightly
ambiguous; we solve this issue pragmatically.%
\footnote{\label{foot:milc} The ``chiral extrapolation'' rating of
  \cite{Bazavov:2009bb} is based on the information on the RMS masses given
  in \cite{Bazavov:2009fk}.}

A graphical summary of the lattice results for the coupling constants $L_4$, $L_5$, 
$L_6$ and $L_8$, which determine the masses and the decay constants of the
pions and kaons at NLO of the chiral SU(3) expansion, is displayed in 
Fig.\,\ref{fig:plot_LECs_SU3}, along with the two phenomenological determinations 
quoted in the above tables. In general, there is a rather convincing overall 
consistency, although some of the lattice results possibly 
underestimate the theoretical error. Since only the calculation by the MILC
collaboration is free of red tags, the gray bands in
Fig.\,\ref{fig:plot_LECs_SU3} mark their results.

In spite of this apparent consistency, there is a point which needs to be
clarified as soon as possible. Some collaborations (RBC/UKQCD and PACS-CS)
find that they are having difficulties in fitting their partially quenched
data to the respective formulas for pion masses above $\simeq$ 400 MeV.
Evidently, this indicates that the data are stretching the regime of
validity of these formulas. To date it is, however, not clear which subset
of the data causes the troubles, whether it is the unitary part extending
to too large values of the quark masses or whether it is due to
$m^\mathrm{val}/m^\mathrm{sea}$ differing too much from one.  In fact,
little is known, in the framework of partially quenched {\Ch}PT, about the
\emph{shape} of the region of applicability in the $m^\mathrm{val}$ versus
$m^\mathrm{sea}$ plane for fixed $\Nf$.

In the large-$N_c$ limit, the Zweig-rule becomes exact, but the quarks have
$N_c=3$.  The work done on the lattice is ideally suited to disprove or
confirm the approximate validity of this rule. Two of the coupling
constants entering the effective SU(3) Lagrangian at NLO disappear when
$N_c$ is sent to infinity: $L_4$ and $L_6$. The upper part of
Tab.\,\ref{tab:SU3_NLO} and the left panels of Fig.~\ref{fig:plot_LECs_SU3}
show that the lattice results for these are quite coherent. At the scale
$\mu=M_\rho$, $L_4$ and $L_6$ are consistent with zero, indicating that
these constants do approximately obey the Zweig-rule. As mentioned above,
the ratios $F/F_0$, $B/B_0$ and $\Sigma/\Sigma_0$ also test the validity of
this rule. Their expansion in powers of $m_s$ starts with unity and the
contributions of first order in $m_s$ are determined by the constants $L_4$
and $L_6$, but they also contain terms of higher order. Quite apart from
measuring the Zweig-rule violations, an accurate determination will thus
also allow us to determine the range of $m_s$ where the first few terms of
the expansion represent an adequate approximation. Unfortunately, at
present, the uncertainties in the lattice data on the ratios are too large
to draw conclusions, both concerning the relative size of the subsequent
terms in the chiral perturbation series and concerning the magnitude of the
Zweig-rule violations. The data do appear to confirm the {\it paramagnetic
  inequalities} \cite{DescotesGenon:1999uh}, which require $F/F_0>1$,
$\Sigma/\Sigma_0>1$ and it appears that the ratio $B/B_0$ is also larger
than unity, but the numerical results need to be improved before further
conclusions can be drawn.

In principle, the matching formulae in 
\cite{Gasser:1984gg} can be used to calculate the SU(2) couplings $\bar{l}_i$ from 
the SU(3) couplings $L_i$.%
\footnote{For instance, for the MILC data this yields
$\bar{l}_3\!=\!3.32(64)(45)$ and $\bar{l}_4=4.03(16)(17)$ \cite{Bazavov:2009fk}.}
This procedure, however, yields less accurate results than a direct determination 
within SU(2), as it relies on the expansion in powers of $m_s$, where the omitted 
higher order contributions generate comparatively large uncertainties. We plead with
every collaboration performing $\Nf=2+1$ simulations to \emph{directly}
analyze their data in the SU(2) framework.
In practice, lattice simulations are performed at values of $m_s$ close to
the physical value and the results are then corrected for the difference of
$m_s$ from its physical value. If simulations with more than one value of
$m_s$ have been performed, then this can be done by
interpolation. Alternatively one can use the technique of
\emph{reweighting} (for a review see \cite{Jung:2010jt}), which applies an
a posteriori correction to shift $m_s$ into its physical value.  


\section{Kaon $B$-parameter $B_K$}
\label{sec:BK}
\subsection{Indirect CP-violation and $\epsilon_{K}$}

The mixing of neutral pseudoscalar mesons plays an important role in
the understanding of the physics of CP-violation. In this section we
will only focus on $K^0 - \bar K^0$ oscillations, which probe the
physics of indirect CP-violation. We collect and comment the
basic formulae; for extended reviews on the subject see, among others,
refs.~\cite{Branco:1999fs,Buchalla:1995vs,Buras:1998raa}. Indirect
CP-violation arises in $K_L \rightarrow \pi \pi$ transitions through
the decay of the $\rm CP=+1$ component of $K_L$ into two pions (which
are also in a $\rm CP=+1$ state). Its measure is defined as
\be 
\epsilon_{K} \,\, = \,\, \dfrac{{\cal A} [ K_L \rightarrow
(\pi\pi)_{I=0}]}{{\cal A} [ K_S \rightarrow (\pi\pi)_{I=0}]} \,\, ,
\ee
with the final state having total isospin zero. The parameter
$\epsilon_{K}$ may also be expressed in terms of $K^0 - \bar K^0$
oscillations. In particular, to lowest order in the electroweak theory, the
contribution to these oscillations arises from so-called box diagrams, in
which two $W$-bosons and two ``up-type'' quarks (i.e. up, charm, top) are
exchanged between the constituent down and strange quarks of the
$K$-mesons. The loop integration of the box diagrams can be performed
exactly. In the limit of vanishing external momenta and external quark
masses, the result can be identified with an effective four-fermion
interaction, expressed in terms of the ``effective Hamiltonian''
\be
  {\cal H}_{\rm eff}^{\Delta S = 2} \,\, = \,\, 
  \frac{G_F^2 M_{\rm{W}}^2}{16\pi^2} {\cal F}^0 Q^{\Delta S=2} \,\, + 
   \,\, {\rm h.c.} \,\,. 
\ee
In this expression, $G_F$ is the Fermi coupling, $M_{\rm{W}}$ the 
$W$-boson mass, and 
\be
   Q^{\Delta S=2} =
   \left[\bar{s}\gamma_\mu(1-\gamma_5)d\right]
   \left[\bar{s}\gamma_\mu(1-\gamma_5)d\right]
   \equiv O_{\rm VV+AA}-O_{\rm VA+AV} \,\, ,
\ee
is a dimension-six, four-fermion operator. The function ${\cal F}^0$
is given by
\be
{\cal F}^0 \,\, = \,\, \lambda_c^2 S_0(x_c) \, + \, \lambda_t^2
S_0(x_t) \, + \, 2 \lambda_c  \lambda_t S_0(x_c,x_t)  \,\, , 
\ee
where $\lambda_a = V^\ast_{as} V_{ad}$, and $a=c\,,t$ denotes a
flavour index. The quantities $S_0(x_c),\,S_0(x_t)$ and $S_0(x_c,x_t)$
with $x_c=m_c^2/M_{\rm{W}}^2$, $x_t=m_t^2/M_{\rm{W}}^2$ are the
Inami-Lim functions \cite{Inami:1980fz}, which express the basic
electroweak loop contributions without QCD corrections. The
contribution of the up quark, which is taken to be massless in this
approach, has been taken into account by imposing the unitarity
constraint $\lambda_u + \lambda_c + \lambda_t = 0$.

When strong interactions are included, $\Delta{S}=2$ transitions can
no longer be discussed at the quark level. Instead, the effective
Hamiltonian must be considered between mesonic initial and final
states. Since the strong coupling constant is large at typical
hadronic scales, the resulting weak matrix element cannot be
calculated in perturbation theory. The operator product expansion
(OPE) does, however, factorize long- and short- distance effects. For
energy scales below the charm threshold, the $K^0-\bar K^0$ transition
amplitude of the effective Hamiltonian can be expressed as
\begin{eqnarray}
\label{eq:Heff}
\langle \bar K^0 \vert {\cal H}_{\rm eff}^{\Delta S = 2} \vert K^0
\rangle  \,\, = \,\, \frac{G_F^2 M_{\rm{W}}^2}{16 \pi^2}  
\Big [ \lambda_c^2 S_0(x_c) \eta_1  \, + \, \lambda_t^2 S_0(x_t)
  \eta_2 \, + \, 2 \lambda_c  \lambda_t S_0(x_c,x_t) \eta_3
  \Big ]  \nn \\ 
\times \left(\frac{\gbar(\mu)^2}{4\pi}\right)^{-\gamma_0/(2\beta_0)}
   \left\{ 1+\dfrac{\gbar(\mu)^2}{(4\pi)^2}\left[
   \frac{\beta_1\gamma_0-\beta_0\gamma_1}{2\beta_0^2} \right]\right\}\,
   \langle \bar K^0 \vert  Q^{\Delta S=2}_{\rm R} (\mu) \vert K^0
   \rangle \,\, + \,\, {\rm h.c.} \,\, ,
\end{eqnarray}
where $\gbar(\mu)$ and $Q^{\Delta S=2}_{\rm R}(\mu)$ are the
renormalized gauge coupling and four-fermion operator in some
renormalization scheme. The factors $\eta_1, \eta_2$ and $\eta_3$
depend on the renormalized coupling $\gbar$, evaluated at the various
flavour thresholds $m_t, m_b, m_c$ and $ M_{\rm{W}}$, as required by
the OPE and RG-running procedure that separates high- and low-energy
contributions. Explicit expressions can be found
in~\cite{Buchalla:1995vs} and references therein. We follow the same
conventions for the RG-equations as in
ref.~\cite{Buchalla:1995vs}. Thus the Callan-Symanzik function and the
anomalous dimension $\gamma(\gbar)$ of $Q^{\Delta S=2}$ are defined by
\be
\dfrac{d \gbar}{d \ln \mu} = \beta(\gbar)\,,\qquad
\dfrac{d Q^{\Delta S=2}_{\rm R}}{d \ln \mu} =
-\gamma(\gbar)\,Q^{\Delta S=2}_{\rm R} \,\,,  
\ee
with perturbative expansions
\begin{eqnarray}
\beta(g)  &=&  -\beta_0 \dfrac{g^3}{(4\pi)^2} \,\, - \,\, \beta_1
\dfrac{g^5}{(4\pi)^4} \,\, - \,\, \cdots 
\\
\gamma(g)  &=&  \gamma_0 \dfrac{g^2}{(4\pi)^2} \,\, + \,\,
\gamma_1 \dfrac{g^4}{(4\pi)^4} \,\, + \,\, \cdots \,.\nn
\end{eqnarray}
We stress that $\beta_0, \beta_1$ and $\gamma_0$ are universal,
i.e. scheme-independent. $K^0-\bar K^0$ mixing is usually considered
in the naive dimensional regularization (NDR) scheme of $\msbar$, and
below we specify the perturbative coefficient $\gamma_1$ in that
scheme:
\begin{eqnarray}
& &\beta_0 = 
         \left\{\frac{11}{3}N-\frac{2}{3}\Nf\right\}, \qquad
   \beta_1 = 
         \left\{\frac{34}{3}N^2-\Nf\left(\frac{13}{3}N-\frac{1}{N}
         \right)\right\}, \\[0.3ex]
& &\gamma_0 = \frac{6(N-1)}{N}, \qquad
         \gamma_1 = \frac{N-1}{2N} 
         \left\{-21 + \frac{57}{N} - \frac{19}{3}N + \frac{4}{3}\Nf
         \right\}\,.\nn
\end{eqnarray}
Note that for QCD the above expressions must be evaluated for $N=3$
colours, while $\Nf$ denotes the number of active quark flavours. As
already stated, eq.~(\ref{eq:Heff}) is valid at scales below the charm
threshold, after all heavier flavours have been integrated out,
i.e. $\Nf = 3$.

In eq.~(\ref{eq:Heff}), the terms proportional to $\eta_1,\,\eta_2$
and $\eta_3$, multiplied by the contributions containing
$\gbar(\mu)^2$, correspond to the Wilson coefficient of the OPE,
estimated at NLO in perturbation theory. Its dependence on the
renormalization scheme and scale $\mu$ is cancelled by that of the
weak matrix element $\langle \bar K^0 \vert Q^{\Delta S=2}_{\rm R}
(\mu) \vert K^0 \rangle$. The latter corresponds to the long-distance
effects of the effective Hamiltonian and must be computed
non-perturbatively. For historical, as well as technical reasons, it
is convenient to express it in terms of the $B$-parameter $B_{K}$,
defined as
\be
   B_{K}(\mu)= \frac{{\left\langle\bar{K}^0\left|
         Q^{\Delta S=2}_{\rm R}(\mu)\right|K^0\right\rangle} }{
         {\frac{8}{3}\fK^2\mK^2}} \,\, .
\ee
The four-quark operator $Q^{\Delta S=2}(\mu)$ is renormalized at scale
$\mu$ in some regularization scheme, usually taken to be NDR. The
renormalization group independent (RGI) $B$-parameter $\hat{B}_{\rm
K}$ is related to $B_{K}(\mu)$ by the exact formula
\be
  \hat{B}_{K} = 
  \left(\frac{\gbar(\mu)^2}{4\pi}\right)^{-\gamma_0/(2\beta_0)}
  \exp\bigg\{ \int_0^{\gbar(\mu)} \, dg \, \bigg(
  \frac{\gamma(g)}{\beta(g)} \, + \, \frac{\gamma_0}{\beta_0g} \bigg)
  \bigg\} \, B_{K}(\mu) \,\,\, .
\ee
At NLO in perturbation theory, the above reduces to
\be
   \hat{B}_{K} =
   \left(\frac{\gbar(\mu)^2}{4\pi}\right)^{- \gamma_0/(2\beta_0)}
   \left\{ 1+\dfrac{\gbar(\mu)^2}{(4\pi)^2}\left[
   \frac{\beta_1\gamma_0-\beta_0\gamma_1}{2\beta_0^2} \right]\right\}\,
   B_{K}(\mu) \,\,\, ,
\label{eq:BKRGI_NLO}
\ee
which, to this order, is the scale-independent product of all
$\mu$-dependent quantities in eq.~(\ref{eq:Heff}).

The implementation of non-perturbative renormalization in lattice
calculations is based on the numerical evaluation of a
non-perturbative matching condition between the bare operator matrix
element and its counterpart in some intermediate renormalization
scheme, where the latter has a controlled perturbative relation to
continuum schemes such as $\msbar$. Examples of intermediate schemes
are the RI/MOM scheme \cite{Martinelli:1994ty} (also dubbed the
``Rome-Southampton method'') and the Schr\"odinger functional (SF)
scheme \cite{Luscher:1992an}. Typical scales for the transition
between lattice regularization and intermediate scheme are
$\mu=\rmO(1\,\gev)$. When applied to the case at hand, this implies
that $B_{\rm{K}}(\mu\approx 1\,\gev)$ in the intermediate scheme must
then be matched to the RGI $B$-parameter via
eq.\,(\ref{eq:BKRGI_NLO}). Obviously, due to asymptotic freedom, the
NLO relation provides a more reliable link if it is applied at scales
far greater than $\mu=\rmO(1\,\gev)$. Therefore, in order to remove
any doubts about the use of perturbation theory at NLO when
determining $\hat{B}_{\rm{K}}$, it is advantageous to run
non-perturbatively to larger values of $\mu$, where
eq.\,(\ref{eq:BKRGI_NLO}) provides an accurate matching
relation. Indeed, neglecting higher orders in perturbation theory has
been identified by the authors of\,\cite{Aubin:2009jh} as one of the
dominant systematic uncertainties in their result for $\hat
B_{\rm{K}}$. We note that the short-distance QCD corrections
in eq.\,(\ref{eq:Heff}) are known in perturbation theory, to NLO 
for $\eta_1$, $\eta_2$ and to NNLO for $\eta_3$ \cite{Brod:2010mj}. 
This implies that an
uncertainty of $\rmO(\alpha_s(m_c))$ pertains to the $K^0 -\bar K^0$
transition amplitude and $\epsilon_{K}$, even if the computation
of the kaon $B$-parameter is essentially free from perturbative
effects. However, since $\eta_1$ and $\eta_2$ are currently
being determined beyond NLO as well \cite{Nierste_private}, the issue of
eliminating systematic uncertainties arising from the use of
perturbation theory in the matching procedure becomes more acute.

The ``master formula'', for $\epsilon_{K}$, which connects the
experimentally observable quantity $\epsilon_{K}$ to the matrix
element of ${\cal H}_{\rm eff}^{\Delta S = 2}$,
is~\cite{Buras:1998raa,Anikeev:2001rk,Nierste:2009wg}
\be
\epsilon_{K} \,\,\, = \,\,\, \exp(i \phi_\epsilon) \,\,
\sin(\phi_\epsilon) \,\, \Big [ \frac{\Im [ \langle \bar K^0 \vert
{\cal H}_{\rm eff}^{\Delta S = 2} \vert K^0 \rangle ]} {\Delta M_K }
\,\,\, + \,\,\, \frac{\Im(A_0)}{\Re(A_0)} \,\, \Big ] \,\,\, ,
\label{eq:epsK}
\ee
for $\lambda_u$ real and positive; the phase of $\epsilon_{K}$ is
given by
\be
\phi_\epsilon \,\,\, = \,\,\, \arctan \frac{\Delta M_{K}}{\Delta
  \Gamma_{K}/2} \,\,\, . 
\ee
The quantities $\Delta M_K$ and $\Delta \Gamma_K$ are the mass- and
decay width-differences between long- and short-lived neutral Kaons,
while $A_0$ is the amplitude of the Kaon decay into an isospin-0 two
pion state. Experimentally known values of the above quantities are:
\begin{eqnarray}
\vert \epsilon_{K} \vert \,\, &=& \,\, 2.280(13) \times 10^{-3} \,\,\, ,
\nn \\
\phi_\epsilon \,\, &=& \,\, 43.51(5)^\circ \,\,\, ,
 \\
\Delta M_{K} \,\, &=& \,\, 3.491(9) \times 10^{-12}\, {\rm MeV} \,\,\, ,
\nn \\
\Delta \Gamma_{K}  \,\, &=& \,\ 7.335(4) \times 10^{-15} \,{\rm GeV} \,\,\,,\nn
\end{eqnarray}
while the last term in the square brackets of eq.~(\ref{eq:epsK}) has
been estimated in ref.\,\cite{Buras:2008nn}. Long distance
contributions, omitted from the first term in the square brackets of
eq.~(\ref{eq:epsK}) have been recently discussed in
ref.\cite{Buras:2010pz}.

\subsection{Lattice computation of $B_{K}$}

The lattice calculation of $B_{K}$ is affected by the same systematic
effects, discussed in previous sections. However, the issue of
renormalization merits special attention. The reason is that the
multiplicative renormalizability of the relevant operator $Q^{\Delta
S=2}$ is lost, once the regularized QCD action ceases to be invariant
under chiral transformations. With Wilson fermions, $Q^{\Delta S=2}$
mixes with four additional dimension-six operators which belong to
different representations of the chiral group, with mixing
coefficients that are finite functions of the gauge coupling. This
complicated renormalization pattern is the main source of systematic
error in computations of $B_{K}$ with Wilson quarks. It can be
bypassed via the implementation of specifically designed methods,
which are either based on Ward identities~\cite{Becirevic:2000cy} or
on a modification of the Wilson quark action, known as twisted mass
QCD~\cite{Frezzotti:2000nk,Dimopoulos:2006dm}. An advantage of
staggered fermions is the presence of a remnant $U(1)$ chiral
symmetry. However, at non-vanishing lattice spacing, the symmetry
among the extra unphysical degrees of freedom (tastes) is broken. As a
result, mixing with other dimension-six operators cannot be avoided in
the staggered formulation, which complicates the determination of the
$B$-parameter.

Fermionic lattice actions based on the Ginsparg-Wilson
relation~\cite{Ginsparg:1981bj} are invariant under the chiral group,
and hence four-quark operators such as $Q^{\Delta S=2}$ renormalize
multiplicatively. However, depending on the particular formulation of
Ginsparg-Wilson fermions, residual chiral symmetry breaking effects
may be present in actual calculations. For instance, in the case of
domain wall fermions, the finiteness of the extra 5th dimension
implies that the decoupling of modes with different chirality is not
exact, which produces a residual non-zero quark mass in the chiral
limit. Whether or not a significant mixing with dimension-six
operators is induced as well must be investigated on a case-by-case
basis.

\begin{figure}[ht]
\centering
\includegraphics[width=7.5cm]{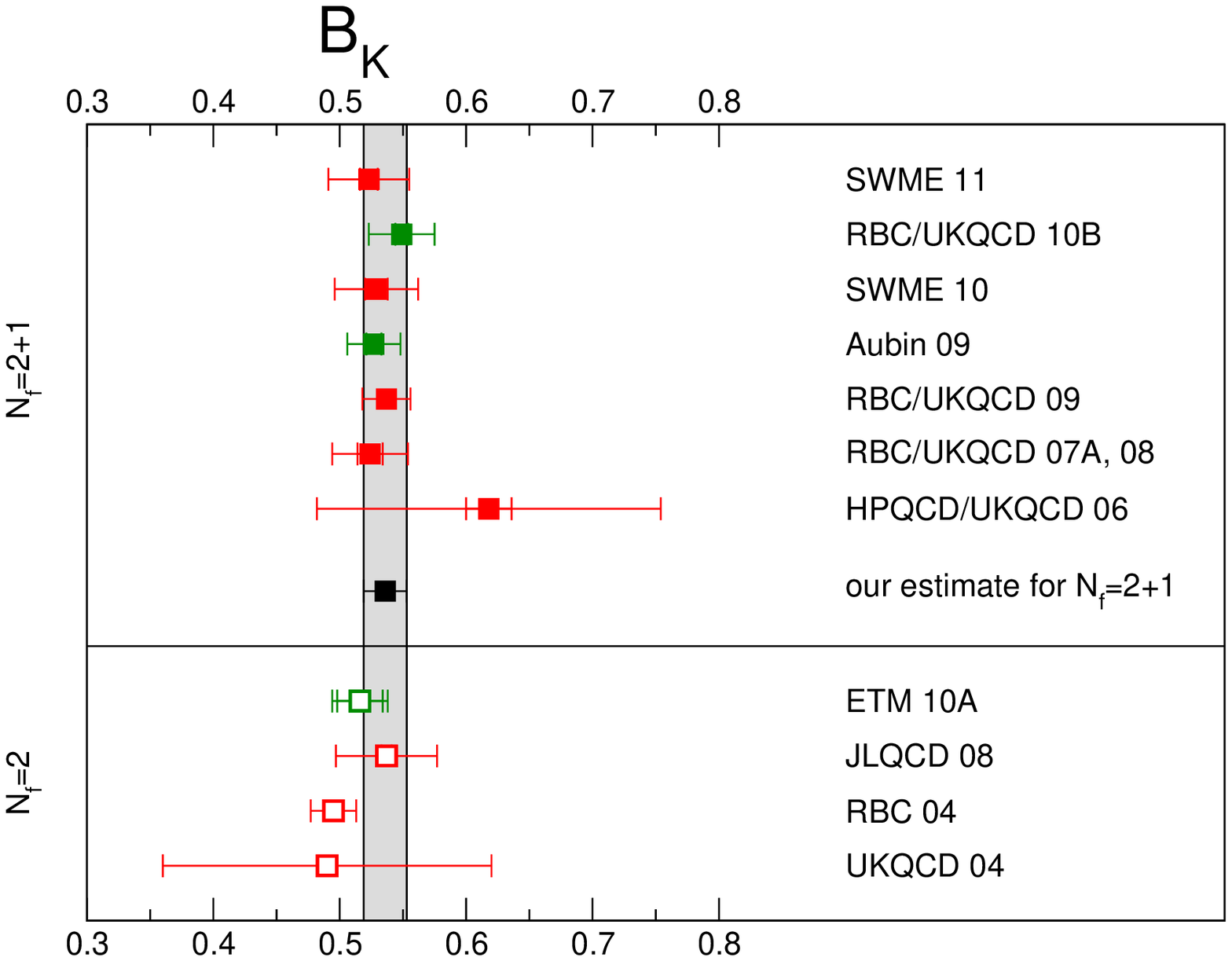}
\hfill
\includegraphics[width=7.5cm]{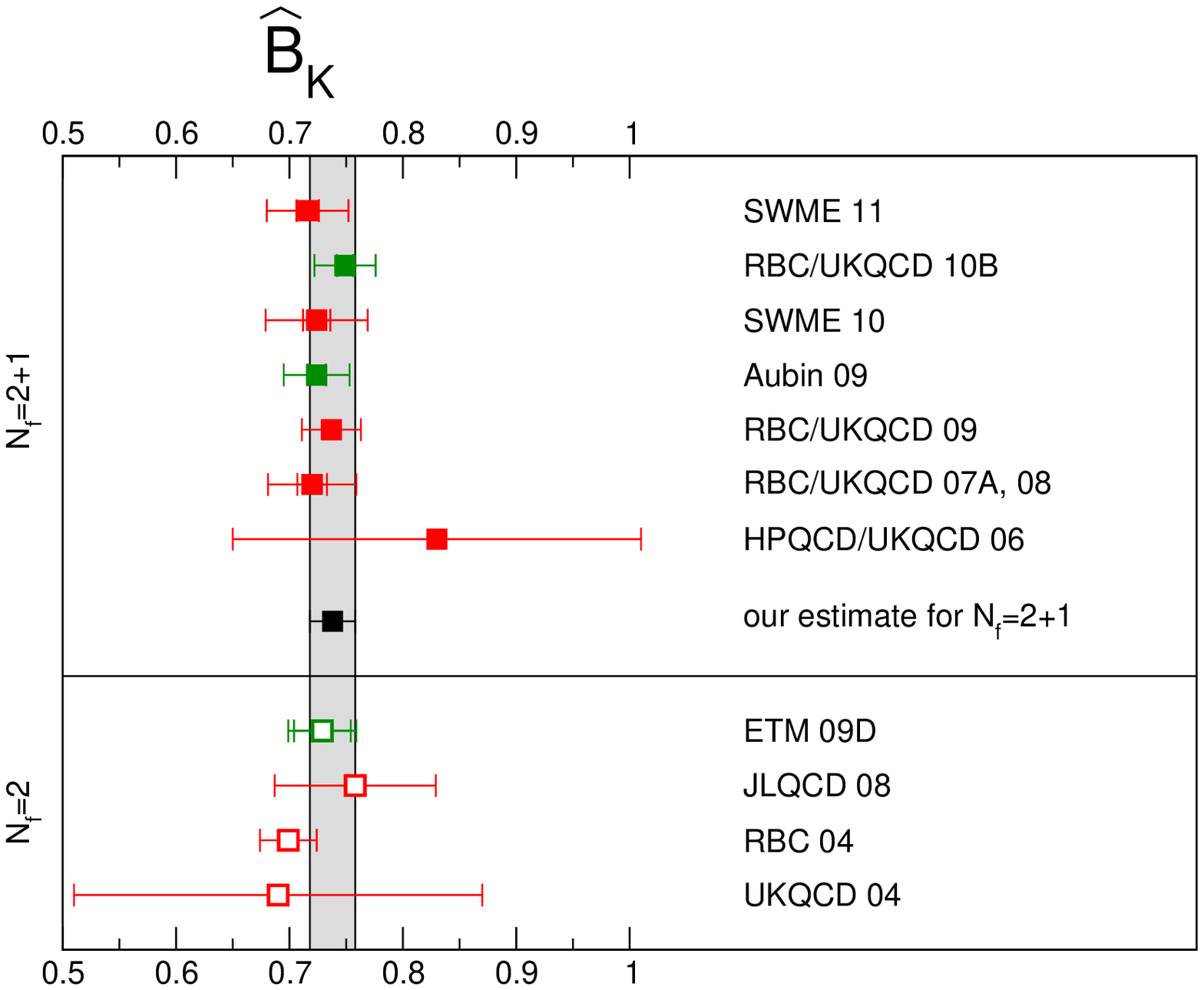}
\caption{Recent unquenched lattice results for $B_K$ ($\msbar$-scheme,
  scale $\mu =2\,\gev$) and for the corresponding RGI parameter
  $\hat{B}_{K}$. The gray band indicates our estimate.
\label{fig_BKsumm}}
\end{figure}

In this section we focus on recent results for $B_{\rm{K}}$, obtained
for $\Nf=2$ and $2+1$ flavours of dynamical quarks. A compilation of
results is shown in Table~\ref{tab_BKsumm} and
Fig.\,\ref{fig_BKsumm}. An overview of the quality of systematic error
studies is represented by the colour coded entries in
Table~\ref{tab_BKsumm}. In Appendix~\ref{app-BK} we gather the
simulation details and results from different collaborations, the
values of the most relevant lattice parameters, and comparative tables
on the various estimates of systematic errors. Note that some
references do not quote results for both
$B_{\rm{K}}$ and $\hat{B}_{K}$. In this case we
have performed the conversion ourselves by evaluating the
proportionality factor in eq.\,(\ref{eq:BKRGI_NLO}) for
$\mu=2\,\gev$. This requires fixing the value of
$\alpha_s(2\,\gev)$ in the two- and three-flavour cases. For
$\Nf=2$ we have inserted the non-perturbative result for the
$\Lambda$-parameter by the ALPHA Collaboration
\cite{DellaMorte:2004bc},
i.e.\,$\Lambda^{(2)}=245(16)(16)\,\mev$, into the NLO
expression for $\alpha_s$, neglecting the quoted error. This gives
$\hat{B}_K/B_K=1.412$, in accordance
with ref.\,\cite{Aoki:2008ss}. For $\Nf=3$ we use the value
$\alpha_s(M_{\rm{Z}})=0.1184$\,\cite{Nakamura:2010zzi} and run it
across the quark thresholds down to $\mu=2\,\gev$, using the four-loop
RG $\beta$-function. We then obtain $\hat{B}_{\rm
K}/B_{\rm{K}}=1.368$ in the three-flavour theory.

\begin{table}[t]
\begin{center}
\mbox{} \\[3.0cm]
\footnotesize
\begin{tabular*}{\textwidth}[l]{l @{\extracolsep{\fill}} r l l l l l l l l l}
Collaboration & Ref. & $\Nf$ & 
\hspace{0.15cm}\begin{rotate}{60}{publication status}\end{rotate}\hspace{-0.15cm} &
\hspace{0.15cm}\begin{rotate}{60}{continuum extrapolation}\end{rotate}\hspace{-0.15cm} &
\hspace{0.15cm}\begin{rotate}{60}{chiral extrapolation}\end{rotate}\hspace{-0.15cm}&
\hspace{0.15cm}\begin{rotate}{60}{finite volume}\end{rotate}\hspace{-0.15cm}&
\hspace{0.15cm}\begin{rotate}{60}{renormalization}\end{rotate}\hspace{-0.15cm}  &
\hspace{0.15cm}\begin{rotate}{60}{running}\end{rotate}\hspace{-0.15cm} & 
\rule{0.3cm}{0cm}$B_{K}$ & \rule{0.3cm}{0cm}$\hat{B}_{K}$ \\
&&&&&&&&&& \\[-0.1cm]
\hline
\hline
&&&&&&&&&& \\[-0.1cm]

SWME 11 & \cite{Kim:2011qg} & 2+1 & \oP & \good & \soso & \soso & \tbr
& $-$ & 0.523(7)(26) & 0.716(10)(35) \\[0.5ex]

RBC/UKQCD 10B & \cite{Aoki:2010pe} & 2+1 & \oP & \soso & \soso & \good &
\good & $\,a$ & 0.549(5)(26) & 0.749(7)(26) \\[0.5ex] 

SWME 10 & \cite{Bae:2010ki} & 2+1 & \gA & \good & \soso & \soso & \tbr
& $-$ & 0.529(9)(32) &  0.724(12)(43) \\[0.5ex] 

Aubin 09 & \cite{Aubin:2009jh} & 2+1 & \gA & \soso & \tbg$^\Box$ &
     \soso & \tbg & $-$ & 0.527(6)(21)& 0.724(8)(29) \\[0.5ex]  

RBC/UKQCD 07A, 08 \rule{1em}{0em}& \cite{Antonio:2007pb,Allton:2008pn} & 2+1 & \gA
                              & \tbr & \soso & \tbg     & \tbg & $-$ &
0.524(10)(28) & 0.720(13)(37) \\[0.5ex]  
HPQCD/UKQCD 06  & \cite{Gamiz:2006sq} & 2+1 & \gA
                              & \tbr & \soso$^\ast$ & \tbg     & \tbr &
$-$ & 0.618(18)(135)& 0.83(18) \\[0.5ex]  
&&&&&&&&&& \\[-0.1cm]
\hline
&&&&&&&&&& \\[-0.1cm]
ETM 10A & \cite{Constantinou:2010qv} & 2 & \gA & \good & \soso & \soso
& \good&  $\,b$ &   0.516(18)(12)  & 0.729(25)(17) \\[0.5ex]
JLQCD 08 & \cite{Aoki:2008ss} & 2 & \gA  & \tbr      & \soso      &
\tbr          &\tbg    & $-$ & 0.537(4)(40) &
0.758(6)(71)\\[0.5ex]  
RBC 04   & \cite{Aoki:2004ht} & 2 & \gA & \tbr      & \tbr      &
\tbr$^\dagger$ & \tbg      &$-$ & 0.495(18)    & 0.699(25)
\\[0.5ex]  
UKQCD 04 & \cite{Flynn:2004au} & 2  & \gA  & \tbr      & \tbr      &
\tbr$^\dagger$ & \tbr      & $-$ & 0.49(13)     & 0.69(18)
\\[0.5ex]  
&&&&&&&&&& \\[-0.1cm]
\hline
\hline
\end{tabular*}
\begin{tabular*}{\textwidth}[l]{l@{\extracolsep{\fill}}lllllllll}
  \multicolumn{10}{l}{\vbox{\begin{flushleft} 
        $^\ast$ This result has been obtained with only two ``light'' sea quark masses. \\
        $^\dagger$ These results have been obtained at  $(M_\pi L)_{\rm
          min} > 4$ in a lattice box with a spatial extension  $L <
        2$~fm.\\ 
        $^\Box$ In this mixed action computation, the lightest valence pion
        weighs $\sim 230$~MeV, while the lightest
        sea\\\hspace{0.3cm}taste-pseudoscalar, used in the chiral fits,
        weighs $\sim 370$~MeV.\\\rule{0cm}{0.3cm}\hspace{-0.1cm}
        $a$ $B_K$ is renormalized non-perturbatively at a scale of 2 GeV
        in a couple of $\Nf = 3$ RI/SMOM schemes. A\\\hspace{0.3cm}careful
        study of perturbative matching uncertainties has been performed by
        comparing results in the two\\\hspace{0.3cm}schemes in the region
        of 2 GeV to 3 GeV \cite{Aoki:2010pe}.\\\rule{0cm}{0.3cm}\hspace{-0.1cm}
        $b$ $B_K$ is renormalized non-perturbatively at scales $1/a \sim 2
        \div 3\,\gev$ in the $\Nf = 2$ RI/MOM scheme. In
        this\\\hspace{0.3cm}scheme, non-perturbative and NLO running for
        the are shown to agree from 4 GeV down 2 GeV to
        better\\\hspace{0.3cm}than 3\%
        \cite{Constantinou:2010gr,Constantinou:2010qv}. 
\end{flushleft}}}
\end{tabular*}

\vspace{-0.5cm}
\caption{Results for the kaon $B$-parameter together with a summary of
  systematic errors. If information about non-perturbative running is
available, this is indicated in the column "running", with details given at the bottom of the
table.\label{tab_BKsumm}}
\end{center}
\end{table}

At present four collaborations have reported estimates of $B_{K}$ for
$\Nf=2+1$. The RBC/UKQCD collaboration uses domain wall fermions,
while the results by HPQCD/UKQCD are based on the configurations
generated by MILC, using Asqtad improved, rooted staggered
fermions. In Aubin~09 a mixed-action approach is adopted in which
staggered sea quarks\footnote{Aubin~09 also use configurations
generated by MILC.} are combined with HYP-smeared
\cite{Hasenfratz:2001hp} domain wall fermions. The SWME result
reported in~\cite{Kim:2011qg} is also based on the Asqtad MILC
ensembles, but uses HYP-smeared staggered quarks in the valence
sector.

In view of our quality criteria we note that one main shortcoming of
the results by HPQCD/UKQCD~06 \cite{Gamiz:2006sq} is that they are
based on data obtained at a single lattice spacing. This is also true
for the results of RBC/UKQCD~07A,\,08
\cite{Antonio:2007pb,Allton:2008pn}, which, however, have recently
been updated by including data at a second value of~$a$ (see
RBC/UKQCD~10B \cite{Aoki:2010pe}). The good agreement between the
earlier results and the latest update indicates that discretization
effects for $B_K$, computed using domain wall fermions appear to be
small at the present level of accuracy.  Aubin\,09 have also data at
two values of the lattice spacing, similar to those of RBC/UKQCD\,10B,
while the SWME\,10 \cite{Bae:2010ki} and SWME\,11 \cite{Kim:2011qg}
results are obtained at three and four lattice spacings,
respectively. Having two smaller lattice spacings at their disposal
compared to other collaborations, SWME~10,\,11 have performed the most
complete study of discretization effects for this quantity so far.

The determination of the renormalization factors merits special
attention, owing to the fact that the renormalization and mixing
patterns of four-quark operators such as $Q^{\Delta{S}=2}$ are
typically much more complicated than for quark bilinears. The
renormalization factors used by HPQCD/UKQCD~06 are based on
perturbation theory at one loop, which is by far the biggest source of
systematic uncertainty quoted by these authors. The same is true for
the SWME~10,\,11 results \cite{Bae:2010ki,Kim:2011qg}: Here the
estimated uncertainty in their latest estimate for $B_{K}$
\cite{Kim:2011qg} arising from neglecting higher orders in
perturbation theory amounts to 4.4\%, which accounts for about 95\% of
the entire systematic error reported by this collaboration. The two
other groups, RBC/UKQCD~07A,\,08,\,10B and Aubin~09, have both
implemented a non-perturbative renormalization procedure based on the
RI/MOM scheme. Nonetheless, Aubin~09 state that their biggest single
systematic uncertainty of 3.2\% in $B_{K}$ (or 85\% of the total
systematic error) is still associated with the uncertainty in the
renormalization factor which links the $B$-parameter of the bare
operator to that in the $\msbar$-scheme at 2\,GeV. 
RBC/UKQCD~10B\,\cite{Aoki:2010pe} have investigated several
different RI/MOM schemes, including so-called non-exceptional
momenta. The resulting spread of results in different schemes is then
included in the uncertainty for the renormalization factor. As observed
in\,\cite{Aoki:2010pe}, this error can be
significantly reduced if the conversion to the $\msbar$-scheme is
performed at 3\,GeV instead of the conventional choice, $\mu=2$\,GeV. 
For this reason, the RBC/UKQCD collaboration prefers to quote 
$B_{K}^{\msbar}(3\,\GeV)$, but also gives the value obtained at 
the conventional scale, for comparison with other results.
All of the numbers listed in the first column of
Table~\ref{tab_BKsumm} refer to the conventional scale. In
ref.\,\cite{Aoki:2007xm} the RBC/UKQCD collaboration also investigated the effects
of residual chiral symmetry breaking induced by the finite extent of
the 5th dimension in the domain wall fermion formulation and found
that the mixing of $Q^{\Delta{S}=2}$ with operators of opposite
chirality was negligibly small.

The rules of section \ref{sec:color-code} stipulate that our averages
only involve results which are free of red tags and are published in a
refereed journal. Papers which, at the time of writing, are still
unpublished but are obvious updates of earlier published results
should also be taken into account. In view of the above, we combine
the results of Aubin~09 and RBC/UKQCD 10B, by applying the following
procedure: in a first step statistical and systematic errors are added
in quadrature, in order to have an overall error for the result of a
particular collaboration. In a second step the results from the two
groups are combined in a weighted average. This yields
\be
\Nf=2+1:\hspace{1cm}B_{K}^\msbar (2 {\rm GeV}) = 0.536(17)\,  \hspace{1cm}
\hat{B}_{K} = 0.738(20)\, .
\label{BKfinal-3}
\ee
Here we have used the RBC/UKQCD\,10B result at the scale of 2\,GeV, in
spite of the bigger uncertainty reported for the associated
renormalization factor.

Due to the use of one-loop perturbation theory for the renormalization
factor, the SWME results are assigned a red tag and are thus not
included in the average. However it must be stressed that they are in
excellent agreement with eq.\,(\ref{BKfinal-3}). Also the earlier
HPQCD/UKQCD estimate, albeit with bigger statistical and systematic
errors, agrees with our average. The observation that independent
calculations produce compatible results -- despite the fact that
according to the colour coding in Table\,\ref{tab_BKsumm} some
systematic effects are not fully controlled -- lends further
confidence to the result quoted in eq.\,(\ref{BKfinal-3}).

We now pass over to the results computed for $\Nf = 2$ flavours of
dynamical quarks. The UKQCD~04 result \cite{Flynn:2004au}, computed
with $\rmO(a)$ improved Wilson fermions, is afflicted by many
systematic errors and has a rather large overall error. Results by the
JLQCD collaboration \cite{Aoki:2008ss} are computed using overlap
fermions in both the sea and valence quark sectors. Besides the usual
systematic uncertainties listed in Table~\ref{tab_BKsumm}, they quote
a 5\% systematic error due to the uncertainty in the physical value of
$r_0$, which sets the scale in their simulation. They also quote a
1.4\% systematic error, owing to the fact that they determine
$B_{\rm{K}}$ on gauge configurations with a fixed value of the
topological charge. In view of the high numerical cost of simulations
with dynamical overlap quarks, it is not surprising that the lattice
sizes of ref.\,\cite{Aoki:2008ss} are smaller than those in present
simulations with $\Nf=2+1$ flavours of dynamical staggered or domain
wall fermions. Consequently, finite-volume effects are less well
controlled in ref.\,\cite{Aoki:2008ss}. The RBC~04 result
\cite{Aoki:2004ht} has been obtained using domain wall fermions at
rather heavy pseudoscalar (pion) masses; otherwise it has similar
systematics as the JLQCD~08 run. The result by ETM~09D
\cite{Bertone:2009bu} is based on a set of ensembles which allows for
an extensive investigation of systematic uncertainties. In particular,
it is so far the only $N_f=2$ calculation involving data computed at three
values of the lattice spacing. Being the only result without red tags,
it is quoted as the currently best overall estimate for $N_f = 2$:
\be
\Nf=2:\hspace{1cm}B_{K}^\msbar (2 {\rm GeV}) = 0.516(18)(12)\,  \hspace{1cm}
\hat{B}_{K} = 0.729(25)(17)\, .
\label{BKfinal-2}
\ee
It is clear that with the present level of accuracy, any dependence of
$B_{K}$ on the number of dynamical quark flavours, $\Nf$, is not
visible.

It is interesting to compare these results to the best available
quenched data, computed for large (valence) quark masses and with
degenerate down and strange quarks. Avoiding a detailed discussion
about the systematics, we show an (incomplete) compilation of
indicative results in Tab.\,\ref{tab_BKquen}. We see that they agree
with the ones obtained with $\Nf = 2$ and $\Nf = 2+1$ at the level of
two standard deviations.
\begin{table}[h]
\begin{center}
\footnotesize
\begin{tabular*}{\textwidth}[l]{l @{\extracolsep{\fill}} l l l l }
\hline\hline
&&& \\[-0.1cm]
Collaboration & Ref. & $\Nf$ &\rule{0.1cm}{0cm} $B_{K}$ &
\rule{0.1cm}{0cm}$\hat{B}_{K}$ \\  
&&& \\[-0.1cm]
\hline
&&& \\[-0.1cm]
ALPHA 09  & \cite{Dimopoulos:2009es} & 0 & 0.532(25) & 0.73(3)  \\[0.5ex]
CP-PACS 08& \cite{Nakamura:2008xz}   & 0 & 0.565(6)  & 0.782(9) \\[0.5ex]
ALPHA 07  & \cite{Dimopoulos:2007cn} & 0 & 0.534(52) & 0.74(7)  \\[0.5ex]
JLQCD 97  & \cite{Aoki:1997nr}       & 0 & 0.628(42) & 0.86(6)  \\[0.5ex]
&&& \\[-0.1cm]
\hline
\hline
\end{tabular*}
\caption{Quenched results for the $B$-parameter $B_{\rm{K}}$ from
  various collaborations. Errors have been combined in quadrature.
\label{tab_BKquen}}
\end{center}
\end{table}

\section*{Acknowledgments}
We wish to thank Sinya Aoki, Silas Beane, Claude Bernard, Joachim Brod,
Petros Dimopoulos, Xining Du, J\"urg Gasser, Peter Hasenfratz, Shoji
Hashimoto, Gregorio Herdoiza, Matthias Jamin, Karl Jansen, Kim Maltman, Ferenc Niedermayer,
Uli Nierste, Carlos Pena, Matts Roos, Gerrit Schierholz, Stephen Sharpe and
Andr\'e Walker-Loud for correspondence and useful comments.  The Albert
Einstein Center for Fundamental Physics at the University of Bern is
supported by the ``Innovations- und Kooperationsprojekt C-13'' of the
``Schweizerische Universit\"atskonferenz SUK/CRUS''. This work was
partially supported by CNRS grants GDR n$^0$ 2921 (Physique subatomique et
calculs sur r\'eseau) and PICS n$^0$ 4707, by UK STFC grant ST/G000557/1,
by the Helmholtz Association through the virtual institute ``Spin and
strong QCD'' (VH-VI-231), by the Swiss National Science Foundation and by
EU contract MRTN-CT-2006-035482 (Flavianet).

\appendix

\begin{appendix}
\section{Glossary}\label{comm}
\subsection{Lattice actions}
In this appendix we give brief descriptions of the lattice actions
used in the simulations and summarize their main features.

\subsubsection{Gauge actions \label{sec_gauge_actions}}

The simplest and most widely used discretization of the Yang-Mills
part of the QCD action is the Wilson plaquette action\,\cite{Wilson:1974sk}:
\be
 S_{\rm G} = \beta\sum_{x} \sum_{\mu<\nu}\Big(
  1-\frac{1}{3}{\rm Re\,\Tr}\,W_{\mu\nu}^{1\times1}(x)\Big),
\label{eq_plaquette}
\ee
where the plaquette, $W_{\mu\nu}^{1\times1}(x)$, is the product of
link variables around an elementary square of the lattice, i.e.
\be
  W_{\mu\nu}^{1\times1}(x) \equiv U_\mu(x)U_\nu(x+a\hat{\mu})
   U_\mu(x+a\hat{\nu})^{-1} U_\nu(x)^{-1}.
\ee
This expression reproduces the Euclidean Yang-Mills action in the
continuum up to corrections of order~$a^2$.  There is a general
formalism, known as the ``Symanzik improvement programme''
\cite{Symanzik:1983dc,Symanzik:1983gh}, which is designed to cancel
the leading lattice artifacts, such that observables have an
accelerated rate of convergence to the continuum limit.  The
improvement programme is implemented by adding higher-dimensional
operators, whose coefficients must be tuned appropriately in order to
cancel the leading lattice artifacts. The effectiveness of this
procedure depends largely on the method with which the coefficients
are determined. The most widely applied methods (in ascending order of
effectiveness) include perturbation theory, tadpole-improved
(partially resummed) perturbation theory, renormalization group
methods, and the non-perturbative evaluation of improvement
conditions.

In the case of Yang-Mills theory, the simplest version of an improved
lattice action is obtained by adding rectangular $1\times2$ loops to
the plaquette action, i.e.
\be
   S_{\rm G}^{\rm imp} = \beta\sum_{x}\left\{ c_0\sum_{\mu<\nu}\Big(
  1-\frac{1}{3}{\rm Re\,\Tr}\,W_{\mu\nu}^{1\times1}(x)\Big) +
   c_1\sum_{\mu,\nu} \Big(
  1-\frac{1}{3}{\rm Re\,\Tr}\,W_{\mu\nu}^{1\times2}(x)\Big) \right\},
\label{eq_Sym}
\ee
where the coefficients $c_0, c_1$ satisfy the normalization condition
$c_0+8c_1=1$. The {\sl Symanzik-improved \cite{Luscher:1984xn},
Iwasaki \cite{Iwasaki:1985we}}, and {\sl DBW2}
\cite{Takaishi:1996xj,deForcrand:1999bi} actions are all defined
through \eq{eq_Sym} via particular choices for $c_0, c_1$. Details are
listed in Table\,\ref{tab_gaugeactions} together with the
abbreviations used in the summary tables.
\vspace{-0.07cm}
\begin{table}[!h]
\begin{center}
{\footnotesize
\begin{tabular*}{\textwidth}[l]{l @{\extracolsep{\fill}} c l}
\hline\hline \\[-1.0ex]
Abbrev. & $c_1$ & Description 
\\[1.0ex] \hline \hline \\[-1.0ex]
Wilson    & 0 & Wilson plaquette action \\[1.0ex] \hline \\[-1.0ex]
tlSym   & $-1/12$ & tree-level Symanzik-improved gauge action \\[1.0ex] \hline \\[-1.0ex]
tadSym  & variable & tadpole Symanzik-improved gauge
action \\[1.0ex] \hline \\[-1.0ex]
Iwasaki & $-0.331$ & Renormalization group improved (``Iwasaki'')
action \\[1.0ex] \hline \\[-1.0ex]
DBW2 & $-1.4088$ & Renormalization group improved (``DBW2'') action 
\\ [1.0ex] 
\hline\hline
\end{tabular*}
}
\caption{Summary of lattice gauge actions. The leading lattice
 artifacts are $O(a^2)$ or better for all
  discretizations. \label{tab_gaugeactions}} 
\end{center}
\end{table}
\clearpage

\subsubsection{Quark actions \label{sec_quark_actions}}

If one attempts to discretize the quark action, one is faced with the
fermion doubling problem: the naive lattice transcription produces a
16-fold degeneracy of the fermion spectrum. \\

\noindent
{\it Wilson fermions}\\
\noindent
Wilson's solution to the
doubling problem is based on adding a dimension-5 operator which
removes the doublers from the low-energy spectrum. The Wilson-Dirac
operator for the massless case reads \cite{Wilson:1974sk}
\be
     D_{\rm w} = \half\gamma_\mu(\nabla_\mu+\nabla_\mu^*)
   +a\nabla_\mu^*\nabla_\mu,
\ee
where $\nabla_\mu,\,\nabla_\mu^*$ denote lattice versions of the
covariant derivative. Adding the Wilson term,
$a\nabla_\mu^*\nabla_\mu$, results in an explicit breaking of chiral
symmetry even in the massless theory. Furthermore, the leading order
lattice artifacts are of order~$a$. With the help of the Symanzik
improvement programme, the leading artifacts can be cancelled by
adding the so-called ``Clover'' or Sheikholeslami-Wohlert (SW) term. The
resulting expression in the massless case reads
\be
   D_{\rm sw} = D_{\rm w}
   +\frac{ia}{4}\,\csw\sigma_{\mu\nu}\widehat{F}_{\mu\nu},
\label{eq_DSW}
\ee
where $\sigma_{\mu\nu}=\frac{i}{2}[\gamma_\mu,\gamma_\nu]$, and
$\widehat{F}_{\mu\nu}$ is a lattice transcription of the gluon field
strength tensor $F_{\mu\nu}$. Provided that the coefficient $\csw$ is
suitably tuned, observables computed using $D_{\rm sw}$ will approach
the continuum limit with a rate proportional to~$a^2$. Chiral symmetry
remains broken, though. The coefficient $\csw$ can be determined
perturbatively at tree-level (tree-level impr., $\csw = 1$ or tlSW in
short), via a mean field approach \cite{Lepage:1992xa} (mean-field
impr. or mfSW) or via a non-perturbative approach
\cite{Luscher:1996ug} (non-perturbativley impr. or npSW).

Finally, we mention ``twisted mass QCD'' as a method which was
originally designed to address another problem of Wilson's
discretization: the Wilson-Dirac operator is not protected against the
occurrence of unphysical zero modes, which manifest themselves as
``exceptional'' configurations. They occur with a certain frequency in
numerical simulations with Wilson quarks and can lead to strong
statistical fluctuations. The problem can be cured by introducing a
so-called ``chirally twisted'' mass term, after which the fermionic
part of the QCD action in the continuum assumes the form
\cite{Frezzotti:2000nk}
\be
   S_{\rm F}^{\rm tm;cont} = \int d^4{x}\, \psibar(x)(\gamma_\mu
   D_\mu +
   m + i\mu_{\rm q}\gamma_5\tau^3)\psi(x).
\ee
Here, $\mu_{\rm q}$ is the twisted mass parameter, and $\tau^3$ is a
Pauli matrix. The standard action in the continuum can be recovered
via a global chiral field rotation.The lattice action of twisted mass
QCD (tmWil) for $\Nf=2$ flavours is defined as
\be
   S_{\rm F}^{\rm tm}[U,\psibar,\psi] = a^4\sum_{x\in\Lambda_{\rm
   E}}\psibar(x)(D_{\rm w}+m_0+i\mu_{\rm q}\gamma_5\tau^3)\psi(x).
\label{eq_tmQCD}
\ee
Although this formulation breaks physical parity and flavour
symmetries, is has a number of advantages over standard Wilson
fermions. In particular, the presence of the twisted mass parameter
$\mu_{\rm q}$ protects the discretized theory against unphysical zero
modes. Another attractive feature of twisted mass lattice QCD is the
fact that the leading lattice artifacts are of order $a^2$ without the
need to add the Sheikholeslami-Wohlert term
\cite{Frezzotti:2003ni}. Although the problem of explicit chiral
symmetry breaking remains, the twisted formulation is particularly
useful to circumvent some of the problems that are encountered in
connection with the renormalization of local operators on the lattice,
such as those required to determine $B_{\rm K}$. \\

\noindent
{\it Staggered fermions}\\
\noindent
An alternative procedure to deal with the doubling problem is based on
so-called ``staggered'' or Kogut-Susskind (KS) fermions
\cite{Susskind:1976jm}. Here the degeneracy is only lifted partially, from 16
down to~4. It has become customary to refer to these residual doublers as
``tastes'' in order to distinguish them from physical flavours. At order~$a^2$
different tastes can interact via gluon exchange, thereby generating large
lattice artifacts. The improvement programme can be used to suppress
taste-changing interactions, leading to ``improved staggered fermions'', with
the so-called ``Asqtad'' and ``HISQ'' actions as the most widely used versions
\cite{Orginos:1999cr,Follana:2006rc}. The standard procedure to remove the
residual doubling of staggered quarks (``four tastes per flavour'') is to take
fractional powers of the quark determinant in the QCD functional integral.
This is usually referred to as the ``fourth root trick''. The validity of this
procedure has not been rigorously proven so far. In fact, it has been
questioned by several authors, and the issue is still hotly debated (for both
sides of the argument see the reviews in refs.\,
\cite{{Durr:2005ax,Sharpe:2006re,Kronfeld:2007ek,Creutz:2007rk,Golterman:2008gt}}).
\\

\noindent
{\it Ginsparg-Wilson fermions}\\
\noindent
Fermionic lattice actions, which do not suffer from the doubling
problem whilst preserving chiral symmetry go under the name of
``Ginsparg-Wilson fermions''. In the continuum the massless Dirac
operator anti-commutes with $\gamma_5$. At non-zero lattice spacing
chiral symmetry can be realized even if this condition is relaxed
according to \cite{Hasenfratz:1998ri,Luscher:1998pqa}
\be
   \left\{D,\gamma_5\right\} = aD\gamma_5 D,
\label{eq_GWrelation}
\ee
which is now known as the Ginsparg-Wilson relation
\cite{Ginsparg:1981bj}. A lattice Dirac operator which satisfies
\eq{eq_GWrelation} can be constructed in several ways. The ``domain
wall'' construction proceeds by introducing a fifth dimension of
length $N_5$ and coupling the fermions to a mass defect (i.e. a
negative mass term) \cite{Kaplan:1992bt}. The five-dimensional action
can be constructed such that modes of opposite chirality are trapped
at the four-dimensional boundaries in the limit of an infinite extent
of the extra dimension \cite{Furman:1994ky}. In any real simulation,
though, one has to work with a finite value of $N_5$, so that the
decoupling of chiral modes is not exact. This leads to a residual
breaking of chiral symmetry, which, however, is exponentially
suppressed. A doubler-free, (approximately) chirally symmetric quark
action can thus be realized at the expense of simulating a
five-dimensional theory.

The so-called ``overlap'' or Neuberger-Dirac operator can be derived
from the domain wall formulation \cite{Neuberger:1997fp}. It acts in
four space-time dimensions and is, in its simplest form, defined by
\be
   D_{\rm N} = \frac{1}{\abar} \left( 1-\frac{A}{\sqrt{A^\dagger{A}}}
   \right), \quad A=1+s-aD_{\rm w},\quad \abar=\frac{a}{1+s},
\label{eq_overlap}
\ee
where $D_{\rm w}$ is the massless Wilson-Dirac operator, and $|s|<1$
is a tunable parameter. The overlap operator $D_{\rm N}$ removes all
doublers from the spectrum, and can easily be shown to satisfy the
Ginsparg-Wilson relation. The occurrence of an inverse square root in
$D_{\rm N}$ renders the application of $D_{\rm N}$ in a computer
program potentially very costly, since it must be implemented using,
for instance, a polynomial approximation.

The third example of an operator which satisfies the Ginsparg-Wilson
relation is the so-called fixed-point action
\cite{Hasenfratz:2000xz,Hasenfratz:2002rp}. This construction proceeds
via a renormalization group approach. A related formalism are the
so-called ``chirally improved'' fermions \cite{Gattringer:2000js}.\\

\begin{table}
\begin{center}
{\footnotesize
\begin{tabular*}{\textwidth}[l]{l @{\extracolsep{\fill}} l l l l}
\hline \hline  \\[-1.0ex]
\parbox[t]{1.5cm}{Abbrev.} & Discretization & \parbox[t]{2.2cm}{Leading lattice \\artifacts} & Chiral symmetry &  Remarks
\\[4.0ex] \hline \hline \\[-1.0ex]
Wilson     & Wilson & $O(a)$ & broken & 
\\[1.0ex] \hline \\[-1.0ex]
tmWil   & Twisted Mass Wilson &  \parbox[t]{2.2cm}{$O(a^2)$
at\\ maximal twist} & broken & \parbox[t]{5cm}{flavour symmetry breaking:\\ $(M_\text{PS}^{0})^2-(M_\text{PS}^\pm)^2\sim O(a^2)$}
\\[4.0ex] \hline \\[-1.0ex]
tlSW      & Sheikholeslami-Wohlert & $O(g^2 a)$ & broken & tree-level
impr., $\csw=1$
\\[1.0ex] \hline \\[-1.0ex]
\parbox[t]{1.0cm}{n-HYP tlSW}      & Sheikholeslami-Wohlert & $O(g^2 a)$ & broken & \parbox[t]{5cm}{tree-level
impr., $\csw=1$,\\
n-HYP smeared gauge links
}
\\[4.0ex] \hline \\[-1.0ex]
\parbox[t]{1.2cm}{n-stout tlSW}      & Sheikholeslami-Wohlert & $O(g^2 a)$ & broken & \parbox[t]{5cm}{tree-level
impr., $\csw=1$,\\
n-stout smeared gauge links
}
\\[4.0ex] \hline \\[-1.0ex]
mfSW      & Sheikholeslami-Wohlert & $O(g^2 a)$ & broken & mean-field impr.
\\[1.0ex] \hline \\[-1.0ex]
npSW      & Sheikholeslami-Wohlert & $O(a^2)$ & broken & non-perturbatively impr.
\\[1.0ex] \hline \\[-1.0ex]
KS      & Staggered & $O(a^2)$ & \parbox[t]{3cm}{$\rm
  U(1)\otimes U(1)$ subgr.\\ unbroken} & rooting for $\Nf<4$
\\[4.0ex] \hline \\[-1.0ex]
Asqtad  & Staggered & $O(a^2)$ & \parbox[t]{3cm}{$\rm
  U(1)\otimes U(1)$ subgr.\\ unbroken}  & \parbox[t]{5cm}{Asqtad
  smeared gauge links, \\rooting for $\Nf<4$}  
\\[4.0ex] \hline \\[-1.0ex]
HISQ  & Staggered & $O(a^2)$ & \parbox[t]{3cm}{$\rm
  U(1)\otimes U(1)$ subgr.\\ unbroken}  & \parbox[t]{5cm}{HISQ
  smeared gauge links, \\rooting for $\Nf<4$}  
\\[4.0ex] \hline \\[-1.0ex]
DW      & Domain Wall & \parbox[t]{2.2cm}{asymptotically \\$O(a^2)$} & \parbox[t]{3cm}{remnant
  breaking \\exponentially suppr.} & \parbox[t]{5cm}{exact chiral symmetry and\\$O(a)$ impr. only in the limit \\
 $L_s\rightarrow \infty$}
\\[7.0ex] \hline \\[-1.0ex]
overlap    & Neuberger & $O(a^2)$ & exact
\\[1.0ex] 
\hline\hline
\end{tabular*}
}
\caption{The most widely used discretizations of the quark action
  and some of their properties. Note that in order to maintain the
  leading lattice artifacts of the action in non-spectral observables
  (like operator matrix elements)
  the corresponding non-spectral operators need to be improved as well. 
\label{tab_quarkactions}}
\end{center}
\end{table}

\noindent
{\it Smearing}\\
\noindent
A simple modification which can help improve the action as well as the
computational performance is the use of smeared gauge fields in the
covariant derivatives of the fermionic action. Any smearing procedure
is acceptable as long as it consists of only adding irrelevant (local)
operators. Moreover, it can be combined with any discretization of the
quark action.  The ``Asqtad'' staggered quark action mentioned above
\cite{Orginos:1999cr} is an example which makes use of so-called
``Asqtad'' smeared (or ``fat'') links. Another example is the use of
n-HYP smeared \cite{Hasenfratz:2001hp,Hasenfratz:2007rf}, n-stout smeared
\cite{Morningstar:2003gk,Durr:2008rw} or n-HEX (hypercubic stout) smeared \cite{Capitani:2006ni} gauge links in the tree-level clover improved
discretization of the quark action, denoted by ``n-HYP tlSW'',
``n-stout tlSW'' and ``n-HEX tlSW'' in the following.\\

\noindent
In Table \ref{tab_quarkactions} we summarize the most widely used
discretizations of the quark action and their main properties together
with the abbreviations used in the summary tables. Note that in order
to maintain the leading lattice artifacts of the actions as given in
the table in non-spectral observables (like operator matrix elements)
the corresponding non-spectral operators need to be improved as well.

\subsection{Setting the scale \label{sec_scale}}

In simulations of lattice QCD quantities such as hadron masses and
decay constants are obtained in ``lattice units'' i.e.~as
dimensionless numbers. In order to convert them into physical units
they must be expressed in terms of some experimentally known,
dimensionful reference quantity $Q$. This procedure is called
``setting the scale''. It amounts to computing the non-perturbative
relation between the bare gauge coupling $g_0$ (which is an input
parameter in any lattice simulation) and the lattice spacing~$a$
expressed in physical units. To this end one chooses a value for $g_0$
and computes the value of the reference quantity in a simulation: This
yields the dimensionless combination, $(aQ)|_{g_0}$, at the chosen
value of $g_0$. The calibration of the lattice spacing is then
achieved via
\be
 a^{-1}\,[{\rm MeV}] = \frac{Q|_{\rm{exp}}\,[{\rm MeV}]}{(aQ)|_{g_0}},
\ee
where $Q|_{\rm{exp}}$ denotes the experimentally known value of the
reference quantity. Common choices for $Q$ are the mass of the
nucleon, the $\Omega$ baryon or the decay constants of the pion and
the kaon. Vector mesons, such as the $\rho$ or $K^\ast$-meson, are
unstable and therefore their masses are not very well suited for
setting the scale, despite the fact that they have been used over many
years for that purpose.

Another widely used quantity to set the scale is the hadronic radius
$r_0$, which can be determined from the force between static quarks
via the relation \cite{Sommer:1993ce}
\be
   F(r_0)r_0^2 = 1.65.
\ee
If the force is derived from potential models describing heavy
quarkonia, the above relation determines the value of $r_0$ as
$r_0\approx0.5$\,fm. A variant of this procedure is obtained
\cite{Bernard:2000gd} by using the definition $F(r_1)r_1^2=1.00$,
which yields $r_1\approx0.32$\,fm. It is important to realize that
both $r_0$ and $r_1$ are not directly accessible in experiment, so
that their values derived from phenomenological potentials are
necessarily model-dependent. Inspite of the inherent ambiguity
whenever hadronic radii are used to calibrate the lattice spacing,
they are very useful quantities for performing scaling tests and
continuum extrapolations of lattice data. Furthermore, they can be
easily computed with good statistical accuracy in lattice simulations.

\subsection{Matching and running \label{sec_match}}

The lattice formulation of QCD amounts to introducing a particular
regularization scheme. Thus, in order to be useful for phenomenology,
hadronic matrix elements computed in lattice simulations must be
related to some continuum reference scheme, such as the
$\msbar$-scheme of dimensional regularization. The matching to the
continuum scheme usually involves running to some reference scale
using the renormalization group. 

In principle, the matching factors which relate lattice matrix
elements to the $\msbar$-scheme, can be computed in perturbation
theory formulated in terms of the bare coupling. It has been known for
a long time, though, that the perturbative expansion is not under good 
control. Several techniques have been developed which allow for a
non-perturbative matching between lattice regularization and continuum
schemes, and are briefly introduced here.\\

\noindent
{\sl Regularization-independent Momentum Subtraction}\\
\noindent
In the {\sl Regularization-independent Momentum Subtraction}
(``RI/MOM'' or ``RI'') scheme \cite{Martinelli:1994ty} a
non-perturbative renormalization condition is formulated in terms of
Green functions involving quark states in a fixed gauge (usually
Landau gauge) at non-zero virtuality. In this way one relates
operators in lattice regularization non-perturbatively to the RI
scheme. In a second step one matches the operator in the RI scheme to
its counterpart in the $\msbar$-scheme. The advantage of this
procedure is that the latter relation involves perturbation theory
formulated in the continuum theory. The uncontrolled use of lattice
perturbation theory can thus be avoided. A technical complication is
associated with the accessible momentum scales (i.e. virtualities),
which must be large enough (typically several $\gev$) in order for the
perturbative relation to $\msbar$ to be reliable. The momentum scales
in simulations must stay well below the cutoff scale (i.e. $2\pi$ over
the lattice spacing), since otherwise large lattice artifacts are
incurred. Thus, the applicability of the RI scheme traditionally relies on the
existence of a ``window'' of momentum scales, which satisfy
\be
   \Lambda_{\rm QCD} \;\lesssim\; p \;\lesssim\; 2\pi a^{-1}.
\ee
However, solutions for mitigating this limitation, which involve
continuum limit, non-perturbative running to higher scales in the
RI/MOM scheme, have recently been proposed and implemented
\cite{Arthur:2010ht,Durr:2010vn,Durr:2010aw,Aoki:2010pe}.

\noindent
{\it Schr\"odinger functional}\\
\noindent
Another example of a non-perturbative matching procedure is provided
by the Schr\"odinger functional (SF) scheme \cite{Luscher:1992an}. It
is based on the formulation of QCD in a finite volume. If all quark
masses are set to zero the box length remains the only scale in the
theory, such that observables like the coupling constant run with the
box size~$L$. The great advantage is that the RG running of
scale-dependent quantities can be computed non-perturbatively using
recursive finite-size scaling techniques. It is thus possible to run
non-perturbatively up to scales of, say, $100\,\gev$, where one is
sure that the perturbative relation between the SF and
$\msbar$-schemes is controlled.\\

\noindent
{\sl Perturbation theory}\\
\noindent
The third matching procedure is based on perturbation theory in which
higher order are effectively resummed \cite{Lepage:1992xa}. Although
this procedure is easier to implement, it is hard to estimate the
uncertainty associated with it.\\

\noindent
In Table~\ref{tab_match} we list the abbreviations used in the
compilation of results together with a short description.

\begin{table}[ht]
{\footnotesize
\begin{tabular*}{\textwidth}[l]{l @{\extracolsep{\fill}} l}
\hline \hline \\[-1.0ex]
Abbrev. & Description
\\[1.0ex] \hline \hline \\[-1.0ex]
RI  &  Regularization-independent momentum subtraction scheme 
\\[1.0ex] \hline \\[-1.0ex]
SF  &  Schr\"odinger functional scheme
\\[1.0ex] \hline \\[-1.0ex]
PT1$\ell$ & matching/running computed in perturbation theory at one loop
\\[1.0ex] \hline \\[-1.0ex]
PT2$\ell$ & matching/running computed in perturbation theory at two loop \\[1.0ex]
\hline\hline
\end{tabular*}
}
\caption{The most widely used matching and running
  techniques. \label{tab_match}} 
\end{table}

\subsection{Chiral extrapolation\label{sec_ChiPT}}
As mentioned in section \ref{sec:scope}, Symanzik's framework can be combined 
with Chiral Perturbation Theory. The well-known terms occurring in the
chiral effective Lagrangian are then supplemented by contributions 
proportional to powers of the lattice spacing $a$. The additional terms are 
constrained by the symmetries of the lattice action and therefore 
depend on the specific choice of the discretization. 
The resulting effective theory can be used to analyze the $a$-dependence of 
the various quantities of interest -- provided the quark masses and the momenta
considered are in the range where the truncated chiral perturbation series yields 
an adequate approximation. Understanding the dependence on the lattice spacing 
is of central importance for a controlled extrapolation to the continuum limit.
 
For staggered fermions, this program has first been carried out for a
single staggered flavor (a single staggered field) \cite{Lee:1999zxa}
at $O(a^2)$. In the following, this effective theory is denoted by
S{\Ch}PT. It was later generalized to an arbitrary number of flavours
\cite{Aubin:2003mg,Aubin:2003uc}, and to next-to-leading order
\cite{Sharpe:2004is}. The corresponding theory is commonly called
Rooted Staggered chiral perturbation theory and is denoted by
RS{\Ch}PT.

For Wilson fermions, the effective theory has been developed in
\cite{Sharpe:1998xm,Rupak:2002sm,Aoki:2003yv}
and is called W{\Ch}PT, while the theory for Wilson twisted mass
fermions \cite{Sharpe:2004ny,Aoki:2004ta,Bar:2010jk} is termed tmW{\Ch}PT.

Another important approach is to consider theories in which the
valence and sea quark masses are chosen to be different. These
theories are called {\it partially quenched}. The acronym for the
corresponding chiral effective theory is PQ{\Ch}PT
\cite{Bernard:1993sv,Golterman:1997st,Sharpe:1997by,Sharpe:2000bc}.

Finally, one can also consider theories where the fermion
discretizations used for the sea and the valence quarks are different. The
effective chiral theories for these ``mixed action'' theories are
referred to as MA{\Ch}PT  \cite{Bar:2002nr,Bar:2003mh,Bar:2005tu,Golterman:2005xa,Chen:2006wf,Chen:2007ug,Chen:2009su}.
 
\newpage
\hspace{-1.5cm}
\subsection{Summary of simulated lattice actions}
In the following two tables we summarize the gauge and quark actions
used in the various calculations. Abbreviations are explained in
section \ref{sec_gauge_actions} and \ref{sec_quark_actions}, and
summarized in tables \ref{tab_gaugeactions} and
\ref{tab_quarkactions}.
\begin{table}[h]
{\footnotesize
\begin{tabular*}{\textwidth}[l]{l @{\extracolsep{\fill}} c c c c}
\hline \hline \\[-1.0ex]
Collab. & Ref. & $\Nf$ & \parbox{1cm}{gauge\\action} & \parbox{1cm}{quark\\action} 
\\[2.0ex] \hline \hline \\[-1.0ex]
{ALPHA 05} & \cite{DellaMorte:2005kg} & 2 & Wilson & npSW  \\
[2.0ex] \hline \\[-1.0ex]
CERN-TOV 06 & \cite{DelDebbio:2006cn} & 2 & Wilson & Wilson/npSW \\
[2.0ex] \hline \\[-1.0ex]
CERN 08 & \cite{Giusti:2008vb} & 2 & Wilson & npSW \\
[2.0ex] \hline \\[-1.0ex]
Bernardoni 10 & \cite{Bernardoni:2010nf} & 2 & Wilson & npSW ${}^\dagger$ \\
[2.0ex] \hline \\[-1.0ex]
{CP-PACS 01} & \cite{AliKhan:2001tx} & 2 & Iwasaki & mfSW  \\
[2.0ex] \hline \\[-1.0ex]
ETM 07, 07A, 08, 09, 09A-D, 10B, D & \parbox[t]{2.5cm}{
\cite{Blossier:2007vv,Boucaud:2007uk,Frezzotti:2008dr,ETM09,Lubicz:2009ht,Jansen:2009tt,Baron:2009wt,Bertone:2009bu,Blossier:2010cr,Lubicz:2010bv}} & 2 &  tlSym & tmWil \\
[7.0ex] \hline \\[-1.0ex]
HHS 08 & \cite{Hasenfratz:2008ce} & 2 & tadSym &
n-HYP tlSW\\
[2.0ex] \hline \\[-1.0ex]
JLQCD 08 & \cite{Aoki:2008ss} & 2 & Iwasaki & overlap \\
[2.0ex] \hline \\[-1.0ex]
{JLQCD 02, 05} & \cite{Aoki:2002uc,Tsutsui:2005cj} & 2 & Wilson & npSW \\
[2.0ex] \hline \\[-1.0ex]
JLQCD/TWQCD 07, 08A, 10 & \cite{Fukaya:2007pn,Noaki:2008iy,Fukaya:2010na} & 2 & Iwasaki & overlap \\
[2.0ex] \hline \\[-1.0ex]
QCDSF 07 & \cite{Brommel:2007wn} & 2 & Wilson &  npSW \\
[2.0ex] \hline \\[-1.0ex]
QCDSF/UKQCD 04, 06, 06A, 07 & \cite{Gockeler:2004rp,Gockeler:2006jt,Brommel:2006ww,QCDSFUKQCD} & 2 & Wilson &  npSW \\
[2.0ex] \hline \\[-1.0ex]
{RBC 07} & \cite{Blum:2007cy} & 2 & DBW2 & DW \\
[2.0ex] \hline \\[-1.0ex]
RBC 04, 06 & \cite{Aoki:2004ht,Dawson:2006qc} & 2 & DBW2 & DW \\
[2.0ex] \hline \\[-1.0ex]
{SPQcdR 05} & \cite{Becirevic:2005ta}
& 2 & Wilson & Wilson \\
[2.0ex] \hline \\ [-1.0ex]
UKQCD 04, 07 & \cite{Flynn:2004au,RBCUKQCD07} & 2 & Wilson & npSW \\ 
[2.0ex]
\hline\hline \\
\end{tabular*}
${}^\dagger$ The calculation uses overlap fermions in the valence quark sector.\\
}
\caption{Summary of simulated lattice actions with $\Nf=2$ quark
  flavours.}
\end{table}
%
%
\begin{table}[h]
{\footnotesize
\begin{tabular*}{\textwidth}[l]{l @{\extracolsep{\fill}} c c c c}
\hline \hline \\[-1.0ex]
Collab. & Ref. & $\Nf$ & \parbox{1cm}{gauge\\action} & \parbox{1cm}{quark\\action} 
\\[2.0ex] \hline \hline \\[-1.0ex]
Aubin 08, 09 & \cite{Aubin:2008ie,Aubin:2009jh} & $2+1$ & tadSym & Asqtad ${}^\dagger$\\
[2.0ex] \hline \\[-1.0ex]
SWME 10, 11 & \cite{Bae:2010ki,Kim:2011qg} & $2+1$ & tadSym & Asqtad ${}^+$ \\
[2.0ex] \hline \\[-1.0ex]
Blum 10 & \cite{Blum:2010ym} & $2+1$ & Iwasaki & DW\\
[2.0ex] \hline \\[-1.0ex]
BMW 10A-C & \cite{Durr:2010vn,Durr:2010aw,Portelli:2010yn} & $2+1$ &  tlSym & 2-HEX tlSW \\
[2.0ex] \hline \\[-1.0ex]
BMW 10 & \cite{BMW10} & $2+1$ & tlSym & 6-stout tlSW\\
[2.0ex] \hline \\[-1.0ex]
CP-PACS/JLQCD 07&  \cite{Ishikawa:2007nn} & $2+1$ & Iwasaki & npSW \\
[2.0ex] \hline \\[-1.0ex]
ETM 10, 10E&  \cite{Baron:2010bv,Farchioni:2010tb} & $2+1+1$ & Iwasaki & tmWil \\
[2.0ex] \hline \\[-1.0ex]
HPQCD 05&  \cite{Mason:2005bj}& $2+1$ & tadSym &  Asqtad \\
[2.0ex] \hline \\[-1.0ex]
HPQCD/UKQCD 06 & \cite{Gamiz:2006sq} & $2+1$ & tadSym & Asqtad \\ 
[2.0ex] \hline \\[-1.0ex]
HPQCD/UKQCD 07 & \cite{Follana:2007uv} & $2+1$ & tadSym & Asqtad ${}^*$\\ 
[2.0ex] \hline \\[-1.0ex]
HPQCD/MILC/UKQCD 04&  \cite{Aubin:2004ck}& $2+1$ & tadSym &  Asqtad \\
[2.0ex] \hline \\[-1.0ex]
JLQCD 09 & \cite{Fukaya:2009fh} & $2+1$ & Iwasaki & overlap \\
[2.0ex] \hline \\[-1.0ex]
JLQCD/TWQCD 08B, 09A & \cite{Chiu:2008kt,JLQCD:2009sk} & $2+1$ & Iwasaki & overlap \\
[2.0ex] \hline \\[-1.0ex]
JLQCD/TWQCD 10 & \cite{Fukaya:2010na} & $2+1, 3$ & Iwasaki & overlap \\
[2.0ex] \hline \\[-1.0ex]
LHP 04 & \cite{Bonnet:2004fr} & $2+1$ & tadSym & Asqtad $^\dagger$ \\
[2.0ex] \hline \\[-1.0ex]
MILC 04, 07, 09, 09A, 10, 10A & \cite{Aubin:2004ck,Aubin:2004fs,Bernard:2007ps,Bazavov:2009bb,Bazavov:2010hj,Bazavov:2010yq}& $2+1$ &
tadSym & Asqtad \\
[2.0ex] \hline \\[-1.0ex]
NPLQCD 06& \cite{Beane:2006kx}& $2+1$ & tadSym & Asqtad $^\dagger$ \\
[2.0ex] \hline \\[-1.0ex]
{PACS-CS 08, 08A, 09, 10} & \cite{Aoki:2008sm,Kuramashi:2008tb,Aoki:2009ix,Aoki:2010wm} & $2+1$ & Iwasaki & npSW \\
[2.0ex] \hline \\[-1.0ex]
{RBC/UKQCD 07, 08, 08A, 10, 10A-B}& \cite{Antonio:2007pb,Allton:2008pn,Boyle:2008yd,RBCUKQCD10,Aoki:2010dy,Aoki:2010pe} & $2+1$ & Iwasaki &  DW \\
[2.0ex] \hline \\[-1.0ex]
TWQCD 08 & \cite{Chiu:2008jq}  & $2+1$ & Iwasaki &  DW \\
[2.0ex] \hline\hline\\
\end{tabular*}
${}^\dagger$ The calculation uses domain wall fermions in the valence quark sector.\\
${}^*$ The calculation uses HISQ staggered fermions in the valence quark sector.\\
${}^+$ The calculation uses HYP smeared improved staggered fermions in the valence quark sector.
}
\caption{Summary of simulated lattice actions with $\Nf=2+1$ or $\Nf=2+1+1$ quark flavours.}
\end{table}

\clearpage

\section{Notes}
\subsection{Notes to section \ref{sec:qmass} on quark masses}

\begin{table}[!ht]
{\footnotesize
\begin{tabular*}{\textwidth}[l]{l @{\extracolsep{\fill}} c c c l}
\hline\hline \\[-1.0ex]
Collab. & Ref. & $\Nf$ & $a$ [fm] & Description
\\[1.0ex] \hline \hline \\[-1.0ex]
PACS-CS 10 & \cite{Aoki:2010wm} &  2+1 &  0.09
& \parbox[t]{5.5cm}{cf.~PACS-CS 08} \\
[1.0ex] \hline \\[-1.0ex]
{MILC 10A}& \cite{Bazavov:2010yq} & 2+1 & & cf.~MILC 09, 09A \\
[1.0ex] \hline \\[-1.0ex]
{BMW 10A, 10B}& \cite{Durr:2010vn,Durr:2010aw}
& 2+1 & \parbox[t]{2.3cm}{0.116,0.093,0.077, 0.065,0.054} &  Scale setting via
$M_\pi, M_K, M_\Omega$.\\
[4.0ex] \hline \\[-1.0ex]
{RBC/UKQCD 10A}& \cite{Aoki:2010dy} & 2+1 & 0.114, 0.087  & Scale set through $M_\Omega$.
\\
[1.0ex] \hline \\[-1.0ex]
{Blum 10}& \cite{Blum:2010ym}
& 2+1 & 0.11 &  Relies on RBC/UKQCD 08 scale setting. \\
[1.0ex] \hline \\[-1.0ex]
{PACS-CS 09}& \cite{Aoki:2009ix}& 2+1 & 0.09 & Scale setting via
$M_\Omega$.\\
[1.0ex] \hline \\[-1.0ex]
{HPQCD 09, 10}&  \cite{Davies:2009ih,McNeile:2010ji}& 2+1 &  & \\
[1.0ex] \hline \\[-1.0ex]
{MILC 09A, 09} & \cite{Bazavov:2009fk,Bazavov:2009bb} & 2+1 & 0.045, 0.06, 0.09 & 
		\parbox[t]{5.5cm}{Scale set through $r_1$ and $\Upsilon$ and 
		continuum extrapolation based on RS{\Ch}PT.} \\ 
[4.0ex] \hline \\[-1.0ex]
PACS-CS 08 &\cite{Aoki:2008sm}& 2+1 &  0.09 & \parbox[t]{5.5cm}{Scale set 
		through $M_\Omega$. Non-perturbatively $O(a)$-improved.} \\
[4.0ex] \hline \\[-1.0ex]
RBC/UKQCD 08	&\cite{Allton:2008pn}	&2+1& 0.11 & \parbox[t]{5.5cm}{Scale
                 set through $M_\Omega$. Automatic  
		$O(a)$-improvement due to appoximate chiral symmetry. 
		$(\Lambda_{\rm QCD}a)^2\approx 4\%$ systematic error
		due to lattice artifacts added.} \\ 
[12.0ex] \hline \\[-1.0ex]
{CP-PACS/JLQCD 07}&  \cite{Ishikawa:2007nn} & 2+1 & 0.07,0.10,0.12 &  \parbox[t]{5.5cm}{Scale set 
		through $M_K$ or $M_\phi$. Non-perturbatively
                $O(a)$-improved.} \\
[4.0ex] \hline \\[-1.0ex]
{HPQCD 05}& \cite{Mason:2005bj}& 2+1 & 0.09,0.12 & \parbox[t]{5.5cm}{Scale set through the $\Upsilon-\Upsilon'$
mass difference.} \\
[4.0ex] \hline \\[-1.0ex]
\parbox [t]{4.5cm}{HPQCD/MILC/UKQCD 04,\\
 MILC 04}
& \cite{Aubin:2004ck,Aubin:2004fs}& 2+1 & 0.09,0.12 & \parbox[t]{5.5cm}{Scale
  set through $r_1$ and $\Upsilon$ and continuum extrapolation based on RS{\Ch}PT.} \\
[4.0ex] \hline \hline \\[-1.0ex]
\end{tabular*}
\caption{Continuum extrapolations/estimation of lattice artifacts in
  determinations of $m_{ud}$ and $m_s$ with $N_f=2+1$ quark flavours.}
}
\end{table}

\begin{table}[!ht]
{\footnotesize
\begin{tabular*}{\textwidth}[l]{l @{\extracolsep{\fill}} c c c l}
\hline\hline \\[-1.0ex]
Collab. & Ref. & $\Nf$ & $a$ [fm] & Description
\\[1.0ex] \hline \hline \\[-1.0ex]
ETM 10B & \cite{Blossier:2010cr} & 2 & \parbox[t]{2.3cm}{0.098, 0.085,\\ 0.067, 0.054} & Scale set through $F_\pi$.\\
[4.0ex] \hline \\[-1.0ex]
{JLQCD/TWQCD 08A} & \cite{Noaki:2008iy} & 2 & 0.12 & Scale set through $r_0$. \\
[1.0ex] \hline \\[-1.0ex]
{RBC 07} & \cite{Blum:2007cy} & 2 & 0.12 & Scale set through $M_\rho$.\\
[1.0ex] \hline \\[-1.0ex]
{ETM 07} & \cite{Blossier:2007vv} & 2 & 0.09 & Scale set through $F_\pi$. \\
[1.0ex] \hline \\[-1.0ex]
{QCDSF/UKQCD 06} & \cite{Gockeler:2006jt} & 2 & 0.065--0.09 & Scale set through $r_0$. \\
[2.0ex] \hline \\[-1.0ex]
{SPQcdR 05} & \cite{Becirevic:2005ta}
& 2 & 0.06,0.08 & Scale set through $M_{K^*}$. \\
[1.0ex] \hline \\[-1.0ex]
{ALPHA 05} & \cite{DellaMorte:2005kg}
& 2 & 0.07-0.12 & Scale set through $r_0$. \\
[1.0ex] \hline \\[-1.0ex]
{QCDSF/UKQCD 04} & \cite{Gockeler:2004rp}
& 2 &  0.07-0.12 & Scale set through $r_0$. \\
[1.0ex] \hline \\[-1.0ex]
{JLQCD 02} & \cite{Aoki:2002uc}
& 2 & 0.09 & Scale set through $M_\rho$. \\
[1.0ex] \hline \\[-1.0ex]
{CP-PACS 01} & \cite{AliKhan:2001tx}
& 2 & 0.11,0.16,0.22 & Scale set through $M_\rho$. \\
[1.0ex] \hline\hline
\end{tabular*}
\caption{Continuum extrapolations/estimation of lattice artifacts in
  determinations of $m_{ud}$ and $m_s$ with $N_f=2$ quark flavours.}
}
\end{table}

\newpage

\begin{table}[!ht]
{\footnotesize
\begin{tabular*}{\textwidth}[l]{l @{\extracolsep{\fill}} c c c l}
\hline \hline  \\[-1.0ex]
Collab. & Ref. & $\Nf$ & $M_{\pi,\text{min}}$ [MeV] & Description
\\[1.0ex] \hline \hline \\[-1.0ex]
PACS-CS 10 & \cite{Aoki:2010wm} &  2+1 & & cf.~PACS-CS 08 \\
[1.0ex] \hline \\[-1.0ex]
{MILC 10A}& \cite{Bazavov:2010yq} & 2+1 & & NLO SU(2)
S{\Ch}PT. Cf.~also MILC 09A,09. \\
[1.0ex] \hline \\[-1.0ex]
{BMW 10A, 10B}& \cite{Durr:2010vn,Durr:2010aw}
& 2+1 & 135 & \parbox[t]{6cm}{Interpolation to the physical point.} \\
[1.0ex] \hline \\[-1.0ex]
{RBC/UKQCD 10A}& \cite{Aoki:2010dy} & 2+1 & 290 & \\
[1.0ex] \hline \\[-1.0ex]
{Blum 10}& \cite{Blum:2010ym,Allton:2008pn}
& 2+1 & \parbox[t]{2cm}{242 (valence), 330 (sea)} &
\parbox[t]{6cm}{Extrapolation done on the basis of PQ$\chi$PT formulae with
virtual photons.} \\
[4.0ex] \hline \\[-1.0ex]
{PACS-CS 09}& \cite{Aoki:2009ix}& 2+1 & 135 & \parbox[t]{6cm}{Physical point reached by
reweighting technique, no chiral extrapolation needed.}\\
[4.0ex] \hline \\[-1.0ex]
{HPQCD 09, 10}&  \cite{Davies:2009ih,McNeile:2010ji}& 2+1 &  & \\
[1.0ex] \hline \\[-1.0ex]
{MILC 09A, 09} & \cite{Bazavov:2009fk,Bazavov:2009bb} & 2+1 & 177, 240 & \parbox[t]{6cm}{
		NLO SU(3) RS{\Ch}PT, continuum {\Ch}PT at NNLO and NNNLO and NNNNLO 
		analytic terms. The lightest Nambu-Goldstone mass
                is 177 MeV (09A) and 224 MeV (09)
		(at $a=$0.09fm) and the lightest RMS mass is 258MeV (at
                $a=$0.06fm).} \\ 
[15.0ex] \hline \\[-1.0ex]
PACS-CS 08 &\cite{Aoki:2008sm}& 2+1 &156& 
		\parbox[t]{6cm}{NLO SU(2) {\Ch}PT and SU(3)
                  (Wilson){\Ch}PT.}\\
[1.0ex] \hline \\[-1.0ex]
RBC/UKQCD 08	&\cite{Allton:2008pn}	&2+1& \parbox[t]{2cm}{242 (valence), 330 (sea)} & \parbox[t]{6cm}{SU(3) PQ{\Ch}PT 
and heavy kaon NLO SU(2) PQ{\Ch}PT fits.} \\
[4.0ex] \hline \\[-1.0ex]
{CP-PACS/JLQCD 07}&  \cite{Ishikawa:2007nn} & 2+1 & 620 & \parbox[t]{6cm}{NLO Wilson {\Ch}PT
fits to meson masses.} \\
[1.0ex] \hline \\[-1.0ex]
{HPQCD 05}& \cite{Mason:2005bj}& 2+1 & 240 & \parbox[t]{6cm}{PQ RS{\Ch}PT fits.} \\
[1.0ex] \hline \\[-1.0ex]
\parbox[t]{4.5cm}{HPQCD/MILC/UKQCD 04,\\ MILC 04}
& \cite{Aubin:2004ck,Aubin:2004fs}& 2+1 & 240 & \parbox[t]{6cm}{PQ RS{\Ch}PT fits.} \\
[4.0ex] \hline \hline \\[-1.0ex]
\end{tabular*}
\caption{Chiral extrapolation/minimum pion mass in determinations of
  $m_{ud}$ and  $m_s$  with $N_f=2+1$ quark flavours.}
}
\end{table}

\begin{table}[!ht]
{\footnotesize
\begin{tabular*}{\textwidth}[l]{l @{\extracolsep{\fill}} c c c l}
\hline \hline  \\[-1.0ex]
Collab. & Ref. & $\Nf$ & $M_{\pi,\text{min}}$ [MeV] & Description
\\[1.0ex] \hline \hline \\[-1.0ex]
ETM 10B & \cite{Blossier:2010cr} & 2 & 270 & \parbox[t]{6cm}{Fits based on NLO {\Ch}PT and Symanzik expansion up to $O(a^2)$}.\\
[4.0ex] \hline \\[-1.0ex]
{JLQCD/TWQCD 08A} & \cite{Noaki:2008iy} & 2 & 290 & NLO {\Ch}PT fits.\\
[1.0ex] \hline \\[-1.0ex]
{RBC 07} & \cite{Blum:2007cy}
& 2 & 440 & \parbox[t]{6cm}{NLO fit including $O(\alpha)$ effects.} \\
[1.0ex] \hline \\[-1.0ex]
{ETM 07} & \cite{Blossier:2007vv}
& 2 & 300 & Polynomial and PQ{\Ch}PT fits. \\
[1.0ex] \hline \\[-1.0ex]
{QCDSF/UKQCD 06} & \cite{Gockeler:2006jt} & 2 & \parbox[t]{2cm}{520 (valence), 620 (sea)} & NLO (PQ){\Ch}PT
fits. \\
[4.0ex] \hline \\[-1.0ex]
{SPQcdR 05} & \cite{Becirevic:2005ta}
& 2 & ~600 & Polynomial fit. \\
[1.0ex] \hline \\[-1.0ex]
{ALPHA 05} & \cite{DellaMorte:2005kg}
& 2 & ~560 & LO {\Ch}PT fit. \\
[1.0ex] \hline \\[-1.0ex]
{QCDSF/UKQCD 04} & \cite{Gockeler:2004rp}
& 2 &  \parbox[t]{2cm}{520 (valence), 620 (sea)} & NLO (PQ){\Ch}PT fits. \\
[4.0ex] \hline \\[-1.0ex]
{JLQCD 02} & \cite{Aoki:2002uc}
& 2 & 560 & Polynomial and {\Ch}PT fits. \\
[1.0ex] \hline \\[-1.0ex]
{CP-PACS 01} & \cite{AliKhan:2001tx}
& 2 & 430 & \parbox{6cm}{Polynomial fits.} \\
[1.0ex]
\hline\hline
\end{tabular*}
\caption{Chiral extrapolation/minimum pion mass in determinations of
  $m_{ud}$ and  $m_s$  with $N_f=2$ quark flavours.}
}
\end{table}

\newpage

\begin{table}
{\footnotesize
\begin{tabular*}{\textwidth}[l]{l @{\extracolsep{\fill}} c c c c l}
\hline\hline  \\[-1.0ex]
Collab. & Ref. & $\Nf$ & $L$ [fm] & $M_{\pi,\text{min}} L$ & Description
\\[1.0ex] \hline \hline \\[-1.0ex]
PACS-CS 10 & \cite{Aoki:2010wm} &  2+1 & & & cf.~PACS-CS 08 \\
[1.0ex] \hline \\[-1.0ex]
{MILC 10A}& \cite{Bazavov:2010yq} & 2+1 & & & cf.~MILC 09A,09 \\
[1.0ex] \hline \\[-1.0ex]
{BMW 10A, 10B}& \cite{Durr:2010vn,Durr:2010aw}
& 2+1 &  $\gtrsim 5.0$ & $\gtrsim 4.0$ & \parbox[t]{4.2cm}{FS corrections
  below 5 per mil on the largest lattices.}\\
[4.0ex] \hline \\[-1.0ex]
{RBC/UKQCD 10A}& \cite{Aoki:2010dy} & 2+1 & 2.7 &  $\gtrsim 4.0$ & \\
[4.0ex] \hline \\[-1.0ex]
{Blum 10}& \cite{Blum:2010ym}
& 2+1 & 1.8, 2.7 & --- & \parbox[t]{4.2cm}{Simulations done with dynamical photons; large
finite volume effects analytically corrected for, but not related to
$M_\pi L$.} \\
[12.0ex] \hline \\[-1.0ex]
{PACS-CS 09}& \cite{Aoki:2009ix}& 2+1 & 2.9 & 2.0 & \parbox[t]{4.2cm}{Only one volume.} \\
[1.0ex] \hline \\[-1.0ex]
{HPQCD 09, 10}&  \cite{Davies:2009ih,McNeile:2010ji}& 2+1 &  &  &  \\
[1.0ex] \hline \\[-1.0ex]
{MILC 09A, 09} & \cite{Bazavov:2009fk,Bazavov:2009bb} & 2+1 &
\parbox[t]{2cm}{2.5, 2.9, 3.4,\\ 3.6, 3.8, 5.8}&4.1, 3.8&  \parbox[t]{4.2cm}{} \\
[4.0ex] \hline \\[-1.0ex]
PACS-CS 08 &\cite{Aoki:2008sm}& 2+1 &2.9&2.3& 
		\parbox[t]{4.2cm}{Correction for FSE from {\Ch}PT using 
		\cite{Colangelo:2005gd}.} \\
[4.0ex] \hline \\[-1.0ex]
RBC/UKQCD 08	&\cite{Allton:2008pn}	&2+1& 1.8, 2.7&4.6&
		\parbox[t]{4.2cm}{Various volumes for comparison and 
		correction for FSE from {\Ch}PT 
		\cite{Gasser:1986vb,Gasser:1987ah,Colangelo:2005gd}.}\\
[6.0ex] \hline \\[-1.0ex]
{CP-PACS/JLQCD 07}&  \cite{Ishikawa:2007nn} & 2+1 & 2.0 & 6.0&
\parbox[t]{4.2cm}{Estimate based on the comparison to a $L=1.6$ fm volume
  assuming powerlike dependence on $L$.} \\
[9.0ex] \hline \\[-1.0ex]
{HPQCD 05}& \cite{Mason:2005bj}& 2+1 & 2.4, 2.9 & 3.5 &  \\
[1.0ex] \hline \\[-1.0ex]
\parbox[t]{4.5cm}{HPQCD/MILC/UKQCD 04,\\ MILC 04}
& \cite{Aubin:2004ck,Aubin:2004fs}& 2+1 & 2.4, 2.9 & 3.5 & NLO S{\Ch}PT. \\
[4.0ex] \hline \hline \\[-1.0ex]
\end{tabular*}
\caption{Finite volume effects in determinations of $m_{ud}$ and $m_s$
   with $N_f=2+1$ quark flavours.}
}
\end{table}

\begin{table}
{\footnotesize
\begin{tabular*}{\textwidth}[l]{l @{\extracolsep{\fill}} c c c c l}
\hline\hline  \\[-1.0ex]
Collab. & Ref. & $\Nf$ & $L$ [fm] & $M_{\pi,\text{min}} L$ & Description
\\[1.0ex] \hline \hline \\[-1.0ex]
ETM 10B & \cite{Blossier:2010cr} & 2 & $\gtrsim 2.0$ & 3.5 & \parbox[t]{4.2cm}{One volume
$L=1.7$ fm at $m_\pi=495, a=0.054$fm.} \\
[4.0ex] \hline \\[-1.0ex]
{JLQCD/TWQCD 08A} & \cite{Noaki:2008iy} & 2 & 1.9 & 2.8 &
\parbox[t]{4.2cm}{Corrections for FSE based on NLO {\Ch}PT.}\\
[4.0ex] \hline \\[-1.0ex]
{RBC 07} & \cite{Blum:2007cy}
& 2 & 1.9 & 4.3 &\parbox[t]{4.2cm}{Estimate of FSE based on a model.}  \\
[4.0ex] \hline \\[-1.0ex]
{ETM 07} & \cite{Blossier:2007vv}
& 2 &  2.1 & 3.2 & NLO PQ{\Ch}PT \\
[1.0ex] \hline \\[-1.0ex]
{QCDSF/UKQCD 06} & \cite{Gockeler:2006jt} & 2 & 1.4--1.9 & 4.7 & \\
[1.0ex] \hline \\[-1.0ex]
{SPQcdR 05} & \cite{Becirevic:2005ta}
& 2 & 1.0--1.5 & 4.3 & \parbox[t]{4.2cm}{Comparison between 1.0 and 1.5 fm.}\\
[4.0ex] \hline \\[-1.0ex]
{ALPHA 05} & \cite{DellaMorte:2005kg}
& 2 & 2.6 & 7.4 & \\
[1.0ex] \hline \\[-1.0ex]
{QCDSF/UKQCD 04} & \cite{Gockeler:2004rp}
& 2 &  1.7--2.0 & 4.7 & \\
[1.0ex] \hline \\[-1.0ex]
{JLQCD 02} & \cite{Aoki:2002uc}
& 2 & 1.8 & 5.1 & \parbox[t]{4.2cm}{Numerical study with three volumes.} \\
[4.0ex] \hline \\[-1.0ex]
{CP-PACS 01} & \cite{AliKhan:2001tx}
& 2 & 2.0--2.6 & 5.7 &  \\
[1.0ex]
\hline\hline
\end{tabular*}
\caption{Finite volume effects in determinations of $m_{ud}$ and $m_s$
   with $N_f=2$ quark flavours.}
}
\end{table}

\newpage

\begin{table}[!ht]
{\footnotesize
\begin{tabular*}{\textwidth}[l]{l @{\extracolsep{\fill}} c c l}
\hline\hline \\[-1.0ex]
Collab. & Ref. & $\Nf$  & Description
\\[1.0ex] \hline \hline \\[-1.0ex]
PACS-CS 10 & \cite{Aoki:2010wm} &  2+1
& \parbox[t]{7cm}{Non-perturbative renormalization and running; Schr\"odinger functional method.} \\
[4.0ex] \hline \\[-1.0ex]
{MILC 10A}& \cite{Bazavov:2010yq} & 2+1 & cf.~MILC 09A,09\\
[1.0ex] \hline \\[-1.0ex]
{BMW 10A, 10B}& \cite{Durr:2010vn,Durr:2010aw}
& 2+1 & \parbox[t]{7cm}{Non-perturbative renormalization (tree-level
  improved RI-MOM), non-perturbative running.} \\
[4.0ex] \hline \\[-1.0ex]
{RBC/UKQCD 10A}& \cite{Aoki:2010dy} & 2+1
& \parbox[t]{7cm}{Non-perturbative renormalization (RI/SMOM).}\\
[1.0ex] \hline \\[-1.0ex]
{Blum 10}& \cite{Blum:2010ym}
& 2+1 & \parbox[t]{7cm}{Relies on non-perturbative renormalization factors calculated
by RBC/UKQCD 08; no QED renormalization.} \\
[6.0ex] \hline \\[-1.0ex]
{PACS-CS 09}& \cite{Aoki:2009ix}& 2+1 & \parbox[t]{7cm}{Non-perturbative
renormalization; Schr\"odinger functional method.} \\
[4.0ex] \hline \\[-1.0ex]
{HPQCD 09, 10}&  \cite{Davies:2009ih,McNeile:2010ji}& 2+1 & \parbox[t]{7cm}{Lattice
  calculation of $m_s/m_c$: $m_s$ derived from the perturbative
  determination of $m_c$.} \\ 
[4.0ex] \hline \\[-1.0ex]
{MILC 09A, 09} & \cite{Bazavov:2009fk,Bazavov:2009bb} & 2+1 & \parbox[t]{7cm}{2-loop
perturbative renormalization.}  \\
[1.0ex] \hline \\[-1.0ex]
PACS-CS 08 &\cite{Aoki:2008sm}& 2+1 & \parbox[t]{7cm}{1-loop perturbative
  renormalization.} \\ 
[1.0ex] \hline \\[-1.0ex]
RBC/UKQCD 08	&\cite{Allton:2008pn}	&2+1& \parbox[t]{7cm}{Non-perturbative
renormalization, 3-loop perturbative matching.} \\
[4.0ex] \hline \\[-1.0ex]
{CP-PACS/JLQCD 07}&  \cite{Ishikawa:2007nn} & 2+1 & \parbox[t]{7cm}{1-loop
  perturbative renormalization, tadpole improved.} \\ 
[4.0ex] \hline \\[-1.0ex]
{HPQCD 05}& \cite{Mason:2005bj}& 2+1 & \parbox[t]{7cm}{2-loop perturbative
  renormalization.} \\ 
[1.0ex] \hline \\[-1.0ex]
\parbox[t]{4.5cm}{HPQCD/MILC/UKQCD 04,\\ MILC 04}
& \cite{Aubin:2004ck,Aubin:2004fs}& 2+1 &  \parbox[t]{7cm}{1-loop perturbative
renormalization.} \\ 
[4.0ex] \hline \hline \\[-1.0ex]
\end{tabular*}
\caption{Renormalization in determinations of $m_{ud}$ and $m_s$  with $N_f=2+1$ quark flavours.}
}
\end{table}

\begin{table}[!ht]
{\footnotesize
\begin{tabular*}{\textwidth}[l]{l @{\extracolsep{\fill}} c c l}
\hline\hline \\[-1.0ex]
Collab. & Ref. & $\Nf$  & Description
\\[1.0ex] \hline \hline \\[-1.0ex]
ETM 10B & \cite{Blossier:2010cr} & 2 &\parbox[t]{7cm}{Non-perturbative
renormalization.} \\
[1.0ex] \hline \\[-1.0ex]
{JLQCD/TWQCD 08A} & \cite{Noaki:2008iy} & 2 & \parbox[t]{7cm}{Non-perturbative
renormalization.} \\
[1.0ex] \hline \\[-1.0ex]
{RBC 07} & \cite{Blum:2007cy}
& 2 & \parbox[t]{7cm}{Non-perturbative renormalization.}  \\
[1.0ex] \hline \\[-1.0ex]
{ETM 07} & \cite{Blossier:2007vv}
& 2 & \parbox[t]{7cm}{Non-perturbative renormalization.}  \\
[1.0ex] \hline \\[-1.0ex]
{QCDSF/UKQCD 06} & \cite{Gockeler:2006jt} & 2 & \parbox[t]{7cm}{Non-perturbative
renormalization.} \\
[1.0ex] \hline \\[-1.0ex]
{SPQcdR 05} & \cite{Becirevic:2005ta}
& 2 &  \parbox[t]{7cm}{Non-perturbative renormalization.} \\
[1.0ex] \hline \\[-1.0ex]
{ALPHA 05} & \cite{DellaMorte:2005kg}
& 2 & \parbox[t]{7cm}{Non-perturbative renormalization.} \\
[1.0ex] \hline \\[-1.0ex]
{QCDSF/UKQCD 04} & \cite{Gockeler:2004rp}
& 2 & \parbox[t]{7cm}{Non-perturbative
renormalization.} \\
[1.0ex] \hline \\[-1.0ex]
{JLQCD 02} & \cite{Aoki:2002uc}
& 2 & \parbox[t]{7cm}{1-loop perturbative
renormalization.} \\
[1.0ex] \hline \\[-1.0ex]
{CP-PACS 01} & \cite{AliKhan:2001tx}
& 2 & \parbox[t]{7cm}{1-loop perturbative
renormalization.} \\
[1.0ex] \hline \hline \\[-1.0ex]
\end{tabular*}
\caption{Renormalization in determinations of $m_{ud}$ and $m_s$  with $N_f=2$ quark flavours.}
}
\end{table}


\clearpage
\subsection{Notes to section \ref{sec:vusvud} on $|V_{ud}|$ and  $|V_{us}|$}

\begin{table}[!h]
{\footnotesize
\begin{tabular*}{\textwidth}[l]{l @{\extracolsep{\fill}} c c c l}
\hline\hline \\[-1.0ex]
Collab. & Ref. & $\Nf$ & $a$ [fm] & Description
\\[1.0ex] \hline \hline \\[-1.0ex]
RBC/UKQCD 07,10 &\cite{RBCUKQCD07,RBCUKQCD10} 	&2+1& 0.114(2) & 
		\parbox[t]{6cm}{
		Scale fixed through $\Omega$ baryon mass. Add 
		$(\Lambda_{\rm QCD}a)^2\approx4$\%
		systematic error for lattice artifacts.
		Fifth dimension with extension $L_s=16$, therefore small
		residual chiral symmetry breaking and approximate $O(a)$-imrovement.}\\
[15.0ex] \hline \\[-2.0ex]
ETM 10D		&\cite{Lubicz:2010bv}		&2  & 
		\parbox[t]{2cm}{\centering 0.054, 0.068, 0.086, 101} & \parbox[t]{6cm}{ Scale set through $F_\pi$. Automatic $O(a)$ 
		impr., flavour symmetry
		breaking:\\ $(M_{PS}^{0})^2 - (M_{PS}^\pm)^2\sim O(a^2)$.} 
\\[7.0ex] \hline \\[-2.0ex]
ETM 09A		&\cite{Lubicz:2009ht}	&2  & 0.069, 0.883(6), 0.103&
		\parbox[t]{6cm}{
		Scale set through $F_\pi$. Automatic $O(a)$ impr., flavour symmetry
		breaking:\\ $(M_{PS}^{0})^2 - (M_{PS}^\pm)^2\sim O(a^2)$. \\
  		Three lattice spacings only for pion mass 470\,MeV.}\\
[12.0ex] \hline \\[-2.0ex]
QCDSF 07	&\cite{Brommel:2007wn}	&2  &0.075& 
		\parbox[t]{6cm}{Scale set with $r_0$. Non-perturbatively
		$O(a)$-improved Wilson fermions, not clear whether 
		currents improved.}\\
[6.0ex] \hline \\[-2.0ex]
RBC 06		&\cite{Dawson:2006qc}	&2  &0.12& 
		\parbox[t]{6cm}{Scale set through $M_\rho$.
		Automatic $O(a)$-improvement due to approximate 
		chiral symmetry of the action. }\\
[6.0ex] \hline \\[-2.0ex]
JLQCD 05	&\cite{Tsutsui:2005cj}	&2  &0.0887&
		\parbox[t]{6cm}{Scale set through $M_\rho$. Non-perturbatively
		$O(a)$-improved Wilson fermions.}\\
[4.0ex]
\hline\hline
\end{tabular*}
\caption{Continuum extrapolations/estimation of lattice artifacts in
  determinations of $f_+(0)$.}
}
\end{table}

\noindent
\begin{table}[!h]
{\footnotesize
\begin{tabular*}{\textwidth}[l]{l @{\extracolsep{\fill}} c c c l}
\hline \hline  \\[-1.0ex]
Collab. & Ref. & $\Nf$ & $M_{\pi,\text{min}}$ [MeV] & Description
\\[1.0ex] \hline \hline \\[-1.0ex]
RBC/UKQCD 07,10 &\cite{RBCUKQCD07,RBCUKQCD10} 	&2+1&330& 
		\parbox[t]{6cm}{NLO SU(3) {\Ch}PT with phenomenological ansatz 
			for higher orders.}\\
[4.0ex] \hline \\[-2.0ex]
ETM 10D         & \cite{Lubicz:2010bv}  &2  &260&
		\parbox[t]{6cm}{NLO heavy kaon SU(2) {\Ch}PT and
		NLO SU(3) {\Ch}PT and phenomenological ansatz for higher orders.
		Average of $f_+(0)$-fit and joint $f_+(0)$-$f_K/f_\pi$-fit.}\\[4.0ex] \hline \\[-2.0ex]
ETM 09A		&\cite{Lubicz:2009ht}	&2  &260& 
		\parbox[t]{6cm}{NLO heavy kaon SU(2) {\Ch}PT and
		NLO SU(3) {\Ch}PT and phenomenological ansatz for higher orders.}\\
[6.0ex] \hline \\[-2.0ex]
QCDSF 07	&\cite{Brommel:2007wn}	&2  &591& 
		\parbox[t]{6cm}{Only one value for the pion mass.}\\
[1.0ex] \hline \\[-2.0ex]
RBC 06		&\cite{Dawson:2006qc}	&2  &490& 
		\parbox[t]{6cm}{NLO SU(3) {\Ch}PT and phenomenological ansatz 
			for higher orders.}\\
[3.0ex] \hline \\[-2.0ex]
JLQCD 05	&\cite{Tsutsui:2005cj}	&2  &550& 
		\parbox[t]{6cm}{NLO SU(3) {\Ch}PT and phenomenological ansatz 
			for higher orders.}\\
[3.0ex]
\hline\hline\\
\end{tabular*}
\caption{Chiral extrapolation/minimum pion mass in determinations of $f_+(0)$.}
}
\end{table}

\noindent
\begin{table}[!h]
{\footnotesize
\begin{tabular*}{\textwidth}[l]{l @{\extracolsep{\fill}} c c c c l}
\hline \hline  \\[-1.0ex]
Collab. & Ref. & $\Nf$ &$L$ [fm]& $M_{\pi,\text{min}}L$ & Description
\\[1.0ex] \hline \hline \\[-1.0ex]
RBC/UKQCD 07,10 &\cite{RBCUKQCD07,RBCUKQCD10} 	&2+1&2.7&4.7& 
		\parbox[t]{5cm}{Two volumes for all but the lightest
			pion mass.}\\
[3.0ex] \hline \\[-2.0ex]
ETM 10D		&\cite{Lubicz:2010bv}		&2  & 1.7-2.8 & 3.7 & \\
[1.0ex] \hline \\[-2.0ex]
ETM 09A		&\cite{Lubicz:2009ht}	&2  &2.1, 2.8&3.2& 
		\parbox[t]{5cm}{Two volumes at $M_\pi=300$\,MeV and {\Ch}PT-motivated 
		estimate of the error due to FSE.}\\
[6.0ex] \hline \\[-2.0ex]
QCDSF 07	&\cite{Brommel:2007wn}	&2  &1.9&5.4& 
		\parbox[t]{5cm}{}\\
[1.0ex] \hline \\[-1.0ex]
RBC 06		&\cite{Dawson:2006qc}	&2
&1.9&4.7& \parbox[t]{5cm}{}\\
[1.0ex] \hline \\[-1.0ex]
JLQCD 05	&\cite{Tsutsui:2005cj}	&2  &1.8&4.9& \parbox[t]{5cm}{}\\
[1.0ex]
\hline\hline
\end{tabular*}
\caption{Finite volume effects in determinations of $f_+(0)$.}
}
\end{table}

\begin{table}[!h]
{\footnotesize
\begin{tabular*}{\textwidth}[l]{l @{\extracolsep{\fill}} c c c l}
\hline \hline  \\[-1.0ex]
Collab. & Ref. & $\Nf$ & $a$ [fm] & Description
\\[1.0ex] \hline \hline \\[-1.0ex]
{ETM 10E}& \cite{Farchioni:2010tb} & 2+1+1 & 0.061, 0.078  &
		\parbox[t]{6cm}{ 
		Scale set through $f_\pi/m_\pi$. 
		Two lattice spacings but
		$a$-dependence ignored in all fits. Finer lattice spacing
		from \cite{Baron:2011sf}.} \\
[6.0ex] \hline  \\[-1.0ex]
{RBC/UKQCD 10A}& \cite{Aoki:2010dy} & 2+1 & 0.114, 0.087  & Scale set
through $M_\Omega$. \\
[1.0ex] \hline  \\[-1.0ex]
MILC 10 & \cite{Bazavov:2010hj} & 2+1 & 0.045 - 0.09 & \parbox[t]{6.0cm}{3 lattice
spacings, continuum extrapolation by means of RS{\Ch}PT.}\\
[4.0ex] \hline  \\[-1.0ex]
BMW 10		&\cite{BMW10}		&2+1&
		\parbox[t]{2.5cm}{\centering 0.065, 0.083,\\ 0.124}&
		\parbox[t]{6cm}{Scale set through $M_{\Omega,\Xi}$. Perturbative
		$O(a)$-improvement.}\\
[4.0ex] \hline \\[-2.0ex]
JLQCD/TWQCD 08A &\cite{Noaki:2008iy}	&2+1& 0.100(2) & 
		\parbox[t]{6cm}{Scale set through $F_\pi$. Automatic 
		$O(a)$-improvement due to chiral symmetry of action.}\\
[6.0ex] \hline \\[-2.0ex]
MILC 09A	&\cite{Bazavov:2009fk}	&2+1& 0.045, 0.06, 0.09 & 
		\parbox[t]{6cm}{Scale set through $r_1$ and $\Upsilon$ and 
		continuum extrapolation based on RS{\Ch}PT.}\\
[4.0ex] \hline \\[-2.0ex]
MILC 09		&\cite{Bazavov:2009bb}	&2+1& 	\parbox[t]{2.5cm}{\centering 0.045, 0.06, \\0.09, 0.12} & 
		\parbox[t]{6cm}{Scale set through $r_1$ and $\Upsilon$ and 
		continuum extrapolation based on RS{\Ch}PT.}\\
[4.0ex] \hline \\[-2.0ex]
Aubin 08	&\cite{Aubin:2008ie}	&2+1& 0.09, 0.12 & 	
		\parbox[t]{6cm}{Scale set through $r_1$ and $\Upsilon$ and 
		continuum extrapolation based on MA{\Ch}PT.}\\
[4.0ex] \hline \\[-2.0ex]
PACS-CS 08, 08A &\cite{Aoki:2008sm,Kuramashi:2008tb}& 2+1 & 0.0907(13) & 
		\parbox[t]{6cm}{Scale set 
		through $M_\Omega$. Non-perturbatively $O(a)$-improved. }\\
[4.0ex] \hline \\[-2.0ex]
HPQCD/UKQCD 07	&\cite{Follana:2007uv}	&2+1& 0.06, 0.09, 0.15 & 
		\parbox[t]{6cm}{Scale set through $r_1$ and $\Upsilon$ and 
		continuum extrapolation on continuum-{\Ch}PT motivated ansatz. Taste
		breaking of sea quarks ignored.}\\
[10.0ex] \hline \\[-2.0ex]
RBC/UKQCD 08	&\cite{Allton:2008pn}	&2+1& 0.114(2) & 
		\parbox[t]{6cm}{Scale set through $M_\Omega$. Automatic 
		$O(a)$-improvement due to appoximate chiral symmetry. 
		$(\Lambda_{\rm QCD}a)^2\approx 4\%$ systematic error
		due to lattice artifacts added.}\\
[10.0ex] \hline \\[-2.0ex]
NPLQCD 06	&\cite{Beane:2006kx}	&2+1& 0.125 & 
		\parbox[t]{6cm}{Scale set through $r_0$ and $F_\pi$. Taste breaking
		of sea quarks ignored.}\\
[4.0ex] \hline \\[-2.0ex]
ETM 10D		&\cite{Lubicz:2010bv}		&2  & 	\parbox[t]{2.5cm}{\centering 0.054, 0.068,\\ 0.086, 0.101} &\parbox[t]{6cm}{Scale set through $F_\pi$.	Automatic $O(a)$ impr., flavour symmetry breaking: $(M_{PS}^{0})^2-(M_{PS}^\pm)^2\sim O(a^2)$.}\\
[3.0ex] \hline \\[-2.0ex]
ETM 09		&\cite{ETM09}		&2  & 0.07, 0.09, 0.10 & 
		\parbox[t]{6cm}{Scale set through $F_\pi$.
		Automatic $O(a)$ impr., flavour symmetry
		breaking: $(M_{PS}^{0})^2-(M_{PS}^\pm)^2\sim O(a^2)$. \\
		  }\\
[3.0ex] \hline \\[-2.0ex]
QCDSF/UKQCD 07 	&\cite{QCDSFUKQCD}	&2  & 0.06, 0.07 & 
		\parbox[t]{6cm}{Scale set through $F_\pi$. Non-perturbative
		$O(a)$-improvement.}\\
[4.0ex]
\hline\hline
\end{tabular*}
\caption{Continuum extrapolations/estimation of lattice artifacts in
  determinations of $f_K/f\pi$.}
}
\end{table}

\noindent
\begin{table}[!h]
{\footnotesize
\begin{tabular*}{\textwidth}[l]{l @{\extracolsep{\fill}} c c c l}
\hline \hline  \\[-1.0ex]
Collab. & Ref. & $\Nf$ & $M_{\pi,\text{min}}$ [MeV] & Description
\\[1.0ex] \hline \hline \\[-1.0ex]
{ETM 10E}& \cite{Farchioni:2010tb} & 2+1+1 & 265  & \parbox[t]{6.7cm}{Pion mass on finer
	lattice from \cite{Herdoiza_private}.}\\
[1.0ex] \hline  \\[-1.0ex]
{RBC/UKQCD 10A}& \cite{Aoki:2010dy} & 2+1 & 290 &\parbox[t]{6.7cm}{
		Results are based on 
		heavy kaon NLO SU(2) PQ{\Ch}PT. }\\
[4.0ex] \hline \\[-2.0ex]
MILC 10       & \cite{Bazavov:2010hj}  & 2+1 & 258 &  \parbox[t]{6.7cm}{Lightest
Nambu-Goldstone mass is 177\,MeV (at 0.09\,fm) and
 lightest RMS mass is 258\,MeV (at 0.06\,fm). NLO rS{\Ch}PT and NNLO {\Ch}PT.} \\
[7.0ex] \hline \\[-2.0ex]
BMW 10		&\cite{BMW10}		&2+1& 190 & 
		\parbox[t]{6.7cm}{Comparison of various fit-ans\"atze:
		SU(3) {\Ch}PT, heavy kaon SU(2) {\Ch}PT, polynomial. }\\
[4.0ex] \hline \\[-2.0ex]
JLQCD/TWQCD 08A &\cite{Noaki:2008iy}	&2+1&340& 
		\parbox[t]{6.7cm}{NNLO SU(3) {\Ch}PT.}\\
[1.0ex] \hline \\[-1.0ex]
MILC 09A	&\cite{Bazavov:2009fk}	&2+1&177& 
		\parbox[t]{6.7cm}{
		NLO SU(3) RS{\Ch}PT, continuum {\Ch}PT at NNLO and NNNLO and NNNNLO 
		analytic terms and heavy kaon SU(2) RS{\Ch}PT with 
		NNLO continuum chiral logs on a smaller 
		sub-set of the lattices. The lightest Nambu-Goldstone mass is 177\,MeV 
		(at $a=0.09$ fm)
		and the lightest RMS mass is 258\,MeV (at $a=0.06$ fm).}\\
[18.0ex] \hline \\[-1.0ex]
MILC 09		&\cite{Bazavov:2009bb}	&2+1&240& 
		\parbox[t]{6.7cm}{
		NLO SU(3) RS{\Ch}PT with continuum {\Ch}PT NNLO and NNNLO analytic
		terms added. According to \cite{Bazavov:2009fk}
		the lightest sea Nambu-Goldstone mass is 224\,MeV
		and the lightest RMS mass is 258\,MeV (at $a=0.06$ fm).}\\
[12.0ex] \hline \\[-1.0ex]
Aubin 08	&\cite{Aubin:2008ie}	&2+1&240& 
		\parbox[t]{6.7cm}{NLO MA{\Ch}PT. According to \cite{Bazavov:2009fk}
		the lightest sea Nambu-Goldstone mass is 246\,MeV (at $a=0.09$ fm)
		and the lightest RMS mass is 329\,MeV (at $a=0.09$ fm).}\\
[9.0ex] \hline \\[-1.0ex]
PACS-CS 08, 08A &\cite{Aoki:2008sm,Kuramashi:2008tb}& 2+1 &156& 
		\parbox[t]{6.7cm}{NLO SU(2) {\Ch}PT and SU(3) (Wilson){\Ch}PT.}\\
[1.0ex] \hline \\[-1.0ex]
HPCD/UKQCD 07	&\cite{Follana:2007uv}	&2+1&240& 
		\parbox[t]{6.7cm}{NLO SU(3) chiral perturbation theory
		with NNLO and NNNLO analytic terms. 
		The sea RMS mass for the employed lattices is heavier
		than 250\,MeV.}\\
[9.0ex] \hline \\[-2.0ex]
RBC/UKQCD 08	&\cite{Allton:2008pn}	&2+1&330& 
		\parbox[t]{6.7cm}{While SU(3) PQ{\Ch}PT fits were studied,
		final results are based on 
		heavy kaon NLO SU(2) PQ{\Ch}PT. }\\
[6.0ex] \hline \\[-2.0ex]
NPLQCD 06	&\cite{Beane:2006kx}	&2+1&300& 
		\parbox[t]{6.7cm}{NLO SU(3) {\Ch}PT and some NNLO terms. 
		The sea RMS mass for the employed lattices is heavier.}\\
[6.0ex] \hline \\[-2.0ex]
ETM 10D         & \cite{Lubicz:2010bv}  &2  &260&\parbox[t]{6.7cm}{
		NLO SU(3) {\Ch}PT and phenomenological ansatz for higher orders.
		Joint $f_+(0)$-$f_K/f_\pi$-fit.}\\
[3.0ex] \hline \\[-1.0ex]
ETM 09		&\cite{ETM09}		&2  &270& 
		\parbox[t]{6.7cm}{NLO heavy meson SU(2) {\Ch}PT and NLO
		SU(3) {\Ch}PT.}\\
[4.0ex] \hline \\[-2.0ex]
QCDSF/UKQCD 07 	&\cite{QCDSFUKQCD}	&2  &300& 
		\parbox[t]{6.7cm}{Linear extrapolation of lattice data.}\\
[1.0ex]
\hline\hline\\
\end{tabular*}
\caption{Chiral extrapolation/minimum pion mass in determinations of $f_K/f\pi$.}
}
\end{table}

\noindent
\begin{table}[!h]
{\footnotesize
\begin{tabular*}{\textwidth}[l]{l @{\extracolsep{\fill}} c c c c l}
\hline \hline  \\[-1.0ex]
Collab. & Ref. & $\Nf$ &$L$ [fm]& $M_{\pi,\text{min}}L$ & Description
\\[1.0ex] \hline \hline \\[-1.0ex]
{ETM 10E}& \cite{Farchioni:2010tb} & 2+1+1 & 1.9 - 2.9 &  $\gtrsim
3.9$ &\parbox[t]{5cm}{Simulation parameters from \cite{Baron:2011sf,
Herdoiza_private}.} \\
[4.0ex] \hline \\[-1.0ex]
{RBC/UKQCD 10A}& \cite{Aoki:2010dy} & 2+1 & 2.7 &  $\gtrsim 4.0$ & \\
[1.0ex] \hline \\[-1.0ex]
MILC 10       & \cite{Bazavov:2010hj}                 & 2+1 & 2.52 &
4.11  & \parbox[t]{5cm}{$L\!\geq\!2.9\,\fm$ for the lighter masses.}\\
[1.0ex] \hline \\[-1.0ex]
BMW 10		&\cite{BMW10}		&2+1&
		\parbox[t]{2cm}{2.0, 2.6, 2.7,\\3.0, 4.0}&4.0& 
		\parbox[t]{5cm}{Various volumes for comparison and
		correction for FSE from {\Ch}PT using \cite{Colangelo:2005gd}.}\\
[7.0ex] \hline \\[-2.0ex]
JLQCD/TWQCD 08A &\cite{Noaki:2008iy}	&2+1&1.6&2.8& 
		\parbox[t]{5cm}{Estimate of FSE using {\Ch}PT 
		\cite{Colangelo:2005gd,Brower:2003yx}}\\
[4.0ex] \hline \\[-1.0ex]
MILC 09A	&\cite{Bazavov:2009fk}	&2+1&
		\parbox[t]{2cm}{2.5, 2.9, 3.4,\\ 3.6, 3.8, 5.8}&4.1& 
		\parbox[t]{5cm}{}\\
[3.0ex] \hline \\[-2.0ex]
MILC 09		&\cite{Bazavov:2009bb}	&2+1&
		\parbox[t]{2cm}{2.4, 2.5, 2.9,\\
				3.0, 3.4, 3.6,\\
				3.8, 5.8}&3.8& 
		\parbox[t]{5cm}{Various volumes for comparison and
		correction for FSEs from (RS){\Ch}PT \cite{Colangelo:2005gd}.}\\
[7.0ex] \hline \\[-1.0ex]
Aubin 08	&\cite{Aubin:2008ie}	&2+1&2.4, 3.4&4.0& 
		\parbox[t]{5cm}{Correction for FSE from MA{\Ch}PT.}\\
[1.0ex] \hline \\[-1.0ex]
PACS-CS 08, 08A &\cite{Aoki:2008sm,Kuramashi:2008tb}& 2+1 &2.9&2.3& 
		\parbox[t]{5cm}{Correction for FSE from {\Ch}PT using 
		\cite{Colangelo:2005gd}.}\\
[4.0ex] \hline \\[-2.0ex]
HPCD/UKQCD 07	&\cite{Follana:2007uv}	&2+1&
		\parbox[t]{2cm}{2.3, 2.4, 2.5,\\ 2.9}&3.8& 
		\parbox[t]{5cm}{Correction for FSE from {\Ch}PT using 
		\cite{Colangelo:2005gd}.}\\
[4.0ex] \hline \\[-2.0ex]
RBC/UKQCD 08	&\cite{Allton:2008pn}	&2+1&1.8, 2.7&4.6&
		\parbox[t]{5cm}{Various volumes for comparison and 
		correction for FSE from {\Ch}PT 
		\cite{Gasser:1986vb,Gasser:1987ah,Colangelo:2005gd}.}\\
[7.0ex] \hline \\[-2.0ex]
NPLQCD 06	&\cite{Beane:2006kx}	&2+1&2.5&3.8& 
		\parbox[t]{5cm}{Correction for FSE from S{\Ch}PT 
		\cite{Aubin:2003uc,Aubin:2003mg}}\\
[4.0ex] \hline \\[-2.0ex]
ETM 10D		&\cite{Lubicz:2010bv}		&2  & 2.0-2.5 & 3.7 & \\
[4.0ex] \hline \\[-2.0ex]
ETM 09		&\cite{ETM09}		&2  &2.0, 2.4, 2.7&3.3& 
		\parbox[t]{5cm}{Correction for FSE from {\Ch}PT 
		\cite{Gasser:1986vb,Gasser:1987ah,Colangelo:2005gd}.}\\
[4.0ex] \hline \\[-2.0ex]
QCDSF/UKQCD 07 	&\cite{QCDSFUKQCD}	&2  &1.4,\dots,2.6&4.2& 
		\parbox[t]{5cm}{Correction for FSE from {\Ch}PT}\\
[1.0ex] \hline\hline
\end{tabular*}
\caption{Finite volume effects in determinations of $f_K/f\pi$.}
}
\end{table}

\clearpage
\newpage
\subsection{Notes to section \ref{sec:LECs} on Low-Energy Constants}

\begin{table}[!ht]
{\footnotesize
\begin{tabular*}{\textwidth}[l]{l @{\extracolsep{\fill}} c c l l}
\hline\hline \\[-1.0ex]
Collab. & Ref. & $\Nf$ & $a$ [fm]& Description 
\\[1.0ex] \hline \hline \\[-1.0ex]
Bernardoni 10 & \cite{Bernardoni:2010nf} & 2 & 0.0784(10) & \parbox[t]{7.5cm}{Scale fixed through $M_K$. Non-perturbative $O(a)$ improvement.  No estimate of systematic error. }\\
[4.0ex] \hline \\[-1.0ex]
JLQCD/TWQCD 10 & \cite{Fukaya:2010na} & 2 & 0.11
& \parbox[t]{7.5cm}{One lattice spacing, scale fixed
through $r_0$. Estimate of 7\% systematic error due to finite lattice spacing.} 
 \\
[4.0ex] \hline  \\[-1.0ex] 
JLQCD/TWQCD 09 & \cite{JLQCD:2009qn} & 2 & 0.1184(21) & \parbox[t]{7.5cm}{Automatic $O(a)$ impr., exact chiral symmetry. Scale fixed through $r_0$. Scale ambiguity added as systematic error.}\\
[6.0ex] \hline  \\[-1.0ex]
ETM 09B & \cite{Jansen:2009tt} & 2 & 0.063, 0.073 & \parbox[t]{7.5cm}{Automatic $O(a)$ impr. Estimate of lattice artifacts ($4\%-10\%$ effect, not added in final result) by comparing two lattice spacings. $r_0=0.49$\,fm used to convert results in physical units.}.\\ 
[9.0ex] \hline  \\[-1.0ex]
ETM 09C & \cite{Baron:2009wt}  & 2 &  0.051 - 0.1 & \parbox[t]{7.5cm}{Automatic $O(a)$ impr. Scale fixed through $F_\pi$. 4 lattice spacings, continuum extrapolation.}\\
[4.0ex] \hline \\[-1.0ex]
ETM 08 & \cite{Frezzotti:2008dr} & 2 & 0.07-0.09             & \parbox[t]{7.5cm}{Automatic $O(a)$ impr. Discretization effects estimated by comparing two lattice spacings. Scale fixed through $F_\pi$.   }              \\
[7.0ex] \hline \\[-1.0ex]
\parbox[t]{3.5 cm}{JLQCD/TWQCD 08A \\JLQCD 08A} & \parbox[t]{0.7
  cm}{\cite{Noaki:2008iy} \\ \cite{Shintani:2008qe}} & 2 & 0.1184(3)(21)
& \parbox[t]{7.5cm}{Automatic $O(a)$ impr., exact chiral symmetry. Scale fixed through $r_0$. Scale ambiguity added as systematic error.}\\
[6.0ex] \hline \\[-1.0ex]
CERN 08 & \cite{Giusti:2008vb} & 2 & 0.0784(10) & \parbox[t]{7.5cm}{Scale fixed through $M_K$. Non-perturbative $O(a)$ improvement.  No estimate of systematic error.}\\
[4.0ex] \hline \\[-1.0ex]
HHS 08 & \cite{Hasenfratz:2008ce} & 2 & 0.1153(5) & \parbox[t]{7.5cm}{Tree level $O(a)$
improvement. Scale fixed through $r_0$. Estimate of lattice artifacts via W{\Ch}PT in \cite{Bar:2008th}.}\\
[4.0ex] \hline \\[-1.0ex]
JLQCD/TWQCD 07  & \cite{Fukaya:2007pn} & 2 & 0.1111(24) & \parbox[t]{7.5cm}{Automatic $O(a)$ impr., exact chiral symmetry. Scale fixed through $r_0$. No estimate of lattice artifacts.} \\
[4.0ex] \hline \\[-1.0ex]
JLQCD/TWQCD 07A & \cite{Aoki:2007pw} & 2 & $\simeq$ 0.12 & \parbox[t]{7.5cm}{Automatic $O(a)$ impr., exact chiral symmetry. Scale fixed through $r_0$. No estimate of lattice artifacts.} \\
[4.0ex] \hline  \\[-1.0ex]
 CERN-TOV 06 & \cite{DelDebbio:2006cn} & 2 & \parbox[t]{1.0cm}{0.0717(15), 0.0521(7),  0.0784(10)} &\parbox[t]{7.5cm}{Scale fixed through $M_K$. The lattice with $a=0.0784(10)$ is obtained with non-perturbative $O(a)$ improvement. No estimate of systematic error.}\\
[7.0ex] \hline \\[-1.0ex]
QCDSF/UKQCD 06A & \cite{Brommel:2006ww} & 2& 0.07-0.115 &  \parbox[t]{7.5cm}{5 lattice spacings. Non-perturbative $O(a)$ improvement. Scale fixed through $r_0$.}\\
[4.0ex] \hline \hline \\[-1.0ex]
\end{tabular*}
\caption{Continuum extrapolations/estimation of lattice artifacts in $\Nf=2$
  determinations of the Low-Energy Constants.}
}
\end{table}

\begin{table}[!ht]
{\footnotesize
\begin{tabular*}{\textwidth}[l]{l @{\extracolsep{\fill}} c c l l}
\hline\hline \\[-1.0ex]
Collab. & Ref. & $\Nf$ & $a$ [fm]& Description 
\\[1.0ex] \hline \hline \\[-1.0ex]
ETM 10 & \cite{Baron:2010bv} & 2+1+1 & 0.078, 0.086 & \parbox[t]{7.0cm}{Scale fixed through $F_\pi/M_\pi$. Systematic error taken into account by considering two lattice spacings.}  \\
[7.0ex] \hline\hline\\[-1.0ex]
 MILC 10, 10A & \cite{Bazavov:2010hj,Bazavov:2010yq} & 2+1 & 0.045 - 0.09 & \parbox[t]{7.0cm}{3 lattice
spacings, continuum extrapolation by means of RS{\Ch}PT.}\\
[4.0ex] \hline  \\[-1.0ex]
JLQCD/TWQCD 10 & \cite{Fukaya:2010na} & 2+1, 3 & 0.11 & \parbox[t]{7.0cm}{One lattice spacing, scale fixed through $m_\Omega$.}
  \\
[1.0ex] \hline\\[-1.0ex]
RBC/UKQCD 10A & \cite{Aoki:2010dy} & 2+1 & \parbox[t]{1.7cm}{0.1106(27), 0.0888(12)} & \parbox[t]{7.0cm}{Two lattice spacings. Data combined in global chiral-continuum fits.}
  \\
[4.0ex] \hline\\[-1.0ex]
JLQCD 09 & \cite{Fukaya:2009fh} & 2+1 & 0.1075(7) &  \parbox[t]{7.0cm}{Scale fixed through $r_0$.  Systematic error associated to lattice artifacts of
 order 7.4\% estimated through mismatch of the lattice spacing obtained from different inputs.} \\
[9.0ex] \hline \\[-1.0ex]
MILC 09, 09A   & \cite{Bazavov:2009bb,Bazavov:2009fk}  & 2+1 & 0.045 - 0.18 & \parbox[t]{7.0cm}
 {Total of 6 lattice spacings, continuum extrapolation by means of RS{\Ch}PT.} \\
[4.0ex] \hline \\[-1.0ex]
TWQCD 08 & \cite{Chiu:2008jq} & 2+1 & 0.122(3) & \parbox[t]{7.0cm}{Scale fixed through $m_\rho$, $r_0$. No estimate of systematic effects.}  \\
[4.0ex] \hline\\[-1.0ex]
JLQCD/TWQCD 08B & \cite{Chiu:2008kt} & 2+1 &  0.1075(7) & \parbox[t]{7.0cm}{Scale fixed through $r_0$. No estimate of systematic effects for the observable adopted in this work.}   \\
[4.0ex] \hline \\[-1.0ex]
PACS-CS 08   & \cite{Aoki:2008sm}     & 2+1 & 0.0907       & \parbox[t]{7.0cm}
 {Only one lattice spacing, no attempt to estimate size of cut-off effects.} \\
[4.0ex] \hline \\[-1.0ex]
RBC/UKQCD 08 & \cite{Allton:2008pn}   & 2+1 & 0.114        & \parbox[t]{7.0cm}
 {Only one lattice spacing, attempt to estimate size of cut-off effects via
 formal argument.} \\
[4.0ex]\hline \\[-1.0ex]
RBC/UKQCD 08A & \cite{Boyle:2008yd} & 2+1 &0.114 & \parbox[t]{7.0cm}
 {Only one lattice spacing, attempt to estimate size of cut-off effects via
 formal argument.} \\
[4.0ex]\hline \\[-1.0ex]
NPLQCD 06 & \cite{Beane:2006kx} & 2+1 & 0.125 & \parbox[t]{7.0cm}{Only one lattice spacing, continuum {\Ch}PT used.} \\
[1.0ex]\hline \\[-1.0ex]
LHP 04 & \cite{Bonnet:2004fr} & 2+1 & $\simeq 0.12 $ & \parbox[t]{7.0cm}{Only one lattice spacing, mixed discretization approach.}\\
[4.0ex]\hline \\[-1.0ex]
\end{tabular*}
\caption{Continuum extrapolations/estimation of lattice artifacts in
  $\Nf=2+1$ and $2+1+1$ determinations of the Low-Energy Constants.}
}
\end{table}

\begin{table}[!ht]
{\footnotesize
\begin{tabular*}{\textwidth}[l]{l @{\extracolsep{\fill}} c c c l}
\hline\hline \\[-1.0ex] 
Collab. & Ref. & $\Nf$ & $M_{\pi,\text{min}}$ [MeV] & Description \\
[3.0ex] \hline \hline \\[-1.0ex]
Bernardoni 10 & \cite{Bernardoni:2010nf}&  2& 297, 377, 426 & \parbox[t]{5.5cm}{NLO SU(2) fit of the topological susceptibility.}\\
[3.0ex] \hline \\[-2.0ex]
JLQCD/TWQCD 10 & \cite{Fukaya:2010na} & 2 &  \parbox[t]{3.5cm}{$m\Sigma V< 1$ ($\epsilon$-reg.)\\ $m\simeq m_s/6-m_s$ ($p$-reg.)} & \parbox[t]{5.5cm}{Quark masses both in the $p$ and in the $\epsilon$-regime. NLO chiral fit of the spectral density interpolating the two regimes. $\binom{+1.2\%}{-1.8\%}$ for $\Sigma^{1/3}$ associated to chiral fit.} \\
[10.0ex]\hline \\[-2.0ex]
JLQCD/TWQCD 09 & \cite{JLQCD:2009qn} & 2 & 290  &  \parbox[t]{5.5cm}{LECs extracted from NNLO chiral fit of vector and scalar radii $\langle r^2\rangle^\pi_{V,S}$.}\\
[4.0ex]\hline \\[-1.0ex]
ETM 09B & \cite{Jansen:2009tt} & 2 & \parbox{3.5cm}{$\mu=m\Sigma V\simeq 0.11-0.35$} &  \parbox[t]{5.5cm}{NLO SU(2) $\epsilon$-regime fit.}\\
[1.0ex]\hline \\[-1.0ex]
ETM 09C & \cite{Baron:2009wt} & 2 & $280$    & \parbox[t]{5.5cm}{NNLO SU(2) fit. Systematic error estimated by performing a large set of different fits.} \\
[6.0ex] \hline \\[-1.0ex]
ETM 08  & \cite{Frezzotti:2008dr} & 2 &   $260$  &   \parbox[t]{5.5cm}{LECs extracted from pion form factor using NNLO {\Ch}PT and fixing the pion scalar radius to the experimental value.  }        \\
[7.0ex] \hline \\[-1.0ex]
\parbox[t]{3.5 cm}{JLQCD/TWQCD 08A \\JLQCD 08A} & \parbox[t]{0.7
  cm}{\cite{Noaki:2008iy} \\ \cite{Shintani:2008qe}} & 2 & 290 &
\parbox[t]{5.5cm}{NNLO SU(2) fit up to $M_\pi=$ 750 MeV. Estimate of
  systematic error due to truncation of chiral series.}\\ 
[7.0ex] \hline \\[-1.0ex]
CERN 08 & \cite{Giusti:2008vb} &  2 & $m_{q,\text{min}}$=13 MeV & \parbox[t]{5.5cm}{NLO SU(2) fit for the mode number of the Dirac operator.}\\
[4.0ex] \hline \\[-1.0ex]
HHS 08 & \cite{Hasenfratz:2008ce} & 2 &  \parbox[t]{3.5cm}{$\mu_1=m\Sigma
V_1\simeq 0.7-2.9$\\ $\mu_2=m\Sigma V_2\simeq 2.1-5.0$}
&  \parbox[t]{5.5cm}{$\epsilon$-regime: higher values might be too large to
  be considered in the $\epsilon$-regime.  NLO SU(2) $\epsilon$-regime fit.}\\
[7.0ex] \hline \\[-1.0ex]
JLQCD/TWQCD 07  & \cite{Fukaya:2007pn} & 2 & $\mu=m\Sigma V\simeq 0.556$ & \parbox[t]{5.5cm}{NLO SU(2) $\epsilon$-regime fit.} \\
[1.0ex]\hline \\[-1.0ex]
JLQCD/TWQCD 07A & \cite{Aoki:2007pw} & 2 & $m_{ud}= m_s/6-m_s$ & \parbox[t]{5.5cm}{Quark condensate extracted from topological susceptibility, LO chiral fit.}\\
[4.0ex] \hline \\[-1.0ex]
 CERN-TOV 06 & \cite{DelDebbio:2006cn} & 2 & 403, 381, 377 & \parbox[t]{5.5cm}{NLO SU(2) fit up to $M_\pi\simeq $ 540 MeV.}\\
[1.0ex] \hline \\[-1.0ex]
QCDSF/UKQCD 06A & \cite{Brommel:2006ww} & 2 & 400 & \parbox[t]{5.5cm}{Several  fit functions to extrapolate the pion form factor.} \\
[4.0ex] \hline \hline \\[-1.0ex]
\end{tabular*}
\caption{Chiral extrapolation/minimum pion mass in $\Nf=2$ determinations of the Low-Energy Constants.}
}
\end{table}

\begin{table}
\vspace{-1.5cm}
{\footnotesize
\begin{tabular*}{\textwidth}[l]{l @{\extracolsep{\fill}} c c c l}
\hline\hline \\[-1.0ex] 
Collab. & Ref. & $\Nf$ & $M_{\pi,\text{min}}$ [MeV] & Description 
\\[1.0ex] \hline \hline \\[-1.0ex]
ETM 10 & \cite{Baron:2010bv} & 2+1+1 &  $270$ &   \parbox[t]{5.9cm}{SU(2) NLO and NNLO fits. }\\
[1.0ex] \hline\hline\\[-1.0ex]
MILC 10, 10A       & \cite{Bazavov:2010hj,Bazavov:2010yq}  & 2+1 &     &  Cf.~MILC 09A. \\
[1.0ex] \hline \\[-1.0ex]
JLQCD/TWQCD 10 & \cite{Fukaya:2010na} & 2+1,3 &  \parbox[t]{2.8cm}{$m_{ud}\Sigma V <1$ ($\epsilon$-reg)\\ $m_{ud}= m_s/6-m_s$ ($p$-reg.), \\ $m_s\sim$ phys.}  & \parbox[t]{5.9cm}{For the $N_f=2+1$ runs, light quark masses both in the $\epsilon$- and in the $p$-regime. For the $N_f=3$ runs, only masses in the $p$-regime. NLO chiral fit of the spectral density interpolating the two regimes. $\binom{+2.2\%}{-0.2\%}$ systematic error associated to $\Sigma^{1/3}$ and $\pm 8.7\%$ to $\Sigma_0^{1/3}$. } 
    \\
[18.0ex] \hline\\[-1.0ex]
RBC/UKQCD 10A & \cite{Aoki:2010dy} & 2+1 &  \parbox[t]{2.8cm}{290-420} &  \parbox[t]{5.9cm}{The range is 225-420 MeV for partially quenched pions. NLO SU(2) $\chi$PT fits are used.} \\
[6.0ex] \hline\\[-1.0ex]
JLQCD 09 & \cite{Fukaya:2009fh} & 2+1 & \parbox[t]{2.8cm}{$m_{ud}\Sigma V <1$ ($\epsilon$-reg)\\ $m_{ud}= m_s/6-m_s$ ($p$-reg.), \\ $m_s\sim$ phys.} &  \parbox[t]{5.9cm}{Quark masses both in the $\epsilon$ and in the $p$-regime.  NLO chiral fit of the spectral density interpolating the two regimes. $\binom{+2.2}{-0.7}$ \% systematic error estimated by varying the fit range and using formulae for SU(2) versus SU(3) {\Ch}PT.   }\\
[14.0ex] \hline \\[-1.0ex]
MILC 09      & \cite{Bazavov:2009bb}  & 2+1 &  $240$ & \parbox[t]{5.9cm}
 {Value taken from text of \cite{Bazavov:2009bb}; according to
 \cite{Bazavov:2009fk}
 lightest Nambu-Goldstone mass is 224\,MeV and
 lightest RMS mass is 258\,MeV (at 0.06\,fm). }\\
[10.0ex] \hline \\[-1.0ex]
MILC 09A      & \cite{Bazavov:2009fk}  & 2+1 & 258 &  \parbox[t]{5.9cm}{Lightest
Nambu-Goldstone mass is 177\,MeV (at 0.09\,fm) and
 lightest RMS mass is 258\,MeV (at 0.06\,fm).} \\
[7.0ex] \hline \\[-1.0ex]
TWQCD 08 & \cite{Chiu:2008jq} & 2+1 &  \parbox[t]{2.8cm}{{$m_{ud}=(m_s-3m_s)/4$},  $m_s\sim$ phys.}  & \parbox[t]{5.9cm}{Quark condensate extracted from topological susceptibility, LO chiral fit.}  \\
[4.0ex] \hline\\[-1.0ex]
JLQCD/TWQCD 08B & \cite{Chiu:2008kt} & 2+1 & \parbox[t]{2.8cm}{$m_{ud}=m_s/6-m_s$, \\ $m_s\sim$ phys.}  &  \parbox[t]{5.9cm}{Quark condensate extracted from topological susceptibility, LO chiral fit.} \\
[4.0ex] \hline \\[-1.0ex]
PACS-CS 08   & \cite{Aoki:2008sm}     & 2+1 &  $156$ & \parbox[t]{5.9cm}
 {To date, lightest published quark mass reached in a direct simulation.} \\
[4.0ex] \hline \\[-1.0ex]
RBC/UKQCD 08 & \cite{Allton:2008pn}   & 2+1 &  $330$ & \parbox[t]{5.9cm}
 {Other unitary masses are 419, 557, 672 MeV,
 lightest non-unitary pion goes down to $242\MeV$.} \\
[6.0ex] \hline \\[-1.0ex]
RBC/UKQCD 08A & \cite{Boyle:2008yd} & 2+1 & 330 & \parbox[t]{5.9cm}{Pion electromagnetic form factor computed at one pion mass. Low values of $q^2$ reached thanks to partially twisted boundary conditions. LECs determined from NLO $\chi$PT fit in $q^2$.}\\ 
[12.0ex] \hline \\[-1.0ex]
NPLQCD 06 & \cite{Beane:2006kx} & 2+1 & $460$  & \parbox[t]{5.9cm}{Value refers to lightest RMS mass at $a=0.125$\,fm as quoted in \cite{Bazavov:2009fk}.} \\
[4.0ex]\hline \\[-1.0ex]
LHP 04 & \cite{Bonnet:2004fr} & 2+1 &  $318$ &  \parbox[t]{5.9cm}{Pion form factor extracted from vector meson dominance fit.}\\
[4.0ex]\hline \\[-1.0ex]
\end{tabular*}
\caption{Chiral extrapolation/minimum pion mass in $\Nf=2+1$ and $2+1+1$
  determinations of the Low-Energy Constants.}
}
\end{table}

\begin{table}[!ht]
{\footnotesize
\begin{tabular*}{\textwidth}[l]{l @{\extracolsep{\fill}} c c c c l}
\hline\hline \\[-1.0ex]
Collab.\ & Ref.\ & $\Nf$ & $L$ [fm] & $M_{\pi,\text{min}}L$ & Description
\\[1.0ex] \hline \hline \\[-1.0ex]
Bernardoni 10 & \cite{Bernardoni:2010nf} & 2 & 1.88 & 2.8 & \parbox[t]{5.25cm}{Finite-volume effects included in the NLO chiral fit.}  \\
[3.0ex] \hline \\[-2.0ex]
JLQCD/TWQCD 10 & \cite{Fukaya:2010na} & 2 & 1.8-1.9 &  & \parbox[t]{5.25cm}{Finite volume
effects estimated through differences between different topological sectors. Systematic error of 3.7\% on $\Sigma^{1/3}$.} \\
[9.0ex] \hline  \\[-1.0ex]
JLQCD/TWQCD 09 & \cite{JLQCD:2009qn} & 2 & 1.89 & 2.9
&  \parbox[t]{5.25cm}{Finite-volume corrections for form factor estimated through NLO  $\chi$PT
  \protect\cite{Bunton:2006va,Borasoy:2004zf}. Additional finite-volume effect for fixed topological charge \protect\cite{Aoki:2007ka}.} \\
[12.0ex] \hline  \\[-2.0ex]
ETM 09B & \cite{Jansen:2009tt} & 2 & 1.3, 1.5 &  $\epsilon$-regime
&  \parbox[t]{5.25cm}{Topology: not fixed. Comparison between 2 volumes, estimate of $8\%$ systematic effect (not quoted in the final result). Volume might be too small for $\epsilon$-expansion.}\\
[12.0ex] \hline  \\[-2.0ex]
ETM 09C & \cite{Baron:2009wt} & 2 & 2.0-2.5   & \parbox[t]{1.7cm}{4.4, 3.7, 3.2,\\ 3.5} & \parbox[t]{5.25cm}{Several volumes. Finite-volume effects estimated through \protect\cite{Colangelo:2005gd}.}\\
[3.0ex] \hline \\[-2.0ex]
ETM 08 & \cite{Frezzotti:2008dr} & 2 & 2.1, 2.8  & 3.4, 3.7 &\parbox[t]{5.25cm}{For pion form factor only data with $M_\pi L\gtrsim 4$ are considered in the final analysis. FSE for pion masses and decay constants are taken into account according to \cite{Colangelo:2005gd}.}\\
[12.0ex] \hline \\[-2.0ex]
\parbox[t]{3.5 cm}{JLQCD/TWQCD 08A \\JLQCD 08A} & \parbox[t]{0.7
  cm}{\cite{Noaki:2008iy} \\ \cite{Shintani:2008qe}} & 2 & 1.89 & 2.9
&  \parbox[t]{5.25cm}{Finite-volume corrections estimates through
  \protect\cite{Colangelo:2005gd}. Additional finite-volume effect for fixed topological charge \protect\cite{Aoki:2007ka}.} \\
[9.0ex] \hline \\[-1.0ex]
CERN 08 & \cite{Giusti:2008vb}  & 2  & 1.88, 2.51  & - & \parbox[t]{5.25cm}{For the specific observable studied here the finite-volume effects $\propto e^{-M_\Lambda L/2}$, with $M_\Lambda \gg M_\pi$.
Finite-volume effects checked by comparing 2 volumes.}  \\
[12.0ex] \hline \\[-1.0ex]
HHS 08 & \cite{Hasenfratz:2008ce} & 2 & 1.84, 2.77 &  $\epsilon$-regime &  \parbox[t]{5.25cm}{Topology: not fixed. Comparison between 2 volumes.}\\
[3.0ex] \hline \\[-2.0ex]
JLQCD/TWQCD 07  & \cite{Fukaya:2007pn} & 2 & 1.78 & $\epsilon$-regime & \parbox[t]{5.25cm}{Topology: fixed to $\nu=0$.}\\
[1.0ex]\hline \\[-1.0ex]
JLQCD/TWQCD 07A & \cite{Aoki:2007pw} & 2 & 1.92 & - & \parbox[t]{5.25cm}{Topology fixed to $\nu=0$ gives finite-volume effects estimated through \protect\cite{Aoki:2007ka}.} \\
[7.0ex] \hline  \\[-1.0ex]
CERN-TOV 06 & \cite{DelDebbio:2006cn} & 2 & 1.72, 1.67, 1.88 & 3.5, 3.2, 3.6 & \parbox[t]{5.25cm}{No estimate of finite-volume effects.}  \\
[1.0ex] \hline \\[-1.0ex]
QCDSF/UKQCD 06A & \cite{Brommel:2006ww} & 2 & 1.4-2.0 & 3.8 & \parbox[t]{5.25cm}{Use NLO $\chi$PT \protect\cite{Bunton:2006va} to investigate finite-volume effects.} \\
[4.0ex] \hline \hline \\[-1.0ex]
\end{tabular*}
\caption{Finite volume effects in $\Nf=2$ determinations of the Low-Energy Constants.}
}
\end{table}

\begin{table}[!ht]
{\footnotesize
\begin{tabular*}{\textwidth}[l]{l @{\extracolsep{\fill}} c c c c l}
\hline\hline \\[-1.0ex]
Collab.\ & Ref.\ & $\Nf$ & $L$ [fm] & $M_{\pi,\text{min}}L$ & Description
\\[1.0ex] \hline \hline \\[-1.0ex]
ETM 10 & \cite{Baron:2010bv} & 2+1+1 & 1.9-2.8 & 3.0 &
\parbox[t]{4.25cm}{Finite-size effects studied using
  \cite{Colangelo:2005gd}. LECs are extrac\-ted from data which satisfy
  $M_{\pi^+} L\gtrsim 4$, but $M_{\pi^0} L \sim 2$. }\\
[10.0ex] \hline\hline\\[-1.0ex]
MILC 10, 10A       & \cite{Bazavov:2010hj,Bazavov:2010yq}                 & 2+1 & 2.52 &
4.11  & \parbox[t]{4.25cm}{$L\!\geq\!2.9\,\fm$ for the lighter masses.}\\
[3.0ex] \hline \\[-1.0ex]
JLQCD/TWQCD 10 & \cite{Fukaya:2010na} & 2+1, 3 & 1.9, 2.7 &
& \parbox[t]{4.25cm}{Finite volume effects estimated by comparing with larger volume for a fixed quark mass. For $F$ estimated to be
  3\%, for $\Sigma^{1/3}$ and  $\Sigma_0^{1/3}$ $\binom{+0.3\%}{-1.0\%}$.} \\
[6.0ex]\hline\\[-1.0ex]
RBC/UKQCD 10A & \cite{Aoki:2010dy} & 2+1 &  2.7   & $\simeq$ 4  & \parbox[t]{4.25cm}{Finite-volume corrections included by means of $\chi$PT.} \\
[4.0ex] \hline\\[-2.0ex]
JLQCD 09 & \cite{Fukaya:2009fh} & 2+1 & 1.72 &  $\epsilon/p$-regime   & \parbox[t]{4.25cm}{Finite-volume effects estimated by comparing with larger volume ($L=2.58$\,fm) for a fixed $m_{ud}$, $m_s$.} 
\\
[9.0ex] \hline \\[-1.0ex]
 MILC 09, 09A  & \cite{Bazavov:2009bb,Bazavov:2009fk}  & 2+1 & 2.4/2.9 & 3.5/4.11 & \parbox[t]{4.25cm}
 {$L_\mathrm{min}\!=\!2.4\,\fm$, but $L\!\geq\!2.9\,\fm$ for lighter masses; one
 coarse ensemble with $\Mpi L\!=\!3.5$, all fine ones have $\Mpi L\!>\!4.11$.} \\
[9.0ex] \hline \\[-1.0ex]
TWQCD 08 & \cite{Chiu:2008jq} & 2+1 &  1.95  &  - &  \parbox[t]{4.25cm}{No estimate of finite-volume effects.} \\
[4.0ex] \hline\\[-1.0ex]
JLQCD/TWQCD 08B & \cite{Chiu:2008kt} & 2+1 &1.72  &  - &  \parbox[t]{4.25cm}{Topology fixed to $\nu=0$ gives finite-volume effects estimated through \protect\cite{Aoki:2007ka}.} \\
[6.0ex]\hline\\[-1.0ex]
 PACS-CS 08   & \cite{Aoki:2008sm}     & 2+1 & 2.9     & 2.3      & \parbox[t]{4.25cm}
 {Possible finite volume effects are the main concern of the authors.} \\
[6.0ex] \hline \\[-1.0ex]
RBC/UKQCD 08 & \cite{Allton:2008pn}   & 2+1 & 2.74    & 4.6      & \parbox[t]{4.25cm}{Good setting to correct residual finite-volume  effects by means of {\Ch}PT.} \\
[6.0ex]\hline \\[-1.0ex]
RBC/UKQCD 08A & \cite{Boyle:2008yd} & 2+1 & 2.74 & 4.6 & \parbox[t]{4.25cm}{Finite-size effects on the form factor and radius estimated to be  $< 1 \%$ using $\chi$PT}\\
[6.0ex] \hline \\[-1.0ex]
NPLQCD 06 & \cite{Beane:2006kx} & 2+1 &  2.5    & 3.7     & \parbox[t]{4.25cm}{Value refers to lightest valence pion mass.} \\
[4.0ex]\hline \\[-1.0ex]
 LHP 04 & \cite{Bonnet:2004fr} & 2+1 & $\simeq 2.4$ & 3.97 &  \parbox[t]{4.25cm}{Value refers to domain-wall valence pion mass.}\\
[4.0ex] \hline \hline\\[-1.0ex]
\end{tabular*}
\caption{Finite volume effects in $\Nf=2+1$ and $2+1+1$ determinations of the Low-Energy Constants.}
}
\end{table}

\begin{table}[!ht]
{\footnotesize
\begin{tabular*}{\textwidth}[l]{l @{\extracolsep{\fill}} c c l}
\hline\hline \\[-1.0ex]
Collab. & Ref. & $\Nf$ &  Description
\\[1.0ex] \hline \hline \\[-1.0ex]
ETM 10 & \cite{Baron:2010bv}  & 2+1+1 & Non-perturbative\\
[1.0ex] \hline\hline \\[-1.0ex]
JLQCD/TWQCD 10 & \cite{Fukaya:2010na} & 2+1, 3 &  Non-perturbative\\
[1.0ex]\hline \\[-1.0ex]
MILC 10, 10A & \cite{Bazavov:2010hj, Bazavov:2010yq}  & 2+1 &  2 loop \\
[1.0ex] \hline \\[-1.0ex]
RBC/UKQCD 10A & \cite{Aoki:2010dy}   & 2+1 &  Non-perturbative\\
[1.0ex]\hline \\[-1.0ex]
JLQCD 09 & \cite{Fukaya:2009fh} & 2+1 & Non-perturbative\\
[1.0ex] \hline \\[-1.0ex]
MILC 09, 09A & \cite{Bazavov:2009bb,Bazavov:2009fk}  & 2+1 &  2 loop \\
[1.0ex] \hline \\[-1.0ex]
TWQCD 08 & \cite{Chiu:2008jq} & 2+1 &  Non-perturbative\\
[1.0ex] \hline \\[-1.0ex]
JLQCD/TWQCD 08B & \cite{Chiu:2008kt} & 2+1 & Non-perturbative\\
[1.0ex]\hline\\[-1.0ex]
PACS-CS 08   & \cite{Aoki:2008sm}     & 2+1 &  1 loop \\
[1.0ex] \hline \\[-1.0ex]
RBC/UKQCD 08, 08A & \cite{Allton:2008pn,Boyle:2008yd}   & 2+1 &  Non-perturbative\\
[1.0ex]\hline \\[-1.0ex]
NPLQCD 06 & \cite{Beane:2006kx} & 2+1 & --- \\
[1.0ex] \hline \\[-1.0ex]
LHP 04 & \cite{Bonnet:2004fr} & 2+1 & --- \\
[1.0ex] \hline \hline\\[-1.0ex]
 All collaborations  &  & 2 & Non-perturbative\\
[1.0ex] \hline\hline \\[-1.0ex]
\end{tabular*}
\caption{Renormalization in determinations of the Low-Energy Constants.}
}
\end{table}

\clearpage
\subsection{Notes to section \ref{sec:BK} on Kaon $B$-parameter $B_K$}
\label{app-BK}

In the following, we summarize the characteristics (lattice actions,
pion masses, lattice spacings, etc.) of the recent $\Nf = 2+1$ and
$\Nf = 2$ runs. We also provide brief descriptions of how systematic
errors are estimated by the various authors.

\begin{table}[!ht]

{\footnotesize
\begin{tabular*}{\textwidth}{l c c c l}
\hline\hline \\[-1.0ex]
Collab. & Ref. & $\Nf$ & $a$ [fm] & Description 
\\[1.0ex] \hline \hline \\[-1.0ex]
SWME 11 & \cite{Kim:2011qg} & 2+1 & 0.12, 0.09, 0.06, 0.04  &
\parbox[t]{5.5cm}{Continuum extrapolation of results obtained at four
lattice spacings; residual discretization error of 0.21\% from
difference to result at smallest lattice spacing.}
\\[12.0ex] \hline \\[-2.0ex]
RBC/UKQCD 10B & \cite{Aoki:2010pe} & 2+1 & 0.114, 0.087  &
\parbox[t]{5.5cm}{Two lattice spacings. Combined chiral and continuum
fits.} 
\\[3.0ex] \hline \\[-2.0ex]
SWME 10   & \cite{Bae:2010ki} & 2+1 &  0.12, 0.09, 0.06 &
\parbox[t]{5.5cm}{Continuum extrapolation of results obtained at four
lattice spacings; residual discretization error of 0.21\% from
difference to result at smallest lattice spacing.}   
\\[12.0ex] \hline \\[-2.0ex]
Aubin 09 & \cite{Aubin:2009jh} & 2+1 & 0.09,\,0.12 & \parbox[t]{5.5cm}{Two
   lattice spacings; quote 0.3\% discretization error, estimated from
   various $a^2$-terms in fit function}
\\[6.0ex] \hline \\[-2.0ex]
RBC/UKQCD 07A, 08 & \cite{Antonio:2007pb,Allton:2008pn} & 2+1 &
0.114(2) & \parbox[t]{5.5cm}{Single lattice spacing; quote 4\%
  discretization error, estimated from the difference between
  computed and experimental values of $f_\pi$.}
\\ [9.0ex] \hline \\[-2.0ex]
HPQCD/UKQCD  06 & \cite{Gamiz:2006sq} & {2+1} & 0.12 & \parbox[t]{5.5cm}{Single
  lattice spacing; 3\% discretization error quoted without providing more
  details.}
\\[6.0ex] \hline \\[-1.0ex]
ETM 10A      & \cite{Constantinou:2010qv} & 2 & 0.07,\, 0.09, \,0.1& 
 \parbox[t]{5.5cm}{Three lattice spacings; 1.2\% error quoted.}
 \\[4.0ex] \hline \\[-1.0ex] 
JLQCD 08 & \cite{Aoki:2008ss} & 2 & 0.118(1) & \parbox[t]{5.5cm}{Single
  lattice spacing; no error quoted.}
\\[1.0ex] \hline \\[-1.0ex]
RBC 04 & \cite{Aoki:2004ht} & 2 & 0.117(4) & \parbox[t]{5.5cm}{Single
   lattice spacing; no error quoted.}
\\[1.0ex] \hline \\[-1.0ex]
UKQCD 04 & \cite{Flynn:2004au} & 2 & 0.10 & \parbox[t]{5.5cm}{Single
   lattice spacing; no error quoted.}
\\[1.0ex]
\hline\hline
\end{tabular*}
\caption{Continuum extrapolations/estimation of lattice artifacts in
  determinations of $B_K$.}
}
\end{table}

\begin{table}[!ht]
{\footnotesize
\begin{tabular*}{\textwidth}{l @{\extracolsep{\fill}} c c c l}
\hline\hline \\[-1.0ex]
{Collab.} & {Ref.} & {$\Nf$} & {$M_{\pi,\rm min}\,[\mev]$} & {Description}  
\\[1.0ex] \hline \hline \\[-1.0ex]
SWME 11 & \cite{Kim:2011qg} & 2+1 & \parbox[t]{2cm}{309, 292, 275, 334}
 & \parbox[t]{7cm}{The $M_{\pi,\rm min}$ entries correspond to the
four lattice spacings. Chiral extrapolations based on SU(2) staggered
{\Ch}PT at NLO, including some analytic NNLO terms. Combined 1.1\%
error from various different variations in the fit procedure.}
\\[12.0ex] \hline \\[-2.0ex]
RBC/UKQCD 10B & \cite{Aoki:2010pe} & 2+1 & 327, 289  &
\parbox[t]{7cm}{The $M_{\pi,\rm min}$ entries correspond to two
lattice spacings. Combined chiral and continuum extrapolations.}
\\[6.0ex] \hline \\[-2.0ex]
SWME 10   & \cite{Bae:2010ki} & 2+1 &  309, 292, 275 &
\parbox[t]{7cm}{The $M_{\pi,\rm min}$ entries correspond to the three
lattice spacings. Chiral extrapolations based on SU(2) staggered {\Ch}PT
at NLO, including some analytic NNLO terms. SU(3) staggered {\Ch}PT as
cross-check. Combined 1.1\% error from various different variations in
the fit procedure.} 
\\[15.0ex] \hline \\[-2.0ex]
Aubin 09 & \cite{Aubin:2009jh} & 2+1 & 240/370 & \parbox[t]{7cm}{The
  $M_{\pi,\rm min}$ entries correspond to the smallest valence/sea quark
  masses. Chiral \& continuum fits based on NLO mixed action {\Ch}PT at
  NLO, including a subset of NNLO terms. Systematic error estimated from
  spread arising from variations in the fit function. Relevant pion mass in
  the sea quark sector is the singlet pion mass of 370\,MeV.} 
\\[18.0ex] \hline \\[-2.0ex]
RBC/UKQCD 07A, 08 & {\cite{Antonio:2007pb,Allton:2008pn}} & {2+1} & 
 {330} & {\parbox[t]{7cm}{Fits based on SU(2) PQ{\Ch}PT at NLO. Effect of
 neglecting higher orders estimated at 6\% via difference between fits
 based on LO and NLO expressions.}} 
\\[9.0ex] \hline \\[-2.0ex]
HPQCD/UKQCD 06 & {\cite{Gamiz:2006sq}} & {2+1} & {360} &
    {\parbox[t]{7cm}{3\% uncertainty from chiral extrapolation 
    quoted, without giving further details.}}
\\[3.0ex] \hline \\[-1.0ex]
ETM 10A & \cite{Constantinou:2010qv} & 2 & 300, \, 270, \,400 &
  \parbox[t]{7cm}{Each $M_{\pi,\rm min}$ entry corresponds to a
  different lattice spacing. Simultaneous chiral \& continuum
  extrapolations, based on {\Ch}PT at NLO, are carried out. Systematic
  error from several sources, including lattice calibration, quark
  mass calibration, chiral and continuum extrapolation etc., estimated
  at 3.1\%.} 
\\ [18.0ex] \hline \\[-1.0ex]
JLQCD 08 & \cite{Aoki:2008ss} & 2 & 290 & \parbox[t]{7cm}{Fits based on
  NLO PQ{\Ch}PT. Range of validity investigated. Fit error included in
  statistical uncertainty.}  
\\[6.0ex] \hline \\[-1.0ex]
RBC 04 & \cite{Aoki:2004ht} & 2 & 490 & \parbox[t]{7cm}{Fits based on NLO
     PQ{\Ch}PT. Fit error included in statistical uncertainty.} 
\\[3.0ex] \hline \\[-1.0ex]
UKQCD 04 & \cite{Flynn:2004au} & 2 & 780 & \parbox[t]{7cm}{Fits to
  continuum chiral behaviour at fixed sea quark mass. Separate
  extrapolation in sea quark mass. Fit error included in overall
  uncertainty.} 
\\[6.0ex]
\hline\hline
\end{tabular*}
\caption{Chiral extrapolation/minimum pion mass in
  determinations of $B_K$.}
}
\end{table}

\begin{table}[!ht]
{\footnotesize
\begin{tabular*}{\textwidth}{l @{\extracolsep{\fill}} c c c c l}
\hline\hline \\[-1.0ex]
Collab. & Ref. & $\Nf$ & $L$ [fm] & ${M_{\pi,\rm min}}L$ & Description 
\\[1.0ex] \hline \hline \\[-1.0ex]
SWME 11 & \cite{Kim:2011qg} & 2+1 & \parbox[t]{1.8cm}{2.4/3.3, 2.4,
2.8, 2.8} & $\gtrsim 3.5$ &
\parbox[t]{5cm}{Each $L$-entry corresponds to a different lattice
spacing, with two volumes at the coarsest lattice. Central value
obtained via finite-volume staggered {\Ch}PT. Residual uncertainty of
0.6\% from difference obtained using {\Ch}PT in infinite volume.}
\\[20.0ex] \hline \\[-1.0ex]
RBC/UKQCD 10B & \cite{Aoki:2010pe} & 2+1 & 2.7 &  $\gtrsim 4.0$ &
\parbox[t]{5cm}{Finite-volume {\Ch}PT applied; residual finite-volume
error estimated from the difference to fit to {\Ch}PT in infinite
volume.} 
\\[9.0ex] \hline \\[-1.0ex]
SWME 10   & \cite{Bae:2010ki} & 2+1 & \parbox[t]{1.8cm}{2.4/3.3, 2.4,
2.8} & $\gtrsim 3.5$ & \parbox[t]{5cm}{Each $L$-entry corresponds to a
different lattice spacing, with two volumes at the coarsest
lattice. Finite-volume error of 0.9\% estimated from difference
obtained on two volumes at the coarsest lattice spacing.}
\\[15.0ex] \hline \\[-1.0ex]
Aubin 09 & \cite{Aubin:2009jh} & 2+1 & 2.4,\,3.4 & 3.5 &
   \parbox[t]{5cm}{Each $L$ entry corresponds to a different
lattice spacing. Keep $m_{\pi}L\;\gtaeq\;3.5$; no
   comparison of results from different volumes; 0.6\% error
   estimated from mixed action {\Ch}PT correction.}
\\[15.0ex] \hline \\[-1.0ex]
RBC/UKQCD 07A, 08 & \cite{Antonio:2007pb,Allton:2008pn} & 2+1 &
  1.83/2.74 & 4.60 & \parbox[t]{5cm}{Each $L$ entry corresponds to a
    different volume at the same lattice spacing; 1\% error from
    difference in results on two volumes.} 
\\[10.0ex] \hline \\[-1.0ex]
HPQCD/UKQCD 06 & \cite{Gamiz:2006sq} & 2+1 & 2.46 & 4.49 & {Single
  volume; no error quoted.}
\\[1.0ex] \hline \\[-1.0ex]
ETM 10A & \cite{Constantinou:2010qv} & 2 & \parbox[t]{1.8cm}{2.2, 2.2/2.9,
  2.1}& \parbox[t]{1.8cm}{3.3, 3.3/4.3, 5} & 
\parbox[t]{5cm}{Each $L$ entry corresponds to a different
lattice spacing, with two volumes at the intermediate lattice
spacing. Results from these two volumes at $M_{\pi} \sim 300$ MeV are
compatible.} 
\\ [15.0ex] \hline \\[-1.0ex]
JLQCD 08 & \cite{Aoki:2008ss} & 2 & 1.89 & 2.75 & \parbox[t]{5cm}{Single
  volume; data points with $m_{\rm val} < m_{\rm sea}$ excluded; 5\%
  error quoted as upper bound of PQ{\Ch}PT estimate of the 
  effect.}
\\[10.0ex] \hline \\[-1.0ex]
RBC 04 & \cite{Aoki:2004ht} & 2 & 1.87 & 4.64 & Single volume; no
   error quoted.
\\[1.0ex] \hline \\[-1.0ex]
UKQCD 04 & \cite{Flynn:2004au} & 2 & 1.6 & 6.51 & Single volume; no
   error quoted.
\\[1.0ex] 
\hline\hline
\end{tabular*}
\caption{Finite volume effects in determinations of $B_K$.}
}
\end{table}

\begin{table}[!ht]
{\footnotesize
\begin{tabular*}{\textwidth}{l @{\extracolsep{\fill}} c c c c l}
\hline\hline \\[-1.0ex]
& & & & running & \\
\rb{Collab.} & \rb{Ref.} & \rb{$\Nf$} & \rb{Ren.} & match. & \rb{Description} 
\\[1.0ex] \hline \hline \\[-1.0ex]
SWME 11 & \cite{Kim:2011qg} & 2+1 & PT1$\ell$ & PT1$\ell$ &
\parbox[t]{6cm}{Uncertainty from neglecting higher orders estimated at
4.4\% by identifying the unknown 2-loop coefficient with result at the
  smallest lattice spacing.}
\\[9.0ex] \hline \\[-1.0ex]
RBC/UKQCD 10B & \cite{Aoki:2010pe} & 2+1 & RI & PT1$\ell$ &
\parbox[t]{6cm}{Variety of different RI-MOM schemes including
non-exceptional momenta. Residual uncertainty of 2\% uncertainty in
running \& matching.}
\\[9.0ex] \hline \\[-1.0ex] 
SWME 10  & \cite{Bae:2010ki} & 2+1 & PT1$\ell$ & PT1$\ell$ &
\parbox[t]{6cm}{Uncertainty from neglecting higher orders estimated at
5.5\% by identifying the unknown 2-loop coefficient with result at the
  smallest lattice spacing.}
\\[9.0ex] \hline \\[-1.0ex]
Aubin 09 & \cite{Aubin:2009jh} & 2+1 & RI & PT1$\ell$ &
  \parbox[t]{6cm}{Total uncertainty in matching \& running of 3.3\%,
  estimated from a number of sources, including chiral extrapolation
  fit ansatz for n.p. determination, strange sea quark mass
  dependence, residual chiral symmetry breaking, perturbative matching
  \& running.}
\\[18.0ex] \hline \\[-1.0ex]
RBC/UKQCD 07A, 08 & {\cite{Antonio:2007pb,Allton:2008pn}} & 
 {2+1} & {RI} & {PT1$\ell$} & {\parbox[t]{6cm}{Uncertainty
 from n.p. determination of ren. factor included in statistical error; 
 2\% systematic error from perturbative matching to $\msbar$
 estimated via size of correction itself.}}
\\[12.0ex] \hline \\[-1.0ex]
HPQCD/UKQCD 06 & {\cite{Gamiz:2006sq}} & {2+1} & {PT1$\ell$}&
    {PT1$\ell$} & {\parbox[t]{6cm}{Uncertainty due to neglecting 2-loop
    order in perturbative matching and running estimated by
    multiplying result by $\alpha^2$.}}
\\[8.0ex] \hline \\[-1.0ex]
ETM 10A & \cite{Constantinou:2010qv} & 2 & RI & PT1$\ell$ &
  \parbox[t]{6cm}{Uncertainty from RI renormalization estimated at 2.5\%.}
\\ [4.0ex] \hline \\[-1.0ex]
JLQCD 08 & \cite{Aoki:2008ss} & 2 & RI & PT1$\ell$ &
 \parbox[t]{6cm}{Uncertainty from n.p. determination of ren. factor 
 included in statistical error; 2.3\% systematic error from
 perturbative matching to $\msbar$ estimated via size of correction
 itself.} 
\\[12.0ex] \hline \\[-1.0ex]
RBC 04 & \cite{Aoki:2004ht} & 2 & RI & PT1$\ell$ &
   \parbox[t]{6cm}{Uncertainty from n.p. determination of ren. factor
 included.}
\\[4.0ex] \hline \\[-1.0ex]
UKQCD 04 & \cite{Flynn:2004au} & 2 & PT1$\ell$ & PT1$\ell$ & No error quoted.
\\[1.0ex] 
\hline\hline
\end{tabular*}
\caption{Running and matching in determinations of $B_K$.}
}
\end{table}

\clearpage

\end{appendix}

\bibliography{FLAG} 
\bibliographystyle{JHEP}   

\end{document}